\documentclass[12pt]{elsarticle}

\usepackage{amsmath}
\usepackage{amssymb}
\usepackage{booktabs}
\usepackage{caption}  
\usepackage{array}
\usepackage{hyperref}
\usepackage{subcaption}
\usepackage{multirow} 
\usepackage{float}
\usepackage{enumitem}
\usepackage{algorithmic}
\usepackage{algorithm}  
\usepackage{adjustbox}
\usepackage{ragged2e} 
\usepackage{pdflscape}
\usepackage{lscape}  
\usepackage{fancyhdr}
\usepackage{etoolbox} 
\usepackage[a4paper, margin=2.5cm]{geometry}

\usepackage{graphicx} 
\usepackage{comment}
\usepackage{afterpage} 
\usepackage{tikz}      
\usepackage{eso-pic}   

\usepackage{setspace}

\renewcommand{\headrulewidth}{0pt}
\renewcommand{\footrulewidth}{0pt}

\newcommand{\landscapePageNumber}{
    \AddToShipoutPicture*{%
        \AtPageLowerLeft{%
            \rotatebox{90}{
                \raisebox{0.5cm}[0pt][0pt]{\makebox[\paperheight][c]{\thepage}}
            }
        }
    }
}

\makeatletter
\def\ps@pprintTitle{%
 \let\@oddhead\@empty
 \let\@evenhead\@empty
 \def\@oddfoot{\reset@font\hfil}%
 \let\@evenfoot\@oddfoot}
\makeatother
\journal{}

\begin{document}

\begin{frontmatter}



\title{Minimal Batch Adaptive Learning Policy Engine for Real-Time Mid-Price Forecasting in High-Frequency Trading}


\author[inst1]{Adamantios Ntakaris}

\affiliation[inst1]{organization={Business School, University of Edinburgh},
            addressline={29 Buccleuch Place}, 
            city={Edinburgh},
            postcode={EH8 9JS}, 
            country={UK}}

\author[inst1]{Gbenga Ibikunle}


\begin{abstract}
High-frequency trading (HFT) has transformed modern financial markets, making reliable short-term price forecasting models essential. In this study, we present a novel approach to mid-price forecasting using Level 1 limit order book (LOB) data from NASDAQ, focusing on 100 U.S. stocks from the S\&P 500 index during the period from September to November 2022. Expanding on our previous work with Radial Basis Function Neural Networks (RBFNN), which leveraged automated feature importance techniques based on mean decrease impurity (MDI) and gradient descent (GD), we introduce the Adaptive Learning Policy Engine (ALPE) - a reinforcement learning (RL)-based agent designed for batch-free, immediate mid-price forecasting. ALPE incorporates adaptive epsilon decay to dynamically balance exploration and exploitation, outperforming a diverse range of highly effective machine learning (ML) and deep learning (DL) models in forecasting performance.
\end{abstract}




\end{frontmatter}


\section{Introduction}
\label{sec:sample1}
\noindent HFT has become a cornerstone of modern financial markets, where large volumes of trades are executed within fractions of a second, This environment's high speed and complexity make accurately forecasting short-term price movements within the LOB a challenging yet crucial task. Traders and market participants constantly seek ways to optimize their strategies and mitigate risks in this highly volatile space. However, the intricate dynamics of the LOB, characterized by its high dimensionality and noisy data, often render traditional statistical models ineffective. These challenges have increasingly prompted researchers to explore ML and DL methods as more robust alternatives for HFT forecasting.

In our previous research \cite{rbfnn}, we introduced a Radial Basis Function Neural Network (RBFNN) model that utilized automated feature clustering and selection methods to enhance mid-price forecasting accuracy based on Level 1 LOB data from NASDAQ, focusing on 20 U.S. stocks from the S\&P 500 index. This model leveraged fully automated feature selection techniques based on MDI and GD to identify and process the most relevant information from HFT data. By applying these techniques, the RBFNN model is able to outperform traditional forecasting methods, showcasing the potential of automated routines for extracting meaningful patterns in the chaotic environment of financial markets. While the RBFNN demonstrates promising results, the non-stationary and adaptive nature of financial markets indicated room for further advancements through more dynamic modeling approaches.

This paper aims to build on that foundation by introducing a novel RL framework for mid-price forecasting in HFT. Unlike traditional supervised learning models that learn from a static dataset (i.e., all the available data is pre-collected, and the model learns from this static pool of examples), RL enables the model to adapt its forecasting strategy based on the rewards and penalties associated with market changes, thereby offering a more flexible solution to the mid-price forecasting task.

To establish the efficacy of the proposed RL-based model, we conduct a comparative analysis against several ML and DL models, including a baseline regressor, autoregressive integrated moving average (ARIMA), multi-layer perceptron (MLP), convolutional neural network (CNN), long short-term memory (LSTM), gated recurrent unit (GRU), and RBFNN. These models are evaluated using different input data variations and feature importance techniques such as MDI and GD, providing a comprehensive benchmark for the new RL approach. This comparative study aims to demonstrate not only the performance advantages of the RL-based model but also its adaptability to varying market conditions, an aspect that is critical in the highly volatile environment of HFT.

The primary contribution of this work is twofold. First, we extend the set of benchmark models and number of stocks from our previous study, including the new RL-based model, providing a broader context for evaluating the mid-price forecasting performance. Second, we investigate the predictive potential of RL within the HFT LOB data, highlighting its ability to dynamically adjust to market shifts and thereby improve mid-price forecasts in a batch-free fashion. Our results indicate that the RL model can outperform traditional statistical, ML, and DL models, particularly in capturing the nuanced and non-linear dependencies present in HFT LOB data.

The remainder of this paper is structured as follows: Section 2 reviews related work in HFT forecasting and RL applications. Section 3 outlines the dataset and the preprocessing techniques employed. Section 4 describes the proposed RL-based model and the experimental design used for evaluation. Section 5 presents the empirical results, followed by a discussion of the findings. Finally, Section 6 concludes with insights into the implications of this research and future research directions.

\section{Literature Review}

\noindent HFT plays a pivotal role in modern financial markets, where the execution of large volumes of orders happens within fractions of a second. As the landscape of electronic trading has evolved, so has the complexity of the data available, especially in LOBs, which capture all the bids and asks in the market at any given time. LOB data is both highly dimensional and noisy, presenting unique challenges for forecasting models. Traditional approaches, such as econometric and time series models like ARIMA \cite{alshawarbeh2023statistical, dong2020predictive, bukhari2020fractional}, struggle to capture the non-linear dependencies and intricate structures in such fast-moving, high-dimensional datasets. Consequently, there has been a growing interest in employing ML and DL methods to improve prediction accuracy for mid-price forecasting, a crucial task in HFT. As we will see in the following subsections, these methods have proven capable of handling the complexity of LOB data and offer superior performance in capturing short-term price movements.

Recent advancements in ML and DL introduced more sophisticated tools for financial forecasting. For instance, the authors in \cite{zhu2024forecasting} utilized a complete ensemble empirical mode decomposition with adaptive noise (CEEMDAN) algorithm together with the Savitzky–Golay (SG) filter for data de-noising and employed a neural network based on three convolutional layers and an LSTM layer, which enabled it to capture complex temporal patterns in CSI 300 index data for price forecasting. Similarly, the authors in \cite{liu2024multimodal} proposed a novel multiscale multimodal dynamic graph convolution network (Melody-GCN) for stock price prediction based on two datasets, one of which was based on 40 US stocks. Additionally, the authors in \cite{YIN2022108209}, using data from the Chinese A-share market, developed a novel ML model named graph attention long short-term memory (GALSTM) to learn correlations between stocks and predict their future prices automatically.

Models such as MLP, CNN, and recurrent neural networks (RNN), such as GRU and LSTMs, demonstrated strong performance in predicting stock prices and other financial variables. More specifically, the authors in \cite{dingli2017financial} achieved higher predictive accuracy by using the TA-Lib library, an open-source library for financial market data analysis, for stock price forecasting and CNNs to estimate weekly and monthly stock movements. Similarly, the authors in \cite{haq2021forecasting} developed a feature selection approach by combining L\_1 logistic regression (LR), support vector machines (SVM), and a random forest (RF)-based technique to improve stock trend prediction using a deep generative model equipped with a temporal attention mechanism. They demonstrated that this ensemble feature selection method enhanced prediction accuracy, outperforming discriminative models.

Apart from daily, weekly, and monthly stock price forecasting, ML and DL models were extensively utilized in the HFT space. For instance, the authors in \cite{arifovic2022machine} explored the effects of HFT speed and ML on market dynamics using a genetic algorithm-based model, showing that HFT improved price efficiency but reduced market liquidity. Similarly, the authors in \cite{mangat2022high} employed ML models, such as SVM and RF, for forecasting stock movements using high-frequency LOB data from the iShares Core S\&P 500 ETF, demonstrating their potential despite challenges related to transaction costs and reliance on the last price change. Additionally, the authors in \cite{moews2020predictive} used deep feed-forward neural networks to exploit lagged correlations in S\&P 500 stocks for predicting intraday and daily price movements. They demonstrated that these models performed well in both stable and volatile market conditions, achieving higher accuracy over longer prediction intervals, particularly during the 2007-2008 financial crisis.

An important area of HFT research are LOBs - a dynamic record of all outstanding buy and sell orders for a particular financial asset at different price levels. This area requires fast information processing and interpretation since trade execution takes place via low-latency, long-haul wireless networks among other fast means of communication with the stock exchange. Much research is conducted in this area. For instance, the author in \cite{sirignano2019deep} developed a DL model called a 'spatial neural network' designed to forecast price movements in LOBs using data from 489 U.S. stocks (S\&P 500 and NASDAQ-100) from 2014-2015. The proposed spatial NN outperformed traditional models by leveraging the local structure of the LOB, providing greater accuracy and efficiency in predicting price distributions. The authors in \cite{nousi2019machine} explored the use of several ML and DL topologies, such as SVMs, single hidden layer feed-forward neural networks (SLFN), and MLPs, to predict mid-price movements using LOB data from five Finnish stocks over a 10-day period. They demonstrated that combining handcrafted and learned features improved prediction accuracy, highlighting the value of ML for stock trend forecasting.

Additionally, the authors in \cite{ibikunle2024can} aimed to model the nonlinear interactions between HFT strategies and financial market dynamics using ML. The authors used NASDAQ HFT data and the Trade and Quote (TAQ) database to develop ML-driven metrics, distinguishing liquidity-demanding and liquidity-supplying HFT activities, ultimately analyzing how these strategies reacted to market events and influenced market quality. The prediction of jump arrivals in stock prices using NASDAQ LOB data was explored by the authors in \cite{makinen2019forecasting}. They focused on five liquid U.S. stocks and introduced a novel attention-based neural network architecture (CNN-LSTM-Attention) that outperformed other models in forecasting price jumps one minute ahead.

These ML and DL architectures are effective at learning hierarchical features from raw data, capturing both short-term patterns (e.g., CNNs) and longer-term trends (e.g., LSTMs and GRUs) in time series data. While they have shown promising results, RL-based methods - another line of research - focus on adjustability to real-time conditions via trial, error, and reward functions, RL has emerged as a promising tool for financial markets, particularly due to its capacity to handle sequential decision-making problems under uncertainty. Unlike traditional supervised learning methods, RL enables the model to learn optimal strategies based on rewards and penalties, making it an ideal fit for environments requiring continuous adaptation, such as HFT. More specifically, the authors in \cite{lee2001stock} proposed using RL to predict stock prices by treating stock price changes as a Markov process. The model was applied to data from the Korean stock market, focusing on learning from price trends and immediate rewards to enhance prediction accuracy. Similarly, the authors in \cite{deng2016deep} introduced a deep direct reinforcement learning (DDR) model for financial signal representation and trading. They used a recurrent deep neural network (RDNN) combined with RL to predict trading decisions in real-time on stock index and commodity futures data.

Deep Q-networks (DQN), an RL variant, and proximal policy optimization (PPO) were used in \cite{wang2023domain} to investigate the application of deep RL (DRL) for trading front-month natural gas futures contracts, focusing on improving agent performance through domain-adapted learning and model explainability. Similarly, RL was implemented extensively in quantitative-based trading (QT) activities, as highlighted by the authors in their survey paper \cite{sun2023reinforcement}. They found that RL was utilized effectively for QT tasks such as algorithmic trading, portfolio management, order execution, and market making. The paper reviewed a wide range of RL models, including Q-learning, DQN, and PPO, applied to different datasets such as stock, cryptocurrency, and futures market data. In a similar fashion, the authors in \cite{karpe2020multi} proposed a multi-agent RL (MARL) approach to optimize order execution strategies in HFT using a realistic LOB simulation. The model employed Double Deep Q-Learning (DDQL) and was trained on the agent-based interactive discrete event simulation (ABIDES), an open-source framework designed primarily for simulating financial markets, with NASDAQ LOB data from US stocks.

Several other studies also focused on the application of RL to LOBs, For instance, the authors in \cite{philip2020estimating} presented a recursive RL model to estimate permanent price impact, addressing the limitations of traditional vector auto-regression (VAR) models. By incorporating LOB data, specifically the depth imbalance between the bid and ask sides, the RL model captured the nonlinear relationship between trade size and price impact. This approach allowed for better estimation of both immediate and long-term price impacts. Similarly, the authors in \cite{cao2022estimating} explored different RL methods in the LOB domain to improve price impact estimation. More specifically, they proposed a deep RL architecture that estimated permanent price impact using LOB data. The model incorporated bid-ask depth imbalance from the LOB to better capture nonlinear relationships between trade size and price impact.

In a similar fashion, the authors in \cite{he2022reinforcement} introduced an information-based RL (IRL) model to study trader behavior in equilibrium limit order markets. They utilized LOB data to examine the dynamics between informed and uninformed traders, focusing on how order chasing and strategic trading influenced price discovery and liquidity. Another deep RL (DRL) model that estimated price impact based on LOB data was introduced by the authors in \cite{tsantekidis2023modeling}. The authors developed a model that used continuous action space and Beta distributions to determine limit prices, optimizing both buy and sell order placements.

Additional applications of RL focused on feature selection, feature clustering tasks, and feature importance tasks. For instance, the authors in \cite{fan2021interactive} explored the effectiveness and efficiency of automated feature selection using an interactive RL (IRL) framework combined with decision tree feedback to improve performance. The model was tested on multiple real-world datasets, such as PRD, forest cover, and spam, and showed enhanced feature selection capabilities. Similarly, the authors in \cite{thajeel2023dynamic} introduced a dynamic feature selection model for detecting evolved cross-site scripting (XSS) attacks using a multi-agent deep Q-learning framework. Their model adapted to feature drift by dynamically updating the selection of relevant features through a fair agent reward distribution (FARD-DFS).

RL-based feature clustering was performed in \cite{liu2023robust}. The authors introduced a novel feature clustering method, named clustering with bisimulation metrics (CBM), to improve visual RL under distractions. CBM was tested on visual control tasks from the distracting control suite (DCS) and showed state-of-the-art performance across multiple and single distraction settings. Another example of a successful application of RL for feature clustering tasks came from the authors in \cite{fathinezhad2023soft}, who introduced a novel clustering method that integrated soft dimensionality reduction with RL. This method used a distance metric based on feature importance, along with a sparsity regularization term, to reduce irrelevant features during dimensionality reduction. The method was evaluated on 14 real-world datasets from the UCI and Keel repositories, showing superior clustering consistency and handling of noisy data.

RL was also utilized for feature importance tasks. More precisely, the authors in \cite{elbaz2023deep} presented a DRL model combining deep Q-learning (DQL) and particle swarm optimization (PSO), applied to optimize the driving performance of shield tunneling machines. The model was validated using field data from a tunneling project in Shenzhen, China. Similarly, the authors in \cite{gauci2018horizon} presented Horizon, a framework that leveraged RL algorithms while incorporating a feature importance mechanism that measured the impact of each feature by observing the increase in model loss when a feature was masked. This approach helped identify the most critical features for predicting state transitions and rewards.

Another research area where RL was applied is time series forecasting. For instance, the authors in \cite{fu2022reinforcement} proposed a novel RL model combination (RLMC) framework designed to dynamically assign weights to ensemble models for various time series forecasting tasks based on electricity transformer temperature data, weather data, electricity load data, and the M4 dataset (a dataset containing demographics, finance, industry, and macro/microeconomic factors). Another example of time series forecasting based on RL is the OEA-RL model, which stands for online ensemble aggregation RL in time series forecasting, developed by the authors in \cite{saadallah2021online}. The model used LOB data from diverse real-world sources, where the RL variant dynamically adapted to concept drifts by learning optimal weights for a pool of forecasting models. Additionally, the authors in \cite{kuremoto2007forecasting} proposed a self-organized fuzzy neural network (SOFNN) combined with a stochastic gradient ascent (SGA) RL algorithm to forecast time series data based on the NN3 forecasting competition, which includes financial, economic, and industrial data.

Several other works relied on RL for time series forecasting, such as the work by the authors in \cite{hirata2018forecasting}, who investigated the application of a deep belief network (DBN) fine-tuned via an RL algorithm named stochastic gradient ascent (SGA) for real-time series forecasting. They utilized datasets such as atmospheric CO$_{2}$ levels, sea level pressure, and sunspot numbers. In a different study, the authors in \cite{zhuang2023data} developed a soft actor-critic (SAC) RL agent integrated with time-series forecasting (TSF) models, such as LSTM-based architectures, to optimize HVAC (heating, ventilation, and air conditioning system) operations in smart buildings. The forecasting task focused on predicting system energy consumption, indoor air temperature, and relative humidity using sensor metadata, Q-learning RL in \cite{liu2005neural} was also utilized to reduce time-complexity issues for forecasting tasks based on a sunspot dataset. The authors in that paper presented an RL method based on a dimension and delay estimator (RLDDE) for optimizing neural network structures in time series prediction by automatically selecting input dimension and time delay.

While much of the existing literature on RL for financial and other forecasting tasks focused on complex ensemble models, advanced preprocessing techniques, and delayed reward structures, these methods introduce significant computational overhead and lack real-time adaptability. Moreover, many RL architectures are not optimized for immediate decision-making, which is critical in HFT environments. In contrast, our approach leverages a streamlined adaptive policy RL-based model (i.e., ALPE) tailored specifically for mid-price forecasting in LOB environments. By directly predicting mid-prices without the need for complex preprocessing or ensemble management, our model achieves real-time (i.e., batch-free) adaptability. Furthermore, the reward structure of the proposed ALPE model focuses on immediate price movements, optimizing decision-making. This approach not only simplifies the forecasting process but also enhances the model’s scalability and performance in rapidly changing financial markets, addressing key limitations observed in the current literature.

\section{Methodology and Experimental Setup}
\noindent This section outlines the experimental methodology used to develop, test, and benchmark the proposed ALPE model for mid-price forecasting using HFT LOB data. We begin by describing the dataset and the preprocessing techniques used to extract key features from the raw LOB data. Next, we explain the custom RL environment and the architecture of the ALPE agent. For a comprehensive evaluation, we compare the performance of the ALPE agent against several models, including RBFNN, as well as other models like the statistical model ARIMA, and ML and DL architectures like CNN, LSTM, and GRU, A simple baseline regressor model is also included for reference.

As part of the experimental setup, we utilize three distinct input feature sets: one created using the MDI feature importance technique, another using GD, and a third consisting of the raw LOB data. Both MDI and GD were developed in our previous work in \cite{rbfnn}, These feature sets are applied to ensure fairness in evaluation across different methodologies. The primary evaluation metrics used are Mean Squared Error (MSE), Root Mean Squared Error (RMSE), and a newly introduced regression metric, Relative Root Mean Squared Error (RRMSE), which we proposed in \cite{rbfnn}. The RRMSE metric is specifically designed to aid HFT traders in making informed decisions across various stocks and can be considered a normalized RMSE, meaning that, in order to compare a wide range of stocks, we divide each stock's RMSE score by its mid-price on an event-by-event basis.

\subsection{Forecasting Objective }
\noindent The objective of this study is to forecast the mid-price of the LOB, and despite the fact that it cannot be traded directly, it serves as a useful proxy for trading activities. For instance, accurately estimating the movement of the mid-price can help in understanding price impact, particularly when large orders are placed. We treat this forecasting objective as an event-by-event regression problem. Our proposed DQR model seeks to minimize forecast error, evaluated through MSE, RMSE, and a novel RRMSE metric, and to compare its performance against a range of benchmark models, including a basic regressor, ARIMA, LSTM, GRU, MLP, CNN, and RBFNN.

The mid-price is defined as the average of the best bid and ask LOB prices:

\begin{equation}
p_t = \frac{p_{Ask_t} + p_{Bid_t}}{2},
\end{equation}

\noindent where $p_{Ask_t}$ and $p_{Bid_t}$ represent the best ask and bid LOB prices at event at time $t$, and the novel RRMSE is definied as:

\begin{equation}
RRMSE_t = \frac{RMSE_t}{p_t},
\end{equation}

\noindent where $RMSE_t$ is the RMSE score at time event $t$. We define time event $t$ as a unique event even if the timestamp is the same for more than one events. That means that our experimental protocol is event-based and no sampling technique was applied. 

\begin{figure}[t!]
    \includegraphics[width=1\textwidth]{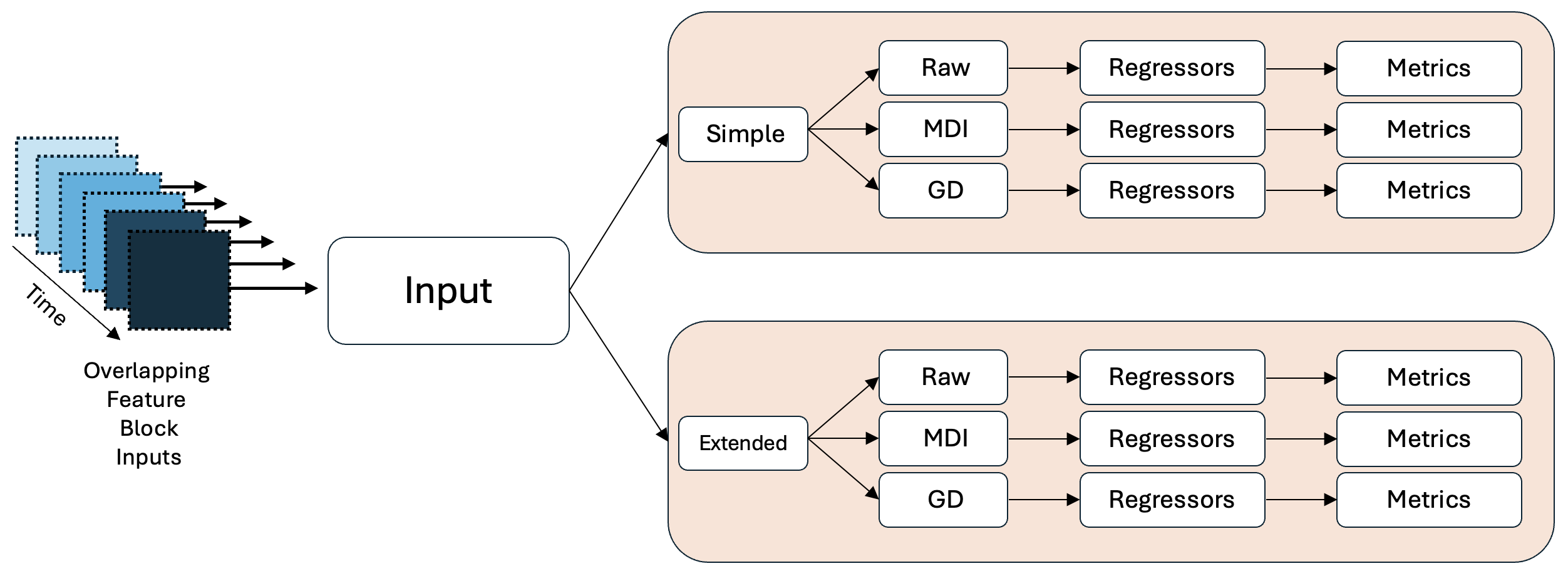}  
    \caption{The figure provides an overview of our experimental protocol, which is structured around two main blocks: \textit{Simple} and \textit{Extended} feature sets, The process begins with the \textit{Input} block, where the HFT trader utilizes each input block consisting of a specific number of LOB states, referred to as \textit{Overlapping Feature Block Inputs}. These blocks sequentially feed raw LOB data into the two feature sets. The \textit{Simple} feature set is based on the best level (i.e., price level 1) of LOB data, while the \textit{Extended} feature set consists of handcrafted kernelized features designed to capture more complex relationships from the data. For both the \textit{Simple} and \textit{Extended} sets, three different inputs are used: Raw data, MDI-adjusted features, and GD-adjusted features. Each input type is passed through various regressors (i.e., baseline regressor, ARIMA, MLP, CNN, LSTM, GRU, RBFNN, and ALPE), and their performance is evaluated using \textit{metrics} such as MSE, RMSE, and RRMSE. This framework is applied to each of the 100 stocks, ensuring a thorough evaluation of the different feature sets and models.}   
  \label{fig:expe_overview}
\end{figure}

This event-based forecasting objective is integrated into our online (i.e., event-by-event forecast) experimental protocol. It includes both batch-based training and batch-free learning settings. More specifically, competitor models such as the baseline regressor, ARIMA, MLP, CNN, LSTM, GRU, and RBFNN are part of the batch-based training, while only the newly developed ALPE model is based on the batch-free setting. This differentiation stems from the fact that the competitor models are not designed to operate under a truly online forecasting setting; therefore, to ensure a fair comparison with the new RL-based model, we developed a rolling window experimental protocol, which considers training based on 10 LOB states (i.e., the current and nine historical LOB states). The motivation for selecting this training window size comes from the work done by the authors in \cite{ntakaris2019feature}, where they observed that, for the selected U.S. stocks, there was an alternation between stationary and non-stationary time series approximately every 10 trading events. The performance of the competitor models dropped significantly when we further reduced the number of training LOB states. For the batch-free RL-based model, we followed exactly the same rolling window protocol but with a window size of one, meaning only the current LOB information was used. An overview of the experimental protocol is shown in \autoref{fig:expe_overview}, where we also present the different feature sets that were utilized as part of our comparative analysis. 

\subsection{Data Preprocessing and Feature Engineering}
\noindent In HFT, the quality of data preprocessing and feature engineering plays a crucial role in model performance, particularly due to the vast amount of noise and high dimensionality typical in such data. Effective preprocessing ensures the features are scaled appropriately to enhance the stability\footnote{Here we refer to numerical stability that ensures computations do not result in extremely large or small values that could lead to numerical issues such as overflow, underflow, or loss of precision.} and convergence of learning algorithms. In this section, we describe the sequence of preprocessing steps that were employed, including feature importance calculations via MDI and GD methods, as well as data normalization based on min-max scaling. More specifically, the MDI and GD feature importance algorithms are developed as follows:

\begin{enumerate}
    \item \textbf{Mean Decrease Impurity} (MDI)\\
     MDI is a feature importance method based on RF. It is calculated as the average reduction in impurity across all trees in which a feature is used for splitting. Since we perform a regression task, it uses variance-based impurity. Specifically, the variance impurity $I$ is used at each node and the variance impurity at a node $j$ is calculated as follows:

     \vspace{-13pt}
     \begin{equation}
     I_j = \frac{1}{N_j} \sum_{i=1}^{N_j} (y_i - \bar{y})^2,
     \end{equation}
     \vspace{-13pt}

     where:
     \vspace{-10pt}
     \begin{itemize}[label=-]
     \item $N_j$: The number of data points at node $j$,
     \item $y_i$: The value of the target variable for the $i$-th data point at the node,
     \item $\bar{y}$: The mean of the target values at node $j$, calculated as: \\$\bar{y} = \frac{1}{N_j} \sum_{i=1}^{N_j} y_i$.
     \end{itemize}
     During training, the algorithm aims to minimize impurity by splitting nodes to create child nodes more homogeneous in their target values (e.g., mid-price), The impurity reduction \(\Delta I_{j,f}\) for feature \(f\) at node \(j\) is given by:
     \begin{equation}
     \Delta I_{j,f} = I_{j,f} - \left( \frac{N_l}{N_j} I_l + \frac{N_r}{N_j}  I_r \right),
     \end{equation}

     where:
     \vspace{-10pt}
     \begin{itemize}[label=-]
     \item \(I_{j,f}\) is the initial impurity at node \(j\) before the split,
     \item \(N_j\), \(N_l\), and \(N_r\) are the numbers of samples at the current, left, and right child nodes, respectively,
     \item \(I_l\) and \(I_r\) are the impurities of the left and right child nodes.
     \end{itemize}
     Having defined the impurity calculation, we proceed with the MDI metric for feature $f$ as:
     
     \vspace{-10pt}
     \begin{equation}
     \text{MDI}_f = \frac{1}{B} \sum_{b=1}^B \left( \sum_{j \in P(b, f)} \Delta I_{j,f} \right),
     \end{equation}
     \vspace{-10pt}
     where:
     \begin{itemize}[label=-]
     \item $B$ is the total number of trees,
     \item $P(b, f)$ is the set of nodes that split on feature $f$ in tree $b$.
     \end{itemize}
    
    \item \textbf{Gradient Descent} (GD)\\
    GD algorithm is a first-order optimization technique commonly used to minimize a loss function by iteratively updating parameters in the direction of the steepest descent. Following \cite{ntakaris2023optimum}, GD is adapted as a method for determining the relative influence of input features by optimizing a weight vector. By adjusting these feature weights, the algorithm minimizes the MSE of predictions. The first step in the GD algorithm is to initialize the input LOB block matrix $X \in \mathbb{R}^{N \times F}$, where $N$ represents the number of samples and $F$ represents the number of features and the target variable $y \in \mathbb{R}^N$, which is the LOB mid-price, GD's objective is to optimize the weight vector $\theta \in \mathbb{R}^F$, representing the relative importance of each feature in minimizing the MSE. We initialize the weight vector to $\theta = [1, 1, \ldots, 1]^T \in \mathbb{R}^F$ which means each feature is given equal initial importance and $\theta$ is updated over $T$ iterations, as follows:

    \begin{itemize}[label=-]
    \item \textit{Predicted Labels}: The predicted target values, denoted as $\hat{y} \in \mathbb{R}^N$, are computed as a linear combination of the features weighted by $\theta$:
    \begin{equation}
    \hat{y} = X \theta.
    \end{equation}
    Then for each sample $i$, the predicted value $\hat{y}_i$ is computed as:
    \begin{equation}
    \hat{y}_i = \sum_{j=1}^F X_{ij} \theta_j,
    \end{equation}
    
    where \( X_{ij} \) is the value of feature \( j \) for sample \( i \), and \( \theta_j \) is the weight associated with feature \( j \).
    
    \item \textit{Error Calculation}: The error term $e_i \in \mathbb{R}^N$ for each sample $i$ is calculated by comparing the predicted value $\hat{y}_i$ with the true target value $y_i$:
    \begin{equation}
    e_i = \hat{y}_i - y_i.
    \end{equation} 
    
    \item \textit{Gradient Calculation}: The objective function $J(\theta)$ represents the MSE across all samples, and it is defined as:
    
    \begin{equation}
    J(\theta) = \frac{1}{N} \sum_{i=1}^N (y_i - X_i \theta)^2,
    \end{equation}
   which denotes the average squared error across all samples, The gradient of $J(\theta)$ with respect to each weight $\theta_j$ is calculated to update the feature weights in the direction of steepest descent, given by:

    \vspace{-10pt}
    \begin{equation}
    \nabla_{\theta} J(\theta) = \frac{2}{N} \sum_{i=1}^N X_{ij} e_i,
    \end{equation}

    where $X_{ij}$ is the value of feature $j$ for sample $i$, and $\nabla_{\theta} J(\theta)$ represents the gradient of the objective function and $e_i$ the error term of sample $i$.
    
    \item \textit{Handling Numerical Instabilities}: To ensure numerical stability (i.e., avoid NaN or infinite values) during the weight update process, we incorporate the following checks:
    
    \begin{itemize}[label=$\diamond$]
        \item Any NaN or infinite values in the gradient vector $\nabla_{\theta} J(\theta)$ are replaced with zeros to ensure valid computations,
        \item Gradient clipping is applied to keep the gradient values within the range $[-1, 1]$:
        
        \vspace{-10pt}
        \begin{equation}
        \nabla_{\theta} J(\theta) = \text{clip}(\nabla_{\theta} J(\theta), -1, 1),
        \end{equation}   

        applied element-wise to the gradient vector.
    \end{itemize}
    
    \item \textit{Weight Update}: The weight vector $\theta$ is updated using the gradient $\nabla_{\theta} J(\theta)$ and a learning rate $\eta$\footnote{A learning rate of $\eta = 0.001$ was found effective for most stocks in this study.}:

    \vspace{-10pt}
    \begin{equation}
    \theta \leftarrow \theta - \eta \nabla_{\theta} J(\theta).
    \end{equation}
    
    \end{itemize}

    \item \textbf{Algorithmic Guidance}\\
    After completing $T$ iterations, the feature importance (FI) vectors for both methods (i.e., MDI and GD) are represented by the absolute values of the weight vectors:

    \vspace{-9pt}
    \begin{equation}
        \begin{aligned}
            FI_{MDI_i} &= |MDI_i|, \quad \text{for every} \quad i \in \{1, 2, \ldots, F\}, \\
            FI_{GD_i} &= |\theta_{i}|, \quad \text{for every} \quad i \in \{1, 2, \ldots, F\}.
        \end{aligned}
    \end{equation}

    where $FI_{MDI_i}$ and $FI_{GD_i}$ are the feature importance vectors of $MDI$ and $GD$ algorithms, respectively. To ensure numerical stability and avoid potential oscillations, a small constant $\delta$ is added to each final feature importance score:

    \vspace{-9pt}
    \begin{equation}
        \begin{aligned}
            FI_{MDI_i} &= |MDI_i| + \delta, \quad \text{for every} \quad i \in \{1, 2, \ldots, F\}, \\
            FI_{GD_i} &= |\theta_{i}| + \delta, \quad \text{for every} \quad i \in \{1, 2, \ldots, F\}.
        \end{aligned}
    \end{equation}

    Here, $\delta$ is defined to be 0,001. The transformed feature matrices for MDI and GD are given by:

    \vspace{-13pt}
    \begin{equation}
    X_{\text{MDI}_i} = X_i \cdot FI_{MDI_i}, \quad \text{and} \quad 
    X_{\text{GD}_i} = X_i \cdot FI_{GD_i}
    \end{equation}
    
    where $X_i$ represents the LOB input matrix, based on either the raw, simple, or extended LOB dataset. We fixed the number of iterations to 10, as convergence was achieved in most cases close to seven iterations, The algorithmic routines for both methods are outlined in \hyperref[algo:mdi]{Algorithm \ref{algo:mdi}} and \hyperref[algo:gd]{Algorithm \ref{algo:gd}}.

\begin{minipage}{0.90\textwidth}
    \scalebox{0.85}{ 
      \parbox{\textwidth}{
        \begin{algorithm}[H]  
          \caption{MDI Feature Importance Algorithm}
              \begin{algorithmic}[1]
                \REQUIRE $RF$ model, Input data $X$, Target variable $y$
                \STATE Train $RF$ using $X$ and $y$
                \FORALL{trees $b$ in $RF$}
                    \FORALL{nodes $j$ in $b$}
                        \STATE Calculate impurity reduction $\Delta I_{j,f}$ for feature $f$ used at node $j$
                        \STATE Accumulate $\Delta I_{j,f}$ for feature $f$
                    \ENDFOR
                \ENDFOR
                \STATE Calculate MDI for each feature $f$: $\text{MDI}_f \gets \frac{1}{B} \sum_{b=1}^B \sum_{j \in P(b,f)} \Delta I_{j,f}$
                \STATE Avoid numerical instabilities: $FI_{MDI_f} \gets |MDI_f| + \delta$\\
                \ENSURE Feature importance scores for all features.
              \end{algorithmic}
          \label{algo:mdi}
        \end{algorithm}
        }
    }
\end{minipage}
        
\begin{minipage}{0.90\textwidth}
    \scalebox{0.85}{ 
      \parbox{\textwidth}{
        \begin{algorithm}[H]
          \caption{GD Feature Importance Algorithm}
          \begin{algorithmic}[1]
            \REQUIRE Input matrix $X$, Target variable $y$, Learning rate $\eta$, $T$ iterations
            \STATE Initialize the weight vector $\theta \gets [1, 1, \ldots, 1]^T$
            \FOR{$t = 1$ to $T$}
                \STATE Compute predicted labels $\hat{y} \gets X \theta$
                \STATE Calculate error vector $e \gets \hat{y} - y$
                \STATE Compute the gradient $\nabla_{\theta} J(\theta) \gets \frac{2}{N} X^T e$
                \STATE Clip the gradient: $\nabla_{\theta} J(\theta) \gets \text{clip}(\nabla_{\theta} J(\theta), -1, 1)$
                \STATE Update weight vector $\theta \gets \theta - \eta \nabla_{\theta} J(\theta)$
            \ENDFOR
            \STATE Avoid numerical instabilities: $FI_{GD_i} \gets |\theta_i| + \delta$
            \ENSURE Feature importance scores for all features.
          \end{algorithmic}
          \label{algo:gd}
        \end{algorithm}
       }
     }
\end{minipage}

        \item \textbf{Feature Engineering}\\
        We utilize a diverse set of features derived from Level 1 LOB to forecast mid-price movements. The features, as outlined in \hyperref[tab:feat]{Table \ref{tab:feat}}, are specifically designed to capture potential relationships between order book characteristics. The feature sets are categorized into two groups, named \textit{Simple} and \textit{Extended} (\textit{Exte}), and they are based on raw LOB data for the \textit{Simple} group and basic, synthesised and kernalised features for the \text{Extended} group.

        \begin{table}[!hbtp]
    \centering
    \captionsetup{width=12.00\textwidth}
    \caption{Feature Sets}
    \scalebox{0.8}{
    \begin{tabular}{r r l }
    \toprule
    \textbf{Group} &\textbf{Set} & \textbf{Feature} \\
    \hline
    Simple &LOB Best Level & $u_1=\{ P^{ask}_1, V^{ask}_1, P^{bid}_1, V^{bid}_1\}$ \\
    &&\\
    Extended & Basic & $u_2=(P^{ask}_1 + P^{bid}_1$)/2 \\
    & & $u_3=P^{ask}_1 - P^{bid}_1$ \\
    & & $u_4=\sin(P^{ask}_1 P^{bid}_1)$ \\
    &&\\ 
    &Synthesized & $u_5=P^{ask}_1 P^{bid}_1$ \\
    & & $u_6=V^{ask}_1 V^{bid}_1$ \\
    & & $u_7={P^{ask}_1}^2 + {P^{bid}_1}^2$ \\
    & & $u_8={V^{ask}_1}^2 + {V^{bid}_1}^2$ \\
    &&\\
    &Linear Kernel & $u_9=P^{ask}_1 P^{bid}_1$ \\
    &Polynomial Kernel & $u_{10}=(P^{ask}_1 P^{bid}_1 + c_0)^d$ \\
    &Sigmoid Kernel & $u_{11}=\tanh(\gamma P^{ask}_1 P^{bid}_1 + c_0)$ \\
    &Exponential Kernel & $u_{12}=e^{-\gamma |P^{ask}_1 - P^{bid}_1|}$ \\
    &RBF Kernel & $u_{13}=e^{-\gamma (P^{ask}_1 - P^{bid}_1)^2}$ \\
    \bottomrule
    \end{tabular}}
    \label{tab:feat}
    \end{table}
        
        \vspace{-0.5em}
        \hspace{1.5em} The first group comprises four key features extracted directly from Level 1 of the LOB, including the best bid and ask prices ($P^{bid}_1$, $P^{ask}_1$) and their corresponding volumes ($V^{bid}_1$, $V^{ask}_1$). This fundamental representation of the LOB provides an unfiltered snapshot of market liquidity and supply-demand equilibrium. The second feature group provides further insights by creating transformations of the best bid and ask prices. These include the mid-price ($u_2$), which is directly connected to our forecasting objective, the bid-ask spread ($u_3$), which captures market friction. Additionally, we incorporate a sinusoidal transformation ($u_4$), which helps model potential cyclical or periodic components that may impact short-term market behavior. The synthesized features ($u_5$ to $u_8$) are designed to capture non-linear interactions between price and volume through products and quadratic forms. For instance, $u_5$ and $u_6 $ represent the product of the best ask and bid prices and volumes, which captures joint price and volume effects. Meanwhile, $u_7$ and $u_8$ introduce second-order dependencies, useful for capturing volatility effects and variances in LOB activity. 
        \vspace{-0.0em}
        \hspace{1.5em} Finally, the \textit{Extended} feature group further includes a variety of kernel transformations, which aim to capture complex, non-linear relationships in the data. More specifically, the linear kernel ($u_9$) provides a simple transformation that preserves proportional interactions between bid and ask prices. The polynomial kernel ($u_{10}$), with degree $d=3$, incorporates higher-order terms to capture more complex interactions between the features. The sigmoid kernel ($u_{11}$), similar to activation functions in neural networks, introduces non-linearity that may capture threshold-based effects in the price interactions. Lastly, the exponential kernel ($u_{12}$) and RBF kernel ($u_{13}$) help capture localized patterns in the data using distance-based metrics. In particular, the RBF kernel is effective for identifying similarities between feature values and for modeling non-linear dynamics through Euclidean distances, which is highly relevant in noisy and rapidly changing HFT environments.
\end{enumerate}

\subsection{Reinforcement Learning – Deep Policy Value Learning }
    \noindent The proposed ALPE RL framework utilize a DL model that approximates the optimal action-value function for decision-making, to forecast the mid-price in an HFT setting. This model falls under the category of model-free, value-based RL approaches. Our RL approach is model-free and value-based, meaning we do not use a model to describe how the environment changes in response to actions. Instead, the agent learns through direct interaction with the environment, which consists of the current state of the LOB, the action bounds, the states represented by the LOB feature sets (i.e., \textit{Simple} and \textit{Extended}), and a reward function that penalizes deviations from actual mid-price movements. The agent operates in an event-driven, online learning manner, continually adapting based on new incoming LOB data. Below, we detail the different components of the agent, including the environment, the action and reward structures, its internal DL model architecture, and the learning process.

\subsubsection{Markov Decision Process Representation}
\noindent The problem is modeled as a Markov decision process (MDP) defined by the tuple $(\mathcal{S}, \mathcal{A}, R, \gamma)$, where:

\begin{itemize}[label=-]
    \item $\mathcal{S}$ is the set of states, where each state $s_t \in \mathbb{R}^n$ represents the features of LOB at time $t$\footnote{We treat the events sequentially as they exist in LOB and the time $t$ is utilised for indexing purposes only,}. The state $s_t$ at time $t$ consists of a vector of features derived from the LOB. Specifically:
     
    \vspace{-12pt}
    \begin{equation}
        s_t = \begin{bmatrix} s_{1,t}, s_{2,t}, \ldots, s_{n,t} \end{bmatrix},
    \end{equation}
    
    where $s_{i,t}$ represents the $i$-th feature at time $t$, In this work, we use features such as bid price, ask price, and other relevant LOB metrics to fully describe the state of the market.
    
    \item $\mathcal{A}$ is the set of actions, representing possible adjustments to the mid-price prediction. The action space $\alpha_t$ is continuous and we also define $a_{e_t} \in [a_{\min}, a_{\max}]$, with $a_{\min} = -0.1$ and $a_{\max} = 0.1$\footnote{These bounds ensure that the agent can adjust its predictions based on the smallest possible market movements defined by the tick size.} the possible exploration-specific parameter. The ALPE agent uses an epsilon-greedy policy for action selection within the action space $\alpha_t$, as follows:
    
    \vspace{-12pt}
    \begin{equation}
        a_t = 
        \begin{cases} 
            p_t + \alpha_{e_t}, &  \xi \leq \epsilon_t \ (\text{exploration}) \\
            f_{\pi}(s_t, \alpha_t; \theta_{ALPE}), & \text{otherwise} \ (\text{exploitation})
        \end{cases}
    \end{equation}

    where $\xi$ is drawn from a uniform distribution\footnote{A uniform distribution ensures that each value within the range $[0, 1]$ has an equal probability of selection, making the exploration decision unbiased.} $U(0, 1)$, $\epsilon_t$ is the adaptive exploration parameter, and $f_{\pi}(s_t; \theta_{ALPE})$ represents the output of the DL model for state $s_t$. A detailed explanation of this model's architecture and functionality can be found in \hyperref[sec:ALPE_architecture]{Section \ref{sec:ALPE_architecture}}. Initially, the agent explores more frequently, with $\epsilon$ decaying over time:
    
    \vspace{-12pt}
    \begin{equation}
        \epsilon_{t+1} = \max(\epsilon_{\text{min}}, \epsilon_t \cdot \epsilon_{\text{decay}}),
    \end{equation}

    where $\epsilon_{\text{min}} = 0.0001$ and $\epsilon_{\text{decay}} = 0.999$\footnote{These values balance exploration and exploitation, allowing the agent to initially explore and learn about LOB dynamics, then exploit actions that yield high rewards. The decay factor reaches the $\epsilon_{\text{min}}$ after approximately 9500 trading events, which is sufficient for many stocks given their daily trading activity. For instance, some stocks, such as ILMN and ORLY, have fewer trading events than the decaying steps of ALPE. We maintained the same decay factor across all stocks for a fair comparison. However, HFT traders should note that the decay factor should ideally be dynamic and stock-specific.}.
    
    \item The environment transitions are modeled as Markovian, where the next state $s_{t+1}$ depends only on the current state $s_t$ and the action $\alpha_t$. These transitions are deterministic and directly derived from the LOB dataset.
    
    \item $R(s_t, a_t)$ is the reward function that evaluates the agent's action based on the deviation from the true mid-price. Specifically, the reward $R_t$ for action $a_t$ at state $s_t$ is defined as:

    \vspace{-12pt}
    \begin{equation}
        R_t = -\left| a_t - p_t \right| \cdot (1 - \epsilon_t),
    \end{equation}
    
    where $p_t$ is the true mid-price at time $t$, $a_t$ is the predicted adjustment to the mid-price and the term $(1 - \epsilon)$ serves as a weighting factor that gradually increases the penalty for errors as the agent's exploration decreases. This reward function encourages the agent to minimize the prediction error while balancing exploration and exploitation.
    
    \item $\gamma \in [0, 1]$ is the discount factor, which, in our model, is set to $\gamma = 0$ to focus on maximizing immediate rewards due to the fast-moving nature of HFT.
\end{itemize}

\subsubsection{Exploitation Network Architecture}
\label{sec:ALPE_architecture}

\noindent The agent uses a DL model, specifically an MLP, as a non-linear regressor that will approximate the policy value function $f_{\pi}(s_t, a_t; \theta_{ALPE})$. This function represents the immediate reward adjustment for taking action $a_t$ in state $s_t$ at the current state $t$ under the current policy. The topology of MLP is as follows,:

\begin{enumerate}
    \item \textit{Input Layer}: The network takes the current state $s_t \in \mathbb{R}^n$ as input,
    \item \textit{Hidden Layers}: The network includes eight hidden layers, with 64 neurons each, using the rectified linear unit (ReLU) to capture non-linearity in the state-action relationships, as follows:
    \begin{equation}
        \textbf{h}_i = \text{ReLU}(\mathbf{W}_i \mathbf{h}_{i-1} + \mathbf{b}_i), \quad i = 1, \ldots, 8,
    \end{equation}
    where $\mathbf{W}_i$ are the weight matrices, and $\mathbf{b}_i$ are the biases.
    \item \textit{Batch Normalization}: Batch normalization is applied after the first hidden layer to stabilize learning and improve convergence. The batch normalization is performed as follows:

    \vspace{-12pt}
    \begin{equation}
        \hat{\mathbf{h}}_1 = \frac{\mathbf{h}_1 - \mu_B}{\sqrt{\sigma_B^2 + \zeta}},
    \end{equation}
    
    where $\mu_B$ is the mean of $\textbf{h}_1$ from the current minimal-batch\footnote{The minimal-batch learning approach is limited to a look-back window of one LOB state (i.e., current LOB state).}, $\sigma_B^2 $ is the variance of the $\textbf{h}_1$ for the current minimal-batch and $\zeta$ is a small constant. Each normalized activation is then scaled and shifted, as follows:
    
    \begin{equation}
        \mathbf{y}_1 = \psi \hat{\mathbf{h}}_1 + \beta,
    \end{equation}
    
    where $\psi$ and $\beta$ are learnable parameters for scaling and shifting the normalized output.
    
    \item \textit{Output Layer}: The output layer that is the ninth layer, has a single neuron representing the predicted policy value:
    \begin{equation}
        f_{\pi}(s_t, a_t; \theta_{ALPE}) = \mathbf{W}_{L} \mathbf{h}_H + b_{L},
    \end{equation}
    where \( \mathbf{W}_{L} \) and \( b_{L} \) represent the weight matrix and bias term for the output layer, respectively. The term \( \mathbf{h}_H \) denotes the output of the final hidden layer, which is activated using the ReLU function. The network comprises \( L = 9 \) total layers, including \( H = 8 \) hidden layers.\footnote{The selected topology was determined through an extensive grid search. The search considered up to 10 hidden layers and various neuron counts per layer (4, 8, 16, 32, 64, 128, 256, 512, and 1024). Additionally, configurations with and without batch normalization were evaluated, as well as two optimizers: Adam and GD. The grid search identified the best-performing architecture on average across the majority of stocks to ensure a fair evaluation.}
\end{enumerate}

\subsubsection{Policy Value Approximation with Minimal Training} \noindent In this study, the Policy Value function $f_{\pi}(s_t, a_t; \theta_{ALPE})$ approximates the expected cumulative reward for taking action $a_t$ in state $s_t$ under the current policy parameterized by $\theta_{ALPE}$. To train this policy network, a Policy Value target $f_{\pi, target}(s_t, a_t; \theta_{ALPE})$ is computed at each step, which reflects the expected reward adjusted by the exploration penalty. The Policy Value target is defined as: 

\begin{equation}
f_{\pi, target}(s_t, a_t; \theta_{ALPE}) = R_t - |\alpha_t - f_{\pi}(s_t, a_t; \theta_{ALPE})| \times (1-\epsilon_t),
\end{equation}

The network's goal is to minimize the squared difference between the predicted Policy Value $f_{\pi}(s_t, a_t; \theta_{ALPE})$ and the target Policy Value $f_{\pi, target}(s_t, a_t;\theta_{ALPE})$:

\begin{equation} 
\mathcal{L}(\theta_{ALPE}) = (f_{\pi, target}(s_t, a_t; \theta_{ALPE}) - f_{\pi}(s_t, a_t; \theta_{ALPE}))^2,
\end{equation}

\noindent where the training process uses the adaptive moment estimation (Adam) optimizer\cite{kingma2014adam}. The training process is minimal, consisting of only two epochs. This is sufficient because the model's input is limited to the current state of the LOB, which provides restricted information for learning. Consequently, the convergence of the optimal weights occurs rapidly, as the DL model's complexity is constrained by the simplicity of the input data.

\subsection{Novelty of the Proposed RL Framework}

\noindent The key innovations in the proposed RL framework are:

\begin{itemize}[label=-]
    \item \textit{Online Adaptation}:
    The agent continuously adapts its policy in an event-driven manner, updating its Policy Value-network with each new LOB state. 
    
    \item \textit{Reward Balancing Mechanism}:   
    The reward function $R_t$ is designed to penalize prediction errors while incorporating the exploration factor $\epsilon_t$. This approach enables the agent to balance exploring new actions and exploiting established strategies.

    \item \textit{Markovian Structure for LOB Dynamics}:
    The framework relies on a Markov framework, where the current state of the LOB provides relevant information for the decision-making of the RL agent.
\end{itemize}

\section{Results and Discussion}

\noindent This study utilises a Level 1 LOB HFT dataset from NASDAQ, provided by London Stock Exchange Group (LSEG), covering the period from September 1$^{st}$ to November 30$^{th}$, 2022. This dataset spans 100 stocks that are listed in \hyperref[tab:stocks]{Table \ref{tab:stocks}}. To benchmark the performance of our proposed ALPE model, we compared it against a range of widely used forecasting models, including traditional statistical models (ARIMA and a Naive regressor), DL models such as MLP, CNN, LSTM, GRU, and RBFNN. Each model was run ten times to calculate the average RMSE and RRMSE scores, thereby reducing the impact of random variations on performance metrics and providing a more reliable estimation of forecasting performance.

\begin{table}[!htb]
\centering
\captionsetup{skip=5pt} 
\caption{List of the 100 S\&P 500 stocks.}
\label{tab:table1a}
\normalsize
\scalebox{0.65}{
\begin{tabular}{ r l r l r l}
\toprule
\textbf{Ticker} & \textbf{Stock} & \textbf{Ticker} & \textbf{Stock} & \textbf{Ticker} & \textbf{Stock} \\
\hline
AMZN & Amazon                   &  LDOS & Leidos Holdings              & PSA & Public Storage   \\
BAC & Bank of America           &  LHX &  L3Harris Technologies        &  PSX & Phillips 66 \\
BRK & Berkshire Hathaway        &  LMT &  Lockheed Martin Corp.        & PXD & Pioneer Natural Res. \\
EIX & Edison International      &  LUV &  Southwest Airlines           & REGN & Regeneron Pharmaceuticals  \\
EMN & Eastman Chemical          &  LVS &  Las Vegas Sands              & ROL & Rollins  \\
ENPH & Enphase Energy           &  MAA & Mid-America Apartment Com.    & ROST & Ross Stores   \\
EOG & EOG Resources             &  MAS & Masco                         & SBAC & SBA Communications  \\
EQIX & Equinix                  &  MCHP & Microchip Technology         & SJM & J. M. Smucker Comp \\
EQR & Equity Residential        &  MCK & McKesson Corp.                & STLD& Steel Dynamics \\
EQT & EQT Corporation           &  MDT & Medtronic                     & TECH & Bio-Techne Corp. \\
EXPD & Expeditors International &  META & Meta Platforms               & TEL &  TE Connectivity  \\ 
EXPE & Expedia                  &  MKC & McCormick \& Company          & TFC & Truist Financial \\
FDX & FedEx                     &  MNST & Monster Beverage             & TGT & Target  \\
FITB & Fifth Third Bancorp      &  MOS & Mosaic                        & TMUS & T-Mobile US \\
FLT & FleetCor Technologies     &  MRK & Merck \& Co                   & TROW & T. Rowe Price Group  \\
GD & General Dynamics           &  MRNA & Moderna                      & TRV & Travelers Companies \\
GILD & Gilead Sciences          &  MSCI & MSCI Inc.                    & TSN & Tyson Foods \\
GIS & General Mills             &  MSFT & Microsoft                    & UHS & Universal Health Services \\
GL & Globe Life                 &  MTB & M\&T Bank Corp.               & UNH & UnitedHealth Group\\
GM & General Motors             &  NFLX & Netflix                      & UPS &  United Parcel Service \\
GOOGL & Alphabet                &  NKE & Nike                          & V & Visa\\
HAS & Hasbro                    &  NTAP & NetApp                       & VRSN & VeriSign  \\
HD & Home Depot                 &  NVDA & NVIDIA                       & WAB & Westinghouse Air Brake Techn.\\
HLT & Hilton Worldwide          &  NXPI & NXP Semiconductors           & WBD & Warner Bros Discovery \\
ILMN & Illumina                 &  ODFL & Old Dominion Freight Line    &  WEC & WEC Energy Group  \\
INCY & Incyte                   &  OMC & Omnicom Group                 & WMT & Walmart \\
IPG & Interpublic               &  ORCL & Oracle                       & WRB & W. R. Berkley Corp, \\
IQV & IQVIA                     &  ORLY & O'Reilly Automotive          & XOM & Exxon Mobil \\
IRM & Iron Mountain             &  OTIS & Otis Worldwide Corp.         & XYL & Xylem \\
J &  Jacobs Solutions           &  PEAK & Healthpeak Properties        & ZBH & Zimmer Biomet Holdings \\
JNJ & Johnson \& Johnson        &  PEP & PepsiCo                       & ZBRA & Zebra Technologies\\ 
JPM & JPMorgan Chase            &  PG & Procter \& Gamble              & ZTS & Zoetis \\
KEYS &   Keysight Technologies  &  PHM & PulteGroup                    &   \\ 
KO &  Coca-Cola                 &  POOL & Pool Corp.                   &   \\       
\bottomrule
\end{tabular}}
\label{tab:stocks}
\end{table}

\begin{table}[H]
    \centering
    \captionsetup{width=12.00\textwidth}
    \caption{RMSE and RRMSE Scores for Amazon.}
    \scalebox{0.62}{
    \begin{small}
    \begin{tabular}{r c c c c c c c}
    \toprule
    \textbf{Stock} & \textbf{Set} & \textbf{Model} & \textbf{RMSE} & \textbf{RRMSE} & \textbf{Set} & \textbf{RMSE} & \textbf{RRMSE} \\
    \hline
   Amazon & Simple & Naive & 6.020E-01 ± 9.518E-03 & 5.287E-03 ± 8.360E-05 & Exte & 6.527E-01 ± 1.032E-03 & 5.732E-03 ± 5.084E-05 \\
           &        & ARIMA & 3.640E-01 ± 5.756E-03 & 3.197E-03 ± 5.055E-05 &      & 1.294E-01 ± 2.046E-03 & 1.136E-03 ± 1.797E-05 \\
           &        & MLP   & 1.902E-01 ± 3.007E-03 & 1.670E-03 ± 2.641E-05 &      & 1.299E-01 ± 1.154E-03 & 6.411E-04 ± 1.014E-05 \\
           &        & CNN   & 1.521E-01 ± 2.404E-03 & 1.335E-03 ± 2.112E-05 &      & 1.449E-01 ± 2.291E-03 & 1.273E-03 ± 2.013E-05 \\
           &        & LSTM  & 1.529E-01 ± 2.417E-03 & 1.343E-03 ± 2.123E-05 &      & 9.405E-02 ± 5.383E-04 & 8.990E-04 ± 9.728E-06 \\
           &        & GRU   & 1.048E-01 ± 1.657E-03 & 9.206E-03 ± 1.456E-05 &      & 2.440E-01 ± 3.857E-03 & 2.143E-03 ± 3.388E-05 \\
           &        & RBFNN & 1.248E-01 ± 1.973E-03 & 1.096E-03 ± 1.732E-05 &      & 1.293E-01 ± 2.046E-03 & 1.106E-03 ± 1.557E-05 \\
           &        & \textbf{ALPE}    & \textbf{5.586E-02 ± 8.832E-04} & \textbf{4.906E-04 ± 7.757E-06} &      & \textbf{2.527E-02 ± 1.012E-03} & \textbf{2.732E-04 ± 5.844E-06} \\
    \cline{2-5} \cline{6-8} 
           & Simple & MLP   & 2.439E-01 ± 1.492E-03 & 4.290E-03 ± 1.311E-05 & Exte & 1.089E-01 ± 1.722E-03 & 9.564E-04 ± 1.512E-05 \\
           & MDI    & CNN   & 1.595E-01 ± 2.522E-03 & 1.401E-03 ± 2.215E-05 & MDI  & 1.390E-01 ± 2.197E-03 & 1.221E-03 ± 1.930E-05 \\
           &        & LSTM  & 1.453E-01 ± 2.298E-03 & 1.276E-03 ± 2.018E-05 &      & 9.795E-02 ± 1.549E-03 & 8.602E-04 ± 1.360E-05 \\
           &        & GRU   & 1.369E-01 ± 2.165E-03 & 1.202E-03 ± 1.901E-05 &      & 1.809E-01 ± 2.861E-03 & 1.589E-03 ± 2.513E-05 \\
           &        & RBFNN & 1.825E-01 ± 2.885E-03 & 1.603E-03 ± 2.534E-05 &      & 1.294E-01 ± 2.116E-03 & 1.136E-03 ± 1.797E-05 \\
           &        & \textbf{ALPE}    & \textbf{6.000E-02 ± 9.487E-04} & \textbf{5.270E-04 ± 8.332E-06} &      & \textbf{3.447E-02 ± 1.032E-03} & \textbf{5.732E-04 ± 9.064E-06} \\
    \cline{2-5} \cline{6-8}
           & Simple & MLP   & 5.113E-01 ± 7.401E-03 & 4.809E-03 ± 7.604E-05 & Exte & 2.046E-01 ± 5.764E-03 & 9.202E-04 ± 5.063E-06 \\
           & GD     & CNN   & 2.587E-01 ± 4.091E-03 & 2.272E-03 ± 3.593E-05 & GD   & 1.528E-01 ± 2.416E-03 & 1.342E-03 ± 2.122E-05 \\
           &        & LSTM  & 2.214E-01 ± 3.501E-03 & 1.991E-03 ± 2.201E-05 &      & 1.808E-01 ± 2.859E-03 & 1.588E-03 ± 2.511E-05 \\
           &        & GRU   & 2.408E-01 ± 3.808E-03 & 2.115E-03 ± 3.344E-05 &      & 1.342E-01 ± 2.121E-03 & 1.178E-03 ± 1.863E-05 \\
           &        & RBFNN & 1.248E-01 ± 1.973E-03 & 1.096E-03 ± 1.732E-05 &      & 1.292E-01 ± 2.046E-03 & 1.139E-03 ± 1.550E-05 \\
           &        & \textbf{ALPE}    & \textbf{5.865E-02 ± 9.274E-04} & \textbf{5.151E-04 ± 8.145E-06} &      & \textbf{2.828E-02 ± 4.472E-04} & \textbf{2.944E-04 ± 3.928E-06} \\
    \bottomrule
    \end{tabular}
    \end{small}}
    \label{tab:amazon_rmse_rrmse}
\end{table}

The forecasting ability of the newly introduced ALPE model is measured based on RMSE and RRMSE.\footnote{We measure the forecasting performance of RMSE and RRMSE based on the six available input datasets that are based on the Simple and Extensive (Exte) feature sets from \hyperref[tab:feat]{Table \ref{tab:feat}}, and each feature set considers additionally the MDI and GD feature importance mechanisms as described in \hyperref[algo:mdi]{Algorithm \ref{algo:mdi}} and \hyperref[algo:gd]{Algorithm \ref{algo:gd}}.} The results indicate that it consistently achieves the lowest forecasting error in predicting mid-price movements during the three-month testing period. We provide results for Amazon stock in \hyperref[tab:amazon_rmse_rrmse]{Table \ref{tab:amazon_rmse_rrmse}} as a representative example, while the complete set of results for all stocks can be found in the Appendix from \hyperref[tab:table_2]{Table \ref{tab:table_2}} to \hyperref[tab:table_14]{Table \ref{tab:table_14}}. Specifically, the ALPE model for Amazon achieved the lowest RMSE and RRMSE scores under the Exte dataset. In the Simple dataset, the ALPE attained an RMSE of 5.586E-02 and an RRMSE of 4.906E-04, with performance improving further under more complex datasets, such as Exte, where it achieved RMSE and RRMSE values of 2.527E-02 and 2.732E-04, respectively. The table also illustrates that the ALPE model provided the lowest RMSE and RRMSE scores across all competitor models for both simple and extensive datasets. This strongly indicates that the proposed model can effectively provide optimal performance even when different datasets carry noisy signals. We also highlight that a noisy signal for one stock may still be useful for another stock. For instance, in the case of WBD (see \hyperref[tab:table_13]{Table \ref{tab:table_13}}), feature engineering and feature importance methods reduced the ALPE model's performance; however, its performance remained significantly better than that of competitor models across all scenarios. 

\begin{figure}[!hbtp]
  \centering
    \hspace*{-1.1cm}  
  \includegraphics[height=9.8cm]{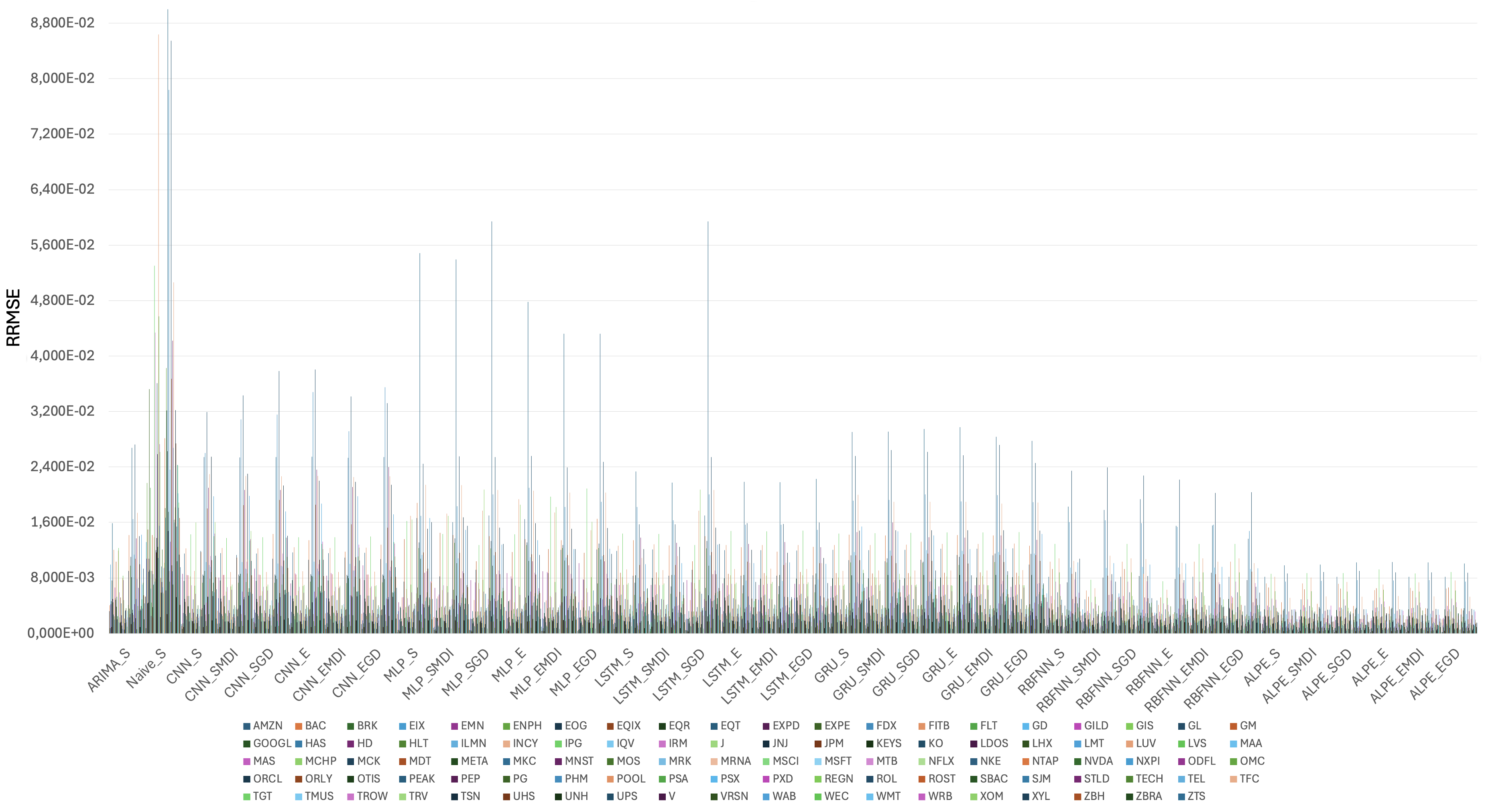}  
  \caption{RRMSE scores for all six input datasets (i.e., \textit{Simple}, \textit{Simple MDI}, \textit{Simple GD}, \textit{Exte}, \textit{Exte MDI}, \textit{Exte GD}) and each model (i.e., ARIMA, Naive regressor, MLP, CNN, LSTM, GRU, RBFNN, and ALPE) for all stocks, except for WBD due to overscaling of the y-axis, are reported. The performance details for WBD can be found in \hyperref[tab:table_13]{Table \ref{tab:table_13}} for performance details of that stock.}
  \label{fig:AMAZON_RRMSE_ALL}
\end{figure}

A general overview of the ALPE error performance across the entire line-up of competitor models and across all the six available input datasets for the 100 stocks can be seen in \hyperref[fig:AMAZON_RRMSE_ALL]{Figure \ref{fig:AMAZON_RRMSE_ALL}}. Notably, the ALPE's performance is enhanced for the majority of stocks by considering a more non-linear input space (i.e., Exte GD), where RRMSE values are lower. This suggests that the combination of the Exte GD dataset with the ALPE architecture improves the model’s ability to minimize error. Building on these findings, we conducted a pairwise statistical comparison using the Conover post-hoc test\footnote{The Conover post-hoc test, following a significant Friedman test, enables precise, pairwise model comparisons in a non-parametric context. The method applies a Bonferroni correction for multiple comparisons, controlling Type I error and rigorously highlighting ALPE’s statistically significant performance.} following a significant Friedman test to assess the consistency of the ALPE’s performance advantage over other competitor models. The statistical analysis, summarized in \hyperref[tab:p_values]{Table \ref{tab:p_values}}, reveals that the ALPE model outperformed all the ML and DL architectures with high statistical significance across various datasets. Specifically, the ALPE’s RMSE improvements over the Naive and ARIMA models were found to be highly significant (p$<$0.001), reinforcing the ALPE's capability to effectively manage the challenges of noisy HFT data. Even against more complex neural networks like CNN and MLP, the ALPE demonstrated statistically significant advantages (p$<$0.01). Although comparisons with LSTM did not yield statistical significance in every dataset, the ALPE consistently achieved lower RMSE and RRMSE scores. 

\begin{table}[H]
    \centering
    \captionsetup{width=0.90\textwidth}
    \caption{RMSE-based adjusted significance levels, derived from pairwise comparisons using the Conover post-hoc test following a significant Friedman test across six datasets.}
    \scalebox{0.60}{
    \begin{small}
    \begin{tabular}{r l l l l l l l l}
    \toprule
    \textbf{Model} & \multicolumn{1}{c}{\textbf{ALPE}} & \multicolumn{1}{c}{\textbf{Naive}} & \multicolumn{1}{c}{\textbf{RBFNN}} & \multicolumn{1}{c}{\textbf{MLP}} & \multicolumn{1}{c}{\textbf{CNN}} & \multicolumn{1}{c}{\textbf{LSTM}} & \multicolumn{1}{c}{\textbf{GRU}} & \multicolumn{1}{c}{\textbf{ARIMA}} \\
    \hline
   \textbf{ALPE}    & \multicolumn{1}{c}{-} & $1.5649e^{-09}***$ & $3.0614e^{-01}$ &  $4.9236e^{-04}***$ &  $2.6234e^{-03}**$ & $6.7727e^{-02}$ &  $4.5304e^{-03}**$ &  $8.0060e^{-07}***$ \\
    \textbf{Naive} & $1.5649e^{-09}***$ & \multicolumn{1}{c}{-} & $7.0985e^{-06}***$ & $7.7680e^{-03}**$ &  $1.5097e^{-03}**$ & $4.4812e^{-05}***$ &  $8.6415e^{-04}***$ & $1.0000e^{+00}$ \\
    \textbf{RBFNN} & $3.0614e^{-01}$ & $7.0985e^{-06}***$ & \multicolumn{1}{c}{-} & $9.2830e^{-01}$ & $1.0000e^{+00}$ & $1.0000e^{+00}$ & $1.0000e^{+00}$ & $4.0637e^{-03}**$ \\
    \textbf{MLP}   & $4.9236e^{-04}***$ & $7.7680e^{-03}**$ & $9.2830e^{-01}$ & \multicolumn{1}{c}{-} & $1.0000e^{+00}$ & $1.0000e^{+00}$ & $1.0000e^{+00}$ & $1.0000e^{+00}$ \\
    \textbf{CNN}   & $2.6234e^{-03}**$ &  $1.5097e^{-03}**$ & $1.0000e^{+00}$ & $1.0000e^{+00}$ & \multicolumn{1}{c}{-} & $1.0000e^{+00}$ & $1.0000e^{+00}$ & $4.3593e^{-01}$ \\
    \textbf{LSTM}  & $6.7727e^{-02}$ &  $4.4812e^{-05}***$ & $1.0000e^{+00}$ & $1.0000e^{+00}$ & $1.0000e^{+00}$ & \multicolumn{1}{c}{-} & $1.0000e^{+00}$ & $2.2272e^{-02}*$ \\
    \textbf{GRU}   & $4.5304e^{-03}**$ & $8.6415e^{-04}***$ & $1.0000e^{+00}$ & $1.0000e^{+00}$ & $1.0000e^{+00}$ & $1.0000e^{+00}$ & \multicolumn{1}{c}{-} & $2.7981e^{-01}$ \\
    \textbf{ARIMA} & $8.0060e^{-07}***$ & $1.0000e^{+00}$ & $4.0637e^{-03}**$ & $1.0000e^{+00}$ & $4.3593e^{-01}$ & $2.2272e^{-02}*$ & $2.7981e^{-01}$ & \multicolumn{1}{c}{-} \\

    \bottomrule
    \end{tabular}
    \end{small}}
    \label{tab:p_values}
    \vspace{2pt} 
    \justifying
    \begin{tiny}
    \textit{$p < 0.05 (*)$, $p < 0.01 (**)$, $p < 0.001 (***)$.}
    \end{tiny}
\end{table}

\begin{figure}[!htbt]
  \centering
    \hspace*{-0.8cm} 
  \includegraphics[height=9.9cm]{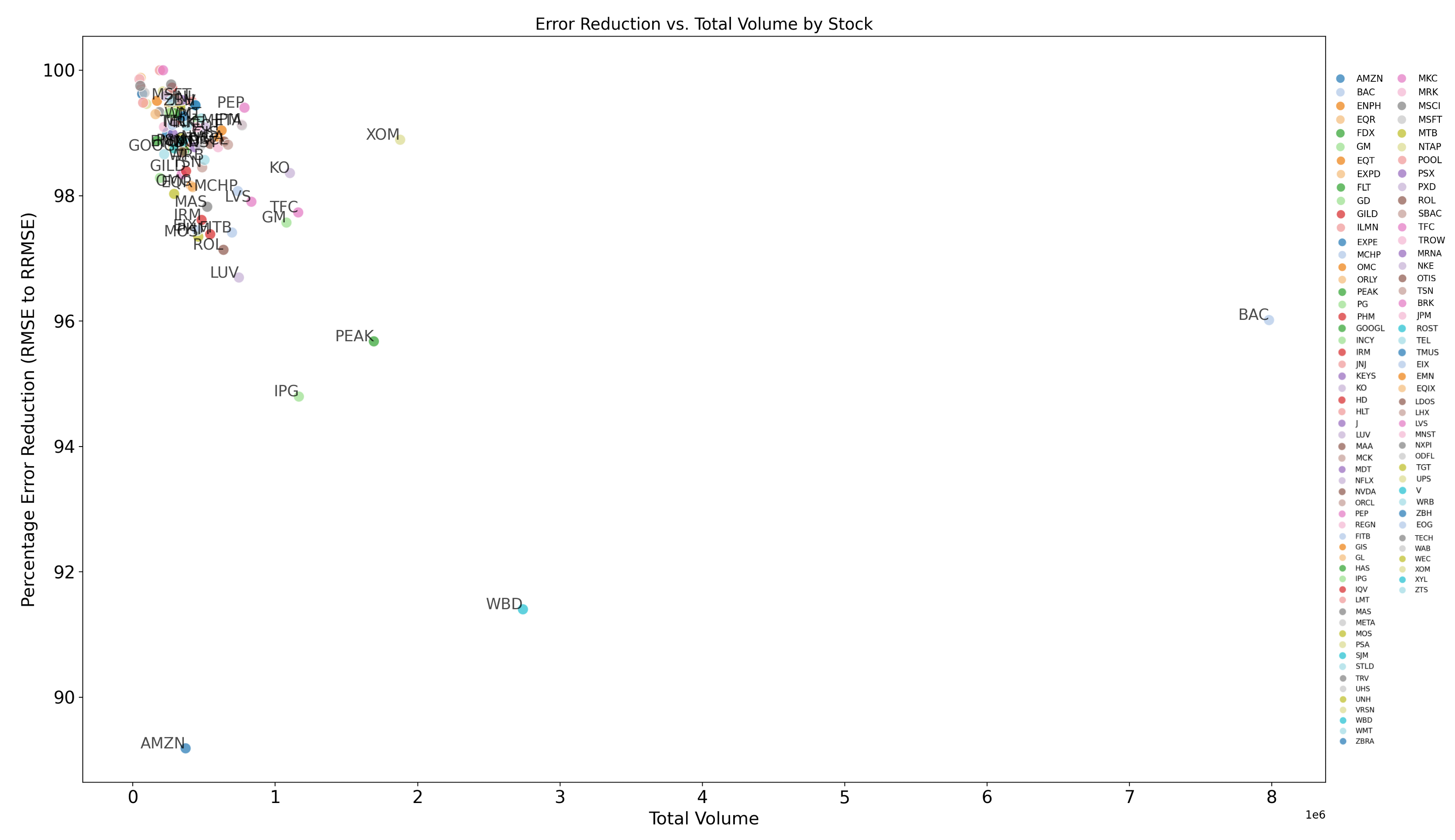}  
  \caption{Percentage error reduction between RMSE and RRMSE across stocks, plotted against total volume. The majority of stocks are concentrated in the top left quadrant, indicating a significant error reduction for lower-volume stocks, which supports the use of RRMSE over RMSE. \textit{Note}: A subset of stock names is annotated to maintain clarity in the scatter plot.}
  \label{fig:error_reduction}
\end{figure}

In analyzing model performance across varying stock volumes, we observe that RRMSE provides a more accurate measure of forecasting error than RMSE, particularly for stocks with lower trading volumes. \hyperref[fig:error_reduction]{Figure \ref{fig:error_reduction}} illustrates the percentage error reduction\footnote{The percentage error reduction is calculated as:
$\text{Error Reduction Percentage} = \frac{\text{RMSE} - \text{RRMSE}}{\text{RMSE}} \times 100$. This formula quantifies the improvement achieved by using RRMSE over RMSE, providing a relative measure of error reduction as a percentage. Lower RRMSE values indicate improved model performance, particularly beneficial for stocks with lower trading volumes where RMSE may overstate error due to scale differences.} achieved when using RRMSE instead of RMSE, with most lower-volume stocks displaying substantial reductions in error. This observation stems from the clustering of lower-volume stocks in the top left quadrant of the scatter plot, where error reduction percentages are highest. This suggests that HFT traders should rely primarily on RRMSE for evaluating performance in lower-volume stocks. Conversely, for stocks that lie on the right side of the plot, the RMSE score can safely replace RRMSE as a reliable alternative indicator of forecasting performance.

\begin{figure}[!htbt]
  \centering
  \scalebox{0.65}{ 
      \hspace*{-0.7cm} 
      \includegraphics[height=9.7cm]{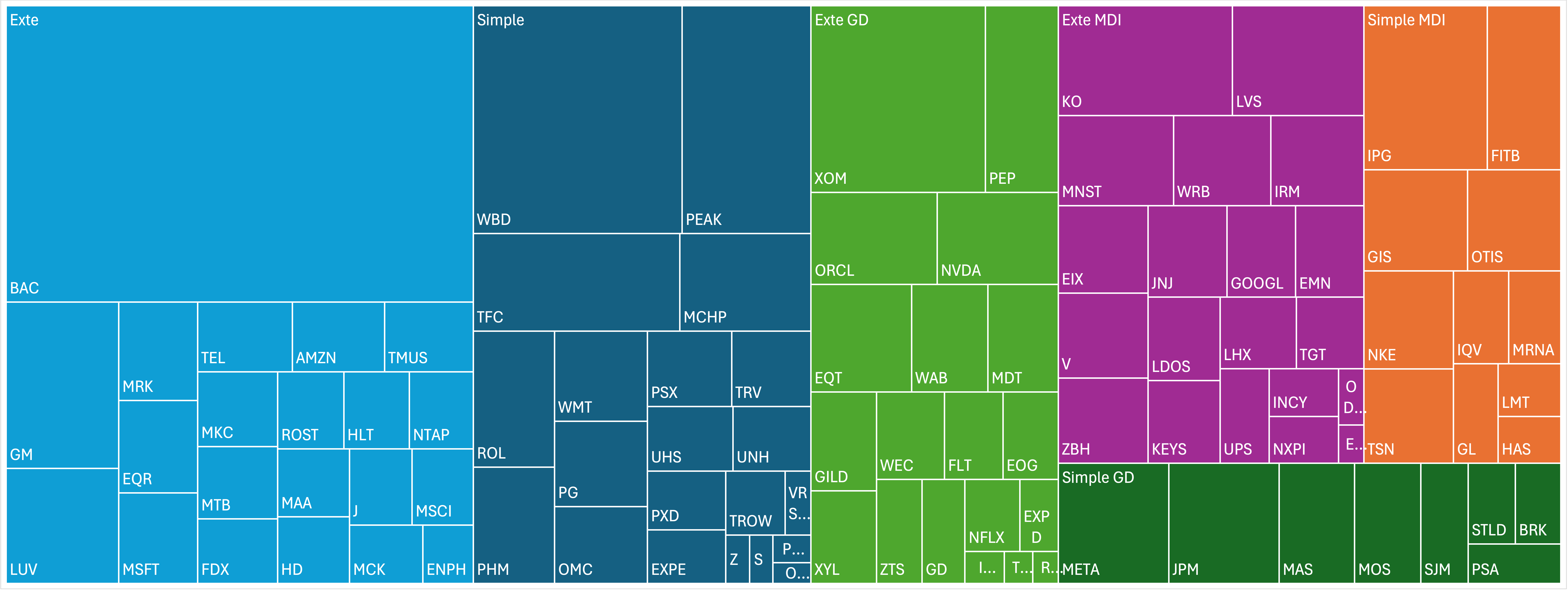}
  }
  \caption{Performance-based volume profiling: The graph shows the volume distribution per stock, classified according to their best ALPE-based RRMSE across the six datasets. \textit{Note}: The following stocks are not fully visible due to image scaling: VRSN, ZBRA, SBAC, POOL, and ORLY for Simple; ILMN, TECH, and REGN for Exte GD; ODFL and EQIX for Exte MDI.}
  \label{fig:vol_profile}
\end{figure}

Investigating further, \hyperref[fig:vol_profile]{Figure \ref{fig:vol_profile}} provides a volume-based performance profile, categorizing stocks by the dataset configuration under which the ALPE engine achieved the lowest RRMSE score. To ensure fairness, the same model hyperparameters were applied across all stocks, creating a consistent baseline for evaluating performance across trading volumes. This profile is particularly valuable for high-frequency traders as it helps identify which dataset configurations are most effective for stocks with specific trading behaviors. For example, BAC (Exte) and XOM (Exte GD) represent the highest-volume stocks in their respective dataset groups, demonstrating that complex feature sets may be critical for accurate forecasting in highly liquid markets. Conversely, WBD (Simple) and IPG (Simple MDI) suggest that simpler input configurations can suffice for lower-volume stocks, providing a pathway for reducing computational costs without sacrificing accuracy.

\section{Conclusion}
\noindent This study introduces a novel, minimal-batch RL-based model named ALPE, designed to rely solely on the current LOB state for HFT mid-price forecasting. By employing adaptive epsilon decay and a finely tuned reward structure, ALPE dynamically balances exploration and exploitation, achieving significant reductions in forecasting error within a true zero-lag, event-by-event experimental protocol. Rigorous empirical evaluation on NASDAQ Level 1 LOB data from 100 S\&P 500 stocks demonstrated the model’s consistent outperformance against several benchmark models. For instance, for Amazon stock under the Extended GD-based dataset, ALPE achieved an RRMSE of 2.484E-04, outperforming GRU (RRMSE of 1.178E-03) by approximately 79\% and MLP (RRMSE of 9.202E-04) by around 73\%. The introduction of the RRMSE metric further revealed the ALPE’s effectiveness across varying trading volumes. Specifically, the error reduction analysis indicated substantial improvements for the majority of the utilised stocks, confirming the metric’s value as a normalized measure for stock comparison. Future research can explore integrating the ALPE model into multi-agent RL frameworks, potentially enhancing its capabilities in cooperative and competitive scenarios. Additionally, adaptations to process Level 2 LOB data may allow the ALPE to capture a broader scope of market dynamics. This work provides a reliable framework for real-time HFT forecasting, establishing the ALPE as a robust and adaptive tool for the complex demands of HFT LOB datasets.

\section*{Acknowledgments}
\noindent We gratefully acknowledge the use of data provided by the London Stock Exchange Group (LSEG).

\appendix
\setcounter{table}{0}

\section{}
\label{sec:sample:appendix}

\begin{table}[!hbtp]
    \centering
    \captionsetup{width=12.00\textwidth}
    \caption{RMSE and RRMSE Scores for BAC, BRK, and EIX.}
    \scalebox{0.45}{
    \begin{small}
    \begin{tabular}{c c r c c c c c}
    \toprule
    \textbf{Stock} & \textbf{Set} & \textbf{Model} & \textbf{RMSE} & \textbf{RRMSE} & \textbf{Set} & \textbf{RMSE} & \textbf{RRMSE} \\
    \hline
     BAC & Simple & Naive & 6.011E-01 ± 7.694E-03 & 5.349E-03 ± 8.458E-05 & Exte     & 6.003E-01 ± 5.234E-03 & 4.356E-03 ± 2.903E-05 \\
           &     & ARIMA & 1.244E-01 ± 1.967E-03 & 4.065E-03 ± 6.428E-05 &          & 1.004E-01 ± 2.570E-03 & 1.982E-03 ± 3.201E-05 \\
           &     & MLP   & 8.565E-02 ± 1.354E-03 & 2.798E-03 ± 4.425E-05 &          & 1.005E-01 ± 1.589E-03 & 3.284E-03 ± 5.193E-05 \\
           &     & CNN   & 5.268E-02 ± 5.166E-04 & 1.868E-03 ± 1.688E-05 &          & 5.695E-02 ± 5.843E-04 & 1.607E-03 ± 1.909E-05 \\
           &     & LSTM  & 9.508E-02 ± 1.503E-03 & 3.107E-03 ± 4.912E-05 &          & 8.538E-02 ± 1.350E-03 & 2.790E-03 ± 4.411E-05 \\
           &     & GRU   & 8.478E-02 ± 1.340E-03 & 2.770E-03 ± 4.380E-05 &          & 9.247E-02 ± 1.462E-03 & 3.021E-03 ± 4.777E-05 \\
           &     & RBFNN & 5.593E-02 ± 2.518E-03 & 3.434E-03 ± 6.554E-05 &          & 6.674E-02 ± 1.746E-03 & 2.690E-03 ± 2.784E-05 \\
           &     & \textbf{ALPE}& \textbf{4.336E-02 ± 6.856E-04} & \textbf{1.417E-03 ± 2.240E-05}&& \textbf{3.404E-02 ± 8.544E-04} & \textbf{1.266E-03 ± 2.792E-05} \\
    \cline{2-5} \cline{6-8} 
        & Simple    & MLP   & 7.242E-02 ± 1.145E-03 & 2.366E-03 ± 3.742E-05 & Exte & 9.890E-03 ± 1.564E-04 & 3.231E-03 ± 5.109E-06 \\
           & MDI    & CNN   & 9.980E-02 ± 1.578E-03 & 3.261E-03 ± 5.156E-05 & MDI  & 8.794E-02 ± 1.390E-03 & 2.873E-03 ± 4.543E-05 \\
           &        & LSTM  & 9.108E-02 ± 1.454E-04 & 3.006E-03 ± 4.752E-06 &      & 9.656E-02 ± 1.527E-03 & 3.155E-03 ± 4.989E-05 \\
           &        & GRU   & 9.129E-02 ± 1.443E-03 & 2.983E-03 ± 4.716E-05 &      & 8.704E-02 ± 1.376E-03 & 2.844E-03 ± 4.496E-05 \\
           &        & RBFNN & 7.783E-02 ± 1.231E-03 & 2.543E-03 ± 4.021E-05 &      & 7.563E-02 ± 1.454E-03 & 2.548E-03 ± 3.994E-05 \\
           &        & \textbf{ALPE}& \textbf{4.980E-02 ± 7.874E-04}& \textbf{1.627E-03 ± 2.573E-05}&& \textbf{6.884E-02 ± 8.887E-04} & \textbf{2.088E-03 ± 2.221E-05} \\
    \cline{2-5} \cline{6-8}
           & Simple & MLP   & 6.080E-02 ± 9.613E-04 & 1.987E-03 ± 3.141E-05 & Exte & 6.846E-02 ± 1.082E-03 & 2.237E-03 ± 3.537E-05 \\
           & GD     & CNN   & 5.117E-02 ± 8.090E-04 & 1.672E-03 ± 2.643E-05 & GD   & 8.714E-02 ± 1.378E-03 & 2.847E-03 ± 4.502E-05 \\
           &        & LSTM  & 5.115E-02 ± 8.087E-04 & 1.671E-03 ± 2.642E-05 &      & 8.824E-02 ± 1.395E-03 & 2.883E-03 ± 4.559E-05 \\
           &        & GRU   & 4.976E-02 ± 7.867E-04 & 1.626E-03 ± 2.571E-05 &      & 8.169E-02 ± 1.292E-03 & 2.669E-03 ± 4.221E-05 \\
           &        & RBFNN & 5.653E-02 ± 2.663E-03 & 4.003E-03 ± 8.228E-05 &      & 7.008E-02 ± 1.402E-03 & 2.411E-03 ± 2.980E-05 \\
           &        & \textbf{ALPE}& \textbf{3.177E-02 ± 8.185E-04}& \textbf{1.292E-03 ± 2.675E-05}&& \textbf{6.309E-02 ± 9.975E-04} & \textbf{2.061E-03 ± 3.259E-05} \\
     \hline
    BRK & Simple & Naive & 3.150E+02 ± 4.981E+00 & 7.493E-04 ± 1.185E-05 & Exte     & 4.108E+02 ± 4.914E+00 & 7.392E-04 ± 1.169E-05 \\
        &        & ARIMA & 3.168E+02 ± 5.009E+00 & 7.534E-04 ± 1.191E-05 &          & 2.850E+02 ± 4.506E+00 & 6.778E-04 ± 1.072E-05 \\
        &        & MLP   & 2.156E+02 ± 3.409E+00 & 5.129E-04 ± 8.109E-06 &          & 2.187E+02 ± 3.460E+00 & 5.204E-04 ± 8.229E-06 \\
        &        & CNN   & 2.811E+02 ± 4.444E+00 & 6.685E-04 ± 1.057E-05 &          & 2.665E+02 ± 4.213E+00 & 6.338E-04 ± 1.002E-05 \\
        &        & LSTM  & 2.793E+02 ± 4.416E+00 & 6.643E-04 ± 1.050E-05 &          & 2.614E+02 ± 4.035E+00 & 5.009E-04 ± 8.430E-06 \\
        &        & GRU   & 2.912E+02 ± 4.604E+00 & 6.925E-04 ± 1.095E-05 &          & 2.750E+02 ± 4.506E+00 & 6.778E-04 ± 1.072E-05 \\
        &        & RBFNN & 2.033E+02 ± 3.530E+00 & 5.310E-04 ± 8.396E-06 &          & 2.108E+02 ± 4.914E+00 & 7.392E-04 ± 1.169E-05 \\
        &        & \textbf{ALPE}    & \textbf{9.434E+01 ± 1.492E+00} & \textbf{2.244E-04 ± 3.548E-06} &          & \textbf{8.230E+01 ± 1.301E+00} & \textbf{1.958E-04 ± 3.095E-06} \\
    \cline{2-5} \cline{6-8} 
     & Simple   & MLP   & 2.137E+02 ± 3.379E+00 & 5.084E-04 ± 8.038E-05 & Exte & 1.509E+02 ± 2.386E+00 & 3.589E-04 ± 5.675E-06 \\
        &   MDI & CNN   & 2.704E+02 ± 4.275E+00 & 6.431E-04 ± 1.017E-05 &  MDI & 2.668E+02 ± 4.218E+00 & 6.346E-04 ± 1.003E-05 \\
        &       & LSTM  & 2.720E+02 ± 4.301E+00 & 6.470E-04 ± 1.023E-05 &      & 2.741E+02 ± 4.387E+00 & 6.600E-04 ± 1.044E-05 \\
        &       & GRU   & 2.957E+02 ± 4.676E+00 & 7.033E-04 ± 1.112E-05 &      & 2.606E+02 ± 4.120E+00 & 6.197E-04 ± 9.799E-06 \\
        &       & RBFNN & 1.940E+02 ± 3.068E+00 & 4.615E-04 ± 7.297E-05 &      & 2.896E+02 ± 4.579E+00 & 6.889E-04 ± 1.089E-05 \\
        &       & \textbf{ALPE}    & \textbf{8.882E+01 ± 1.404E+00} & \textbf{2.113E-04 ± 3.340E-06} &      & \textbf{7.677E+01 ± 1.214E+00} & \textbf{1.826E-04 ± 2.887E-06} \\
    \cline{2-5} \cline{6-8}
     & Simple   & MLP   & 1.844E+02 ± 2.915E+00 & 5.385E-04 ± 6.933E-05 & Exte & 2.167E+02 ± 3.426E+00 & 5.154E-04 ± 8.149E-06 \\
        &   GD  & CNN   & 2.811E+02 ± 4.444E+00 & 6.685E-04 ± 1.057E-05 & GD   & 2.782E+02 ± 4.398E+00 & 6.616E-04 ± 1.046E-05 \\
        &       & LSTM  & 1.961E+02 ± 4.681E+00 & 7.042E-04 ± 1.113E-05 &      & 2.846E+02 ± 4.500E+00 & 6.769E-04 ± 1.070E-05 \\
        &       & GRU   & 2.015E+02 ± 4.506E+00 & 6.778E-04 ± 1.072E-05 &      & 2.811E+02 ± 4.444E+00 & 6.685E-04 ± 1.057E-05 \\
        &       & RBFNN & 1.382E+02 ± 3.766E+00 & 4.665E-04 ± 8.957E-06 &      & 3.082E+02 ± 4.873E+00 & 7.331E-04 ± 1.159E-05 \\
        &       & \textbf{ALPE}   & \textbf{7.601E+01 ± 1.202E+00} & \textbf{1.808E-04 ± 2.859E-06} &      & \textbf{6.559E+01 ± 1.037E+00} & \textbf{1.560E-04 ± 2.467E-06} \\
    \hline
    EIX & Simple & Naive & 7.617E-01 ± 1.204E-02 & 1.318E-02 ± 2.084E-04 & Exte     & 7.746E-01 ± 2.761E-03 & 8.021E-03 ± 4.777E-05 \\
        &        & ARIMA & 5.749E-01 ± 9.090E-03 & 9.948E-03 ± 1.573E-04 &          & 1.133E-01 ± 1.586E-03 & 1.935E-03 ± 2.744E-05 \\
        &        & MLP   & 1.858E-01 ± 2.938E-03 & 3.215E-03 ± 5.083E-05 &          & 1.843E-01 ± 2.914E-03 & 3.189E-03 ± 5.042E-05 \\
        &        & CNN   & 1.056E-01 ± 1.670E-03 & 1.828E-03 ± 2.890E-05 &          & 3.147E-01 ± 4.976E-03 & 5.445E-03 ± 8.610E-05 \\
        &        & LSTM  & 9.926E-02 ± 1.411E-03 & 1.545E-03 ± 2.442E-05 &          & 1.005E-01 ± 1.589E-03 & 1.739E-03 ± 2.749E-05 \\
        &        & GRU   & 2.976E-01 ± 4.706E-03 & 5.150E-03 ± 8.142E-05 &          & 2.985E-01 ± 4.719E-03 & 5.165E-03 ± 8.166E-05 \\
        &        & RBFNN & 3.339E-01 ± 5.279E-03 & 5.778E-03 ± 9.136E-05 &          & 2.012E-01 ± 3.181E-03 & 3.482E-03 ± 5.505E-05 \\
        &        & \textbf{ALPE}    & \textbf{7.733E-02 ± 1.223E-03} & \textbf{1.238E-03 ± 2.116E-05} &          & \textbf{4.123E-02 ± 6.519E-04} & \textbf{7.135E-04 ± 1.128E-05} \\
    \cline{2-5} \cline{6-8} 
     & Simple & MLP   & 1.592E-01 ± 8.842E-04 & 9.677E-04 ± 1.530E-05 & Exte    & 1.268E-01 ± 2.005E-03 & 2.194E-03 ± 3.469E-05 \\
        & MDI & CNN   & 2.932E-01 ± 4.635E-03 & 5.073E-03 ± 8.021E-05 & MDI     & 2.924E-01 ± 4.623E-03 & 5.059E-03 ± 7.999E-05 \\
        &     & LSTM  & 2.622E-01 ± 4.146E-03 & 4.537E-03 ± 7.173E-05 &         & 2.821E-01 ± 4.461E-03 & 4.882E-03 ± 7.719E-05 \\
        &     & GRU   & 2.883E-01 ± 4.558E-03 & 4.988E-03 ± 7.887E-05 &         & 3.013E-01 ± 4.764E-03 & 5.214E-03 ± 8.244E-05 \\
        &     & RBFNN & 2.004E-01 ± 3.168E-03 & 3.467E-03 ± 5.482E-05 &         & 1.449E-01 ± 3.114E-03 & 2.342E-03 ± 4.030E-05 \\
        &     & \textbf{ALPE}    & \textbf{4.919E-02 ± 7.778E-04} & \textbf{4.512E-04 ± 1.346E-05} &         & \textbf{6.083E-02 ± 9.618E-04} & \textbf{1.053E-03 ± 1.664E-05} \\
    \cline{2-5} \cline{6-8}
     & Simple& MLP   & 1.250E-01 ± 1.976E-03 & 2.163E-03 ± 3.420E-05 & Exte  & 2.593E-01 ± 4.099E-03 & 4.486E-03 ± 7.094E-05 \\
        & GD & CNN   & 1.966E-01 ± 3.109E-03 & 3.402E-03 ± 5.379E-05 &   GD  & 1.089E-01 ± 1.722E-03 & 1.885E-03 ± 2.980E-05 \\
        &    & LSTM  & 1.689E-01 ± 2.671E-03 & 2.923E-03 ± 4.622E-05 &       & 2.947E-01 ± 4.660E-03 & 5.100E-03 ± 8.063E-05 \\
        &    & GRU   & 1.687E-01 ± 2.668E-03 & 2.920E-03 ± 4.617E-05 &       & 2.975E-01 ± 3.886E-03 & 5.140E-03 ± 7.511E-05 \\
        &    & RBFNN & 2.546E-01 ± 4.001E-03 & 4.608E-03 ± 8.101E-05 &       & 1.423E-01 ± 2.250E-03 & 2.462E-03 ± 3.893E-05 \\
        &    & \textbf{ALPE}    & \textbf{4.266E-02 ± 6.745E-04} & \textbf{7.382E-04 ± 1.167E-05} &       & \textbf{6.434E-02 ± 1.017E-03} & \textbf{1.113E-03 ± 1.760E-05}\\
    \bottomrule
    \end{tabular}
    \end{small}}
    \label{tab:table_2}
\end{table}


\begin{landscape}
    \fancyhf{}  
    \fancyfoot[R]{\rotatebox{90}{\thepage}}  
    \thispagestyle{fancy}  

    \begin{table}[!hbtp]
        \centering
        \caption{RMSE and RRMSE scores for EMN, ENPH, EOG, EQIX, EQR, EQT, EXPD, and EXPE.}
        \hspace*{-4cm} 
        \begin{minipage}[t]{0.48\textwidth}
            \centering
            \scalebox{0.42}{
    \begin{small}
    \begin{tabular}{c c r c c c c c}
    \toprule
    \textbf{Stock} & \textbf{Set} & \textbf{Model} & \textbf{RMSE} & \textbf{RRMSE} & \textbf{Set} & \textbf{RMSE} & \textbf{RRMSE} \\
    \hline
        EMN & Simple & Naive & 7.898E-01 ± 1.249E-02 & 1.077E-02 ± 1.702E-04 & Exte     & 7.286E-01 ± 8.358E-03 & 9.206E-03 ± 1.139E-04 \\
        &            & ARIMA & 3.114E-01 ± 4.924E-03 & 4.246E-03 ± 6.713E-05 &          & 5.994E-01 ± 3.567E-03 & 8.111E-03 ± 5.401E-05  \\
        &            & MLP   & 3.298E-01 ± 5.214E-03 & 4.496E-03 ± 7.109E-05 &          & 6.363E-01 ± 1.006E-02 & 8.674E-03 ± 1.372E-04 \\
        &            & CNN   & 6.291E-01 ± 9.947E-03 & 8.577E-03 ± 1.356E-04 &          & 5.286E-01 ± 8.358E-03 & 7.206E-03 ± 1.139E-04 \\
        &            & LSTM  & 5.178E-01 ± 8.187E-03 & 7.059E-03 ± 1.116E-04 &          & 5.266E-01 ± 8.327E-03 & 7.180E-03 ± 1.135E-04 \\
        &            & GRU   & 4.995E-01 ± 7.898E-03 & 6.810E-03 ± 1.077E-04 &          & 4.953E-01 ± 7.831E-03 & 6.752E-03 ± 1.068E-04 \\
        &            & RBFNN & 3.318E-01 ± 5.247E-03 & 4.524E-03 ± 7.153E-05 &          & 1.877E-01 ± 2.189E-03 & 2.652E-03 ± 4.193E-05 \\
        &            & \textbf{ALPE}    & \textbf{1.770E-01 ± 2.799E-03} & \textbf{2.413E-03 ± 3.816E-05} &          & \textbf{1.620E-01 ± 2.561E-03} & \textbf{2.208E-03 ± 3.491E-05} \\
    \cline{2-5} \cline{6-8} 
     & Simple   & MLP   & 4.817E-01 ± 7.616E-03 & 6.567E-03 ± 1.038E-04 & Exte     & 6.572E-01 ± 1.039E-02 & 8.960E-03 ± 1.417E-04 \\
     & MDI      & CNN   & 5.903E-01 ± 9.334E-03 & 8.048E-03 ± 1.272E-04 &  MDI     & 6.122E-01 ± 9.680E-03 & 8.347E-03 ± 1.320E-04 \\
        &       & LSTM  & 5.117E-01 ± 8.091E-03 & 6.976E-03 ± 1.103E-04 &          & 5.143E-01 ± 8.132E-03 & 7.012E-03 ± 1.109E-04 \\
        &       & GRU   & 4.939E-01 ± 7.810E-03 & 6.734E-03 ± 1.065E-04 &          & 5.081E-01 ± 8.033E-03 & 6.927E-03 ± 1.095E-04 \\
        &       & RBFNN & 2.004E-01 ± 3.168E-03 & 2.732E-03 ± 4.319E-05 &          & 1.455E-01 ± 1.826E-03 & 1.975E-03 ± 2.490E-05 \\
        &       & \textbf{ALPE}    & \textbf{1.801E-01 ± 2.848E-03} & \textbf{2.456E-03 ± 3.883E-05} &          & \textbf{1.224E-01 ± 2.884E-03} & \textbf{1.487E-03 ± 3.932E-05} \\
    \cline{2-5} \cline{6-8}
     & Simple  & MLP   & 5.639E-01 ± 8.916E-03 & 7.688E-03 ± 1.216E-04 & Exte     & 7.437E-01 ± 1.176E-02 & 1.014E-02 ± 1.603E-04 \\
     & GD      & CNN   & 6.846E-01 ± 1.083E-02 & 9.334E-03 ± 1.476E-04 &  GD      & 7.214E-01 ± 1.141E-02 & 9.835E-03 ± 1.555E-04 \\
     &         & LSTM  & 5.084E-01 ± 8.038E-03 & 6.931E-03 ± 1.096E-04 &          & 5.227E-01 ± 8.265E-03 & 7.126E-03 ± 1.127E-04 \\
     &         & GRU   & 5.039E-01 ± 7.967E-03 & 6.870E-03 ± 1.086E-04 &          & 4.945E-01 ± 7.819E-03 & 6.742E-03 ± 1.066E-04 \\
     &         & RBFNN & 3.863E-01 ± 6.108E-03 & 5.267E-03 ± 8.328E-05 &          & 1.356E-01 ± 1.670E-03 & 1.940E-03 ± 2.276E-05 \\
     &         & \textbf{ALPE}    & \textbf{1.758E-01 ± 2.780E-03} & \textbf{2.397E-03 ± 3.790E-05} &          & \textbf{1.645E-01 ± 2.601E-03} & \textbf{2.243E-03 ± 3.546E-05}\\
    \hline
    ENPH & Simple & Naive & 6.101E+00 ± 9.646E-02 & 2.169E-02 ± 3.430E-04 & Exte     & 6.201E-01 ± 3.310E-03 & 5.034E-03 ± 3.357E-05 \\
         &        & ARIMA & 3.308E-01 ± 5.231E-03 & 1.176E-03 ± 1.860E-05 &          & 4.679E-01 ± 6.666E-03 & 4.643E-03 ± 1.231E-05 \\
         &        & MLP   & 4.351E-01 ± 2.136E-03 & 2.804E-03 ± 7.596E-05 &          & 1.959E-01 ± 3.097E-03 & 2.965E-03 ± 1.101E-05 \\
         &        & CNN   & 4.027E-01 ± 6.367E-03 & 1.432E-03 ± 2.264E-05 &          & 4.231E-01 ± 6.690E-03 & 1.504E-03 ± 2.379E-05 \\
         &        & LSTM  & 4.265E-01 ± 6.744E-03 & 1.517E-03 ± 2.398E-05 &          & 4.424E-01 ± 6.995E-03 & 1.573E-03 ± 2.487E-05 \\
         &        & GRU   & 4.150E-01 ± 6.562E-03 & 1.476E-03 ± 2.333E-05 &          & 4.216E-01 ± 6.666E-03 & 1.499E-03 ± 2.370E-05 \\
         &        & RBFNN & 3.887E-01 ± 3.898E-03 & 1.187E-03 ± 1.877E-05 &          & 1.899E-01 ± 4.788E-03 & 7.915E-04 ± 1.093E-05 \\
         &        & \textbf{ALPE}    & \textbf{2.290E-01 ± 3.621E-03} & \textbf{8.142E-04 ± 1.287E-05} &          & \textbf{1.183E-01 ± 3.452E-03} & \textbf{5.762E-04 ± 1.227E-05} \\
    \cline{2-5} \cline{6-8} 
         & Simple & MLP   & 3.641E-01 ± 3.595E-03 & 5.835E-03 ± 9.226E-06 & Exte     & 3.096E-01 ± 3.314E-03 & 3.452E-03 ± 1.178E-05 \\
         & MDI    & CNN   & 4.040E-01 ± 6.388E-03 & 1.436E-03 ± 2.271E-05 & MDI      & 4.050E-01 ± 6.404E-03 & 1.440E-03 ± 2.277E-05 \\
         &        & LSTM  & 4.564E-01 ± 7.216E-03 & 1.623E-03 ± 2.566E-05 &          & 4.308E-01 ± 6.812E-03 & 1.532E-03 ± 2.422E-05 \\
         &        & GRU   & 4.105E-01 ± 6.491E-03 & 1.460E-03 ± 2.308E-05 &          & 4.270E-01 ± 6.751E-03 & 1.518E-03 ± 2.400E-05 \\
         &        & RBFNN & 2.304E-01 ± 3.168E-03 & 7.824E-04 ± 1.126E-05 &          & 2.155E-01 ± 1.826E-03 & 1.575E-03 ± 2.490E-05 \\
         &        & \textbf{ALPE}    & \textbf{2.099E-01 ± 3.636E-03} & \textbf{7.276E-04 ± 1.293E-05} &          & \textbf{1.372E-01 ± 3.751E-03} & \textbf{8.434E-04 ± 1.334E-05} \\
    \cline{2-5} \cline{6-8}
         & Simple & MLP   & 4.256E-01 ± 1.986E-03 & 4.466E-03 ± 7.061E-05 & Exte     & 3.932E-01 ± 2.739E-03 & 2.160E-03 ± 9.739E-05 \\
         & GD     & CNN   & 4.074E-01 ± 6.441E-03 & 1.448E-03 ± 2.290E-05 & GD       & 4.067E-01 ± 6.430E-03 & 1.477E-03 ± 3.588E-05 \\
         &        & LSTM  & 4.196E-01 ± 6.635E-03 & 1.492E-03 ± 2.359E-05 &          & 4.334E-01 ± 6.852E-03 & 1.541E-03 ± 2.436E-05 \\
         &        & GRU   & 4.154E-01 ± 6.568E-03 & 1.477E-03 ± 2.335E-05 &          & 4.394E-01 ± 6.947E-03 & 1.562E-03 ± 2.470E-05 \\
         &        & RBFNN & 3.545E-01 ± 2.009E-03 & 1.187E-03 ± 1.039E-05 &          & 3.737E-01 ± 4.029E-03 & 1.185E-03 ± 1.877E-05 \\
         &        & \textbf{ALPE}    & \textbf{1.251E-01 ± 3.559E-03} & \textbf{8.003E-04 ± 1.265E-05} &          & \textbf{2.881E-01 ± 2.001E-03} & \textbf{8.320E-04 ± 2.236E-05} \\
    \hline
    EOG & Simple & Naive & 5.053E-01 ± 7.989E-03 & 4.319E-03 ± 6.829E-05 & Exte     & 7.002E-01 ± 5.153E-03 & 4.228E-03 ± 4.402E-05 \\
        &        & ARIMA & 5.458E-01 ± 8.630E-03 & 4.665E-03 ± 7.376E-05 &          & 3.008E-01 ± 5.993E-03 & 3.008E-03 ± 4.021E-05 \\
        &        & MLP   & 2.300E-01 ± 3.636E-03 & 1.965E-03 ± 3.108E-05 &          & 3.583E-01 ± 4.084E-03 & 2.208E-03 ± 3.491E-05 \\
        &        & CNN   & 3.462E-01 ± 5.473E-03 & 2.959E-03 ± 4.678E-05 &          & 3.621E-01 ± 5.725E-03 & 3.095E-03 ± 4.893E-05 \\
        &        & LSTM  & 3.649E-01 ± 5.770E-03 & 3.119E-03 ± 4.932E-05 &          & 3.773E-01 ± 5.973E-03 & 3.229E-03 ± 5.105E-05 \\
        &        & GRU   & 3.554E-01 ± 5.620E-03 & 3.038E-03 ± 4.804E-05 &          & 3.605E-01 ± 5.700E-03 & 3.081E-03 ± 4.872E-05 \\
        &        & RBFNN & 2.557E-01 ± 1.672E-03 & 1.837E-04 ± 1.429E-05 &          & 2.577E-01 ± 4.202E-03 & 1.662E-03 ± 2.629E-05 \\
        &        & \textbf{ALPE}   & \textbf{2.009E-01 ± 3.651E-03} & \textbf{1.001E-03 ± 3.121E-05} &          & \textbf{1.213E-01 ± 3.499E-03} & \textbf{1.392E-03 ± 2.991E-05} \\
    \cline{2-5} \cline{6-8} 
     & Simple   & MLP   & 3.616E-01 ± 4.137E-03 & 2.236E-03 ± 3.536E-05 & Exte     & 2.447E-01 ± 3.869E-03 & 3.092E-03 ± 3.307E-05 \\
     & MDI      & CNN   & 3.479E-01 ± 5.501E-03 & 2.974E-03 ± 4.702E-05 &  MDI     & 3.496E-01 ± 5.527E-03 & 2.988E-03 ± 4.724E-05 \\
        &       & LSTM  & 3.884E-01 ± 6.140E-03 & 3.320E-03 ± 5.249E-05 &          & 3.686E-01 ± 5.828E-03 & 3.151E-03 ± 4.982E-05 \\
        &       & GRU   & 3.522E-01 ± 5.568E-03 & 3.010E-03 ± 4.760E-05 &          & 3.647E-01 ± 5.767E-03 & 3.118E-03 ± 4.929E-05 \\
        &       & RBFNN & 2.604E-01 ± 3.168E-03 & 1.713E-03 ± 2.708E-05 &          & 1.365E-01 ± 2.159E-03 & 1.167E-03 ± 1.845E-05 \\
        &       & \textbf{ALPE}    & \textbf{2.234E-01 ± 3.691E-03} & \textbf{1.295E-03 ± 3.155E-05} &          & \textbf{1.270E-01 ± 3.747E-03} & \textbf{1.026E-03 ± 3.203E-05} \\
    \cline{2-5} \cline{6-8}
     & Simple  & MLP   & 3.409E-01 ± 3.904E-03 & 2.110E-03 ± 3.337E-05 & Exte      & 2.666E-01 ± 4.215E-03 & 2.278E-03 ± 3.603E-05 \\
     & GD      & CNN   & 3.497E-01 ± 5.529E-03 & 2.989E-03 ± 4.726E-05 &  GD       & 3.492E-01 ± 5.522E-03 & 2.985E-03 ± 4.720E-05 \\
     &         & LSTM  & 3.596E-01 ± 5.685E-03 & 3.073E-03 ± 4.859E-05 &           & 3.702E-01 ± 5.854E-03 & 3.165E-03 ± 5.004E-05 \\
     &         & GRU   & 3.557E-01 ± 5.625E-03 & 3.041E-03 ± 4.808E-05 &           & 3.744E-01 ± 5.920E-03 & 3.200E-03 ± 5.060E-05 \\
     &         & RBFNN & 1.333E-01 ± 2.107E-03 & 1.739E-03 ± 1.801E-05 &           & 1.806E-01 ± 1.591E-03 & 3.600E-03 ± 1.360E-05 \\
     &         & \textbf{ALPE}    & \textbf{1.150E-01 ± 3.557E-03} & \textbf{1.023E-03 ± 3.041E-05} &           & \textbf{1.364E-01 ± 3.738E-03} & \textbf{1.001E-03 ± 3.195E-05} \\
     \hline
    EQIX & Simple& Naive & 9.448E+00 ± 1.494E-01 & 1.499E-02 ± 2.371E-04 & Exte     & 8.314E-01 ± 4.563E-03 & 9.986E-04 ± 8.905E-05 \\
        &        & ARIMA & 5.755E-01 ± 9.099E-03 & 9.132E-04 ± 1.444E-05 &          & 4.773E-01 ± 5.002E-03 & 7.896E-04 ± 3.003E-06 \\
        &        & MLP   & 3.840E-01 ± 2.909E-03 & 8.920E-04 ± 4.617E-06 &          & 5.323E-01 ± 3.673E-03 & 8.686E-04 ± 5.829E-06 \\
        &        & CNN   & 5.706E-01 ± 9.022E-03 & 9.055E-04 ± 1.432E-05 &          & 5.786E-01 ± 9.148E-03 & 9.181E-04 ± 1.452E-05 \\
        &        & LSTM  & 5.883E-01 ± 9.302E-03 & 9.336E-04 ± 1.476E-05 &          & 5.955E-01 ± 9.416E-03 & 7.450E-04 ± 1.494E-05 \\
        &        & GRU   & 5.982E-01 ± 9.458E-03 & 9.493E-04 ± 1.501E-05 &          & 6.059E-01 ± 9.580E-03 & 8.615E-04 ± 1.520E-05 \\
        &        & RBFNN & 3.480E-01 ± 2.736E-03 & 3.181E-04 ± 7.645E-06 &          & 3.840E-01 ± 3.903E-03 & 3.111E-04 ± 6.554E-06 \\
        &        & \textbf{ALPE}    & \textbf{2.311E-01 ± 3.653E-03} & \textbf{3.667E-04 ± 5.798E-06} &          & \textbf{2.335E-01 ± 3.693E-03} & \textbf{3.706E-04 ± 5.860E-06} \\
    \cline{2-5} \cline{6-8} 
     & Simple   & MLP   & 2.506E-01 ± 3.963E-03 & 6.977E-04 ± 6.289E-06 & Exte     & 4.225E-01 ± 3.519E-03 & 3.532E-04 ± 5.584E-06 \\
     & MDI      & CNN   & 5.724E-01 ± 9.050E-03 & 9.084E-04 ± 1.436E-05 &  MDI     & 5.734E-01 ± 9.067E-03 & 9.100E-04 ± 1.439E-05 \\
        &       & LSTM  & 6.001E-01 ± 9.488E-03 & 9.523E-04 ± 1.506E-05 &          & 5.946E-01 ± 9.401E-03 & 8.436E-04 ± 1.492E-05 \\
        &       & GRU   & 5.960E-01 ± 9.424E-03 & 9.458E-04 ± 1.495E-05 &          & 5.070E-01 ± 9.598E-03 & 7.633E-04 ± 1.523E-05 \\
        &       & RBFNN & 2.304E-01 ± 3.168E-03 & 3.180E-04 ± 5.027E-06 &          & 3.681E-01 ± 2.658E-03 & 3.668E-04 ± 4.218E-06 \\
        &       & \textbf{ALPE}    & \textbf{2.003E-01 ± 3.641E-03} & \textbf{3.055E-04 ± 5.778E-06} &          & \textbf{2.332E-01 ± 3.687E-03} & \textbf{2.435E-04 ± 4.342E-06} \\
    \cline{2-5} \cline{6-8}
     & Simple  & MLP   & 3.587E-01 ± 2.509E-03 & 5.518E-04 ± 3.981E-06 & Exte     & 5.681E-01 ± 2.658E-03 & 6.668E-04 ± 4.218E-06 \\
     & GD      & CNN   & 5.819E-01 ± 8.736E-03 & 8.122E-04 ± 2.736E-05 &  GD      & 5.800E-01 ± 8.998E-03 & 7.604E-04 ± 2.736E-05 \\
     &         & LSTM  & 5.862E-01 ± 9.269E-03 & 8.303E-04 ± 1.471E-05 &          & 5.934E-01 ± 9.382E-03 & 8.416E-04 ± 1.489E-05 \\
     &         & GRU   & 5.977E-01 ± 9.450E-03 & 9.484E-04 ± 1.500E-05 &          & 6.084E-01 ± 9.620E-03 & 7.655E-04 ± 1.527E-05 \\
     &         & RBFNN & 3.390E-01 ± 2.001E-03 & 4.109E-04 ± 6.932E-06 &          & 3.874E-01 ± 4.004E-03 & 5.299E-04 ± 8.378E-06 \\
     &         & \textbf{ALPE}    & \textbf{2.264E-01 ± 3.579E-03} & \textbf{3.593E-04 ± 5.680E-06} &          & \textbf{2.364E-01 ± 3.738E-03} & \textbf{3.701E-04 ± 5.851E-06} \\
    \bottomrule
    \end{tabular}
    \end{small}}
            \label{tab:table1}
        \end{minipage}%
        \hspace{0.25\textwidth} 
        \begin{minipage}[t]{0.48\textwidth}
            \centering
            \scalebox{0.42}{
    \begin{small}
    \begin{tabular}{c c r c c c c c}
    \toprule
    \textbf{Stock} & \textbf{Set} & \textbf{Model} & \textbf{RMSE} & \textbf{RRMSE} & \textbf{Set} & \textbf{RMSE} & \textbf{RRMSE} \\
    \hline
    EQR & Simple & Naive & 7.985E-01 ± 4.720E-03 & 9.421E-03 ± 6.991E-04 & Exte     & 8.714E-01 ± 2.135E-03 & 9.881E-03 ± 4.555E-05 \\
        &        & ARIMA & 5.163E-01 ± 8.163E-03 & 7.646E-03 ± 1.209E-04 &          & 4.873E-01 ± 4.985E-03 & 4.670E-03 ± 7.384E-05 \\
        &        & MLP   & 3.599E-01 ± 4.109E-03 & 5.849E-03 ± 6.086E-05 &          & 3.153E-01 ± 4.985E-03 & 4.670E-03 ± 7.384E-05 \\
        &        & CNN   & 5.131E-01 ± 8.113E-03 & 7.600E-03 ± 1.202E-04 &          & 5.202E-01 ± 8.225E-03 & 7.705E-03 ± 1.218E-04 \\
        &        & LSTM  & 5.291E-01 ± 8.366E-03 & 6.437E-03 ± 1.239E-04 &          & 5.358E-01 ± 8.472E-03 & 7.936E-03 ± 1.255E-04 \\
        &        & GRU   & 5.377E-01 ± 8.502E-03 & 6.965E-03 ± 1.259E-04 &          & 5.446E-01 ± 8.610E-03 & 8.066E-03 ± 1.275E-04 \\
        &        & RBFNN & 3.019E-01 ± 2.001E-03 & 4.946E-03 ± 7.820E-05 &          & 2.820E-01 ± 3.661E-03 & 3.822E-03 ± 3.902E-05 \\
        &        & \textbf{ALPE}    & \textbf{2.343E-01 ± 3.705E-03} & \textbf{3.471E-03 ± 5.488E-05} &          & \textbf{1.253E-01 ± 3.562E-03} & \textbf{2.336E-03 ± 5.275E-05} \\
    \cline{2-5} \cline{6-8} 
     & Simple   & MLP   & 3.246E-01 ± 3.551E-03 & 6.327E-03 ± 5.260E-05 & Exte     & 4.003E-01 ± 3.167E-03 & 5.967E-03 ± 4.691E-05 \\
     & MDI      & CNN   & 5.147E-01 ± 8.138E-03 & 7.624E-03 ± 1.205E-04 &  MDI     & 5.157E-01 ± 8.154E-03 & 7.638E-03 ± 1.208E-04 \\
        &       & LSTM  & 5.396E-01 ± 8.532E-03 & 6.993E-03 ± 1.264E-04 &          & 5.347E-01 ± 8.454E-03 & 7.919E-03 ± 1.252E-04 \\
        &       & GRU   & 5.359E-01 ± 8.473E-03 & 6.938E-03 ± 1.255E-04 &          & 5.456E-01 ± 8.626E-03 & 8.081E-03 ± 1.278E-04 \\
        &       & RBFNN & 2.204E-01 ± 3.168E-03 & 3.968E-03 ± 4.692E-05 &          & 3.202E-01 ± 4.009E-03 & 3.887E-03 ± 6.873E-05 \\
        &       & \textbf{ALPE}    & \textbf{2.194E-01 ± 3.627E-03} & \textbf{2.397E-03 ± 5.372E-05} &          & \textbf{2.335E-01 ± 3.692E-03} & \textbf{3.459E-03 ± 3.469E-05} \\
    \cline{2-5} \cline{6-8}
     & Simple  & MLP   & 4.428E-01 ± 2.259E-03 & 7.116E-03 ± 3.345E-05 & Exte     & 4.153E-01 ± 4.985E-03 & 5.670E-03 ± 7.384E-05 \\
     & GD      & CNN   & 5.132E-01 ± 8.114E-03 & 8.601E-03 ± 1.202E-04 &  GD      & 5.180E-01 ± 8.190E-03 & 7.672E-03 ± 1.213E-04 \\
     &         & LSTM  & 5.272E-01 ± 8.336E-03 & 6.809E-03 ± 1.235E-04 &          & 4.336E-01 ± 8.437E-03 & 7.904E-03 ± 1.250E-04 \\
     &         & GRU   & 5.372E-01 ± 8.494E-03 & 6.957E-03 ± 1.258E-04 &          & 5.468E-01 ± 8.646E-03 & 8.099E-03 ± 1.281E-04 \\
     &         & RBFNN & 3.231E-01 ± 3.938E-03 & 4.946E-03 ± 7.820E-05 &          & 3.773E-01 ± 4.992E-03 & 4.946E-03 ± 7.820E-05 \\
     &         & \textbf{ALPE}    & \textbf{2.264E-01 ± 3.580E-03} & \textbf{3.353E-03 ± 5.302E-05} &          & \textbf{2.340E-01 ± 3.700E-03} & \textbf{3.466E-03 ± 5.480E-05}\\
     \hline
     EQT & Simple & Naive & 3.659E-01 ± 5.785E-03 & 8.935E-02 ± 1.413E-04 & Exte     & 4.099E-01 ± 3.887E-03 & 7.775E-02 ± 2.115E-04 \\
        &        & ARIMA & 6.507E-01 ± 1.029E-02 & 1.589E-02 ± 2.513E-04 &          & 4.001E-01 ± 2.158E-02 & 3.568E-02 ± 4.678E-04  \\
        &        & MLP   & 1.519E-01 ± 2.401E-03 & 3.709E-03 ± 5.865E-05 &          & 3.906E-01 ± 3.014E-03 & 4.655E-03 ± 7.361E-05 \\
        &        & CNN   & 4.713E-01 ± 7.453E-03 & 1.151E-02 ± 1.820E-04 &          & 4.778E-01 ± 7.554E-03 & 1.167E-02 ± 1.845E-04 \\
        &        & LSTM  & 4.859E-01 ± 7.683E-03 & 1.187E-02 ± 1.877E-04 &          & 4.921E-01 ± 7.781E-03 & 1.202E-02 ± 1.900E-04 \\
        &        & GRU   & 4.937E-01 ± 7.806E-03 & 1.206E-02 ± 1.906E-04 &          & 5.008E-01 ± 7.918E-03 & 1.223E-02 ± 1.740E-04 \\
        &        & RBFNN & 3.379E-01 ± 5.279E-03 & 8.155E-03 ± 1.289E-04 &          & 4.887E-01 ± 4.665E-03 & 6.750E-03 ± 7.511E-05 \\
        &        & \textbf{ALPE}    & \textbf{3.302E-01 ± 4.310E-03} & \textbf{8.004E-03 ± 2.003E-04} &          & \textbf{2.760E-01 ± 4.701E-03} & \textbf{3.810E-03 ± 1.773E-04} \\
    \cline{2-5} \cline{6-8} 
     & Simple   & MLP   & 2.055E-01 ± 3.249E-03 & 5.018E-03 ± 7.934E-05 & Exte     & 1.832E-01 ± 2.896E-03 & 1.073E-02 ± 7.073E-05 \\
     & MDI      & CNN   & 4.728E-01 ± 7.476E-03 & 1.155E-02 ± 1.826E-04 &  MDI     & 4.738E-01 ± 7.491E-03 & 1.157E-02 ± 1.830E-04 \\
        &       & LSTM  & 4.955E-01 ± 7.834E-03 & 1.210E-02 ± 1.913E-04 &          & 4.911E-01 ± 7.765E-03 & 1.199E-02 ± 1.896E-04 \\
        &       & GRU   & 4.920E-01 ± 7.779E-03 & 1.202E-02 ± 1.900E-04 &          & 5.008E-01 ± 7.918E-03 & 1.223E-02 ± 1.934E-04 \\
        &       & RBFNN & 3.034E-01 ± 3.168E-03 & 6.893E-03 ± 7.737E-05 &          & 3.386E-01 ± 2.191E-03 & 5.385E-03 ± 5.352E-05 \\
        &       & \textbf{ALPE}    & \textbf{2.384E-01 ± 3.997E-03} & \textbf{4.101E-03 ± 1.555E-04} &          & \textbf{2.247E-01 ± 2.598E-03} & \textbf{4.009E-03 ± 3.303E-04} \\
    \cline{2-5} \cline{6-8}
     & Simple    & MLP   & 4.309E-01 ± 2.070E-03 & 9.197E-03 ± 5.055E-05 & Exte    & 3.886E-01 ± 2.191E-03 & 5.385E-03 ± 5.352E-05 \\
     &    GD     & CNN   & 4.714E-01 ± 7.454E-03 & 1.151E-02 ± 1.820E-04 &   GD    & 4.757E-01 ± 7.522E-03 & 1.162E-02 ± 1.837E-04 \\
     &           & LSTM  & 4.842E-01 ± 7.656E-03 & 1.183E-02 ± 1.870E-04 &         & 4.901E-01 ± 7.749E-03 & 1.197E-02 ± 1.893E-04 \\
     &           & GRU   & 4.932E-01 ± 7.798E-03 & 1.205E-02 ± 1.905E-04 &         & 5.019E-01 ± 7.936E-03 & 1.226E-02 ± 1.938E-04 \\
     &           & RBFNN & 3.632E-01 ± 4.939E-03 & 8.155E-03 ± 1.289E-04 &         & 3.939E-01 ± 5.736E-03 & 5.155E-03 ± 1.289E-04 \\
     &           & \textbf{ALPE}    & \textbf{3.038E-01 ± 4.229E-03} & \textbf{6.155E-03 ± 2.3999E-04} &         & \textbf{3.337E-01 ± 4.003E-03} & \textbf{2.155E-03 ± 1.289E-04} \\
    \hline
    EXPD & Simple& Naive & 9.822E-01 ± 1.553E-02 & 1.078E-02 ± 1.705E-04 & Exte     & 6.678E-01 ± 2.425E-02 & 2.045E-02 ± 4.203E-04 \\
        &        & ARIMA & 4.518E-01 ± 7.143E-03 & 4.960E-03 ± 7.843E-05 &          & 3.248E-01 ± 8.063E-03 & 5.032E-03 ± 3.256E-05  \\
        &        & MLP   & 3.542E-01 ± 2.438E-03 & 2.893E-03 ± 2.676E-05 &          & 2.895E-01 ± 4.577E-03 & 3.179E-03 ± 5.026E-05 \\
        &        & CNN   & 4.529E-01 ± 7.160E-03 & 4.972E-03 ± 7.862E-05 &          & 4.587E-01 ± 7.253E-03 & 5.037E-03 ± 7.964E-05 \\
        &        & LSTM  & 4.671E-01 ± 7.386E-03 & 5.129E-03 ± 8.110E-05 &          & 4.720E-01 ± 7.463E-03 & 5.183E-03 ± 8.195E-05 \\
        &        & GRU   & 4.735E-01 ± 7.486E-03 & 5.199E-03 ± 8.220E-05 &          & 4.790E-01 ± 7.574E-03 & 5.260E-03 ± 8.317E-05 \\
        &        & RBFNN & 3.665E-01 ± 4.949E-03 & 4.093E-03 ± 4.264E-05 &          & 3.093E-01 ± 3.830E-03 & 3.636E-03 ± 4.003E-05 \\
        &        & \textbf{ALPE}    & \textbf{2.348E-01 ± 3.713E-03} & \textbf{2.578E-03 ± 4.076E-05} &          & \textbf{2.233E-01 ± 3.530E-03} & \textbf{2.451E-03 ± 3.876E-05} \\
    \cline{2-5} \cline{6-8} 
     & Simple   & MLP   & 2.010E-01 ± 3.178E-03 & 2.807E-03 ± 3.489E-05 & Exte     & 2.748E-01 ± 3.554E-03 & 2.468E-03 ± 3.903E-05 \\
     & MDI      & CNN   & 4.556E-01 ± 7.203E-03 & 5.002E-03 ± 7.909E-05 &  MDI     & 4.550E-01 ± 7.194E-03 & 4.995E-03 ± 7.899E-05 \\
        &       & LSTM  & 4.754E-01 ± 7.516E-03 & 5.220E-03 ± 8.253E-05 &          & 4.709E-01 ± 7.446E-03 & 5.171E-03 ± 8.176E-05 \\
        &       & GRU   & 4.721E-01 ± 7.465E-03 & 5.184E-03 ± 8.196E-05 &          & 4.796E-01 ± 7.584E-03 & 5.266E-03 ± 8.327E-05 \\
        &       & RBFNN & 2.004E-01 ± 3.168E-03 & 2.600E-03 ± 3.478E-05 &          & 3.574E-01 ± 2.939E-03 & 3.990E-03 ± 5.797E-05 \\
        &       & \textbf{ALPE}    & \textbf{2.315E-01 ± 3.661E-03} & \textbf{2.142E-03 ± 4.019E-05} &          & \textbf{2.340E-01 ± 3.699E-03} & \textbf{2.006E-03 ± 4.062E-05} \\
    \cline{2-5} \cline{6-8}
     & Simple  & MLP   & 3.302E-01 ± 2.058E-03 & 3.429E-03 ± 2.260E-05 & Exte     & 1.645E-01 ± 2.127E-03 & 2.477E-03 ± 2.335E-05 \\
     & GD      & CNN   & 4.533E-01 ± 7.167E-03 & 4.977E-03 ± 7.869E-05 &  GD      & 4.572E-01 ± 7.228E-03 & 5.020E-03 ± 7.937E-05 \\
     &         & LSTM  & 4.646E-01 ± 7.346E-03 & 5.101E-03 ± 8.066E-05 &          & 4.706E-01 ± 7.441E-03 & 5.167E-03 ± 8.170E-05 \\
     &         & GRU   & 4.732E-01 ± 7.482E-03 & 5.196E-03 ± 8.215E-05 &          & 4.807E-01 ± 7.600E-03 & 5.278E-03 ± 8.345E-05 \\
     &         & RBFNN & 3.309E-01 ± 1.432E-03 & 3.909E-03 ± 2.143E-05 &          & 3.464E-01 ± 5.279E-03 & 3.090E-03 ± 3.647E-05 \\
     &         & \textbf{ALPE}    & \textbf{2.293E-01 ± 4.055E-03} & \textbf{2.517E-03 ± 3.973E-05} &          & \textbf{1.335E-01 ± 3.692E-03} & \textbf{1.564E-03 ± 4.054E-05} \\
    \hline
    EXPE & Simple & Naive & 3.353E+00 ± 5.301E-02 & 3.521E-02 ± 5.568E-04 & Exte     & 5.636E+00 ± 2.567E-02 & 5.413E-02 ± 2.467E-04 \\
        &         & ARIMA & 4.292E-01 ± 6.786E-03 & 4.508E-03 ± 7.128E-05 &          & 7.054E-01 ± 2.466E-03 & 6.567E-03 ± 3.256E-05  \\
        &         & MLP   & 2.548E-01 ± 3.238E-03 & 2.951E-03 ± 3.401E-05 &          & 3.291E-01 ± 3.545E-03 & 3.355E-03 ± 3.724E-05 \\
        &         & CNN   & 4.338E-01 ± 6.859E-03 & 4.556E-03 ± 7.204E-05 &          & 4.402E-01 ± 6.960E-03 & 4.623E-03 ± 7.310E-05 \\
        &         & LSTM  & 4.478E-01 ± 7.080E-03 & 4.703E-03 ± 7.437E-05 &          & 4.522E-01 ± 7.150E-03 & 4.749E-03 ± 7.509E-05 \\
        &         & GRU   & 4.533E-01 ± 7.167E-03 & 4.761E-03 ± 7.528E-05 &          & 4.580E-01 ± 7.242E-03 & 4.810E-03 ± 7.606E-05 \\
        &         & RBFNN & 3.314E-01 ± 4.873E-03 & 3.884E-03 ± 4.003E-05 &          & 3.634E-01 ± 3.928E-03 & 3.993E-03 ± 4.838E-05 \\
        &         & \textbf{ALPE}    & \textbf{2.017E-01 ± 3.663E-03} & \textbf{2.003E-03 ± 3.847E-05} &          & \textbf{2.254E-01 ± 3.564E-03} & \textbf{2.367E-03 ± 3.743E-05} \\
    \cline{2-5} \cline{6-8} 
     & Simple   & MLP   & 4.152E-01 ± 3.403E-03 & 3.060E-03 ± 3.574E-05 & Exte     & 3.192E-01 ± 3.466E-03 & 4.302E-03 ± 3.640E-05 \\
     & MDI      & CNN   & 4.364E-01 ± 6.900E-03 & 4.583E-03 ± 7.247E-05 &  MDI     & 4.359E-01 ± 6.893E-03 & 4.579E-03 ± 7.240E-05 \\
        &       & LSTM  & 4.556E-01 ± 7.204E-03 & 4.785E-03 ± 7.566E-05 &          & 4.518E-01 ± 7.144E-03 & 4.745E-03 ± 7.503E-05 \\
        &       & GRU   & 4.520E-01 ± 7.146E-03 & 4.747E-03 ± 7.506E-05 &          & 4.588E-01 ± 7.254E-03 & 4.819E-03 ± 7.619E-05 \\
        &       & RBFNN & 2.831E-01 ± 4.476E-03 & 2.973E-03 ± 4.701E-05 &          & 3.635E-01 ± 4.403E-03 & 3.443E-03 ± 4.114E-05 \\
        &       & \textbf{ALPE}    & \textbf{2.315E-01 ± 3.660E-03} & \textbf{2.431E-03 ± 3.844E-05} &          & \textbf{2.341E-01 ± 3.701E-03} & \textbf{2.459E-03 ± 3.887E-05} \\
    \cline{2-5} \cline{6-8}
     & Simple  & MLP   & 3.154E-01 ± 3.405E-03 & 3.962E-03 ± 3.577E-05 & Exte     & 4.168E-01 ± 3.427E-03 & 3.277E-03 ± 3.600E-05 \\
     & GD      & CNN   & 4.344E-01 ± 6.868E-03 & 4.563E-03 ± 7.214E-05 &  GD      & 4.381E-01 ± 6.926E-03 & 4.601E-03 ± 7.275E-05 \\
     &         & LSTM  & 4.458E-01 ± 7.049E-03 & 4.682E-03 ± 7.404E-05 &          & 4.521E-01 ± 7.148E-03 & 4.748E-03 ± 7.508E-05 \\
     &         & GRU   & 4.542E-01 ± 7.182E-03 & 4.771E-03 ± 7.543E-05 &          & 4.600E-01 ± 7.273E-03 & 4.831E-03 ± 7.639E-05 \\
     &         & RBFNN & 3.432E-01 ± 2.013E-03 & 3.673E-03 ± 3.001E-05 &          & 3.637E-01 ± 2.978E-03 & 3.603E-03 ± 4.332E-05 \\
     &         & \textbf{ALPE}    & \textbf{2.281E-01 ± 3.607E-03} & \textbf{2.396E-03 ± 3.789E-05} &          & \textbf{2.348E-01 ± 3.712E-03} & \textbf{2.466E-03 ± 3.899E-05} \\

    \bottomrule
    \end{tabular}
    \end{small}}
            \label{tab:table_3}
        \end{minipage}
    \end{table}
\end{landscape}


\begin{landscape}
    \fancyhf{}  
    \fancyfoot[R]{\rotatebox{90}{\thepage}}  
    \thispagestyle{fancy}  

    \begin{table}[!hbtp]
        \centering
        \caption{RMSE and RRMSE scores for FDX, FITB, FLT, GD, GILD, GIS, GL, and GM.}
        \hspace*{-4cm} 
        \begin{minipage}[t]{0.48\textwidth}
            \centering
            \scalebox{0.42}{
    \begin{small}
    \begin{tabular}{c c r c c c c c}
    \toprule
    \textbf{Stock} & \textbf{Set} & \textbf{Model} & \textbf{RMSE} & \textbf{RRMSE} & \textbf{Set} & \textbf{RMSE} & \textbf{RRMSE} \\
    \hline
        FDX & Simple  & Naive & 1.374E+00 ± 2.173E-02 & 9.155E-03 ± 1.447E-04 & Exte & 3.456E+00 ± 5.303E-02 & 8.456E-03 ± 2.409E-04 \\
         &        & ARIMA & 4.089E-01 ± 6.465E-03 & 2.724E-03 ± 4.307E-05 &      & 5.145E-01 ± 4.356E-03 & 3.656E-03 ± 3.567E-05  \\
         &        & MLP   & 2.643E-01 ± 4.179E-03 & 1.761E-03 ± 2.784E-05 &      & 2.837E-01 ± 3.853E-03 & 2.623E-03 ± 2.567E-05 \\
         &        & CNN   & 4.154E-01 ± 6.569E-03 & 2.767E-03 ± 4.376E-05 &      & 4.218E-01 ± 6.669E-03 & 2.810E-03 ± 4.443E-05 \\
         &        & LSTM  & 4.288E-01 ± 6.780E-03 & 2.856E-03 ± 4.516E-05 &      & 4.331E-01 ± 6.848E-03 & 2.885E-03 ± 4.562E-05 \\
         &        & GRU   & 4.335E-01 ± 6.855E-03 & 2.888E-03 ± 4.566E-05 &      & 4.379E-01 ± 6.924E-03 & 2.917E-03 ± 4.612E-05 \\
         &        & RBFNN & 3.110E-01 ± 2.432E-03 & 2.224E-03 ± 3.517E-05 &      & 2.831E-01 ± 4.771E-03 & 1.883E-03 ± 3.192E-05 \\
         &        & \textbf{ALPE}    & \textbf{2.317E-01 ± 3.663E-03} & \textbf{1.543E-03 ± 2.440E-05} &      & \textbf{2.501E-01 ± 3.998E-03} & \textbf{1.046E-03 ± 2.476E-05} \\
    \cline{2-5} \cline{6-8} 
     & Simple     & MLP   & 2.696E-01 ± 4.263E-03 & 1.796E-03 ± 2.840E-05 & Exte & 2.996E-01 ± 3.836E-03 & 2.016E-03 ± 2.555E-05 \\
     &  MDI       & CNN   & 4.181E-01 ± 6.611E-03 & 2.785E-03 ± 4.404E-05 &  MDI & 4.174E-01 ± 6.600E-03 & 2.781E-03 ± 4.397E-05 \\
     &            & LSTM  & 4.182E-01 ± 6.612E-03 & 2.802E-03 ± 4.430E-05 &      & 4.325E-01 ± 6.839E-03 & 2.881E-03 ± 4.556E-05 \\
     &            & GRU   & 4.323E-01 ± 6.835E-03 & 2.879E-03 ± 4.553E-05 &      & 4.386E-01 ± 6.935E-03 & 2.922E-03 ± 4.620E-05 \\
     &            & RBFNN & 2.604E-01 ± 3.168E-03 & 1.335E-03 ± 2.110E-05 &      & 3.166E-01 ± 5.005E-03 & 2.121E-03 ± 3.353E-05 \\
     &            & \textbf{ALPE}    & \textbf{2.338E-01 ± 3.697E-03} & \textbf{1.158E-03 ± 2.463E-05} &      & \textbf{2.351E-01 ± 3.717E-03} & \textbf{1.566E-03 ± 2.476E-05} \\
    \cline{2-5} \cline{6-8}
     & Simple    & MLP   & 2.551E-01 ± 4.034E-03 & 1.699E-03 ± 2.687E-05 & Exte & 2.559E-01 ± 4.046E-03 & 1.705E-03 ± 2.695E-05 \\
     &   GD      & CNN   & 4.165E-01 ± 6.585E-03 & 2.774E-03 ± 4.386E-05 &   GD & 4.199E-01 ± 6.639E-03 & 2.797E-03 ± 4.423E-05 \\
     &           & LSTM  & 4.166E-01 ± 6.587E-03 & 2.791E-03 ± 4.413E-05 &      & 4.236E-01 ± 6.698E-03 & 2.838E-03 ± 4.487E-05 \\
     &           & GRU   & 4.343E-01 ± 6.867E-03 & 2.893E-03 ± 4.575E-05 &      & 4.397E-01 ± 6.953E-03 & 2.929E-03 ± 4.631E-05 \\
     &           & RBFNN & 3.762E-01 ± 3.920E-03 & 2.224E-03 ± 3.517E-05 &      & 3.202E-01 ± 4.898E-03 & 2.224E-03 ± 3.517E-05 \\
     &           & \textbf{ALPE}    & \textbf{2.292E-01 ± 3.623E-03} & \textbf{1.527E-03 ± 2.414E-05} &      & \textbf{2.348E-01 ± 3.712E-03} & \textbf{1.564E-03 ± 2.473E-05}\\
     \hline
     FITB & Simple& Naive & 2.909E-01 ± 4.599E-03 & 9.009E-02 ± 1.424E-04 & Exte     & 3.056E-01 ± 5.463E-03 & 8.765E-02 ± 5.554E-04 \\
        &        & ARIMA & 3.888E-01 ± 6.147E-03 & 1.204E-02 ± 1.904E-04 &          & 4.456E-01 ± 6.147E-03 & 2.545E-02 ± 4.564E-04  \\
        &        & MLP   & 3.674E-01 ± 2.647E-03 & 5.185E-03 ± 8.198E-05 &          & 3.572E-01 ± 2.486E-03 & 4.869E-02 ± 7.699E-05 \\
        &        & CNN   & 3.952E-01 ± 6.249E-03 & 1.224E-02 ± 1.935E-04 &          & 4.012E-01 ± 6.344E-03 & 1.242E-02 ± 1.965E-04 \\
        &        & LSTM  & 4.079E-01 ± 6.449E-03 & 1.263E-02 ± 1.997E-04 &          & 4.120E-01 ± 6.515E-03 & 1.276E-02 ± 2.018E-04 \\
        &        & GRU   & 4.124E-01 ± 6.520E-03 & 1.277E-02 ± 2.019E-04 &          & 4.165E-01 ± 6.585E-03 & 1.290E-02 ± 2.039E-04 \\
        &        & RBFNN & 3.777E-01 ± 4.039E-03 & 1.146E-02 ± 2.566E-04 &          & 3.536E-01 ± 3.034E-03 & 1.111E-02 ± 1.536E-04 \\
        &        & \textbf{ALPE}    & \textbf{2.318E-01 ± 3.665E-03} & \textbf{7.178E-03 ± 1.135E-04} &          & \textbf{2.018E-01 ± 3.191E-03} & \textbf{6.251E-03 ± 9.883E-05} \\
    \cline{2-5} \cline{6-8} 
     & Simple   & MLP   & 4.720E-01 ± 2.720E-03 & 5.327E-02 ± 8.423E-05 & Exte     & 3.529E-01 ± 2.418E-03 & 1.736E-02 ± 7.488E-05 \\
     & MDI      & CNN   & 3.977E-01 ± 6.288E-03 & 1.232E-02 ± 1.947E-04 &  MDI     & 3.971E-01 ± 6.279E-03 & 1.230E-02 ± 1.943E-04 \\
        &       & LSTM  & 4.147E-01 ± 6.557E-03 & 1.284E-02 ± 2.031E-04 &          & 4.114E-01 ± 6.505E-03 & 1.274E-02 ± 2.015E-04 \\
        &       & GRU   & 4.112E-01 ± 6.502E-03 & 1.274E-02 ± 2.014E-04 &          & 4.171E-01 ± 6.595E-03 & 1.292E-02 ± 2.043E-04 \\
        &       & RBFNN & 2.884E-01 ± 3.168E-03 & 7.205E-03 ± 9.811E-05 &          & 3.849E-01 ± 3.930E-03 & 1.231E-02 ± 1.543E-04 \\
        &       & \textbf{ALPE}    & \textbf{2.232E-01 ± 3.688E-03} & \textbf{5.223E-03 ± 1.142E-04} &          & \textbf{2.116E-01 ± 3.346E-03} & \textbf{6.553E-03 ± 1.033E-04} \\
    \cline{2-5} \cline{6-8}
     & Simple  & MLP   & 3.554E-01 ± 2.458E-03 & 2.814E-02 ± 7.612E-05 & Exte     & 2.550E-01 ± 2.450E-03 & 1.800E-01 ± 7.589E-05 \\
     & GD      & CNN   & 3.962E-01 ± 6.264E-03 & 1.227E-02 ± 1.940E-04 &  GD      & 3.995E-01 ± 6.317E-03 & 1.237E-02 ± 1.956E-04 \\
     &         & LSTM  & 4.063E-01 ± 6.424E-03 & 1.258E-02 ± 1.989E-04 &          & 4.118E-01 ± 6.511E-03 & 1.275E-02 ± 2.016E-04 \\
     &         & GRU   & 4.131E-01 ± 6.532E-03 & 1.279E-02 ± 2.023E-04 &          & 4.182E-01 ± 6.612E-03 & 1.295E-02 ± 2.048E-04 \\
     &         & RBFNN & 3.430E-01 ± 4.023E-03 & 1.320E-02 ± 1.554E-04 &          & 3.637E-01 ± 3.839E-03 & 1.322E-02 ± 2.004E-04 \\
     &         & \textbf{ALPE}    & \textbf{2.310E-01 ± 3.652E-03} & \textbf{7.152E-03 ± 1.131E-04} &          & \textbf{2.083E-01 ± 3.294E-03} & \textbf{6.451E-03 ± 1.020E-04} \\
     \hline
    FLT & Simple & Naive & 3.742E+00 ± 5.917E-02 & 2.101E-02 ± 3.321E-04 & Exte     & 4.932E+00 ± 4.325E-02 & 3.432E-02 ± 2.343E-04 \\
        &        & ARIMA & 4.373E-01 ± 6.914E-03 & 2.454E-03 ± 3.881E-05 &          & 5.343E-01 ± 2.343E-03 & 3.345E-03 ± 4.345E-05  \\
        &        & MLP   & 1.960E-01 ± 3.099E-03 & 1.500E-03 ± 1.739E-05 &          & 3.984E-01 ± 3.137E-03 & 2.113E-03 ± 1.760E-05 \\
        &        & CNN   & 4.453E-01 ± 6.129E-03 & 2.496E-03 ± 3.947E-05 &          & 4.512E-01 ± 7.134E-03 & 2.532E-03 ± 4.004E-05 \\
        &        & LSTM  & 4.628E-01 ± 7.318E-03 & 2.598E-03 ± 4.107E-05 &          & 4.589E-01 ± 7.256E-03 & 2.576E-03 ± 4.073E-05 \\
        &        & GRU   & 4.584E-01 ± 7.248E-03 & 2.573E-03 ± 4.033E-05 &          & 4.618E-01 ± 7.302E-03 & 2.592E-03 ± 4.098E-05 \\
        &        & RBFNN & 3.449E-01 ± 3.020E-03 & 1.563E-03 ± 2.738E-05 &          & 3.092E-01 ± 4.282E-03 & 1.839E-03 ± 2.667E-05 \\
        &        & \textbf{ALPE}    & \textbf{1.903E-01 ± 3.009E-03} & \textbf{1.068E-03 ± 1.689E-05} &          & \textbf{2.205E-01 ± 3.486E-03} & \textbf{1.237E-03 ± 1.956E-05} \\
    \cline{2-5} \cline{6-8} 
     & Simple   & MLP   & 3.752E-01 ± 4.223E-03 & 1.873E-03 ± 1.447E-05 & Exte     & 3.653E-01 ± 4.192E-03 & 2.111E-03 ± 1.757E-05 \\
     & MDI      & CNN   & 4.468E-01 ± 7.065E-03 & 2.508E-03 ± 3.965E-05 &  MDI     & 4.525E-01 ± 7.155E-03 & 2.540E-03 ± 4.016E-05 \\
        &       & LSTM  & 4.682E-01 ± 7.403E-03 & 2.628E-03 ± 4.155E-05 &          & 4.614E-01 ± 7.295E-03 & 2.590E-03 ± 4.094E-05 \\
        &       & GRU   & 4.570E-01 ± 7.225E-03 & 2.565E-03 ± 4.055E-05 &          & 4.646E-01 ± 7.346E-03 & 2.608E-03 ± 4.123E-05 \\
        &       & RBFNN & 2.704E-01 ± 3.168E-03 & 1.925E-03 ± 1.778E-05 &          & 3.930E-01 ± 2.838E-03 & 2.432E-03 ± 2.894E-05 \\
        &       & \textbf{ALPE}    & \textbf{2.081E-01 ± 4.082E-03} & \textbf{1.449E-03 ± 2.291E-05} &          & \textbf{2.494E-01 ± 3.944E-03} & \textbf{1.400E-03 ± 2.214E-05} \\
    \cline{2-5} \cline{6-8}
     & Simple  & MLP   & 2.931E-01 ± 4.632E-03 & 2.083E-03 ± 1.713E-05 & Exte     & 3.168E-01 ± 3.428E-03 & 2.217E-03 ± 1.924E-05 \\
     & GD      & CNN   & 4.474E-01 ± 7.075E-03 & 2.511E-03 ± 3.971E-05 &  GD      & 4.489E-01 ± 7.098E-03 & 2.520E-03 ± 3.984E-05 \\
     &         & LSTM  & 4.537E-01 ± 7.174E-03 & 2.546E-03 ± 4.026E-05 &          & 4.686E-01 ± 7.409E-03 & 2.630E-03 ± 4.159E-05 \\
     &         & GRU   & 4.597E-01 ± 7.269E-03 & 2.580E-03 ± 4.080E-05 &          & 4.634E-01 ± 7.327E-03 & 2.601E-03 ± 4.112E-05 \\
     &         & RBFNN & 3.534E-01 ± 3.020E-03 & 1.738E-03 ± 3.453E-05 &          & 3.839E-01 ± 3.930E-03 & 1.993E-03 ± 4.893E-05 \\
     &         & \textbf{ALPE}    & \textbf{1.788E-01 ± 2.827E-03} & \textbf{1.003E-03 ± 1.586E-05} &          & \textbf{2.192E-01 ± 3.466E-03} & \textbf{1.229E-03 ± 1.746E-05} \\
     \hline
    GD & Simple & Naive  & 2.383E+00 ± 2.187E-02 & 6.440E-03 ± 1.018E-04 & Exte     & 2.234E+00 ± 5.456E-02 & 5.643E-03 ± 2.453E-04 \\
        &        & ARIMA & 4.229E-01 ± 6.686E-03 & 1.969E-03 ± 3.113E-05 &          & 5.401E-01 ± 4.055E-03 & 3.455E-03 ± 6.234E-05 \\
        &        & MLP   & 3.301E-01 ± 3.637E-03 & 1.971E-03 ± 1.694E-05 &          & 3.195E-01 ± 3.471E-03 & 2.022E-03 ± 1.616E-05 \\
        &        & CNN   & 4.311E-01 ± 6.816E-03 & 2.007E-03 ± 3.174E-05 &          & 4.372E-01 ± 6.913E-03 & 2.036E-03 ± 3.219E-05 \\
        &        & LSTM  & 4.484E-01 ± 7.090E-03 & 2.088E-03 ± 3.301E-05 &          & 4.450E-01 ± 7.037E-03 & 2.072E-03 ± 3.276E-05 \\
        &        & GRU   & 4.440E-01 ± 7.021E-03 & 2.067E-03 ± 3.269E-05 &          & 4.471E-01 ± 7.070E-03 & 2.082E-03 ± 3.292E-05 \\
        &        & RBFNN & 3.340E-01 ± 3.663E-03 & 1.609E-03 ± 2.553E-05 &          & 3.588E-01 ± 3.750E-03 & 1.700E-03 ± 2.101E-05 \\
        &        & \textbf{ALPE}    & \textbf{1.432E-01 ± 3.846E-03} & \textbf{1.133E-03 ± 1.791E-05} &          & \textbf{2.510E-01 ± 3.968E-03} & \textbf{1.168E-03 ± 1.848E-05} \\
    \cline{2-5} \cline{6-8} 
     & Simple   & MLP   & 3.101E-01 ± 3.323E-03 & 2.018E-03 ± 1.547E-05 & Exte     & 3.071E-01 ± 3.274E-03 & 2.643E-03 ± 1.525E-05 \\
     & MDI      & CNN   & 4.330E-01 ± 6.846E-03 & 2.016E-03 ± 3.187E-05 &  MDI     & 4.387E-01 ± 6.936E-03 & 2.043E-03 ± 3.230E-05 \\
        &       & LSTM  & 4.537E-01 ± 7.173E-03 & 2.112E-03 ± 3.340E-05 &          & 4.470E-01 ± 7.068E-03 & 2.082E-03 ± 3.291E-05 \\
        &       & GRU   & 4.427E-01 ± 6.999E-03 & 2.061E-03 ± 3.259E-05 &          & 4.433E-01 ± 5.204E-03 & 2.094E-03 ± 3.311E-05 \\
        &       & RBFNN & 2.904E-01 ± 3.168E-03 & 1.629E-03 ± 1.475E-05 &          & 3.463E-01 ± 2.887E-03 & 1.943E-03 ± 2.563E-05 \\
        &       & \textbf{ALPE}    & \textbf{2.164E-01 ± 3.579E-03} & \textbf{1.054E-03 ± 1.666E-05} &          & \textbf{2.440E-01 ± 3.859E-03} & \textbf{1.136E-03 ± 1.797E-05} \\
    \cline{2-5} \cline{6-8}
     & Simple  & MLP   & 3.065E-01 ± 3.265E-03 & 2.615E-03 ± 1.520E-05 & Exte     & 2.926E-01 ± 3.361E-03 & 2.598E-03 ± 1.565E-05 \\
     & GD      & CNN   & 4.336E-01 ± 6.856E-03 & 2.019E-03 ± 3.192E-05 &  GD      & 4.351E-01 ± 6.879E-03 & 2.026E-03 ± 3.203E-05 \\
     &         & LSTM  & 4.397E-01 ± 6.953E-03 & 2.047E-03 ± 3.237E-05 &          & 4.540E-01 ± 7.178E-03 & 2.114E-03 ± 3.342E-05 \\
     &         & GRU   & 4.452E-01 ± 7.040E-03 & 2.073E-03 ± 3.278E-05 &          & 4.486E-01 ± 7.093E-03 & 2.089E-03 ± 3.303E-05 \\
     &         & RBFNN & 3.603E-01 ± 3.112E-03 & 1.738E-03 ± 3.093E-05 &          & 3.453E-01 ± 4.453E-03 & 1.903E-03 ± 2.563E-05 \\
     &         & \textbf{ALPE}    & \textbf{2.189E-01 ± 3.461E-03} & \textbf{1.019E-03 ± 1.612E-05} &          & \textbf{2.099E-01 ± 3.319E-03} & \textbf{9.773E-04 ± 1.545E-05} \\
    \bottomrule
    \end{tabular}
    \end{small}}
            \label{tab:table1}
        \end{minipage}%
        \hspace{0.25\textwidth} 
        \begin{minipage}[t]{0.48\textwidth}
            \centering
            \scalebox{0.42}{
    \begin{small}
    \begin{tabular}{c c r c c c c c}
    \toprule
    \textbf{Stock} & \textbf{Set} & \textbf{Model} & \textbf{RMSE} & \textbf{RRMSE} & \textbf{Set} & \textbf{RMSE} & \textbf{RRMSE} \\
    \hline
        GILD & Simple& Naive & 3.567E-01 ± 2.478E-03 & 8.519E-03 ± 3.983E-05 & Exte     & 6.062E-01 ± 4.405E-03 & 7.253E-03 ± 2.459E-05 \\
        &        & ARIMA & 4.106E-01 ± 6.492E-03 & 6.600E-03 ± 1.044E-04 &          & 5.513E-01 ± 5.645E-03 & 5.541E-03 ± 2.323E-04  \\
        &        & MLP   & 3.259E-01 ± 5.153E-03 & 5.238E-03 ± 8.283E-05 &          & 5.122E-01 ± 8.099E-03 & 8.233E-03 ± 1.302E-04 \\
        &        & CNN   & 5.244E-01 ± 8.291E-03 & 8.429E-03 ± 1.333E-04 &          & 5.384E-01 ± 8.513E-03 & 8.654E-03 ± 1.368E-04 \\
        &        & LSTM  & 4.349E-01 ± 6.876E-03 & 6.990E-03 ± 1.105E-04 &          & 4.316E-01 ± 6.825E-03 & 6.938E-03 ± 1.097E-04 \\
        &        & GRU   & 4.307E-01 ± 6.809E-03 & 6.922E-03 ± 1.094E-04 &          & 4.337E-01 ± 6.857E-03 & 6.971E-03 ± 1.102E-04 \\
        &        & RBFNN & 3.302E-01 ± 3.093E-03 & 5.367E-03 ± 8.486E-05 &          & 3.938E-01 ± 3.028E-03 & 5.897E-03 ± 8.948E-05 \\
        &        & \textbf{ALPE}    & \textbf{2.491E-01 ± 3.938E-03} & \textbf{4.004E-03 ± 6.330E-05} &          & \textbf{2.307E-01 ± 3.647E-03} & \textbf{3.707E-03 ± 5.862E-05} \\
    \cline{2-5} \cline{6-8} 
     & Simple   & MLP   & 3.481E-01 ± 5.504E-03 & 5.595E-03 ± 8.847E-05 & Exte     & 5.525E-01 ± 8.736E-03 & 8.881E-03 ± 1.404E-04 \\
     & MDI      & CNN   & 5.258E-01 ± 8.314E-03 & 8.452E-03 ± 1.336E-04 &  MDI     & 5.395E-01 ± 8.531E-03 & 8.672E-03 ± 1.371E-04 \\
        &       & LSTM  & 4.399E-01 ± 6.955E-03 & 7.070E-03 ± 1.118E-04 &          & 4.335E-01 ± 6.855E-03 & 6.968E-03 ± 1.102E-04 \\
        &       & GRU   & 4.294E-01 ± 6.789E-03 & 6.901E-03 ± 1.091E-04 &          & 4.361E-01 ± 6.896E-03 & 7.010E-03 ± 1.108E-04 \\
        &       & RBFNN & 2.574E-01 ± 3.168E-03 & 4.220E-03 ± 5.092E-05 &          & 3.673E-01 ± 4.984E-03 & 5.367E-03 ± 7.490E-05 \\
        &       & \textbf{ALPE}    & \textbf{2.534E-01 ± 4.006E-03} & \textbf{4.073E-03 ± 6.439E-05} &          & \textbf{2.447E-01 ± 3.870E-03} & \textbf{3.934E-03 ± 6.220E-05} \\
    \cline{2-5} \cline{6-8}
     & Simple  & MLP   & 4.591E-01 ± 7.259E-03 & 7.379E-03 ± 1.167E-04 & Exte     & 4.841E-01 ± 7.654E-03 & 7.781E-03 ± 1.230E-04 \\
     & GD      & CNN   & 5.263E-01 ± 8.321E-03 & 8.459E-03 ± 1.337E-04 &  GD      & 5.274E-01 ± 8.339E-03 & 8.477E-03 ± 1.340E-04 \\
     &         & LSTM  & 4.266E-01 ± 6.745E-03 & 6.856E-03 ± 1.084E-04 &          & 4.401E-01 ± 6.959E-03 & 7.074E-03 ± 1.119E-04 \\
     &         & GRU   & 4.318E-01 ± 6.827E-03 & 6.940E-03 ± 1.097E-04 &          & 4.350E-01 ± 6.879E-03 & 6.992E-03 ± 1.106E-04 \\
     &         & RBFNN & 3.748E-01 ± 3.839E-03 & 6.093E-03 ± 6.983E-05 &          & 3.599E-01 ± 4.528E-03 & 5.003E-03 ± 7.001E-05 \\
     &         & \textbf{ALPE}    & \textbf{2.324E-01 ± 3.674E-03} & \textbf{3.735E-03 ± 5.906E-05} &          & \textbf{2.249E-01 ± 3.557E-03} & \textbf{3.616E-03 ± 5.717E-05}\\
     \hline
     GIS & Simple & Naive & 3.426E-01 ± 5.417E-03 & 6.446E-03 ± 7.030E-05 & Exte     & 4.568E-01 ± 7.353E-03 & 7.456E-03 ± 6.575E-05 \\
        &        & ARIMA & 3.960E-01 ± 6.262E-03 & 5.140E-03 ± 8.128E-05 &          & 2.431E-01 ± 3.121E-03 & 6.454E-03 ± 5.676E-05 \\
        &        & MLP   & 3.617E-01 ± 4.139E-03 & 4.397E-03 ± 5.372E-05 &          & 2.539E-01 ± 4.014E-03 & 3.295E-03 ± 5.210E-05 \\
        &        & CNN   & 4.159E-01 ± 6.576E-03 & 5.398E-03 ± 8.535E-05 &          & 4.188E-01 ± 6.622E-03 & 5.436E-03 ± 8.594E-05 \\
        &        & LSTM  & 4.200E-01 ± 6.641E-03 & 5.452E-03 ± 8.620E-05 &          & 4.241E-01 ± 5.664E-03 & 4.009E-03 ± 6.040E-05 \\
        &        & GRU   & 4.159E-01 ± 6.576E-03 & 5.398E-03 ± 8.535E-05 &          & 4.188E-01 ± 6.622E-03 & 5.436E-03 ± 8.594E-05 \\
        &        & RBFNN & 3.440E-01 ± 4.930E-03 & 4.473E-03 ± 5.994E-05 &          & 3.894E-01 ± 5.004E-03 & 4.355E-03 ± 5.702E-05 \\
        &        & \textbf{ALPE}    & \textbf{2.594E-01 ± 4.101E-03} & \textbf{3.367E-03 ± 5.323E-05} &          & \textbf{2.032E-01 ± 3.687E-03} & \textbf{3.027E-03 ± 4.786E-05} \\
    \cline{2-5} \cline{6-8} 
     & Simple   & MLP   & 2.828E-01 ± 4.472E-03 & 3.671E-03 ± 5.804E-05 & Exte     & 2.800E-01 ± 4.428E-03 & 3.635E-03 ± 5.747E-05 \\
     & MDI      & CNN   & 4.146E-01 ± 6.556E-03 & 5.382E-03 ± 8.509E-05 &  MDI     & 4.212E-01 ± 6.659E-03 & 5.467E-03 ± 8.643E-05 \\
        &       & LSTM  & 4.248E-01 ± 6.717E-03 & 5.514E-03 ± 8.719E-05 &          & 4.188E-01 ± 6.621E-03 & 5.435E-03 ± 8.594E-05 \\
        &       & GRU   & 4.146E-01 ± 6.556E-03 & 5.382E-03 ± 8.509E-05 &          & 4.212E-01 ± 6.659E-03 & 5.467E-03 ± 8.643E-05 \\
        &       & RBFNN & 2.704E-01 ± 3.168E-03 & 3.601E-03 ± 4.112E-05 &          & 3.648E-01 ± 3.009E-03 & 4.393E-03 ± 6.852E-05 \\
        &       & \textbf{ALPE}    & \textbf{2.352E-01 ± 3.877E-03} & \textbf{2.182E-03 ± 5.032E-05} &          & \textbf{2.456E-01 ± 3.883E-03} & \textbf{3.188E-03 ± 5.040E-05} \\
    \cline{2-5} \cline{6-8}
     & Simple  & MLP   & 3.442E-01 ± 3.861E-03 & 4.350E-03 ± 5.011E-05 & Exte     & 2.862E-01 ± 4.525E-03 & 3.715E-03 ± 5.874E-05 \\
     & GD      & CNN   & 4.200E-01 ± 5.404E-03 & 4.312E-03 ± 7.302E-05 &  GD      & 4.071E-01 ± 5.321E-03 & 3.202E-03 ± 7.001E-05 \\
     &         & LSTM  & 4.121E-01 ± 6.515E-03 & 5.348E-03 ± 8.457E-05 &          & 4.251E-01 ± 6.721E-03 & 5.517E-03 ± 8.723E-05 \\
     &         & GRU   & 4.170E-01 ± 6.593E-03 & 5.412E-03 ± 8.557E-05 &          & 4.201E-01 ± 3.232E-03 & 3.003E-03 ± 5.009E-05 \\
     &         & RBFNN & 3.883E-01 ± 3.921E-03 & 4.334E-03 ± 6.852E-05 &          & 3.444E-01 ± 3.942E-03 & 4.334E-03 ± 4.610E-05 \\
     &         & \textbf{ALPE}    & \textbf{2.469E-01 ± 3.904E-03} & \textbf{3.205E-03 ± 5.067E-05} &          & \textbf{2.249E-01 ± 3.557E-03} & \textbf{2.920E-03 ± 4.616E-05} \\
     \hline
    GL & Simple  & Naive & 9.076E-01 ± 1.435E-02 & 9.006E-03 ± 1.424E-04 & Exte     & 8.765E-01 ± 3.245E-02 & 8.577E-03 ± 2.341E-04 \\
        &        & ARIMA & 4.880E-01 ± 7.716E-03 & 4.842E-03 ± 7.656E-05 &          & 3.691E-01 ± 4.405E-03 & 3.540E-03 ± 6.434E-05 \\
        &        & MLP   & 7.555E-01 ± 1.195E-02 & 7.497E-03 ± 1.185E-04 &          & 7.949E-01 ± 1.257E-02 & 7.888E-03 ± 1.247E-04 \\
        &        & CNN   & 3.930E-01 ± 6.214E-03 & 3.900E-03 ± 6.167E-05 &          & 3.982E-01 ± 6.296E-03 & 3.952E-03 ± 6.248E-05 \\
        &        & LSTM  & 4.084E-01 ± 6.457E-03 & 4.052E-03 ± 6.407E-05 &          & 4.055E-01 ± 6.412E-03 & 4.024E-03 ± 6.363E-05 \\
        &        & GRU   & 4.044E-01 ± 6.394E-03 & 4.013E-03 ± 6.345E-05 &          & 4.072E-01 ± 6.438E-03 & 4.040E-03 ± 6.388E-05 \\
        &        & RBFNN & 3.910E-01 ± 3.023E-03 & 3.313E-03 ± 5.239E-05 &          & 3.003E-01 ± 2.920E-03 & 3.455E-03 ± 3.303E-05 \\
        &        & \textbf{ALPE}    & \textbf{2.375E-01 ± 3.755E-03} & \textbf{2.357E-03 ± 3.727E-05} &          & \textbf{2.421E-01 ± 3.828E-03} & \textbf{2.403E-03 ± 3.799E-05} \\
    \cline{2-5} \cline{6-8} 
     & Simple   & MLP   & 8.261E-01 ± 1.306E-02 & 8.197E-03 ± 1.296E-04 & Exte     & 7.787E-01 ± 1.231E-02 & 7.727E-03 ± 1.222E-04 \\
     & MDI      & CNN   & 3.945E-01 ± 6.237E-03 & 3.914E-03 ± 6.189E-05 &  MDI     & 3.995E-01 ± 6.316E-03 & 3.964E-03 ± 6.268E-05 \\
        &       & LSTM  & 4.129E-01 ± 6.529E-03 & 4.098E-03 ± 6.479E-05 &          & 4.073E-01 ± 6.440E-03 & 4.042E-03 ± 6.390E-05 \\
        &       & GRU   & 4.054E-01 ± 6.410E-03 & 4.023E-03 ± 6.361E-05 &          & 4.094E-01 ± 6.473E-03 & 4.063E-03 ± 6.424E-05 \\
        &       & RBFNN & 2.994E-01 ± 3.168E-03 & 2.988E-03 ± 3.144E-05 &          & 3.736E-01 ± 2.004E-03 & 3.837E-03 ± 2.028E-05 \\
        &       & \textbf{ALPE}    & \textbf{2.299E-01 ± 3.636E-03} & \textbf{1.282E-03 ± 3.608E-05} &          & \textbf{2.278E-01 ± 3.602E-03} & \textbf{2.261E-03 ± 3.575E-05} \\
    \cline{2-5} \cline{6-8}
     & Simple  & MLP   & 9.233E-01 ± 1.460E-02 & 9.162E-03 ± 1.449E-04 & Exte     & 8.762E-01 ± 1.385E-02 & 8.694E-03 ± 1.375E-04 \\
     & GD      & CNN   & 3.950E-01 ± 6.245E-03 & 3.919E-03 ± 6.197E-05 &  GD      & 3.963E-01 ± 6.266E-03 & 3.933E-03 ± 6.218E-05 \\
     &         & LSTM  & 4.009E-01 ± 6.338E-03 & 3.978E-03 ± 6.290E-05 &          & 4.133E-01 ± 6.535E-03 & 4.102E-03 ± 6.485E-05 \\
     &         & GRU   & 4.032E-01 ± 6.375E-03 & 4.001E-03 ± 6.326E-05 &          & 4.085E-01 ± 6.460E-03 & 4.054E-03 ± 6.410E-05 \\
     &         & RBFNN & 3.224E-01 ± 2.936E-03 & 3.210E-03 ± 3.992E-05 &          & 3.928E-01 ± 3.832E-03 & 3.645E-03 ± 3.837E-05 \\
     &         & \textbf{ALPE}    & \textbf{2.379E-01 ± 3.761E-03} & \textbf{2.360E-03 ± 3.732E-05} &          & \textbf{2.503E-01 ± 3.958E-03} & \textbf{2.484E-03 ± 3.928E-05} \\
     \hline
    GM & Simple  & Naive & 5.358E-01 ± 8.471E-03 & 1.420E-02 ± 2.245E-04 & Exte     & 4.563E-01 ± 5.352E-03 & 2.456E-02 ± 5.345E-04 \\
        &        & ARIMA & 3.899E-01 ± 6.165E-03 & 1.033E-02 ± 1.634E-04 &          & 4.350E-01 ± 5.523E-03 & 2.160E-02 ± 2.303E-04 \\
        &        & MLP   & 5.135E-01 ± 8.120E-03 & 1.361E-02 ± 2.152E-04 &          & 5.466E-01 ± 8.643E-03 & 1.449E-02 ± 2.291E-04 \\
        &        & CNN   & 3.168E-01 ± 5.009E-03 & 8.396E-03 ± 1.328E-04 &          & 3.196E-01 ± 5.053E-03 & 8.471E-03 ± 1.339E-04 \\
        &        & LSTM  & 3.301E-01 ± 5.219E-03 & 8.748E-03 ± 1.383E-04 &          & 3.234E-01 ± 5.114E-03 & 8.573E-03 ± 1.355E-04 \\
        &        & GRU   & 3.276E-01 ± 5.180E-03 & 8.683E-03 ± 1.373E-04 &          & 3.302E-01 ± 5.221E-03 & 8.751E-03 ± 1.384E-04 \\
        &        & RBFNN & 3.309E-01 ± 3.399E-03 & 9.266E-03 ± 1.235E-04 &          & 2.990E-01 ± 4.754E-03 & 5.230E-03 ± 7.009E-05 \\
        &        & \textbf{ALPE}    & \textbf{2.498E-01 ± 3.950E-03} & \textbf{6.621E-03 ± 1.047E-04} &          & \textbf{1.859E-01 ± 3.888E-03} & \textbf{4.517E-03 ± 1.030E-04} \\
    \cline{2-5} \cline{6-8} 
     & Simple   & MLP   & 3.916E-01 ± 6.192E-03 & 1.038E-02 ± 1.641E-04 & Exte     & 4.586E-01 ± 7.251E-03 & 1.215E-02 ± 1.922E-04 \\
     & MDI      & CNN   & 3.172E-01 ± 5.015E-03 & 8.407E-03 ± 1.329E-04 &  MDI     & 3.230E-01 ± 5.108E-03 & 8.562E-03 ± 1.354E-04 \\
        &       & LSTM  & 3.319E-01 ± 5.248E-03 & 8.798E-03 ± 1.391E-04 &          & 3.271E-01 ± 5.172E-03 & 8.670E-03 ± 1.371E-04 \\
        &       & GRU   & 3.264E-01 ± 5.161E-03 & 8.651E-03 ± 1.368E-04 &          & 3.321E-01 ± 5.251E-03 & 8.802E-03 ± 1.392E-04 \\
        &       & RBFNN & 2.004E-01 ± 3.169E-03 & 6.312E-03 ± 8.400E-05 &          & 2.901E-01 ± 3.204E-03 & 6.546E-03 ± 4.510E-05 \\
        &       & \textbf{ALPE}    & \textbf{2.397E-01 ± 3.790E-03} & \textbf{5.352E-03 ± 1.004E-04} &          & \textbf{1.896E-01 ± 3.946E-03} & \textbf{5.615E-03 ± 1.046E-04} \\
    \cline{2-5} \cline{6-8}
     & Simple  & MLP   & 4.428E-01 ± 7.001E-03 & 1.174E-02 ± 1.856E-04 & Exte     & 4.389E-01 ± 6.939E-03 & 1.163E-02 ± 1.839E-04 \\
     & GD      & CNN   & 3.184E-01 ± 5.034E-03 & 8.439E-03 ± 1.334E-04 &  GD      & 3.192E-01 ± 5.047E-03 & 8.460E-03 ± 1.338E-04 \\
     &         & LSTM  & 3.209E-01 ± 5.075E-03 & 8.506E-03 ± 1.345E-04 &          & 3.351E-01 ± 5.299E-03 & 8.882E-03 ± 1.404E-04 \\
     &         & GRU   & 3.282E-01 ± 5.189E-03 & 8.698E-03 ± 1.375E-04 &          & 3.300E-01 ± 5.218E-03 & 8.747E-03 ± 1.383E-04 \\
     &         & RBFNN & 3.556E-01 ± 4.112E-03 & 8.773E-03 ± 1.007E-04 &          & 3.500E-01 ± 2.330E-03 & 8.040E-03 ± 1.504E-04 \\
     &         & \textbf{ALPE}    & \textbf{2.427E-01 ± 3.838E-03} & \textbf{6.433E-03 ± 1.017E-04} &          & \textbf{2.322E-01 ± 3.671E-03} & \textbf{6.153E-03 ± 9.729E-05} \\

    \bottomrule
    \end{tabular}
    \end{small}}
            \label{tab:table_4}
        \end{minipage}
    \end{table}
\end{landscape}


\begin{landscape}
    \fancyhf{}  
    \fancyfoot[R]{\rotatebox{90}{\thepage}}  
    \thispagestyle{fancy}  

    \begin{table}[!hbtp]
        \centering
        \caption{RMSE and RRMSE scores for GOOGL, HAS, HD, HLT, ILMN, INCY, IPG, and IQV.}
        \hspace*{-5cm} 
        \begin{minipage}[t]{0.48\textwidth}
            \centering
            \scalebox{0.42}{
    \begin{small}
    \begin{tabular}{c c r c c c c c}
    \toprule
    \textbf{Stock} & \textbf{Set} & \textbf{Model} & \textbf{RMSE} & \textbf{RRMSE} & \textbf{Set} & \textbf{RMSE} & \textbf{RRMSE} \\
    \hline     
    GOOGL& Simple& Naive & 3.639E-01 ± 5.754E-03 & 5.785E-03 ± 5.984E-05 & Exte     & 5.567E-01 ± 4.567E-03 & 4.673E-03 ± 3.644E-05 \\
        &        & ARIMA & 3.679E-01 ± 5.817E-03 & 3.826E-03 ± 6.049E-05 &          & 5.651E-01 ± 4.523E-03 & 4.450E-03 ± 7.646E-05  \\
        &        & MLP   & 2.103E-01 ± 3.325E-03 & 3.187E-03 ± 3.457E-05 &          & 2.156E-01 ± 3.409E-03 & 2.242E-03 ± 3.546E-05 \\
        &        & CNN   & 3.755E-01 ± 5.937E-03 & 3.905E-03 ± 6.174E-05 &          & 3.806E-01 ± 6.017E-03 & 3.958E-03 ± 6.257E-05 \\
        &        & LSTM  & 3.897E-01 ± 6.162E-03 & 4.053E-03 ± 6.408E-05 &          & 3.872E-01 ± 6.122E-03 & 4.027E-03 ± 6.367E-05 \\
        &        & GRU   & 3.860E-01 ± 6.104E-03 & 4.015E-03 ± 6.348E-05 &          & 3.886E-01 ± 6.145E-03 & 4.042E-03 ± 6.390E-05 \\
        &        & RBFNN & 3.534E-01 ± 3.929E-03 & 3.472E-03 ± 5.490E-05 &          & 2.873E-01 ± 4.673E-03 & 2.923E-03 ± 3.198E-05 \\
        &        & \textbf{ALPE}    & \textbf{2.418E-01 ± 3.823E-03} & \textbf{2.514E-03 ± 3.975E-05} &          & \textbf{1.417E-01 ± 3.822E-03} & \textbf{2.014E-03 ± 3.975E-05} \\
    \cline{2-5} \cline{6-8} 
     & Simple   & MLP   & 3.133E-01 ± 3.372E-03 & 3.218E-03 ± 3.399E-05 & Exte     & 3.093E-01 ± 3.310E-03 & 2.177E-03 ± 3.442E-05 \\
     & MDI      & CNN   & 3.768E-01 ± 5.958E-03 & 3.919E-03 ± 6.196E-05 &  MDI     & 3.816E-01 ± 6.034E-03 & 3.969E-03 ± 6.275E-05 \\
        &       & LSTM  & 3.940E-01 ± 6.229E-03 & 4.097E-03 ± 6.478E-05 &          & 3.944E-01 ± 6.237E-03 & 4.102E-03 ± 6.486E-05 \\
        &       & GRU   & 3.850E-01 ± 6.087E-03 & 4.004E-03 ± 6.330E-05 &          & 3.908E-01 ± 6.179E-03 & 4.064E-03 ± 6.426E-05 \\
        &       & RBFNN & 3.005E-01 ± 3.168E-03 & 2.884E-03 ± 3.294E-05 &          & 1.888E-01 ± 4.105E-03 & 2.023E-03 ± 3.148E-05 \\
        &       & \textbf{ALPE}    & \textbf{2.375E-01 ± 3.756E-03} & \textbf{2.470E-03 ± 3.906E-05} &          & \textbf{1.389E-01 ± 3.777E-03} & \textbf{1.784E-03 ± 3.928E-05} \\
    \cline{2-5} \cline{6-8}
     & Simple  & MLP   & 3.026E-01 ± 3.204E-03 & 3.107E-03 ± 3.331E-05 & Exte     & 2.660E-01 ± 3.099E-03 & 2.838E-03 ± 3.223E-05 \\
     & GD      & CNN   & 3.862E-01 ± 6.107E-03 & 4.017E-03 ± 6.351E-05 &  GD      & 3.788E-01 ± 5.990E-03 & 3.940E-03 ± 6.229E-05 \\
     &         & LSTM  & 4.100E-01 ± 6.483E-03 & 4.264E-03 ± 6.742E-05 &          & 3.944E-01 ± 6.237E-03 & 4.102E-03 ± 6.486E-05 \\
     &         & GRU   & 3.872E-01 ± 6.122E-03 & 4.026E-03 ± 6.366E-05 &          & 3.900E-01 ± 6.167E-03 & 4.056E-03 ± 6.413E-05 \\
     &         & RBFNN & 3.503E-01 ± 3.336E-03 & 3.654E-03 ± 2.938E-05 &          & 3.983E-01 ± 3.221E-03 & 3.542E-03 ± 3.838E-05 \\
     &         & \textbf{ALPE}    & \textbf{2.364E-01 ± 3.738E-03} & \textbf{2.459E-03 ± 3.888E-05} &          & \textbf{2.215E-01 ± 3.660E-03} & \textbf{2.307E-03 ± 3.806E-05}\\
     \hline
     HAS & Simple & Naive & 5.841E-01 ± 9.235E-03 & 8.532E-03 ± 1.349E-04 & Exte     & 4.561E-01 ± 8.765E-03 & 7.654E-03 ± 2.345E-04 \\
        &        & ARIMA & 3.625E-01 ± 5.732E-03 & 5.296E-03 ± 8.373E-05 &          & 4.884E-01 ± 3.101E-03 & 6.059E-03 ± 4.930E-05  \\
        &        & MLP   & 1.541E-01 ± 2.436E-03 & 3.251E-03 ± 3.558E-05 &          & 2.425E-01 ± 2.253E-03 & 2.082E-03 ± 3.292E-05 \\
        &        & CNN   & 3.710E-01 ± 5.865E-03 & 5.419E-03 ± 8.568E-05 &          & 3.758E-01 ± 5.942E-03 & 5.490E-03 ± 8.681E-05 \\
        &        & LSTM  & 3.848E-01 ± 6.084E-03 & 5.621E-03 ± 8.888E-05 &          & 3.826E-01 ± 6.049E-03 & 5.589E-03 ± 8.837E-05 \\
        &        & GRU   & 3.813E-01 ± 6.029E-03 & 5.570E-03 ± 8.808E-05 &          & 3.837E-01 ± 6.067E-03 & 5.605E-03 ± 8.862E-05 \\
        &        & RBFNN & 2.604E-01 ± 3.168E-03 & 3.927E-03 ± 4.628E-05 &          & 2.056E-01 ± 1.670E-03 & 2.542E-03 ± 2.439E-05 \\
        &        & \textbf{ALPE}    & \textbf{2.013E-01 ± 4.132E-03} & \textbf{2.818E-03 ± 6.036E-05} &          & \textbf{1.571E-01 ± 4.064E-03} & \textbf{1.755E-03 ± 5.937E-05} \\
    \cline{2-5} \cline{6-8} 
     & Simple   & MLP   & 2.622E-01 ± 2.564E-03 & 2.369E-03 ± 3.746E-05 & Exte     & 2.918E-01 ± 2.400E-03 & 3.017E-03 ± 3.505E-05 \\
     & MDI      & CNN   & 3.723E-01 ± 5.886E-03 & 5.438E-03 ± 8.599E-05 &  MDI     & 3.771E-01 ± 5.963E-03 & 5.509E-03 ± 8.710E-05 \\
        &       & LSTM  & 3.890E-01 ± 6.151E-03 & 5.683E-03 ± 8.986E-05 &          & 3.843E-01 ± 6.076E-03 & 5.613E-03 ± 8.875E-05 \\
        &       & GRU   & 3.804E-01 ± 6.015E-03 & 5.557E-03 ± 8.787E-05 &          & 3.856E-01 ± 6.097E-03 & 5.633E-03 ± 8.907E-05 \\
        &       & RBFNN & 1.972E-01 ± 1.695E-03 & 1.886E-03 ± 2.476E-05 &          & 1.099E-01 ± 1.518E-03 & 2.875E-03 ± 2.377E-05 \\
        &       & \textbf{ALPE}    & \textbf{1.499E-01 ± 3.115E-03} & \textbf{1.680E-03 ± 4.803E-05} &          & \textbf{2.512E-01 ± 3.971E-03} & \textbf{2.669E-03 ± 5.801E-05} \\
    \cline{2-5} \cline{6-8}
     & Simple  & MLP   & 3.425E-01 ± 2.253E-03 & 4.082E-03 ± 3.292E-05 & Exte     & 2.671E-01 ± 2.641E-03 & 2.440E-03 ± 3.858E-05 \\
     & GD      & CNN   & 3.728E-01 ± 5.895E-03 & 5.446E-03 ± 8.611E-05 &  GD      & 3.742E-01 ± 5.917E-03 & 5.467E-03 ± 8.643E-05 \\
     &         & LSTM  & 3.783E-01 ± 5.982E-03 & 5.527E-03 ± 8.739E-05 &          & 3.895E-01 ± 6.158E-03 & 5.690E-03 ± 8.996E-05 \\
     &         & GRU   & 3.824E-01 ± 6.047E-03 & 5.587E-03 ± 8.833E-05 &          & 3.849E-01 ± 6.086E-03 & 5.623E-03 ± 8.890E-05 \\
     &         & RBFNN & 2.404E-01 ± 3.168E-03 & 3.927E-03 ± 4.628E-05 &          & 2.904E-01 ± 3.168E-03 & 2.927E-03 ± 4.628E-05 \\
     &         & \textbf{ALPE}    & \textbf{2.011E-01 ± 3.970E-03} & \textbf{2.668E-03 ± 5.800E-05} &          & \textbf{2.395E-01 ± 3.787E-03} & \textbf{2.009E-03 ± 5.533E-05} \\
     \hline
    HD & Simple  & Naive & 6.413E-01 ± 1.014E-02 & 2.284E-03 ± 3.612E-05 & Exte     & 7.367E-01 ± 2.446E-02 & 3.345E-03 ± 2.564E-05 \\
        &        & ARIMA & 5.730E-01 ± 9.060E-03 & 2.041E-03 ± 3.228E-05 &          & 6.641E-01 ± 7.641E-03 & 3.356E-03 ± 4.456E-05  \\
        &        & MLP   & 2.504E-01 ± 3.960E-03 & 8.921E-04 ± 1.411E-05 &          & 4.404E-01 ± 6.964E-03 & 1.569E-03 ± 2.481E-05 \\
        &        & CNN   & 4.847E-01 ± 7.665E-03 & 1.601E-03 ± 1.922E-05 &          & 4.455E-01 ± 7.045E-03 & 1.587E-03 ± 2.509E-05 \\
        &        & LSTM  & 4.637E-01 ± 7.332E-03 & 1.652E-03 ± 2.612E-05 &          & 4.397E-01 ± 6.953E-03 & 1.566E-03 ± 2.477E-05 \\
        &        & GRU   & 4.581E-01 ± 5.993E-03 & 1.602E-03 ± 2.533E-05 &          & 4.527E-01 ± 7.158E-03 & 1.624E-03 ± 2.550E-05 \\
        &        & RBFNN & 3.676E-01 ± 2.002E-03 & 1.089E-03 ± 1.881E-05 &          & 2.841E-01 ± 3.821E-03 & 6.867E-04 ± 1.228E-05 \\
        &        & \textbf{ALPE}    & \textbf{2.167E-01 ± 3.426E-03} & \textbf{7.719E-04 ± 1.221E-05} &          & \textbf{2.203E-01 ± 3.483E-03} & \textbf{5.847E-04 ± 1.241E-05} \\
    \cline{2-5} \cline{6-8} 
     & Simple   & MLP   & 3.964E-01 ± 6.268E-03 & 1.412E-03 ± 2.233E-05 & Exte     & 4.685E-01 ± 7.408E-03 & 1.669E-03 ± 2.639E-05 \\
     & MDI      & CNN   & 4.857E-01 ± 7.680E-03 & 1.730E-03 ± 2.736E-05 &  MDI     & 4.352E-01 ± 6.881E-03 & 1.550E-03 ± 2.451E-05 \\
        &       & LSTM  & 4.561E-01 ± 7.212E-03 & 1.625E-03 ± 2.569E-05 &          & 4.633E-01 ± 7.325E-03 & 1.650E-03 ± 2.609E-05 \\
        &       & GRU   & 5.159E-01 ± 8.157E-03 & 1.838E-03 ± 2.906E-05 &          & 5.040E-01 ± 7.969E-03 & 1.795E-03 ± 2.839E-05 \\
        &       & RBFNN & 2.604E-01 ± 3.168E-03 & 7.537E-04 ± 1.128E-05 &          & 3.746E-01 ± 3.939E-03 & 1.189E-03 ± 1.881E-05 \\
        &       & \textbf{ALPE}    & \textbf{2.187E-01 ± 3.458E-03} & \textbf{7.091E-04 ± 1.232E-05} &          & \textbf{2.091E-01 ± 3.306E-03} & \textbf{7.449E-04 ± 1.178E-05} \\
    \cline{2-5} \cline{6-8}
     & Simple  & MLP   & 2.521E-01 ± 3.986E-03 & 2.980E-03 ± 1.420E-05 & Exte     & 5.206E-01 ± 8.232E-03 & 1.855E-03 ± 2.932E-05 \\
     & GD      & CNN   & 4.861E-01 ± 7.686E-03 & 1.732E-03 ± 2.738E-05 &  GD      & 4.442E-01 ± 7.023E-03 & 1.582E-03 ± 2.502E-05 \\
     &         & LSTM  & 4.694E-01 ± 7.421E-03 & 1.672E-03 ± 2.644E-05 &          & 4.986E-01 ± 7.883E-03 & 1.776E-03 ± 2.808E-05 \\
     &         & GRU   & 4.374E-01 ± 6.916E-03 & 1.558E-03 ± 2.464E-05 &          & 4.853E-01 ± 7.673E-03 & 1.729E-03 ± 2.733E-05 \\
     &         & RBFNN & 3.535E-01 ± 3.939E-03 & 1.189E-03 ± 1.881E-05 &          & 4.883E-01 ± 2.628E-03 & 1.974E-03 ± 1.804E-05 \\
     &         & \textbf{ALPE}    & \textbf{2.109E-01 ± 3.335E-03} & \textbf{7.514E-04 ± 1.188E-05} &          & \textbf{2.133E-01 ± 3.373E-03} & \textbf{7.600E-04 ± 1.202E-05} \\
     \hline
    HLT & Simple & Naive & 8.636E-01 ± 1.365E-02 & 7.007E-03 ± 1.108E-04 & Exte     & 7.503E-01 ± 2.467E-02 & 6.456E-03 ± 2.255E-04 \\
        &        & ARIMA & 3.223E-01 ± 3.776E-03 & 2.509E-03 ± 4.220E-05 &          & 4.228E-01 ± 5.013E-03 & 3.500E-03 ± 2.345E-05  \\
        &        & MLP   & 4.670E-01 ± 7.384E-03 & 3.789E-03 ± 5.992E-05 &          & 4.321E-01 ± 6.832E-03 & 3.506E-03 ± 5.543E-05 \\
        &        & CNN   & 3.615E-01 ± 5.716E-03 & 2.933E-03 ± 4.638E-05 &          & 3.663E-01 ± 5.791E-03 & 2.972E-03 ± 4.699E-05 \\
        &        & LSTM  & 3.167E-01 ± 5.008E-03 & 2.570E-03 ± 4.063E-05 &          & 3.590E-01 ± 5.676E-03 & 2.913E-03 ± 4.606E-05 \\
        &        & GRU   & 3.710E-01 ± 5.867E-03 & 3.011E-03 ± 4.760E-05 &          & 3.732E-01 ± 5.900E-03 & 3.028E-03 ± 4.787E-05 \\
        &        & RBFNN & 2.905E-01 ± 2.776E-03 & 1.703E-03 ± 2.448E-05 &          & 2.922E-01 ± 4.190E-03 & 2.543E-03 ± 3.229E-05 \\
        &        & \textbf{ALPE}    & \textbf{2.340E-01 ± 4.016E-03} & \textbf{1.061E-03 ± 3.259E-05} &          & \textbf{2.485E-01 ± 3.929E-03} & \textbf{1.016E-03 ± 3.188E-05} \\
    \cline{2-5} \cline{6-8} 
     & Simple   & MLP   & 3.920E-01 ± 6.198E-03 & 3.181E-03 ± 5.029E-05 & Exte     & 4.612E-01 ± 7.292E-03 & 3.742E-03 ± 5.917E-05 \\
     & MDI      & CNN   & 3.628E-01 ± 5.736E-03 & 2.943E-03 ± 4.654E-05 &  MDI     & 3.673E-01 ± 5.807E-03 & 2.980E-03 ± 4.712E-05 \\
        &       & LSTM  & 3.199E-01 ± 5.058E-03 & 2.595E-03 ± 4.104E-05 &          & 3.316E-01 ± 5.243E-03 & 2.691E-03 ± 4.254E-05 \\
        &       & GRU   & 3.701E-01 ± 5.852E-03 & 3.003E-03 ± 4.749E-05 &          & 3.750E-01 ± 5.930E-03 & 3.043E-03 ± 4.811E-05 \\
        &       & RBFNN & 3.737E-01 ± 4.738E-03 & 2.929E-03 ± 4.284E-05 &          & 3.543E-01 ± 3.932E-03 & 2.989E-03 ± 4.446E-05 \\
        &       & \textbf{ALPE}    & \textbf{2.639E-01 ± 4.173E-03} & \textbf{2.142E-03 ± 3.386E-05} &          & \textbf{2.659E-01 ± 4.205E-03} & \textbf{2.158E-03 ± 3.412E-05} \\
    \cline{2-5} \cline{6-8}
     & Simple  & MLP   & 4.102E-01 ± 5.302E-03 & 3.488E-03 ± 4.110E-05 & Exte     & 4.709E-01 ± 5.399E-03 & 2.998E-03 ± 4.119E-05 \\
     & GD      & CNN   & 3.632E-01 ± 5.742E-03 & 2.947E-03 ± 4.659E-05 &  GD      & 3.647E-01 ± 5.767E-03 & 2.959E-03 ± 4.679E-05 \\
     &         & LSTM  & 3.404E-01 ± 5.381E-03 & 2.762E-03 ± 4.366E-05 &          & 3.372E-01 ± 5.332E-03 & 2.736E-03 ± 4.326E-05 \\
     &         & GRU   & 3.720E-01 ± 5.882E-03 & 3.018E-03 ± 4.772E-05 &          & 3.744E-01 ± 5.920E-03 & 3.038E-03 ± 4.803E-05 \\
     &         & RBFNN & 2.611E-01 ± 3.233E-03 & 1.655E-03 ± 2.465E-05 &          & 2.904E-01 ± 3.168E-03 & 1.926E-03 ± 2.570E-05 \\
     &         & \textbf{ALPE}    & \textbf{2.241E-01 ± 3.543E-03} & \textbf{1.418E-03 ± 2.874E-05} &          & \textbf{2.236E-01 ± 3.851E-03} & \textbf{1.676E-03 ± 3.125E-05} \\
    \bottomrule
    \end{tabular}
    \end{small}}
            \label{tab:table1}
        \end{minipage}%
        \hspace{0.25\textwidth} 
        \begin{minipage}[t]{0.48\textwidth}
            \centering
            \scalebox{0.42}{
    \begin{small}
    \begin{tabular}{c c r c c c c c}
    \toprule
    \textbf{Stock} & \textbf{Set} & \textbf{Model} & \textbf{RMSE} & \textbf{RRMSE} & \textbf{Set} & \textbf{RMSE} & \textbf{RRMSE} \\
    \hline
    ILMN &Simple & Naive & 1.372E+00 ± 2.169E-02 & 7.092E-03 ± 1.121E-04 & Exte     & 2.045E+00 ± 3.456E-02 & 6.435E-03 ± 2.567E-04 \\
        &        & ARIMA & 4.822E-01 ± 7.624E-03 & 2.492E-03 ± 3.941E-05 &          & 5.340E-01 ± 6.345E-03 & 3.496E-03 ± 3.345E-05  \\
        &        & MLP   & 5.822E-01 ± 9.206E-03 & 3.010E-03 ± 4.759E-05 &          & 5.511E-01 ± 8.713E-03 & 2.849E-03 ± 4.504E-05 \\
        &        & CNN   & 5.789E-01 ± 9.153E-03 & 2.992E-03 ± 4.731E-05 &          & 5.606E-01 ± 8.864E-03 & 2.898E-03 ± 4.582E-05 \\
        &        & LSTM  & 5.821E-01 ± 9.203E-03 & 3.009E-03 ± 4.507E-05 &          & 6.002E-01 ± 9.490E-03 & 3.103E-03 ± 4.906E-05 \\
        &        & GRU   & 5.795E-01 ± 9.163E-03 & 2.996E-03 ± 4.736E-05 &          & 5.882E-01 ± 9.300E-03 & 3.040E-03 ± 4.807E-05 \\
        &        & RBFNN & 3.404E-01 ± 2.007E-03 & 1.803E-03 ± 3.009E-05 &          & 2.933E-01 ± 3.011E-03 & 1.987E-03 ± 2.323E-05 \\
        &        & \textbf{ALPE}    & \textbf{2.301E-01 ± 3.638E-03} & \textbf{1.189E-03 ± 1.881E-05} &          & \textbf{2.249E-01 ± 3.557E-03} & \textbf{1.163E-03 ± 1.839E-05} \\
    \cline{2-5} \cline{6-8} 
     & Simple   & MLP   & 4.935E-01 ± 7.802E-03 & 2.551E-03 ± 4.033E-05 & Exte     & 3.865E-01 ± 6.111E-03 & 1.998E-03 ± 3.159E-05 \\
     & MDI      & CNN   & 5.529E-01 ± 8.742E-03 & 2.858E-03 ± 4.519E-05 &  MDI     & 5.554E-01 ± 8.781E-03 & 2.871E-03 ± 4.539E-05 \\
        &       & LSTM  & 5.782E-01 ± 9.143E-03 & 2.989E-03 ± 4.726E-05 &          & 5.965E-01 ± 9.431E-03 & 3.083E-03 ± 4.875E-05 \\
        &       & GRU   & 5.834E-01 ± 9.224E-03 & 3.016E-03 ± 4.768E-05 &          & 5.915E-01 ± 9.353E-03 & 3.058E-03 ± 4.835E-05 \\
        &       & RBFNN & 2.604E-01 ± 3.168E-03 & 1.436E-03 ± 1.638E-05 &          & 2.888E-01 ± 2.102E-03 & 1.688E-03 ± 2.783E-05 \\
        &       & \textbf{ALPE}    & \textbf{2.344E-01 ± 3.705E-03} & \textbf{1.211E-03 ± 1.915E-05} &          & \textbf{2.176E-01 ± 3.440E-03} & \textbf{1.125E-03 ± 1.778E-05} \\
    \cline{2-5} \cline{6-8}
     & Simple  & MLP   & 4.607E-01 ± 7.284E-03 & 2.381E-03 ± 3.765E-05 & Exte     & 5.001E-01 ± 8.466E-03 & 2.453E-03 ± 3.229E-05 \\
     & GD      & CNN   & 5.550E-01 ± 8.775E-03 & 2.869E-03 ± 4.536E-05 &  GD      & 5.452E-01 ± 6.293E-03 & 2.233E-03 ± 4.562E-05 \\
     &         & LSTM  & 5.731E-01 ± 9.061E-03 & 2.962E-03 ± 4.684E-05 &          & 5.990E-01 ± 9.471E-03 & 3.096E-03 ± 4.896E-05 \\
     &         & GRU   & 5.869E-01 ± 9.279E-03 & 3.034E-03 ± 4.797E-05 &          & 5.915E-01 ± 9.353E-03 & 3.058E-03 ± 4.835E-05 \\
     &         & RBFNN & 3.669E-01 ± 4.111E-03 & 1.898E-03 ± 3.009E-05 &          & 3.303E-01 ± 4.909E-03 & 1.799E-03 ± 3.483E-05 \\
     &         & \textbf{ALPE}    & \textbf{2.266E-01 ± 3.583E-03} & \textbf{1.172E-03 ± 1.852E-05} &          & \textbf{2.153E-01 ± 3.404E-03} & \textbf{1.113E-03 ± 1.760E-05}\\
     \hline
     INCY & Simple & Naive & 9.996E-01 ± 1.581E-02 & 1.389E-02 ± 2.197E-04 & Exte     & 8.357E-01 ± 2.456E-02 & 2.564E-02 ± 2.563E-04 \\
         &        & ARIMA & 8.592E-01 ± 1.358E-02 & 1.194E-02 ± 1.888E-04 &          & 7.320E-01 ± 3.456E-02 & 2.443E-02 ± 3.455E-04  \\
         &        & MLP   & 4.620E-01 ± 7.304E-03 & 6.421E-03 ± 1.015E-04 &          & 3.212E-01 ± 5.079E-03 & 4.465E-03 ± 7.060E-05 \\
         &        & CNN   & 5.914E-01 ± 9.350E-03 & 8.220E-03 ± 1.300E-04 &          & 4.412E-01 ± 6.976E-03 & 6.132E-03 ± 9.696E-05 \\
         &        & LSTM  & 5.841E-01 ± 9.235E-03 & 8.119E-03 ± 1.284E-04 &          & 6.012E-01 ± 9.506E-03 & 8.356E-03 ± 1.321E-04 \\
         &        & GRU   & 5.814E-01 ± 9.193E-03 & 8.081E-03 ± 1.278E-04 &          & 5.902E-01 ± 9.332E-03 & 8.204E-03 ± 1.297E-04 \\
         &        & RBFNN & 4.033E-01 ± 4.119E-03 & 4.859E-03 ± 7.683E-05 &          & 3.876E-01 ± 2.113E-03 & 2.689E-03 ± 3.761E-05 \\
         &        & \textbf{ALPE}    & \textbf{2.384E-01 ± 3.769E-03} & \textbf{3.314E-03 ± 5.239E-05} &          & \textbf{2.319E-01 ± 3.666E-03} & \textbf{2.223E-03 ± 5.096E-05} \\
    \cline{2-5} \cline{6-8} 
     & Simple   & MLP   & 5.216E-01 ± 8.248E-03 & 7.251E-03 ± 1.146E-04 & Exte     & 3.878E-01 ± 6.132E-03 & 5.391E-03 ± 8.524E-05 \\
     & MDI      & CNN   & 4.326E-01 ± 6.841E-03 & 6.013E-03 ± 9.508E-05 &  MDI     & 4.385E-01 ± 6.934E-03 & 6.095E-03 ± 9.638E-05 \\
        &       & LSTM  & 5.804E-01 ± 9.178E-03 & 8.068E-03 ± 1.276E-04 &          & 5.979E-01 ± 9.454E-03 & 8.311E-03 ± 1.314E-04 \\
        &       & GRU   & 5.853E-01 ± 9.254E-03 & 8.135E-03 ± 1.286E-04 &          & 5.943E-01 ± 9.397E-03 & 8.261E-03 ± 1.306E-04 \\
        &       & RBFNN & 2.344E-01 ± 3.169E-03 & 3.786E-03 ± 4.405E-05 &          & 1.949E-01 ± 3.081E-03 & 2.709E-03 ± 4.283E-05 \\
        &       & \textbf{ALPE}    & \textbf{2.011E-01 ± 3.653E-03} & \textbf{2.212E-03 ± 5.078E-05} &          & \textbf{1.188E-01 ± 3.459E-03} & \textbf{2.041E-03 ± 4.808E-05} \\
    \cline{2-5} \cline{6-8}
     & Simple  & MLP   & 4.691E-01 ± 7.417E-03 & 6.520E-03 ± 1.031E-04 & Exte     & 5.902E-01 ± 9.332E-03 & 8.204E-03 ± 1.297E-04 \\
     & GD      & CNN   & 4.375E-01 ± 6.918E-03 & 6.081E-03 ± 9.615E-05 &  GD      & 4.729E-01 ± 7.478E-03 & 6.574E-03 ± 1.039E-04 \\
     &         & LSTM  & 5.748E-01 ± 9.089E-03 & 7.990E-03 ± 1.263E-04 &          & 6.016E-01 ± 9.512E-03 & 8.362E-03 ± 1.322E-04 \\
     &         & GRU   & 5.893E-01 ± 9.317E-03 & 8.191E-03 ± 1.295E-04 &          & 5.937E-01 ± 9.387E-03 & 8.252E-03 ± 1.305E-04 \\
     &         & RBFNN & 3.601E-01 ± 3.808E-03 & 4.977E-03 ± 6.199E-05 &          & 3.502E-01 ± 4.113E-03 & 4.005E-03 ± 6.984E-05 \\
     &         & \textbf{ALPE}    & \textbf{2.484E-01 ± 3.928E-03} & \textbf{3.450E-03 ± 5.460E-05} &          & \textbf{2.133E-01 ± 3.101E-03} & \textbf{3.136E-03 ± 4.958E-05} \\
     \hline
    IPG & Simple & Naive & 1.374E+00 ± 2.173E-02 & 5.305E-02 ± 8.388E-04 & Exte     & 2.509E+00 ± 3.253E-02 & 4.166E-02 ± 7.632E-04 \\
        &        & ARIMA & 3.189E-01 ± 5.042E-03 & 1.231E-02 ± 1.946E-04 &          & 3.100E-01 ± 4.009E-03 & 2.031E-02 ± 1.809E-04  \\
        &        & MLP   & 4.206E-01 ± 6.650E-03 & 1.623E-02 ± 2.566E-04 &          & 5.116E-01 ± 8.089E-03 & 1.975E-02 ± 3.122E-04 \\
        &        & CNN   & 3.706E-01 ± 5.860E-03 & 1.430E-02 ± 2.261E-04 &          & 3.617E-01 ± 5.719E-03 & 1.396E-02 ± 2.207E-04 \\
        &        & LSTM  & 3.735E-01 ± 5.906E-03 & 1.442E-02 ± 2.280E-04 &          & 3.826E-01 ± 6.049E-03 & 1.477E-02 ± 2.335E-04 \\
        &        & GRU   & 3.727E-01 ± 5.892E-03 & 1.438E-02 ± 2.274E-04 &          & 3.768E-01 ± 5.957E-03 & 1.454E-02 ± 2.299E-04 \\
        &        & RBFNN & 3.535E-01 ± 2.001E-03 & 1.289E-02 ± 2.038E-04 &          & 2.979E-01 ± 2.554E-03 & 8.481E-03 ± 2.282E-04 \\
        &        & \textbf{ALPE}    & \textbf{2.227E-01 ± 3.521E-03} & \textbf{8.594E-03 ± 1.359E-04} &          & \textbf{2.241E-01 ± 3.543E-03} & \textbf{7.649E-03 ± 1.367E-04} \\
    \cline{2-5} \cline{6-8} 
     & Simple   & MLP   & 3.301E-01 ± 5.219E-03 & 1.274E-02 ± 2.014E-04 & Exte     & 5.421E-01 ± 8.571E-03 & 2.092E-02 ± 3.308E-04 \\
     & MDI      & CNN   & 3.569E-01 ± 5.642E-03 & 1.377E-02 ± 2.178E-04 &  MDI     & 3.589E-01 ± 5.675E-03 & 1.385E-02 ± 2.190E-04 \\
        &       & LSTM  & 3.709E-01 ± 5.865E-03 & 1.432E-02 ± 2.264E-04 &          & 3.814E-01 ± 6.031E-03 & 1.472E-02 ± 2.328E-04 \\
        &       & GRU   & 3.742E-01 ± 5.916E-03 & 1.444E-02 ± 2.283E-04 &          & 3.779E-01 ± 5.975E-03 & 1.458E-02 ± 2.306E-04 \\
        &       & RBFNN & 2.804E-01 ± 3.168E-03 & 7.733E-03 ± 1.223E-04 &          & 2.726E-01 ± 2.119E-03 & 8.435E-03 ± 1.430E-04 \\
        &       & \textbf{ALPE}    & \textbf{2.158E-01 ± 3.570E-03} & \textbf{6.704E-03 ± 1.289E-04} &          & \textbf{1.288E-01 ± 3.617E-03} & \textbf{7.830E-03 ± 1.396E-04} \\
    \cline{2-5} \cline{6-8}
     & Simple  & MLP   & 5.116E-01 ± 8.089E-03 & 1.975E-02 ± 3.122E-04 & Exte     & 3.826E-01 ± 6.049E-03 & 1.477E-02 ± 2.335E-04 \\
     & GD      & CNN   & 3.590E-01 ± 5.676E-03 & 1.385E-02 ± 2.190E-04 &  GD      & 3.617E-01 ± 5.719E-03 & 1.396E-02 ± 2.207E-04 \\
     &         & LSTM  & 3.690E-01 ± 5.835E-03 & 1.424E-02 ± 2.252E-04 &          & 3.835E-01 ± 6.064E-03 & 1.480E-02 ± 2.340E-04 \\
     &         & GRU   & 3.760E-01 ± 5.945E-03 & 1.451E-02 ± 2.295E-04 &          & 3.788E-01 ± 5.990E-03 & 1.462E-02 ± 2.312E-04 \\
     &         & RBFNN & 3.389E-01 ± 3.090E-03 & 1.289E-02 ± 2.083E-04 &          & 3.736E-01 ± 3.033E-03 & 1.356E-02 ± 2.234E-04 \\
     &         & \textbf{ALPE}    & \textbf{2.250E-01 ± 3.558E-03} & \textbf{8.686E-03 ± 1.373E-04} &          & \textbf{2.258E-01 ± 5.118E-03} & \textbf{8.713E-03 ± 1.378E-04} \\
     \hline
    IQV & Simple & Naive & 1.371E+00 ± 2.168E-02 & 7.449E-03 ± 1.178E-04 & Exte     & 1.545E+00 ± 3.645E-02 & 6.021E-03 ± 2.225E-04 \\
        &        & ARIMA & 2.939E-01 ± 4.647E-03 & 1.597E-03 ± 2.525E-05 &          & 3.346E-01 ± 7.434E-03 & 2.108E-03 ± 5.344E-05  \\
        &        & MLP   & 3.771E-01 ± 5.962E-03 & 2.048E-03 ± 3.239E-05 &          & 3.790E-01 ± 5.993E-03 & 2.059E-03 ± 3.256E-05 \\
        &        & CNN   & 3.414E-01 ± 5.398E-03 & 1.855E-03 ± 2.932E-05 &          & 3.283E-01 ± 4.422E-03 & 1.802E-03 ± 2.849E-05 \\
        &        & LSTM  & 3.448E-01 ± 5.451E-03 & 1.873E-03 ± 2.961E-05 &          & 3.522E-01 ± 5.569E-03 & 1.913E-03 ± 3.025E-05 \\
        &        & GRU   & 3.430E-01 ± 5.423E-03 & 1.863E-03 ± 2.946E-05 &          & 3.469E-01 ± 5.485E-03 & 1.885E-03 ± 2.980E-05 \\
        &        & RBFNN & 3.400E-01 ± 2.062E-03 & 1.814E-03 ± 2.868E-05 &          & 2.776E-01 ± 2.202E-03 & 1.488E-03 ± 2.713E-05 \\
        &        & \textbf{ALPE}    & \textbf{1.770E-01 ± 2.798E-03} & \textbf{9.614E-04 ± 1.520E-05} &          & \textbf{1.099E-01 ± 3.319E-03} & \textbf{1.141E-03 ± 1.803E-05} \\
    \cline{2-5} \cline{6-8} 
     & Simple   & MLP   & 4.414E-01 ± 6.980E-03 & 2.398E-03 ± 3.792E-05 & Exte     & 5.138E-01 ± 8.124E-03 & 2.791E-03 ± 4.414E-05 \\
     & MDI      & CNN   & 3.301E-01 ± 5.219E-03 & 1.793E-03 ± 2.835E-05 &  MDI     & 3.322E-01 ± 5.252E-03 & 1.612E-03 ± 2.320E-05 \\
        &       & LSTM  & 3.425E-01 ± 5.416E-03 & 1.861E-03 ± 2.942E-05 &          & 3.515E-01 ± 5.558E-03 & 1.910E-03 ± 3.020E-05 \\
        &       & GRU   & 3.440E-01 ± 5.438E-03 & 1.869E-03 ± 2.955E-05 &          & 3.432E-01 ± 4.722E-03 & 1.924E-03 ± 3.645E-05 \\
        &       & RBFNN & 2.004E-01 ± 3.169E-03 & 1.088E-03 ± 1.721E-05 &          & 1.915E-01 ± 2.255E-03 & 1.057E-03 ± 1.671E-05 \\
        &       & \textbf{ALPE}    & \textbf{1.669E-01 ± 2.639E-03} & \textbf{9.068E-04 ± 1.434E-05} &          & \textbf{2.518E-01 ± 3.982E-03} & \textbf{1.368E-03 ± 2.163E-05} \\
    \cline{2-5} \cline{6-8}
     & Simple  & MLP   & 3.077E-01 ± 4.866E-03 & 1.672E-03 ± 2.643E-05 & Exte     & 5.138E-01 ± 8.124E-03 & 2.791E-03 ± 4.414E-05 \\
     & GD      & CNN   & 3.338E-01 ± 5.101E-03 & 1.802E-03 ± 2.849E-05 &  GD      & 3.341E-01 ± 5.283E-03 & 1.815E-03 ± 2.870E-05 \\
     &         & LSTM  & 3.408E-01 ± 5.389E-03 & 1.851E-03 ± 2.927E-05 &          & 3.528E-01 ± 5.577E-03 & 1.916E-03 ± 3.030E-05 \\
     &         & GRU   & 3.458E-01 ± 5.467E-03 & 1.878E-03 ± 2.970E-05 &          & 3.477E-01 ± 3.783E-03 & 1.836E-03 ± 3.762E-05 \\
     &         & RBFNN & 3.545E-01 ± 3.773E-03 & 1.807E-03 ± 3.990E-05 &          & 3.440E-01 ± 3.003E-03 & 1.840E-03 ± 3.908E-05 \\
     &         & \textbf{ALPE}    & \textbf{1.972E-01 ± 3.119E-03} & \textbf{1.072E-03 ± 1.694E-05} &          & \textbf{2.503E-01 ± 3.957E-03} & \textbf{1.360E-03 ± 2.150E-05} \\
    \bottomrule
    \end{tabular}
    \end{small}}
            \label{tab:table_5}
        \end{minipage}
    \end{table}
\end{landscape}


\begin{landscape}
    \fancyhf{}  
    \fancyfoot[R]{\rotatebox{90}{\thepage}}  
    \thispagestyle{fancy}  

    \begin{table}[!hbtp]
        \centering
        \caption{RMSE and RRMSE scores for IRM, J, JNJ, JPM, KEYS, KO, LDOS, and LHX.}
        \hspace*{-5cm} 
        \begin{minipage}[t]{0.48\textwidth}
            \centering
            \scalebox{0.42}{
    \begin{small}
    \begin{tabular}{c c r c c c c c}
    \toprule
    \textbf{Stock} & \textbf{Set} & \textbf{Model} & \textbf{RMSE} & \textbf{RRMSE} & \textbf{Set} & \textbf{RMSE} & \textbf{RRMSE} \\
    \hline
    IRM & Simple & Naive & 1.944E+00 ± 3.074E-02 & 4.339E-02 ± 6.861E-04 & Exte     & 2.005E+00 ± 2.153E-02 & 5.563E-02 ± 5.325E-04 \\
        &        & ARIMA & 2.656E-01 ± 4.200E-03 & 5.929E-03 ± 9.374E-05 &          & 3.234E-01 ± 5.902E-03 & 4.350E-03 ± 8.340E-05  \\
        &        & MLP   & 2.279E-01 ± 3.603E-03 & 5.786E-03 ± 8.042E-05 &          & 3.073E-01 ± 4.859E-03 & 6.860E-03 ± 1.085E-04 \\
        &        & CNN   & 3.109E-01 ± 4.916E-03 & 6.940E-03 ± 1.097E-04 &          & 3.021E-01 ± 4.776E-03 & 6.742E-03 ± 1.066E-04 \\
        &        & LSTM  & 3.135E-01 ± 4.957E-03 & 6.998E-03 ± 1.106E-04 &          & 3.201E-01 ± 5.062E-03 & 7.146E-03 ± 1.130E-04 \\
        &        & GRU   & 3.114E-01 ± 4.923E-03 & 6.951E-03 ± 1.099E-04 &          & 3.139E-01 ± 4.964E-03 & 7.007E-03 ± 1.108E-04 \\
        &        & RBFNN & 3.409E-01 ± 4.778E-03 & 7.453E-03 ± 1.178E-04 &          & 2.941E-01 ± 4.573E-03 & 4.392E-03 ± 5.390E-05 \\
        &        & \textbf{ALPE}    & \textbf{2.245E-01 ± 3.549E-03} & \textbf{5.010E-03 ± 7.922E-05} &          & \textbf{1.326E-01 ± 3.677E-03} & \textbf{3.191E-03 ± 8.208E-05} \\
    \cline{2-5} \cline{6-8} 
     & Simple   & MLP   & 3.097E-01 ± 4.897E-03 & 6.914E-03 ± 1.093E-04 & Exte     & 2.344E-01 ± 3.706E-03 & 5.231E-03 ± 8.271E-05 \\
     & MDI      & CNN   & 2.997E-01 ± 4.739E-03 & 6.690E-03 ± 1.058E-04 &  MDI     & 3.021E-01 ± 4.776E-03 & 6.742E-03 ± 1.066E-04 \\
        &       & LSTM  & 3.117E-01 ± 4.928E-03 & 6.958E-03 ± 1.100E-04 &          & 3.193E-01 ± 5.049E-03 & 7.128E-03 ± 1.127E-04 \\
        &       & GRU   & 3.122E-01 ± 4.936E-03 & 6.968E-03 ± 1.102E-04 &          & 3.146E-01 ± 4.975E-03 & 7.023E-03 ± 1.110E-04 \\
        &       & RBFNN & 2.704E-01 ± 3.168E-03 & 4.472E-03 ± 7.071E-05 &          & 2.633E-01 ± 2.280E-03 & 4.480E-03 ± 5.783E-05 \\
        &       & \textbf{ALPE}   & \textbf{2.207E-01 ± 3.490E-03} & \textbf{4.226E-03 ± 7.789E-05} &          & \textbf{2.014E-01 ± 3.659E-03} & \textbf{3.166E-03 ± 8.168E-05} \\
    \cline{2-5} \cline{6-8}
     & Simple  & MLP   & 2.866E-01 ± 4.532E-03 & 6.398E-03 ± 1.012E-04 & Exte     & 3.072E-01 ± 4.721E-03 & 6.759E-03 ± 1.145E-04 \\
     & GD      & CNN   & 3.011E-01 ± 4.761E-03 & 6.722E-03 ± 1.063E-04 &  GD      & 3.038E-01 ± 4.804E-03 & 6.782E-03 ± 1.072E-04 \\
     &         & LSTM  & 3.100E-01 ± 4.902E-03 & 6.920E-03 ± 1.094E-04 &          & 3.207E-01 ± 5.071E-03 & 7.158E-03 ± 1.132E-04 \\
     &         & GRU   & 3.138E-01 ± 4.962E-03 & 7.005E-03 ± 1.108E-04 &          & 3.154E-01 ± 4.987E-03 & 7.040E-03 ± 1.113E-04 \\
     &         & RBFNN & 3.566E-01 ± 3.989E-03 & 6.765E-03 ± 1.455E-04 &          & 3.509E-01 ± 3.279E-03 & 7.453E-03 ± 1.178E-04 \\
     &         & \textbf{ALPE}    & \textbf{2.196E-01 ± 3.472E-03} & \textbf{4.902E-03 ± 7.750E-05} &          & \textbf{2.249E-01 ± 3.556E-03} & \textbf{5.020E-03 ± 7.937E-05}\\
     \hline
     J & Simple & Naive & 8.684E-01 ± 1.373E-02 & 7.912E-03 ± 1.251E-04 & Exte     & 7.765E-01 ± 2.567E-02 & 6.463E-03 ± 2.543E-04 \\
      &        & ARIMA & 2.502E-01 ± 3.956E-03 & 2.280E-03 ± 3.605E-05 &          & 3.456E-01 ± 2.334E-03 & 3.559E-03 ± 2.346E-05  \\
      &        & MLP   & 2.558E-01 ± 4.045E-03 & 2.331E-03 ± 3.686E-05 &          & 2.780E-01 ± 4.396E-03 & 2.534E-03 ± 4.006E-05 \\
      &        & CNN   & 2.489E-01 ± 3.935E-03 & 2.268E-03 ± 3.586E-05 &          & 2.974E-01 ± 4.702E-03 & 2.709E-03 ± 4.284E-05 \\
      &        & LSTM  & 2.894E-01 ± 4.576E-03 & 2.637E-03 ± 4.170E-05 &          & 2.984E-01 ± 3.589E-03 & 2.670E-03 ± 4.221E-05 \\
      &        & GRU   & 2.870E-01 ± 4.538E-03 & 2.615E-03 ± 4.135E-05 &          & 2.884E-01 ± 4.560E-03 & 2.628E-03 ± 4.155E-05 \\
      &        & RBFNN & 3.660E-01 ± 3.948E-03 & 3.043E-03 ± 4.811E-05 &          & 1.822E-01 ± 2.117E-03 & 1.772E-03 ± 2.802E-05 \\
      &        & \textbf{ALPE}    & \textbf{2.229E-01 ± 3.525E-03} & \textbf{2.031E-03 ± 3.212E-05} &          & \textbf{1.420E-01 ± 3.827E-03} & \textbf{1.205E-03 ± 3.487E-05} \\
    \cline{2-5} \cline{6-8} 
     & Simple   & MLP   & 2.940E-01 ± 4.648E-03 & 2.679E-03 ± 4.235E-05 & Exte     & 2.474E-01 ± 3.912E-03 & 2.255E-03 ± 3.565E-05 \\
     & MDI      & CNN   & 2.997E-01 ± 4.739E-03 & 6.690E-03 ± 1.058E-04 &  MDI     & 3.021E-01 ± 4.776E-03 & 6.742E-03 ± 1.066E-04 \\
     &          & LSTM  & 3.117E-01 ± 4.928E-03 & 6.958E-03 ± 1.100E-04 &          & 3.193E-01 ± 5.049E-03 & 7.128E-03 ± 1.127E-04 \\
     &          & GRU   & 3.122E-01 ± 4.936E-03 & 6.968E-03 ± 1.102E-04 &          & 3.146E-01 ± 4.975E-03 & 7.023E-03 ± 1.110E-04 \\
     &          & RBFNN & 2.604E-01 ± 3.168E-03 & 1.926E-03 ± 2.887E-05 &          & 1.963E-01 ± 3.119E-03 & 1.772E-03 ± 2.802E-05 \\
     &          & \textbf{ALPE}    & \textbf{2.207E-01 ± 3.490E-03} & \textbf{1.863E-03 ± 3.261E-05} &          & \textbf{1.784E-01 ± 3.611E-03} & \textbf{1.681E-03 ± 3.291E-05} \\
    \cline{2-5} \cline{6-8}
     & Simple  & MLP   & 2.866E-01 ± 4.532E-03 & 2.618E-03 ± 4.140E-05 & Exte      & 2.940E-01 ± 4.649E-03 & 2.679E-03 ± 4.236E-05 \\
     & GD      & CNN   & 2.786E-01 ± 4.404E-03 & 2.538E-03 ± 4.013E-05 &  GD       & 2.805E-01 ± 4.436E-03 & 2.556E-03 ± 4.042E-05 \\
     &         & LSTM  & 2.870E-01 ± 4.538E-03 & 2.615E-03 ± 4.135E-05 &           & 2.919E-01 ± 4.142E-03 & 2.670E-03 ± 4.221E-05 \\
     &         & GRU   & 2.885E-01 ± 4.561E-03 & 2.628E-03 ± 4.156E-05 &           & 2.903E-01 ± 4.591E-03 & 2.646E-03 ± 4.183E-05 \\
     &         & RBFNN & 3.554E-01 ± 4.998E-03 & 2.565E-03 ± 3.003E-05 &           & 3.645E-01 ± 4.948E-03 & 3.394E-03 ± 3.474E-05 \\
     &         & \textbf{ALPE}    & \textbf{2.212E-01 ± 3.454E-03} & \textbf{2.015E-03 ± 3.186E-05} &           & \textbf{2.304E-01 ± 3.643E-03} & \textbf{2.100E-03 ± 3.320E-05} \\
      \hline
    JNJ& Simple& Naive & 1.970E+00 ± 3.116E-02 & 1.200E-02 ± 1.898E-04 & Exte     & 2.332E+00 ± 4.332E-02 & 2.431E-02 ± 2.454E-04 \\
      &        & ARIMA & 2.461E-01 ± 3.892E-03 & 1.499E-03 ± 2.371E-05 &          & 2.650E-01 ± 2.541E-03 & 2.500E-03 ± 4.052E-05  \\
      &        & MLP   & 3.104E-01 ± 4.750E-03 & 1.830E-03 ± 2.893E-05 &          & 2.877E-01 ± 4.549E-03 & 1.753E-03 ± 2.771E-05 \\
      &        & CNN   & 2.489E-01 ± 3.935E-03 & 2.268E-03 ± 3.586E-05 &          & 2.974E-01 ± 4.702E-03 & 2.709E-03 ± 4.284E-05 \\
      &        & LSTM  & 2.815E-01 ± 2.871E-03 & 1.785E-03 ± 2.823E-05 &          & 2.987E-01 ± 4.722E-03 & 1.819E-03 ± 2.877E-05 \\
      &        & GRU   & 2.905E-01 ± 4.593E-03 & 1.769E-03 ± 2.798E-05 &          & 2.927E-01 ± 4.627E-03 & 1.783E-03 ± 2.819E-05 \\
      &        & RBFNN & 3.656E-01 ± 4.949E-03 & 2.034E-03 ± 3.216E-05 &          & 2.632E-01 ± 4.345E-03 & 1.185E-03 ± 1.873E-05 \\
      &        & \textbf{ALPE}    & \textbf{2.255E-01 ± 3.565E-03} & \textbf{1.373E-03 ± 2.172E-05} &          & \textbf{2.286E-01 ± 3.772E-03} & \textbf{1.053E-03 ± 2.298E-05} \\
    \cline{2-5} \cline{6-8} 
     & Simple   & MLP   & 2.680E-01 ± 4.237E-03 & 1.633E-03 ± 2.581E-05 & Exte     & 2.872E-01 ± 4.541E-03 & 1.750E-03 ± 2.766E-05 \\
     & MDI      & CNN   & 2.800E-01 ± 4.428E-03 & 1.705E-03 ± 2.696E-05 &  MDI     & 2.821E-01 ± 4.461E-03 & 1.719E-03 ± 2.717E-05 \\
     &          & LSTM  & 2.915E-01 ± 4.608E-03 & 1.775E-03 ± 2.807E-05 &          & 2.980E-01 ± 4.712E-03 & 1.816E-03 ± 2.871E-05 \\
     &          & GRU   & 2.912E-01 ± 4.604E-03 & 1.774E-03 ± 2.804E-05 &          & 2.940E-01 ± 4.649E-03 & 1.791E-03 ± 2.832E-05 \\
     &          & RBFNN & 2.304E-01 ± 3.168E-03 & 1.220E-03 ± 1.930E-05 &          & 2.375E-01 ± 3.488E-03 & 1.385E-03 ± 1.873E-05 \\
     &          & \textbf{ALPE}    & \textbf{2.063E-01 ± 3.579E-03} & \textbf{1.179E-03 ± 2.180E-05} &          & \textbf{2.034E-01 ± 3.691E-03} & \textbf{1.022E-03 ± 2.248E-05} \\
    \cline{2-5} \cline{6-8}
     & Simple  & MLP   & 2.877E-01 ± 4.549E-03 & 1.753E-03 ± 2.771E-05 & Exte      & 2.987E-01 ± 4.722E-03 & 1.819E-03 ± 2.877E-05 \\
     & GD      & CNN   & 2.786E-01 ± 4.404E-03 & 1.538E-03 ± 2.793E-05 &  GD       & 2.837E-01 ± 4.486E-03 & 1.728E-03 ± 2.733E-05 \\
     &         & LSTM  & 2.771E-01 ± 4.480E-03 & 1.785E-03 ± 2.823E-05 &           & 2.998E-01 ± 4.740E-03 & 1.826E-03 ± 2.887E-05 \\
     &         & GRU   & 2.928E-01 ± 4.629E-03 & 1.783E-03 ± 2.820E-05 &           & 2.933E-01 ± 4.637E-03 & 1.786E-03 ± 2.825E-05 \\
     &         & RBFNN & 3.454E-01 ± 4.764E-03 & 2.276E-03 ± 2.737E-05 &           & 3.837E-01 ± 4.119E-03 & 2.233E-03 ± 3.515E-05 \\
     &         & \textbf{ALPE}    & \textbf{2.196E-01 ± 3.473E-03} & \textbf{1.338E-03 ± 2.116E-05} &           & \textbf{2.278E-01 ± 3.602E-03} & \textbf{1.388E-03 ± 2.194E-05} \\
      \hline
    JPM & Simple & Naive & 1.231E+00 ± 1.947E-02 & 1.166E-02 ± 1.843E-04 & Exte     & 2.001E+00 ± 2.253E-02 & 2.341E-02 ± 2.409E-04 \\
        &        & ARIMA & 5.445E-01 ± 8.609E-03 & 5.155E-03 ± 8.150E-05 &          & 4.346E-01 ± 7.468E-03 & 4.334E-03 ± 6.241E-05  \\
        &        & MLP   & 4.110E-01 ± 6.498E-03 & 3.891E-03 ± 6.152E-05 &          & 3.656E-01 ± 5.781E-03 & 3.462E-03 ± 5.473E-05 \\
        &        & CNN   & 3.704E-01 ± 5.857E-03 & 3.773E-03 ± 4.303E-05 &          & 3.518E-01 ± 5.562E-03 & 3.331E-03 ± 5.266E-05 \\
        &        & LSTM  & 3.667E-01 ± 5.799E-03 & 3.472E-03 ± 5.490E-05 &          & 3.765E-01 ± 5.953E-03 & 3.564E-03 ± 5.636E-05 \\
        &        & GRU   & 3.650E-01 ± 5.771E-03 & 3.455E-03 ± 5.464E-05 &          & 3.699E-01 ± 5.849E-03 & 3.502E-03 ± 5.537E-05 \\
        &        & RBFNN & 2.062E-01 ± 3.108E-03 & 2.700E-03 ± 3.867E-05 &          & 2.943E-01 ± 4.165E-03 & 2.831E-03 ± 3.845E-05 \\
        &        & \textbf{ALPE}    & \textbf{1.340E-01 ± 3.700E-03} & \textbf{1.816E-03 ± 3.503E-05} &          & \textbf{2.496E-01 ± 3.946E-03} & \textbf{2.363E-03 ± 3.736E-05} \\
    \cline{2-5} \cline{6-8} 
     & Simple   & MLP   & 3.925E-01 ± 6.205E-03 & 3.716E-03 ± 5.875E-05 & Exte     & 3.533E-01 ± 5.586E-03 & 3.345E-03 ± 5.289E-05 \\
     & MDI      & CNN   & 3.493E-01 ± 5.523E-03 & 3.307E-03 ± 5.229E-05 &  MDI     & 3.513E-01 ± 5.555E-03 & 3.326E-03 ± 5.259E-05 \\
        &       & LSTM  & 3.644E-01 ± 5.762E-03 & 3.450E-03 ± 5.455E-05 &          & 3.749E-01 ± 5.928E-03 & 3.550E-03 ± 5.612E-05 \\
        &       & GRU   & 3.671E-01 ± 5.804E-03 & 3.475E-03 ± 5.495E-05 &          & 3.721E-01 ± 5.883E-03 & 3.523E-03 ± 5.570E-05 \\
        &       & RBFNN & 2.804E-01 ± 3.169E-03 & 2.598E-03 ± 3.000E-05 &          & 2.349E-01 ± 4.121E-03 & 2.361E-03 ± 3.657E-05 \\
        &       & \textbf{ALPE}    & \textbf{2.231E-01 ± 3.686E-03} & \textbf{2.207E-03 ± 3.490E-05} &          & \textbf{1.844E-01 ± 3.707E-03} & \textbf{1.919E-03 ± 3.509E-05} \\
    \cline{2-5} \cline{6-8}
     & Simple  & MLP   & 3.348E-01 ± 5.294E-03 & 3.170E-03 ± 5.012E-05 & Exte     & 3.500E-01 ± 5.534E-03 & 3.313E-03 ± 5.239E-05 \\
     & GD      & CNN   & 3.509E-01 ± 5.548E-03 & 3.322E-03 ± 5.252E-05 &  GD      & 3.540E-01 ± 5.597E-03 & 3.352E-03 ± 5.299E-05 \\
     &         & LSTM  & 3.613E-01 ± 5.713E-03 & 3.421E-03 ± 5.409E-05 &          & 3.772E-01 ± 5.964E-03 & 3.571E-03 ± 5.646E-05 \\
     &         & GRU   & 3.694E-01 ± 5.841E-03 & 3.497E-03 ± 5.530E-05 &          & 3.720E-01 ± 5.882E-03 & 3.522E-03 ± 5.569E-05 \\
     &         & RBFNN & 3.011E-01 ± 2.663E-03 & 2.861E-03 ± 2.943E-05 &          & 2.966E-01 ± 3.234E-03 & 3.443E-03 ± 2.943E-05 \\
     &         & \textbf{ALPE}    & \textbf{1.948E-01 ± 3.602E-03} & \textbf{1.157E-03 ± 3.410E-05} &          & \textbf{2.390E-01 ± 3.779E-03} & \textbf{2.263E-03 ± 3.578E-05} \\
    \bottomrule
    \end{tabular}
    \end{small}}
            \label{tab:table1}
        \end{minipage}%
        \hspace{0.25\textwidth} 
        \begin{minipage}[t]{0.48\textwidth}
            \centering
            \scalebox{0.42}{
    \begin{small}
    \begin{tabular}{c c r c c c c c}
    \toprule
    \textbf{Stock} & \textbf{Set} & \textbf{Model} & \textbf{RMSE} & \textbf{RRMSE} & \textbf{Set} & \textbf{RMSE} & \textbf{RRMSE} \\
    \hline     
    KEYS & Simple & Naive & 2.185E+00 ± 3.455E-02 & 1.377E-02 ± 2.177E-04 & Exte     & 3.253E+00 ± 5.611E-02 & 2.342E-02 ± 3.342E-04 \\
         &        & ARIMA & 2.627E-01 ± 4.154E-03 & 1.656E-03 ± 2.618E-05 &          & 3.456E-01 ± 3.201E-03 & 2.349E-03 ± 1.300E-05 \\
         &        & MLP   & 2.536E-01 ± 4.010E-03 & 1.598E-03 ± 2.527E-05 &          & 2.570E-01 ± 4.063E-03 & 1.620E-03 ± 2.561E-05 \\
         &        & CNN   & 2.774E-01 ± 4.386E-03 & 1.748E-03 ± 2.764E-05 &          & 2.718E-01 ± 4.298E-03 & 1.713E-03 ± 2.709E-05 \\
         &        & LSTM  & 2.807E-01 ± 4.439E-03 & 1.769E-03 ± 2.798E-05 &          & 2.543E-01 ± 3.310E-03 & 1.719E-03 ± 2.853E-05 \\
         &        & GRU   & 2.782E-01 ± 4.398E-03 & 1.753E-03 ± 2.772E-05 &          & 2.793E-01 ± 4.416E-03 & 1.760E-03 ± 2.783E-05 \\
         &        & RBFNN & 3.555E-01 ± 3.363E-03 & 2.104E-03 ± 3.327E-05 &          & 2.531E-01 ± 3.100E-03 & 1.198E-03 ± 2.366E-05 \\
         &        & \textbf{ALPE}    & \textbf{2.272E-01 ± 3.593E-03} & \textbf{1.432E-03 ± 2.264E-05} &          & \textbf{2.384E-01 ± 3.770E-03} & \textbf{1.103E-03 ± 2.376E-05} \\
    \cline{2-5} \cline{6-8} 
     & Simple   & MLP   & 2.811E-01 ± 4.444E-03 & 1.771E-03 ± 2.801E-05 & Exte     & 2.788E-01 ± 4.409E-03 & 1.357E-03 ± 2.778E-05 \\
     & MDI      & CNN   & 2.702E-01 ± 4.272E-03 & 1.703E-03 ± 2.692E-05 &  MDI     & 2.716E-01 ± 4.295E-03 & 1.712E-03 ± 2.707E-05 \\
     &          & LSTM  & 2.798E-01 ± 4.424E-03 & 1.763E-03 ± 2.788E-05 &          & 2.840E-01 ± 4.490E-03 & 1.790E-03 ± 2.830E-05 \\
     &          & GRU   & 2.807E-01 ± 4.438E-03 & 1.769E-03 ± 2.797E-05 &          & 2.811E-01 ± 4.444E-03 & 1.771E-03 ± 2.801E-05 \\
     &          & RBFNN & 2.704E-01 ± 3.168E-03 & 1.663E-03 ± 1.996E-05 &          & 2.581E-01 ± 4.654E-03 & 1.226E-03 ± 1.938E-05 \\
     &          & \textbf{ALPE}    & \textbf{2.268E-01 ± 3.586E-03} & \textbf{1.429E-03 ± 2.260E-05} &          & \textbf{2.161E-01 ± 3.574E-03} & \textbf{1.025E-03 ± 2.253E-05} \\
    \cline{2-5} \cline{6-8}
     & Simple  & MLP   & 2.853E-01 ± 4.511E-03 & 1.798E-03 ± 2.843E-05 & Exte      & 2.851E-01 ± 4.508E-03 & 1.797E-03 ± 2.841E-05 \\
     & GD      & CNN   & 2.702E-01 ± 4.272E-03 & 1.703E-03 ± 2.692E-05 &  GD       & 2.726E-01 ± 4.310E-03 & 1.718E-03 ± 2.716E-05 \\
     &         & LSTM  & 2.784E-01 ± 4.403E-03 & 1.755E-03 ± 2.775E-05 &           & 2.863E-01 ± 4.527E-03 & 1.701E-03 ± 2.475E-05 \\
     &         & GRU   & 2.788E-01 ± 4.409E-03 & 1.647E-03 ± 2.778E-05 &           & 2.811E-01 ± 4.444E-03 & 1.771E-03 ± 2.801E-05 \\
     &         & RBFNN & 3.737E-01 ± 5.279E-03 & 2.121E-03 ± 3.454E-05 &           & 3.839E-01 ± 3.017E-03 & 2.364E-03 ± 3.998E-05 \\
     &         & \textbf{ALPE}    & \textbf{2.254E-01 ± 3.564E-03} & \textbf{1.420E-03 ± 2.246E-05} &           & \textbf{2.297E-01 ± 3.631E-03} & \textbf{1.447E-03 ± 2.288E-05}\\
     \hline
     KO & Simple& Naive & 2.032E+00 ± 3.213E-02 & 3.606E-02 ± 5.702E-04 & Exte     & 3.943E+00 ± 2.002E-02 & 4.252E-02 ± 3.242E-04 \\
      &        & ARIMA & 2.429E-01 ± 3.841E-03 & 4.311E-03 ± 6.817E-05 &          & 3.340E-01 ± 3.870E-03 & 5.390E-03 ± 3.890E-05  \\
      &        & MLP   & 2.306E-01 ± 3.646E-03 & 4.092E-03 ± 6.470E-05 &          & 2.040E-01 ± 3.225E-03 & 3.620E-03 ± 5.724E-05 \\
      &        & CNN   & 2.774E-01 ± 4.386E-03 & 4.923E-03 ± 7.783E-05 &          & 2.716E-01 ± 4.295E-03 & 4.821E-03 ± 7.622E-05 \\
      &        & LSTM  & 2.807E-01 ± 4.439E-03 & 4.983E-03 ± 7.878E-05 &          & 2.573E-01 ± 3.401E-03 & 5.082E-03 ± 8.035E-05 \\
      &        & GRU   & 2.782E-01 ± 4.398E-03 & 4.937E-03 ± 7.806E-05 &          & 2.793E-01 ± 4.416E-03 & 4.957E-03 ± 7.838E-05 \\
      &        & RBFNN & 3.887E-01 ± 4.100E-03 & 5.926E-03 ± 9.370E-05 &          & 2.478E-01 ± 3.191E-03 & 3.450E-03 ± 6.228E-05 \\
      &        & \textbf{ALPE}    & \textbf{2.262E-01 ± 3.576E-03} & \textbf{4.014E-03 ± 6.347E-05} &          & \textbf{2.066E-01 ± 3.741E-03} & \textbf{3.199E-03 ± 6.640E-05} \\
    \cline{2-5} \cline{6-8} 
     & Simple   & MLP   & 2.236E-01 ± 3.535E-03 & 3.968E-03 ± 6.274E-05 & Exte     & 2.116E-01 ± 3.346E-03 & 3.755E-03 ± 5.938E-05 \\
     & MDI      & CNN   & 2.702E-01 ± 4.272E-03 & 4.795E-03 ± 7.582E-05 &  MDI     & 2.716E-01 ± 4.295E-03 & 4.821E-03 ± 7.622E-05 \\
     &          & LSTM  & 2.798E-01 ± 4.424E-03 & 4.965E-03 ± 7.851E-05 &          & 2.840E-01 ± 4.490E-03 & 5.040E-03 ± 7.968E-05 \\
     &          & GRU   & 2.807E-01 ± 4.438E-03 & 4.981E-03 ± 7.876E-05 &          & 2.794E-01 ± 4.418E-03 & 4.959E-03 ± 7.840E-05 \\
     &          & RBFNN & 2.304E-01 ± 3.168E-03 & 3.556E-03 ± 5.622E-05 &          & 2.199E-01 ± 3.243E-03 & 3.752E-03 ± 5.458E-05 \\
     &          & \textbf{ALPE}    & \textbf{2.133E-01 ± 3.531E-03} & \textbf{3.364E-03 ± 6.267E-05} &          & \textbf{1.976E-01 ± 3.599E-03} & \textbf{3.240E-03 ± 6.388E-05} \\
    \cline{2-5} \cline{6-8}
     & Simple  & MLP   & 2.424E-01 ± 3.358E-03 & 4.370E-03 ± 5.960E-05 & Exte      & 2.210E-01 ± 3.494E-03 & 3.922E-03 ± 6.201E-05 \\
     & GD      & CNN   & 2.702E-01 ± 4.272E-03 & 4.795E-03 ± 7.582E-05 &  GD       & 2.726E-01 ± 4.310E-03 & 4.838E-03 ± 7.649E-05 \\
     &         & LSTM  & 2.784E-01 ± 4.403E-03 & 4.942E-03 ± 7.814E-05 &           & 2.620E-01 ± 2.009E-03 & 5.082E-03 ± 8.035E-05 \\
     &         & GRU   & 2.788E-01 ± 4.409E-03 & 4.949E-03 ± 7.824E-05 &           & 2.811E-01 ± 4.444E-03 & 4.988E-03 ± 7.887E-05 \\
     &         & RBFNN & 3.443E-01 ± 4.363E-03 & 5.848E-03 ± 8.430E-05 &           & 3.398E-01 ± 2.992E-03 & 4.406E-03 ± 8.430E-05 \\
     &         & \textbf{ALPE}    & \textbf{2.261E-01 ± 3.576E-03} & \textbf{4.014E-03 ± 6.346E-05} &           & \textbf{2.095E-01 ± 3.628E-03} & \textbf{3.073E-03 ± 6.439E-05} \\
      \hline
    LDOS & Simple & Naive & 2.279E+00 ± 3.603E-02 & 2.586E-02 ± 4.088E-04 & Exte     & 3.444E+00 ± 2.353E-02 & 3.302E-02 ± 5.325E-04 \\
         &        & ARIMA & 2.309E-01 ± 3.652E-03 & 2.620E-03 ± 4.143E-05 &          & 3.580E-01 ± 4.990E-03 & 3.701E-03 ± 5.102E-05  \\
         &        & MLP   & 2.073E-01 ± 3.278E-03 & 2.352E-03 ± 3.719E-05 &          & 2.110E-01 ± 3.336E-03 & 2.394E-03 ± 3.785E-05 \\
         &        & CNN   & 2.640E-01 ± 4.175E-03 & 2.996E-03 ± 4.737E-05 &          & 2.588E-01 ± 4.092E-03 & 2.936E-03 ± 4.642E-05 \\
         &        & LSTM  & 2.037E-01 ± 3.220E-03 & 2.394E-03 ± 3.654E-05 &          & 2.113E-01 ± 3.341E-03 & 2.398E-03 ± 3.791E-05 \\
         &        & GRU   & 2.452E-01 ± 3.877E-03 & 2.782E-03 ± 4.399E-05 &          & 2.576E-01 ± 4.073E-03 & 2.923E-03 ± 4.621E-05 \\
         &        & RBFNN & 3.444E-01 ± 3.838E-03 & 3.789E-03 ± 5.990E-05 &          & 1.913E-01 ± 2.324E-03 & 2.388E-03 ± 4.576E-05 \\
         &        & \textbf{ALPE}    & \textbf{2.069E-01 ± 3.271E-03} & \textbf{2.347E-03 ± 3.711E-05} &          & \textbf{1.548E-01 ± 3.396E-03} & \textbf{2.037E-03 ± 3.853E-05} \\
    \cline{2-5} \cline{6-8} 
     & Simple   & MLP   & 2.391E-01 ± 3.780E-03 & 2.712E-03 ± 4.289E-05 & Exte     & 2.108E-01 ± 3.332E-03 & 2.391E-03 ± 3.781E-05 \\
     & MDI      & CNN   & 2.572E-01 ± 4.067E-03 & 2.919E-03 ± 4.615E-05 &  MDI     & 2.586E-01 ± 4.089E-03 & 2.934E-03 ± 4.639E-05 \\
     &          & LSTM  & 2.026E-01 ± 3.203E-03 & 2.299E-03 ± 3.634E-05 &          & 2.076E-01 ± 3.283E-03 & 2.356E-03 ± 3.725E-05 \\
     &          & GRU   & 2.477E-01 ± 3.917E-03 & 2.811E-03 ± 4.444E-05 &          & 2.464E-01 ± 3.896E-03 & 2.796E-03 ± 4.421E-05 \\
     &          & RBFNN & 2.204E-01 ± 3.168E-03 & 2.373E-03 ± 3.594E-05 &          & 1.838E-01 ± 4.127E-03 & 2.188E-03 ± 4.673E-05 \\
     &          & \textbf{ALPE}    & \textbf{2.030E-01 ± 3.209E-03} & \textbf{2.103E-03 ± 3.641E-05} &          & \textbf{1.785E-01 ± 2.823E-03} & \textbf{2.026E-03 ± 3.203E-05} \\
    \cline{2-5} \cline{6-8}
     & Simple  & MLP   & 2.447E-01 ± 3.174E-03 & 2.578E-03 ± 3.601E-05 & Exte      & 2.109E-01 ± 2.461E-03 & 2.207E-03 ± 4.320E-05 \\
     & GD      & CNN   & 2.573E-01 ± 4.068E-03 & 2.919E-03 ± 4.615E-05 &  GD       & 2.595E-01 ± 4.103E-03 & 2.944E-03 ± 4.655E-05 \\
     &         & LSTM  & 2.208E-01 ± 3.175E-03 & 2.278E-03 ± 3.603E-05 &           & 2.090E-01 ± 3.304E-03 & 2.371E-03 ± 3.749E-05 \\
     &         & GRU   & 2.459E-01 ± 3.887E-03 & 2.790E-03 ± 4.411E-05 &           & 2.644E-01 ± 4.181E-03 & 3.000E-03 ± 4.744E-05 \\
     &         & RBFNN & 3.399E-01 ± 2.783E-03 & 3.878E-03 ± 3.838E-05 &           & 3.787E-01 ± 2.092E-03 & 3.688E-03 ± 4.900E-05 \\
     &         & \textbf{ALPE}    & \textbf{2.041E-01 ± 3.227E-03} & \textbf{2.116E-03 ± 3.662E-05} &           & \textbf{1.833E-01 ± 2.898E-03} & \textbf{2.080E-03 ± 3.288E-05} \\
     \hline
    LHX& Simple& Naive & 2.629E+00 ± 4.157E-02 & 1.250E-02 ± 1.976E-04 & Exte     & 3.450E+00 ± 4.335E-02 & 1.632E-02 ± 4.132E-04 \\
      &        & ARIMA & 2.395E-01 ± 3.786E-03 & 1.138E-03 ± 1.799E-05 &          & 3.340E-01 ± 4.333E-03 & 1.550E-03 ± 2.250E-05  \\
      &        & MLP   & 3.532E-01 ± 5.584E-03 & 1.678E-03 ± 2.654E-05 &          & 4.117E-01 ± 6.510E-03 & 1.957E-03 ± 3.094E-05 \\
      &        & CNN   & 3.044E-01 ± 4.813E-03 & 1.447E-03 ± 2.287E-05 &          & 2.983E-01 ± 4.717E-03 & 1.487E-03 ± 3.346E-05 \\
      &        & LSTM  & 2.749E-01 ± 4.347E-03 & 1.307E-03 ± 2.066E-05 &          & 2.779E-01 ± 4.393E-03 & 1.320E-03 ± 2.088E-05 \\
      &        & GRU   & 2.699E-01 ± 4.268E-03 & 1.283E-03 ± 2.028E-05 &          & 2.726E-01 ± 4.311E-03 & 1.296E-03 ± 2.049E-05 \\
      &        & RBFNN & 2.219E-01 ± 3.509E-03 & 1.055E-03 ± 1.667E-05 &          & 1.990E-01 ± 2.639E-03 & 9.277E-04 ± 3.827E-05 \\
      &        & \textbf{ALPE}    & \textbf{1.549E-01 ± 2.449E-03} & \textbf{7.361E-04 ± 1.164E-05} &          & \textbf{1.630E-01 ± 2.578E-03} & \textbf{7.749E-04 ± 1.225E-05} \\
    \cline{2-5} \cline{6-8} 
     & Simple   & MLP   & 3.190E-01 ± 5.044E-03 & 1.516E-03 ± 2.397E-05 & Exte     & 4.253E-01 ± 6.725E-03 & 2.021E-03 ± 3.196E-05 \\
     & MDI      & CNN   & 3.154E-01 ± 4.987E-03 & 1.499E-03 ± 2.370E-05 &  MDI     & 3.160E-01 ± 4.996E-03 & 1.502E-03 ± 2.374E-05 \\
     &          & LSTM  & 2.717E-01 ± 4.295E-03 & 1.291E-03 ± 2.041E-05 &          & 2.798E-01 ± 4.424E-03 & 1.330E-03 ± 2.102E-05 \\
     &          & GRU   & 2.722E-01 ± 4.304E-03 & 1.294E-03 ± 2.046E-05 &          & 2.709E-01 ± 4.283E-03 & 1.287E-03 ± 2.035E-05 \\
     &          & RBFNN & 2.004E-01 ± 3.168E-03 & 9.522E-04 ± 1.506E-05 &          & 1.775E-01 ± 2.806E-03 & 8.434E-04 ± 1.333E-05 \\
     &          & \textbf{ALPE}    & \textbf{1.442E-01 ± 2.279E-03} & \textbf{6.851E-04 ± 1.083E-05} &          & \textbf{1.457E-01 ± 2.303E-03} & \textbf{6.922E-04 ± 1.094E-05} \\
    \cline{2-5} \cline{6-8}
     & Simple  & MLP   & 3.215E-01 ± 5.084E-03 & 1.528E-03 ± 2.416E-05 & Exte      & 3.334E-01 ± 5.271E-03 & 1.584E-03 ± 2.505E-05 \\
     & GD      & CNN   & 3.160E-01 ± 4.996E-03 & 1.502E-03 ± 2.374E-05 &  GD       & 2.984E-01 ± 4.718E-03 & 1.566E-03 ± 3.008E-05 \\
     &         & LSTM  & 2.702E-01 ± 4.273E-03 & 1.284E-03 ± 2.031E-05 &           & 2.765E-01 ± 4.371E-03 & 1.314E-03 ± 2.077E-05 \\
     &         & GRU   & 2.709E-01 ± 4.283E-03 & 1.287E-03 ± 2.035E-05 &           & 2.323E-01 ± 2.990E-03 & 1.324E-03 ± 2.131E-05 \\
     &         & RBFNN & 2.219E-01 ± 3.509E-03 & 1.055E-03 ± 1.667E-05 &           & 2.466E-01 ± 3.318E-03 & 9.423E-04 ± 1.435E-05 \\
     &         & \textbf{ALPE}    & \textbf{1.501E-01 ± 2.374E-03} & \textbf{7.136E-04 ± 1.128E-05} &           & \textbf{1.530E-01 ± 2.420E-03} & \textbf{7.272E-04 ± 1.150E-05} \\
    \bottomrule
    \end{tabular}
    \end{small}}
            \label{tab:table_6}
        \end{minipage}
    \end{table}
\end{landscape}


\begin{landscape}
    \fancyhf{}  
    \fancyfoot[R]{\rotatebox{90}{\thepage}}  
    \thispagestyle{fancy}  

    \begin{table}[!hbtp]
        \centering
        \caption{RMSE and RRMSE scores for LMT, LUV, LVS, MAA, MAS, MCHP, MCK, and MDT.}
        \hspace*{-5cm} 
        \begin{minipage}[t]{0.48\textwidth}
            \centering
            \scalebox{0.42}{
    \begin{small}
    \begin{tabular}{c c r c c c c c}
    \toprule
    \textbf{Stock} & \textbf{Set} & \textbf{Model} & \textbf{RMSE} & \textbf{RRMSE} & \textbf{Set} & \textbf{RMSE} & \textbf{RRMSE} \\
    \hline     
    LMT & Simple & Naive & 4.232E+00 ± 6.582E-02 & 1.087E-02 ± 1.718E-04 & Exte     & 5.300E+00 ± 4.432E-02 & 1.144E-02 ± 2.432E-04 \\
        &        & ARIMA & 2.469E-01 ± 3.903E-03 & 6.338E-04 ± 1.002E-05 &          & 3.347E-01 ± 5.340E-03 & 6.349E-04 ± 2.334E-05  \\
        &        & MLP   & 3.041E-01 ± 4.808E-03 & 7.807E-04 ± 1.234E-05 &          & 2.932E-01 ± 4.635E-03 & 7.527E-04 ± 1.190E-05 \\
        &        & CNN   & 2.780E-01 ± 4.395E-03 & 7.136E-04 ± 1.128E-05 &          & 2.739E-01 ± 4.331E-03 & 7.033E-04 ± 1.112E-05 \\
        &        & LSTM  & 2.839E-01 ± 4.488E-03 & 7.288E-04 ± 1.152E-05 &          & 2.864E-01 ± 4.529E-03 & 7.354E-04 ± 1.163E-05 \\
        &        & GRU   & 2.783E-01 ± 4.400E-03 & 7.144E-04 ± 1.130E-05 &          & 2.796E-01 ± 4.421E-03 & 7.179E-04 ± 1.135E-05 \\
        &        & RBFNN & 1.886E-01 ± 2.111E-03 & 4.976E-04 ± 6.131E-06 &          & 1.947E-01 ± 2.920E-03 & 4.742E-04 ± 7.498E-06 \\
        &        & \textbf{ALPE}    & \textbf{1.533E-01 ± 2.423E-03} & \textbf{3.935E-04 ± 6.222E-06} &          & \textbf{1.611E-01 ± 3.021E-03} & \textbf{4.206E-04 ± 7.757E-06} \\
    \cline{2-5} \cline{6-8} 
     & Simple   & MLP   & 3.071E-01 ± 4.855E-03 & 7.884E-04 ± 1.247E-05 & Exte     & 2.969E-01 ± 4.694E-03 & 7.622E-04 ± 1.205E-05 \\
     & MDI      & CNN   & 2.732E-01 ± 4.320E-03 & 7.014E-04 ± 1.109E-05 &  MDI     & 2.813E-01 ± 3.009E-03 & 6.821E-04 ± 1.231E-05 \\
     &          & LSTM  & 2.809E-01 ± 4.442E-03 & 7.213E-04 ± 1.140E-05 &          & 2.881E-01 ± 4.555E-03 & 7.397E-04 ± 1.170E-05 \\
     &          & GRU   & 2.800E-01 ± 4.427E-03 & 7.189E-04 ± 1.137E-05 &          & 2.794E-01 ± 4.418E-03 & 7.174E-04 ± 1.134E-05 \\
     &          & RBFNN & 2.004E-01 ± 3.168E-03 & 5.144E-04 ± 8.134E-06 &          & 1.847E-01 ± 2.920E-03 & 5.001E-04 ± 6.587E-06 \\
     &          & \textbf{ALPE}    & \textbf{1.494E-01 ± 2.362E-03} & \textbf{3.799E-04 ± 3.346E-06} &          & \textbf{1.762E-01 ± 2.786E-03} & \textbf{4.524E-04 ± 7.153E-06} \\
    \cline{2-5} \cline{6-8}
     & Simple  & MLP   & 3.127E-01 ± 4.944E-03 & 8.029E-04 ± 1.269E-05 & Exte      & 2.985E-01 ± 4.719E-03 & 7.663E-04 ± 1.212E-05 \\
     & GD      & CNN   & 2.739E-01 ± 4.330E-03 & 7.031E-04 ± 1.112E-05 &  GD       & 2.749E-01 ± 4.347E-03 & 7.058E-04 ± 1.116E-05 \\
     &         & LSTM  & 2.811E-01 ± 4.445E-03 & 7.218E-04 ± 1.141E-05 &           & 2.856E-01 ± 4.516E-03 & 7.333E-04 ± 1.160E-05 \\
     &         & GRU   & 2.789E-01 ± 4.410E-03 & 7.161E-04 ± 1.132E-05 &           & 2.807E-01 ± 4.439E-03 & 7.208E-04 ± 1.140E-05 \\
     &         & RBFNN & 2.261E-01 ± 3.575E-03 & 5.783E-04 ± 8.203E-06 &           & 2.261E-01 ± 3.575E-03 & 5.805E-04 ± 9.178E-06 \\
     &         & \textbf{ALPE}    & \textbf{1.557E-01 ± 2.461E-03} & \textbf{3.996E-04 ± 6.319E-06} &           & \textbf{1.859E-01 ± 2.940E-03} & \textbf{4.773E-04 ± 7.547E-06}\\
     \hline
      LUV & Simple & Naive & 2.663E+00 ± 4.211E-02 & 8.636E-02 ± 1.366E-03 & Exte     & 3.456E+00 ± 4.501E-02 & 7.425E-02 ± 2.125E-03 \\
        &        & ARIMA & 2.469E-01 ± 3.903E-03 & 8.004E-03 ± 1.266E-04 &          & 3.847E-01 ± 2.554E-03 & 8.342E-03 ± 2.889E-04  \\
        &        & MLP   & 5.220E-01 ± 8.253E-03 & 1.693E-02 ± 2.676E-04 &          & 5.966E-01 ± 9.433E-03 & 1.935E-02 ± 3.059E-04 \\
        &        & CNN   & 2.780E-01 ± 4.395E-03 & 9.013E-03 ± 1.425E-04 &          & 2.739E-01 ± 4.331E-03 & 8.882E-03 ± 1.404E-04 \\
        &        & LSTM  & 2.839E-01 ± 4.488E-03 & 9.204E-03 ± 1.455E-04 &          & 2.864E-01 ± 4.529E-03 & 9.287E-03 ± 1.468E-04 \\
        &        & GRU   & 2.783E-01 ± 4.400E-03 & 9.023E-03 ± 1.427E-04 &          & 2.796E-01 ± 4.421E-03 & 9.067E-03 ± 1.434E-04 \\
        &        & RBFNN & 1.885E-01 ± 2.187E-03 & 6.873E-03 ± 8.098E-05 &          & 2.404E-01 ± 3.168E-03 & 6.497E-03 ± 1.027E-04 \\
        &        & \textbf{ALPE}    & \textbf{1.821E-01 ± 3.987E-03} & \textbf{6.076E-03 ± 1.293E-04} &          & \textbf{2.165E-01 ± 3.423E-03} & \textbf{6.020E-03 ± 1.110E-04} \\
    \cline{2-5} \cline{6-8} 
     & Simple   & MLP   & 5.332E-01 ± 8.430E-03 & 1.729E-02 ± 2.734E-04 & Exte     & 5.368E-01 ± 8.487E-03 & 1.741E-02 ± 2.752E-04 \\
     & MDI      & CNN   & 2.732E-01 ± 4.320E-03 & 8.858E-03 ± 1.401E-04 &  MDI     & 2.733E-01 ± 4.330E-03 & 8.880E-03 ± 1.400E-04 \\
     &          & LSTM  & 2.809E-01 ± 4.442E-03 & 9.109E-03 ± 1.440E-04 &          & 2.881E-01 ± 4.555E-03 & 9.342E-03 ± 1.477E-04 \\
     &          & GRU   & 2.800E-01 ± 4.427E-03 & 9.079E-03 ± 1.436E-04 &          & 2.794E-01 ± 4.418E-03 & 9.060E-03 ± 1.433E-04 \\
     &          & RBFNN & 3.500E-01 ± 2.343E-03 & 1.083E-02 ± 1.712E-04 &          & 2.604E-01 ± 2.208E-03 & 6.496E-03 ± 1.027E-04 \\
     &          & \textbf{ALPE}    & \textbf{2.486E-01 ± 3.930E-03} & \textbf{6.060E-03 ± 1.274E-04} &          & \textbf{2.272E-01 ± 3.593E-03} & \textbf{6.368E-03 ± 1.165E-04} \\
    \cline{2-5} \cline{6-8}
     & Simple  & MLP   & 5.457E-01 ± 8.628E-03 & 1.769E-02 ± 2.798E-04 & Exte      & 4.591E-01 ± 7.259E-03 & 1.489E-02 ± 2.354E-04 \\
     & GD      & CNN   & 2.739E-01 ± 4.330E-03 & 8.880E-03 ± 1.404E-04 &  GD       & 2.749E-01 ± 4.347E-03 & 8.914E-03 ± 1.409E-04 \\
     &         & LSTM  & 2.811E-01 ± 4.445E-03 & 9.116E-03 ± 1.441E-04 &           & 2.856E-01 ± 4.516E-03 & 9.261E-03 ± 1.464E-04 \\
     &         & GRU   & 2.789E-01 ± 4.410E-03 & 9.043E-03 ± 1.430E-04 &           & 2.807E-01 ± 4.439E-03 & 9.103E-03 ± 1.439E-04 \\
     &         & RBFNN & 3.670E-01 ± 2.730E-03 & 1.123E-02 ± 1.893E-04 &           & 3.460E-01 ± 3.939E-03 & 1.145E-02 ± 1.939E-04 \\
     &         & \textbf{ALPE}    & \textbf{2.289E-01 ± 3.619E-03} & \textbf{7.421E-03 ± 1.173E-04} &           & \textbf{2.350E-01 ± 3.715E-03} & \textbf{7.619E-03 ± 1.205E-04} \\
     \hline
    LVS & Simple & Naive & 1.732E+00 ± 2.739E-02 & 4.575E-02 ± 7.233E-04 & Exte     & 2.009E+00 ± 3.123E-02 & 5.801E-02 ± 6.001E-04 \\
        &        & ARIMA & 3.152E-01 ± 4.984E-03 & 8.323E-03 ± 1.316E-04 &          & 3.001E-01 ± 5.990E-03 & 7.810E-03 ± 2.901E-04  \\
        &        & MLP   & 2.664E-01 ± 4.213E-03 & 7.036E-03 ± 1.112E-04 &          & 2.731E-01 ± 3.330E-03 & 6.541E-03 ± 1.234E-04 \\
        &        & CNN   & 2.608E-01 ± 4.123E-03 & 6.886E-03 ± 1.089E-04 &          & 2.570E-01 ± 4.064E-03 & 6.786E-03 ± 1.073E-04 \\
        &        & LSTM  & 2.665E-01 ± 4.214E-03 & 7.037E-03 ± 1.113E-04 &          & 2.690E-01 ± 4.253E-03 & 7.103E-03 ± 1.123E-04 \\
        &        & GRU   & 2.611E-01 ± 4.128E-03 & 6.895E-03 ± 1.090E-04 &          & 2.624E-01 ± 4.148E-03 & 6.928E-03 ± 1.095E-04 \\
        &        & RBFNN & 2.119E-01 ± 3.443E-03 & 6.136E-03 ± 8.121E-05 &          & 2.673E-01 ± 3.327E-03 & 5.998E-03 ± 6.030E-05 \\
        &        & \textbf{ALPE}    & \textbf{1.914E-01 ± 3.659E-03} & \textbf{5.110E-03 ± 9.661E-05} &          & \textbf{2.076E-01 ± 3.756E-03} & \textbf{4.273E-03 ± 9.919E-05} \\
    \cline{2-5} \cline{6-8} 
     & Simple   & MLP   & 2.578E-01 ± 4.076E-03 & 6.808E-03 ± 1.076E-04 & Exte     & 2.712E-01 ± 3.542E-03 & 7.500E-03 ± 1.240E-04 \\
     & MDI      & CNN   & 2.564E-01 ± 4.054E-03 & 6.770E-03 ± 1.071E-04 &  MDI     & 2.462E-01 ± 3.893E-03 & 6.501E-03 ± 1.028E-04 \\
     &          & LSTM  & 2.638E-01 ± 4.171E-03 & 6.966E-03 ± 1.101E-04 &          & 2.705E-01 ± 4.278E-03 & 7.144E-03 ± 1.130E-04 \\
     &          & GRU   & 2.629E-01 ± 4.157E-03 & 6.942E-03 ± 1.098E-04 &          & 2.621E-01 ± 4.145E-03 & 6.922E-03 ± 1.094E-04 \\
     &          & RBFNN & 2.804E-01 ± 3.168E-03 & 5.291E-03 ± 8.365E-05 &          & 2.347E-01 ± 2.920E-03 & 5.742E-03 ± 7.498E-05 \\
     &          & \textbf{ALPE}    & \textbf{2.001E-01 ± 3.638E-03} & \textbf{4.076E-03 ± 9.607E-05} &          & \textbf{2.008E-01 ± 3.601E-03} & \textbf{4.014E-03 ± 9.509E-05} \\
    \cline{2-5} \cline{6-8}
     & Simple  & MLP   & 2.934E-01 ± 4.639E-03 & 7.748E-03 ± 1.225E-04 & Exte      & 3.073E-01 ± 4.859E-03 & 8.115E-03 ± 1.283E-04 \\
     & GD      & CNN   & 2.570E-01 ± 4.064E-03 & 6.786E-03 ± 1.073E-04 &  GD       & 2.580E-01 ± 4.079E-03 & 6.812E-03 ± 1.077E-04 \\
     &         & LSTM  & 2.640E-01 ± 4.174E-03 & 6.971E-03 ± 1.102E-04 &           & 2.683E-01 ± 4.242E-03 & 7.084E-03 ± 1.120E-04 \\
     &         & GRU   & 2.618E-01 ± 4.139E-03 & 6.912E-03 ± 1.093E-04 &           & 2.634E-01 ± 4.165E-03 & 6.956E-03 ± 1.100E-04 \\
     &         & RBFNN & 3.646E-01 ± 2.839E-03 & 7.009E-03 ± 2.783E-04 &           & 3.776E-01 ± 3.030E-03 & 7.003E-03 ± 1.938E-04 \\
     &         & \textbf{ALPE}    & \textbf{2.282E-01 ± 3.609E-03} & \textbf{6.026E-03 ± 9.529E-05} &           & \textbf{2.346E-01 ± 3.709E-03} & \textbf{6.194E-03 ± 9.793E-05} \\
     \hline
    MAA & Simple & Naive & 2.427E+00 ± 3.837E-02 & 1.551E-02 ± 2.452E-04 & Exte     & 3.944E+00 ± 4.336E-02 & 2.901E-02 ± 3.051E-04 \\
        &        & ARIMA & 2.309E-01 ± 3.650E-03 & 1.475E-03 ± 2.333E-05 &          & 3.402E-01 ± 4.303E-03 & 1.989E-03 ± 3.010E-05  \\
        &        & MLP   & 3.315E-01 ± 5.242E-03 & 2.119E-03 ± 3.350E-05 &          & 4.225E-01 ± 6.681E-03 & 2.700E-03 ± 4.270E-05 \\
        &        & CNN   & 3.131E-01 ± 4.950E-03 & 2.001E-03 ± 3.163E-05 &          & 2.970E-01 ± 4.697E-03 & 1.898E-03 ± 3.002E-05 \\
        &        & LSTM  & 2.665E-01 ± 4.214E-03 & 1.703E-03 ± 2.693E-05 &          & 2.690E-01 ± 4.253E-03 & 1.719E-03 ± 2.718E-05 \\
        &        & GRU   & 2.611E-01 ± 4.128E-03 & 1.669E-03 ± 2.638E-05 &          & 2.624E-01 ± 4.148E-03 & 1.677E-03 ± 2.651E-05 \\
        &        & RBFNN & 1.809E-01 ± 4.129E-03 & 1.501E-03 ± 2.776E-05 &          & 2.876E-01 ± 3.112E-03 & 1.438E-03 ± 2.763E-05 \\
        &        & \textbf{ALPE}    & \textbf{1.722E-01 ± 3.672E-03} & \textbf{1.284E-03 ± 2.347E-05} &          & \textbf{1.869E-01 ± 3.746E-03} & \textbf{1.014E-03 ± 2.394E-05} \\
    \cline{2-5} \cline{6-8} 
     & Simple   & MLP   & 4.010E-01 ± 6.341E-03 & 2.563E-03 ± 4.052E-05 & Exte     & 4.410E-01 ± 6.972E-03 & 2.818E-03 ± 4.456E-05 \\
     & MDI      & CNN   & 3.228E-01 ± 5.763E-03 & 1.986E-03 ± 3.140E-05 &  MDI     & 2.580E-01 ± 4.079E-03 & 1.649E-03 ± 2.607E-05 \\
     &          & LSTM  & 2.638E-01 ± 4.171E-03 & 1.686E-03 ± 2.666E-05 &          & 2.705E-01 ± 4.278E-03 & 1.729E-03 ± 2.734E-05 \\
     &          & GRU   & 2.629E-01 ± 4.157E-03 & 1.680E-03 ± 2.656E-05 &          & 2.621E-01 ± 4.145E-03 & 1.675E-03 ± 2.649E-05 \\
     &          & RBFNN & 2.404E-01 ± 3.168E-03 & 1.280E-03 ± 2.025E-05 &          & 2.267E-01 ± 3.066E-03 & 1.473E-03 ± 1.750E-05 \\
     &          & \textbf{ALPE}    & \textbf{2.117E-01 ± 3.664E-03} & \textbf{1.081E-03 ± 2.341E-05} &          & \textbf{1.930E-01 ± 3.610E-03} & \textbf{1.159E-03 ± 2.307E-05} \\
    \cline{2-5} \cline{6-8}
     & Simple  & MLP   & 4.225E-01 ± 6.681E-03 & 2.700E-03 ± 4.270E-05 & Exte      & 4.425E-01 ± 5.758E-03 & 2.746E-03 ± 3.032E-05 \\
     & GD      & CNN   & 3.117E-01 ± 3.887E-03 & 1.986E-03 ± 3.141E-05 &  GD       & 2.970E-01 ± 4.697E-03 & 1.898E-03 ± 3.002E-05 \\
     &         & LSTM  & 2.640E-01 ± 4.174E-03 & 1.687E-03 ± 2.668E-05 &           & 2.683E-01 ± 4.242E-03 & 1.715E-03 ± 2.711E-05 \\
     &         & GRU   & 2.618E-01 ± 4.139E-03 & 1.673E-03 ± 2.645E-05 &           & 2.634E-01 ± 4.165E-03 & 1.684E-03 ± 2.662E-05 \\
     &         & RBFNN & 2.504E-01 ± 3.168E-03 & 1.580E-03 ± 2.025E-05 &           & 3.551E-01 ± 4.453E-03 & 2.134E-03 ± 3.374E-05 \\
     &         & \textbf{ALPE}    & \textbf{2.302E-01 ± 3.640E-03} & \textbf{1.171E-03 ± 2.327E-05} &           & \textbf{2.348E-01 ± 3.713E-03} & \textbf{1.501E-03 ± 2.373E-05} \\
    \bottomrule
    \end{tabular}
    \end{small}}
            \label{tab:table1}
        \end{minipage}%
        \hspace{0.25\textwidth} 
        \begin{minipage}[t]{0.48\textwidth}
            \centering
            \scalebox{0.42}{
    \begin{small}
    \begin{tabular}{c c r c c c c c}
    \toprule
    \textbf{Stock} & \textbf{Set} & \textbf{Model} & \textbf{RMSE} & \textbf{RRMSE} & \textbf{Set} & \textbf{RMSE} & \textbf{RRMSE} \\
    \hline    
    MAS & Simple & Naive & 1.288E+00 ± 2.036E-02 & 2.729E-02 ± 4.315E-04 & Exte     & 2.402E+00 ± 3.873E-02 & 3.303E-02 ± 5.104E-04 \\
        &        & ARIMA & 3.623E-01 ± 5.728E-03 & 7.679E-03 ± 1.214E-04 &          & 4.009E-01 ± 4.474E-03 & 6.553E-03 ± 2.009E-04  \\
        &        & MLP   & 2.807E-01 ± 4.439E-03 & 5.950E-03 ± 9.408E-05 &          & 2.624E-01 ± 4.149E-03 & 5.562E-03 ± 8.795E-05 \\
        &        & CNN   & 2.687E-01 ± 4.248E-03 & 5.695E-03 ± 9.004E-05 &          & 2.783E-01 ± 4.400E-03 & 5.899E-03 ± 9.327E-05 \\
        &        & LSTM  & 2.293E-01 ± 3.615E-03 & 4.860E-03 ± 7.684E-05 &          & 2.326E-01 ± 3.677E-03 & 4.930E-03 ± 7.795E-05 \\
        &        & GRU   & 2.982E-01 ± 4.715E-03 & 6.321E-03 ± 9.994E-05 &          & 2.873E-01 ± 4.543E-03 & 6.090E-03 ± 9.629E-05 \\
        &        & RBFNN & 1.877E-01 ± 4.191E-03 & 4.122E-03 ± 6.518E-05 &          & 1.453E-01 ± 2.298E-03 & 4.080E-03 ± 4.870E-05 \\
        &        & \textbf{ALPE}    & \textbf{1.300E-01 ± 3.637E-03} & \textbf{3.876E-03 ± 7.709E-05} &          & \textbf{1.161E-01 ± 3.416E-03} & \textbf{3.580E-03 ± 7.241E-05} \\
    \cline{2-5} \cline{6-8} 
     & Simple   & MLP   & 2.789E-01 ± 4.409E-03 & 5.910E-03 ± 9.345E-05 & Exte     & 2.799E-01 ± 4.426E-03 & 5.933E-03 ± 9.381E-05 \\
     & MDI      & CNN   & 2.979E-01 ± 4.710E-03 & 6.313E-03 ± 9.982E-05 &  MDI     & 3.154E-01 ± 4.986E-03 & 6.684E-03 ± 1.057E-04 \\
     &          & LSTM  & 2.304E-01 ± 3.643E-03 & 4.883E-03 ± 7.721E-05 &          & 2.331E-01 ± 3.685E-03 & 4.940E-03 ± 7.811E-05 \\
     &          & GRU   & 2.897E-01 ± 4.581E-03 & 6.141E-03 ± 9.710E-05 &          & 2.970E-01 ± 4.697E-03 & 6.296E-03 ± 9.955E-05 \\
     &          & RBFNN & 2.004E-01 ± 3.168E-03 & 4.247E-03 ± 6.715E-05 &          & 2.579E-01 ± 2.496E-03 & 4.346E-03 ± 5.291E-05 \\
     &          & \textbf{ALPE}    & \textbf{1.552E-01 ± 2.453E-03} & \textbf{3.289E-03 ± 5.200E-05} &          & \textbf{1.782E-01 ± 3.292E-03} & \textbf{3.414E-03 ± 6.979E-05} \\
    \cline{2-5} \cline{6-8}
     & Simple  & MLP   & 2.855E-01 ± 4.514E-03 & 6.051E-03 ± 9.568E-05 & Exte      & 2.861E-01 ± 4.523E-03 & 6.063E-03 ± 9.587E-05 \\
     & GD      & CNN   & 2.914E-01 ± 4.608E-03 & 6.177E-03 ± 9.767E-05 &  GD       & 2.493E-01 ± 3.941E-03 & 5.283E-03 ± 8.354E-05 \\
     &         & LSTM  & 2.296E-01 ± 3.630E-03 & 4.866E-03 ± 7.694E-05 &           & 2.528E-01 ± 3.997E-03 & 5.358E-03 ± 8.472E-05 \\
     &         & GRU   & 3.161E-01 ± 4.998E-03 & 6.700E-03 ± 1.059E-04 &           & 3.121E-01 ± 4.935E-03 & 6.616E-03 ± 1.046E-04 \\
     &         & RBFNN & 1.356E-01 ± 1.670E-03 & 3.038E-03 ± 3.539E-05 &           & 2.736E-01 ± 3.119E-03 & 4.323E-03 ± 3.730E-05 \\
     &         & \textbf{ALPE}    & \textbf{1.025E-01 ± 2.411E-03} & \textbf{2.232E-03 ± 5.110E-05} &           & \textbf{1.921E-01 ± 3.353E-03} & \textbf{3.495E-03 ± 7.107E-05}\\
     \hline
     MCHP& Simple & Naive & 1.627E+00 ± 2.573E-02 & 2.618E-02 ± 4.140E-04 & Exte     & 1.352E+00 ± 3.241E-02 & 3.234E-02 ± 5.998E-04 \\
        &        & ARIMA & 3.001E-01 ± 4.746E-03 & 4.829E-03 ± 7.635E-05 &          & 4.382E-01 ± 5.009E-03 & 5.435E-03 ± 6.134E-05  \\
        &        & MLP   & 2.321E+00 ± 1.614E-02 & 1.642E-02 ± 2.597E-04 &          & 1.134E+00 ± 1.793E-02 & 1.825E-02 ± 2.885E-04 \\
        &        & CNN   & 9.955E-01 ± 1.574E-02 & 1.602E-02 ± 2.532E-04 &          & 4.292E-01 ± 6.786E-03 & 6.905E-03 ± 1.092E-04 \\
        &        & LSTM  & 4.521E-01 ± 7.148E-03 & 7.273E-03 ± 1.150E-04 &          & 4.562E-01 ± 7.212E-03 & 7.339E-03 ± 1.160E-04 \\
        &        & GRU   & 4.379E-01 ± 6.924E-03 & 7.045E-03 ± 1.114E-04 &          & 4.299E-01 ± 6.797E-03 & 6.916E-03 ± 1.094E-04 \\
        &        & RBFNN & 2.774E-01 ± 2.847E-03 & 3.392E-03 ± 3.736E-05 &          & 2.504E-01 ± 3.168E-03 & 3.223E-03 ± 5.097E-05 \\
        &        & \textbf{ALPE}    & \textbf{1.405E-01 ± 3.645E-03} & \textbf{2.709E-03 ± 5.865E-05} &          & \textbf{2.313E-01 ± 3.657E-03} & \textbf{2.922E-03 ± 5.884E-05} \\
    \cline{2-5} \cline{6-8} 
     & Simple   & MLP   & 1.290E+00 ± 2.040E-02 & 2.076E-02 ± 3.282E-04 & Exte     & 1.155E+00 ± 1.826E-02 & 1.858E-02 ± 2.938E-04 \\
     & MDI      & CNN   & 4.375E-01 ± 6.917E-03 & 7.039E-03 ± 1.113E-04 &  MDI     & 4.325E-01 ± 6.838E-03 & 6.958E-03 ± 1.100E-04 \\
     &          & LSTM  & 4.490E-01 ± 7.099E-03 & 7.224E-03 ± 1.142E-04 &          & 4.609E-01 ± 7.287E-03 & 7.415E-03 ± 1.172E-04 \\
     &          & GRU   & 4.403E-01 ± 6.961E-03 & 7.083E-03 ± 1.120E-04 &          & 4.303E-01 ± 6.803E-03 & 6.923E-03 ± 1.095E-04 \\
     &          & RBFNN & 2.904E-01 ± 3.168E-03 & 3.923E-03 ± 5.097E-05 &          & 1.874E-01 ± 2.199E-03 & 3.209E-03 ± 5.388E-05 \\
     &          & \textbf{ALPE}    & \textbf{2.359E-01 ± 3.698E-03} & \textbf{3.063E-03 ± 5.949E-05} &          & \textbf{1.655E-01 ± 3.723E-03} & \textbf{2.889E-03 ± 5.991E-05} \\
    \cline{2-5} \cline{6-8}
     & Simple  & MLP   & 1.055E+00 ± 1.668E-02 & 1.505E-02 ± 2.376E-04 & Exte      & 1.001E+00 ± 1.583E-02 & 1.611E-02 ± 2.547E-04 \\
     & GD      & CNN   & 4.385E-01 ± 6.934E-03 & 7.055E-03 ± 1.116E-04 &  GD       & 4.325E-01 ± 6.838E-03 & 6.958E-03 ± 1.100E-04 \\
     &         & LSTM  & 5.523E-01 ± 5.438E-03 & 7.202E-03 ± 1.139E-04 &           & 4.533E-01 ± 7.167E-03 & 7.293E-03 ± 1.153E-04 \\
     &         & GRU   & 4.403E-01 ± 6.961E-03 & 7.083E-03 ± 1.120E-04 &           & 4.302E-01 ± 5.939E-03 & 5.231E-03 ± 2.887E-04 \\
     &         & RBFNN & 3.664E-01 ± 2.383E-03 & 6.837E-03 ± 7.837E-05 &           & 3.474E-01 ± 4.389E-03 & 4.989E-03 ± 7.343E-05 \\
     &         & \textbf{ALPE}    & \textbf{2.307E-01 ± 3.648E-03} & \textbf{3.712E-03 ± 5.870E-05} &           & \textbf{2.365E-01 ± 3.740E-03} & \textbf{3.805E-03 ± 6.017E-05} \\
     \hline
    MCK & Simple & Naive & 8.141E-01 ± 1.287E-02 & 2.382E-03 ± 3.767E-05 & Exte     & 7.812E-01 ± 2.553E-02 & 3.909E-03 ± 5.453E-05 \\
        &        & ARIMA & 1.929E-01 ± 3.051E-03 & 5.646E-04 ± 8.927E-06 &          & 2.010E-01 ± 4.665E-03 & 5.010E-04 ± 7.992E-06  \\
        &        & MLP   & 3.161E-01 ± 4.998E-03 & 9.250E-04 ± 1.463E-05 &          & 2.719E-01 ± 4.265E-03 & 7.977E-04 ± 2.403E-05 \\
        &        & CNN   & 2.142E-01 ± 3.387E-03 & 6.269E-04 ± 9.911E-06 &          & 2.149E-01 ± 3.397E-03 & 6.287E-04 ± 9.941E-06 \\
        &        & LSTM  & 2.181E-01 ± 3.449E-03 & 6.383E-04 ± 1.009E-05 &          & 2.273E-01 ± 4.367E-03 & 6.498E-04 ± 2.189E-05 \\
        &        & GRU   & 2.133E-01 ± 3.373E-03 & 6.243E-04 ± 9.871E-06 &          & 2.130E-01 ± 3.368E-03 & 6.109E-04 ± 9.855E-06 \\
        &        & RBFNN & 3.233E-01 ± 5.112E-03 & 9.460E-04 ± 1.496E-05 &          & 1.615E-01 ± 2.920E-03 & 6.054E-04 ± 6.007E-06 \\
        &        & \textbf{ALPE}    & \textbf{1.911E-01 ± 3.655E-03} & \textbf{5.764E-04 ± 1.069E-05} &          & \textbf{1.355E-01 ± 3.647E-03} & \textbf{4.562E-04 ± 1.220E-05} \\
    \cline{2-5} \cline{6-8} 
     & Simple   & MLP   & 3.002E-01 ± 4.746E-03 & 8.784E-04 ± 1.389E-05 & Exte     & 2.783E-01 ± 4.292E-03 & 7.324E-04 ± 2.920E-05 \\
     & MDI      & CNN   & 2.141E-01 ± 3.385E-03 & 6.265E-04 ± 9.905E-06 &  MDI     & 2.155E-01 ± 3.408E-03 & 6.307E-04 ± 9.973E-06 \\
     &          & LSTM  & 2.171E-01 ± 3.433E-03 & 6.353E-04 ± 1.005E-05 &          & 2.221E-01 ± 3.511E-03 & 6.498E-04 ± 1.027E-05 \\
     &          & GRU   & 2.138E-01 ± 3.381E-03 & 6.257E-04 ± 9.893E-06 &          & 2.130E-01 ± 3.368E-03 & 6.233E-04 ± 9.843E-06 \\
     &          & RBFNN & 2.004E-01 ± 3.168E-03 & 5.863E-04 ± 9.270E-06 &          & 1.701E-01 ± 3.997E-03 & 5.054E-04 ± 7.991E-06 \\
     &          & \textbf{ALPE}    & \textbf{1.842E-01 ± 3.704E-03} & \textbf{5.255E-04 ± 1.084E-05} &          & \textbf{1.376E-01 ± 3.646E-03} & \textbf{4.754E-04 ± 1.100E-05} \\
    \cline{2-5} \cline{6-8}
     & Simple  & MLP   & 2.984E-01 ± 4.718E-03 & 8.732E-04 ± 1.381E-05 & Exte      & 2.701E-01 ± 4.270E-03 & 7.813E-04 ± 1.221E-05 \\
     & GD      & CNN   & 2.143E-01 ± 3.389E-03 & 6.272E-04 ± 9.917E-06 &  GD       & 2.169E-01 ± 3.776E-03 & 6.820E-04 ± 9.091E-06 \\
     &         & LSTM  & 2.168E-01 ± 3.428E-03 & 6.344E-04 ± 1.003E-05 &           & 2.207E-01 ± 3.489E-03 & 6.458E-04 ± 1.021E-05 \\
     &         & GRU   & 2.138E-01 ± 3.383E-03 & 6.260E-04 ± 9.899E-06 &           & 2.162E-01 ± 2.488E-03 & 6.233E-04 ± 9.855E-06 \\
     &         & RBFNN & 1.958E-01 ± 3.096E-03 & 5.731E-04 ± 9.061E-06 &           & 1.609E-01 ± 3.683E-03 & 6.254E-04 ± 6.620E-06 \\
     &         & \textbf{ALPE}    & \textbf{1.312E-01 ± 3.656E-03} & \textbf{5.066E-04 ± 1.070E-05} &           & \textbf{1.356E-01 ± 3.726E-03} & \textbf{5.895E-04 ± 1.090E-05} \\
     \hline
    MDT & Simple & Naive & 6.534E-01 ± 1.033E-02 & 8.018E-03 ± 1.268E-04 & Exte     & 5.776E-01 ± 2.001E-02 & 7.770E-03 ± 2.434E-04 \\
        &        & ARIMA & 1.728E-01 ± 2.732E-03 & 2.120E-03 ± 3.352E-05 &          & 1.553E-01 ± 2.335E-03 & 3.998E-03 ± 4.701E-05  \\
        &        & MLP   & 2.513E-01 ± 3.973E-03 & 3.083E-03 ± 4.875E-05 &          & 2.488E-01 ± 3.933E-03 & 3.053E-03 ± 4.827E-05 \\
        &        & CNN   & 1.948E-01 ± 3.080E-03 & 2.391E-03 ± 3.780E-05 &          & 1.943E-01 ± 3.073E-03 & 2.385E-03 ± 3.770E-05 \\
        &        & LSTM  & 1.976E-01 ± 3.125E-03 & 2.425E-03 ± 3.835E-05 &          & 2.011E-01 ± 2.766E-03 & 2.563E-03 ± 2.703E-05 \\
        &        & GRU   & 1.941E-01 ± 3.069E-03 & 2.381E-03 ± 3.765E-05 &          & 1.876E-01 ± 3.338E-03 & 2.298E-03 ± 3.432E-05 \\
        &        & RBFNN & 3.776E-01 ± 4.101E-03 & 3.847E-03 ± 5.909E-05 &          & 3.994E-01 ± 4.001E-03 & 4.433E-03 ± 5.345E-05 \\
        &        & \textbf{ALPE}    & \textbf{1.598E-01 ± 3.277E-03} & \textbf{2.203E-03 ± 3.883E-05} &          & \textbf{1.376E-01 ± 3.757E-03} & \textbf{2.054E-03 ± 3.898E-05} \\
    \cline{2-5} \cline{6-8} 
     & Simple   & MLP   & 2.602E-01 ± 4.114E-03 & 3.193E-03 ± 5.049E-05 & Exte     & 2.290E-01 ± 4.302E-03 & 3.337E-03 ± 5.919E-05 \\
     & MDI      & CNN   & 1.937E-01 ± 3.062E-03 & 2.376E-03 ± 3.757E-05 &  MDI     & 1.956E-01 ± 3.092E-03 & 2.400E-03 ± 3.794E-05 \\
     &          & LSTM  & 1.981E-01 ± 3.132E-03 & 2.430E-03 ± 3.843E-05 &          & 2.163E-01 ± 3.948E-03 & 2.337E-03 ± 2.784E-05 \\
     &          & GRU   & 1.941E-01 ± 3.069E-03 & 2.381E-03 ± 3.765E-05 &          & 1.928E-01 ± 3.049E-03 & 2.366E-03 ± 3.741E-05 \\
     &          & RBFNN & 2.004E-01 ± 3.168E-03 & 2.459E-03 ± 3.887E-05 &          & 3.605E-01 ± 4.999E-03 & 4.197E-03 ± 6.478E-05 \\
     &          & \textbf{ALPE}    & \textbf{1.711E-01 ± 2.706E-03} & \textbf{2.100E-03 ± 3.320E-05} &          & \textbf{1.613E-01 ± 4.772E-03} & \textbf{2.207E-03 ± 5.070E-05} \\
    \cline{2-5} \cline{6-8}
     & Simple  & MLP   & 2.518E-01 ± 3.981E-03 & 3.089E-03 ± 4.885E-05 & Exte      & 2.548E-01 ± 4.029E-03 & 3.127E-03 ± 4.944E-05 \\
     & GD      & CNN   & 1.949E-01 ± 3.082E-03 & 2.392E-03 ± 3.782E-05 &  GD       & 1.890E-01 ± 2.114E-03 & 2.597E-03 ± 3.576E-05 \\
     &         & LSTM  & 1.983E-01 ± 3.135E-03 & 2.309E-03 ± 2.343E-05 &           & 2.004E-01 ± 3.168E-03 & 2.459E-03 ± 3.888E-05 \\
     &         & GRU   & 1.879E-01 ± 3.449E-03 & 2.502E-03 ± 3.773E-05 &           & 1.944E-01 ± 3.074E-03 & 2.386E-03 ± 3.772E-05 \\
     &         & RBFNN & 1.790E-01 ± 3.324E-03 & 2.387E-03 ± 3.774E-05 &           & 3.440E-01 ± 3.838E-03 & 3.989E-03 ± 5.595E-05 \\
     &         & \textbf{ALPE}    & \textbf{1.554E-01 ± 2.457E-03} & \textbf{1.907E-03 ± 3.014E-05} &           & \textbf{1.512E-01 ± 2.391E-03} & \textbf{1.855E-03 ± 2.933E-05} \\
    \bottomrule
    \end{tabular}
    \end{small}}
            \label{tab:table_7}
        \end{minipage}
    \end{table}
\end{landscape}


\begin{landscape}
    \fancyhf{}  
    \fancyfoot[R]{\rotatebox{90}{\thepage}}  
    \thispagestyle{fancy}  

    \begin{table}[!hbtp]
        \centering
        \caption{RMSE and RRMSE scores for META, MKC, MNST, MOS, MRK, MRNA, MSCI, and MSFT.}
        \hspace*{-5cm} 
        \begin{minipage}[t]{0.48\textwidth}
            \centering
            \scalebox{0.42}{
    \begin{small}
    \begin{tabular}{c c r c c c c c}
    \toprule
    \textbf{Stock} & \textbf{Set} & \textbf{Model} & \textbf{RMSE} & \textbf{RRMSE} & \textbf{Set} & \textbf{RMSE} & \textbf{RRMSE} \\
    \hline     
    META & Simple& Naive & 8.030E-01 ± 1.270E-02 & 5.867E-03 ± 9.276E-05 & Exte     & 7.998E-01 ± 2.787E-02 & 4.090E-03 ± 8.263E-05 \\
        &        & ARIMA & 2.089E-01 ± 3.303E-03 & 1.526E-03 ± 2.413E-05 &          & 3.129E-01 ± 4.404E-03 & 2.778E-03 ± 3.339E-05  \\
        &        & MLP   & 2.189E-01 ± 3.462E-03 & 1.600E-03 ± 2.529E-05 &          & 2.034E-01 ± 3.288E-03 & 1.500E-03 ± 2.152E-05 \\
        &        & CNN   & 2.070E-01 ± 3.273E-03 & 1.512E-03 ± 2.391E-05 &          & 2.072E-01 ± 3.276E-03 & 1.514E-03 ± 2.394E-05 \\
        &        & LSTM  & 2.319E-01 ± 3.667E-03 & 1.694E-03 ± 2.679E-05 &          & 2.360E-01 ± 3.734E-03 & 1.725E-03 ± 2.728E-05 \\
        &        & GRU   & 2.296E-01 ± 3.630E-03 & 1.677E-03 ± 2.652E-05 &          & 2.288E-01 ± 3.618E-03 & 1.672E-03 ± 2.644E-05 \\
        &        & RBFNN & 1.453E-01 ± 2.298E-03 & 1.562E-03 ± 1.679E-05 &          & 1.539E-01 ± 2.558E-03 & 1.301E-03 ± 1.381E-05 \\
        &        & \textbf{ALPE}    & \textbf{2.234E-01 ± 3.532E-03} & \textbf{1.292E-03 ± 2.581E-05} &          & \textbf{1.342E-01 ± 3.702E-03} & \textbf{1.011E-03 ± 2.705E-05} \\
    \cline{2-5} \cline{6-8} 
     & Simple   & MLP   & 2.026E-01 ± 3.204E-03 & 1.481E-03 ± 2.341E-05 & Exte     & 2.124E-01 ± 2.201E-03 & 1.551E-03 ± 3.632E-05 \\
     & MDI      & CNN   & 2.055E-01 ± 3.249E-03 & 1.501E-03 ± 2.374E-05 &  MDI     & 2.114E-01 ± 3.938E-03 & 1.564E-03 ± 3.983E-05 \\
     &          & LSTM  & 2.321E-01 ± 3.670E-03 & 1.696E-03 ± 2.682E-05 &          & 2.209E-01 ± 2.657E-03 & 1.499E-03 ± 2.694E-05 \\
     &          & GRU   & 2.288E-01 ± 3.618E-03 & 1.672E-03 ± 2.644E-05 &          & 2.301E-01 ± 3.637E-03 & 1.681E-03 ± 2.658E-05 \\
     &          & RBFNN & 2.404E-01 ± 3.168E-03 & 1.464E-03 ± 2.315E-05 &          & 1.788E-01 ± 3.008E-03 & 1.532E-03 ± 1.873E-05 \\
     &          & \textbf{ALPE}    & \textbf{1.719E-01 ± 3.667E-03} & \textbf{1.294E-03 ± 2.679E-05} &          & \textbf{1.350E-01 ± 3.721E-03} & \textbf{1.125E-03 ± 2.728E-05} \\
    \cline{2-5} \cline{6-8}
     & Simple  & MLP   & 2.026E-01 ± 3.204E-03 & 1.481E-03 ± 2.341E-05 & Exte      & 2.551E-01 ± 3.904E-03 & 1.511E-03 ± 2.281E-05 \\
     & GD      & CNN   & 2.506E-01 ± 3.962E-03 & 1.831E-03 ± 2.895E-05 &  GD       & 2.667E-01 ± 4.290E-03 & 1.647E-03 ± 3.110E-05 \\
     &         & LSTM  & 2.357E-01 ± 3.727E-03 & 1.722E-03 ± 2.723E-05 &           & 2.532E-01 ± 3.625E-03 & 1.690E-03 ± 2.543E-05 \\
     &         & GRU   & 2.292E-01 ± 3.624E-03 & 1.675E-03 ± 2.648E-05 &           & 2.381E-01 ± 2.543E-03 & 1.593E-03 ± 2.490E-05 \\
     &         & RBFNN & 1.764E-01 ± 2.789E-03 & 1.289E-03 ± 2.038E-05 &           & 1.631E-01 ± 2.421E-03 & 1.519E-03 ± 1.769E-05 \\
     &         & \textbf{ALPE}    & \textbf{1.191E-01 ± 3.781E-03} & \textbf{1.047E-03 ± 2.762E-05} &           & \textbf{1.361E-01 ± 3.734E-03} & \textbf{1.225E-03 ± 2.728E-05}\\
     \hline
     MKC & Simple & Naive & 6.962E-01 ± 1.101E-02 & 9.663E-03 ± 1.528E-04 & Exte     & 5.736E-01 ± 2.001E-02 & 8.703E-03 ± 2.338E-04 \\
        &        & ARIMA & 2.349E-01 ± 3.713E-03 & 3.260E-03 ± 5.154E-05 &          & 3.778E-01 ± 4.038E-03 & 4.980E-03 ± 1.997E-05  \\
        &        & MLP   & 2.383E-01 ± 3.768E-03 & 3.308E-03 ± 5.230E-05 &          & 2.317E-01 ± 4.291E-03 & 3.102E-03 ± 3.736E-05 \\
        &        & CNN   & 2.545E-01 ± 4.024E-03 & 3.533E-03 ± 5.585E-05 &          & 2.604E-01 ± 5.113E-03 & 3.609E-03 ± 4.943E-05 \\
        &        & LSTM  & 2.559E-01 ± 4.046E-03 & 3.552E-03 ± 5.617E-05 &          & 2.588E-01 ± 4.092E-03 & 3.592E-03 ± 5.680E-05 \\
        &        & GRU   & 2.541E-01 ± 4.018E-03 & 3.527E-03 ± 5.577E-05 &          & 2.548E-01 ± 4.028E-03 & 3.536E-03 ± 5.591E-05 \\
        &        & RBFNN & 3.883E-01 ± 2.339E-03 & 4.853E-03 ± 7.673E-05 &          & 1.949E-01 ± 3.081E-03 & 2.705E-03 ± 4.277E-05 \\
        &        & \textbf{ALPE}    & \textbf{2.218E-01 ± 3.506E-03} & \textbf{3.079E-03 ± 4.868E-05} &          & \textbf{1.201E-01 ± 2.400E-03} & \textbf{1.995E-03 ± 3.088E-05} \\
    \cline{2-5} \cline{6-8} 
     & Simple   & MLP   & 2.479E-01 ± 3.920E-03 & 3.441E-03 ± 5.441E-05 & Exte     & 2.291E-01 ± 3.372E-03 & 3.152E-03 ± 4.792E-05 \\
     & MDI      & CNN   & 2.531E-01 ± 4.002E-03 & 3.513E-03 ± 5.555E-05 &  MDI     & 2.549E-01 ± 4.331E-03 & 3.251E-03 ± 4.755E-05 \\
     &          & LSTM  & 2.554E-01 ± 4.038E-03 & 3.545E-03 ± 5.605E-05 &          & 2.983E-01 ± 3.112E-03 & 3.352E-03 ± 4.776E-05 \\
     &          & GRU   & 2.531E-01 ± 4.002E-03 & 3.513E-03 ± 5.555E-05 &          & 2.302E-01 ± 3.110E-03 & 3.424E-03 ± 4.784E-05 \\
     &          & RBFNN & 2.004E-01 ± 3.169E-03 & 2.782E-03 ± 4.399E-05 &          & 1.832E-01 ± 4.113E-03 & 2.681E-03 ± 5.309E-05 \\
     &          & \textbf{ALPE}    & \textbf{1.956E-01 ± 3.409E-03} & \textbf{2.293E-03 ± 4.732E-05} &          & \textbf{1.393E-01 ± 3.310E-03} & \textbf{2.005E-03 ± 4.594E-05} \\
    \cline{2-5} \cline{6-8}
     & Simple  & MLP   & 4.130E-01 ± 6.530E-03 & 5.732E-03 ± 9.064E-05 & Exte      & 2.255E-01 ± 3.565E-03 & 3.130E-03 ± 4.949E-05 \\
     & GD      & CNN   & 2.544E-01 ± 4.022E-03 & 3.531E-03 ± 5.582E-05 &  GD       & 2.326E-01 ± 5.132E-03 & 3.478E-03 ± 4.393E-05 \\
     &         & LSTM  & 2.553E-01 ± 4.036E-03 & 3.543E-03 ± 5.603E-05 &           & 2.596E-01 ± 4.104E-03 & 3.603E-03 ± 5.496E-05 \\
     &         & GRU   & 2.538E-01 ± 4.013E-03 & 3.523E-03 ± 5.570E-05 &           & 2.411E-01 ± 4.084E-03 & 3.093E-03 ± 4.673E-05 \\
     &         & RBFNN & 3.505E-01 ± 3.112E-03 & 4.853E-03 ± 7.673E-05 &           & 1.834E-01 ± 4.667E-03 & 2.677E-03 ± 4.205E-05 \\
     &         & \textbf{ALPE}    & \textbf{2.116E-01 ± 3.346E-03} & \textbf{2.937E-03 ± 4.644E-05} &           & \textbf{1.487E-01 ± 3.299E-03} & \textbf{1.833E-03 ± 3.340E-05} \\
     \hline
    MNST & Simple & Naive & 6.569E-01 ± 1.039E-02 & 7.443E-03 ± 1.177E-04 & Exte     & 5.887E-01 ± 1.899E-02 & 8.339E-03 ± 2.087E-04\\
         &        & ARIMA & 2.964E-01 ± 4.686E-03 & 3.358E-03 ± 5.309E-05 &          & 3.493E-01 ± 5.448E-03 & 4.990E-03 ± 4.202E-05 \\
         &        & MLP   & 3.152E-01 ± 4.985E-03 & 3.572E-03 ± 5.648E-05 &          & 2.815E-01 ± 4.452E-03 & 3.190E-03 ± 5.044E-05 \\
         &        & CNN   & 2.472E-01 ± 3.909E-03 & 2.651E-03 ± 4.569E-05 &          & 2.483E-01 ± 3.926E-03 & 2.755E-03 ± 4.448E-05 \\
         &        & LSTM  & 2.487E-01 ± 3.932E-03 & 2.818E-03 ± 4.455E-05 &          & 2.519E-01 ± 3.983E-03 & 2.854E-03 ± 4.513E-05 \\
         &        & GRU   & 2.469E-01 ± 3.904E-03 & 2.798E-03 ± 4.424E-05 &          & 2.472E-01 ± 3.909E-03 & 2.570E-03 ± 3.349E-05 \\
         &        & RBFNN & 2.318E-01 ± 3.666E-03 & 2.627E-03 ± 4.153E-05 &          & 2.219E-01 ± 3.509E-03 & 2.514E-03 ± 3.975E-05 \\
         &        & \textbf{ALPE}    & \textbf{2.122E-01 ± 3.671E-03} & \textbf{2.131E-03 ± 4.160E-05} &          & \textbf{2.007E-01 ± 3.892E-03} & \textbf{2.171E-03 ± 4.021E-05} \\
    \cline{2-5} \cline{6-8} 
     & Simple   & MLP   & 3.058E-01 ± 4.835E-03 & 3.465E-03 ± 5.478E-05 & Exte     & 2.745E-01 ± 3.009E-03 & 3.320E-03 ± 4.119E-05 \\
     & MDI      & CNN   & 2.465E-01 ± 3.898E-03 & 2.793E-03 ± 4.417E-05 &  MDI     & 2.403E-01 ± 3.776E-03 & 2.766E-03 ± 3.202E-05 \\
     &          & LSTM  & 2.481E-01 ± 3.923E-03 & 2.811E-03 ± 4.445E-05 &          & 2.412E-01 ± 4.532E-03 & 2.594E-03 ± 5.773E-05 \\
     &          & GRU   & 2.462E-01 ± 3.892E-03 & 2.789E-03 ± 4.410E-05 &          & 2.590E-01 ± 2.028E-03 & 2.609E-03 ± 3.763E-05 \\
     &          & RBFNN & 2.452E-01 ± 3.877E-03 & 2.778E-03 ± 4.392E-05 &          & 2.338E-01 ± 3.509E-03 & 2.487E-03 ± 4.323E-05 \\
     &          & \textbf{ALPE}    & \textbf{2.202E-01 ± 3.956E-03} & \textbf{2.135E-03 ± 4.483E-05} &          & \textbf{1.768E-01 ± 2.365E-03} & \textbf{2.170E-03 ± 3.249E-05} \\
    \cline{2-5} \cline{6-8}
     & Simple  & MLP   & 3.152E-01 ± 4.985E-03 & 3.572E-03 ± 5.644E-05 & Exte      & 2.683E-01 ± 4.391E-03 & 3.160E-03 ± 4.129E-05 \\
     & GD      & CNN   & 2.474E-01 ± 3.912E-03 & 2.803E-03 ± 4.432E-05 &  GD       & 2.502E-01 ± 3.870E-03 & 2.667E-03 ± 4.636E-05 \\
     &         & LSTM  & 2.483E-01 ± 3.925E-03 & 2.813E-03 ± 4.447E-05 &           & 2.411E-01 ± 3.440E-03 & 2.848E-03 ± 4.303E-05 \\
     &         & GRU   & 2.466E-01 ± 3.899E-03 & 2.794E-03 ± 4.418E-05 &           & 2.538E-01 ± 3.739E-03 & 2.601E-03 ± 3.483E-05 \\
     &         & RBFNN & 2.318E-01 ± 3.666E-03 & 2.627E-03 ± 4.153E-05 &           & 2.277E-01 ± 3.262E-03 & 2.514E-03 ± 3.778E-05 \\
     &         & \textbf{ALPE}    & \textbf{2.096E-01 ± 3.788E-03} & \textbf{2.314E-03 ± 4.292E-05} &           & \textbf{2.068E-01 ± 3.702E-03} & \textbf{2.161E-03 ± 3.224E-05} \\
     \hline
    MOS & Simple & Naive & 6.014E-01 ± 9.509E-03 & 1.209E-02 ± 1.911E-04 & Exte     & 5.998E-01 ± 8.998E-03 & 3.001E-02 ± 2.008E-04 \\
        &        & ARIMA & 1.807E-01 ± 2.857E-03 & 3.631E-03 ± 5.742E-05 &          & 2.776E-01 ± 3.389E-03 & 4.786E-03 ± 4.665E-05  \\
        &        & MLP   & 2.133E-01 ± 3.373E-03 & 4.287E-03 ± 6.778E-05 &          & 1.733E-01 ± 2.740E-03 & 3.482E-03 ± 5.506E-05 \\
        &        & CNN   & 2.022E-01 ± 3.197E-03 & 4.063E-03 ± 6.424E-05 &          & 2.022E-01 ± 3.234E-03 & 3.324E-03 ± 5.566E-05 \\
        &        & LSTM  & 2.044E-01 ± 3.231E-03 & 4.107E-03 ± 6.494E-05 &          & 2.068E-01 ± 3.270E-03 & 4.156E-03 ± 6.572E-05 \\
        &        & GRU   & 2.020E-01 ± 3.194E-03 & 4.060E-03 ± 6.420E-05 &          & 2.199E-01 ± 4.848E-03 & 3.776E-03 ± 4.006E-05 \\
        &        & RBFNN & 1.454E-01 ± 2.299E-03 & 2.923E-03 ± 4.621E-05 &          & 1.929E-01 ± 3.050E-03 & 3.876E-03 ± 6.129E-05 \\
        &        & \textbf{ALPE}    & \textbf{1.188E-01 ± 2.353E-03} & \textbf{2.390E-03 ± 4.728E-05} &          & \textbf{1.703E-01 ± 3.444E-03} & \textbf{3.298E-03 ± 5.215E-05} \\
    \cline{2-5} \cline{6-8} 
     & Simple   & MLP   & 1.903E-01 ± 3.009E-03 & 3.824E-03 ± 6.046E-05 & Exte     & 1.745E-01 ± 2.644E-03 & 3.089E-03 ± 4.481E-05 \\
     & MDI      & CNN   & 2.022E-01 ± 3.197E-03 & 4.064E-03 ± 6.425E-05 &  MDI     & 2.302E-01 ± 3.304E-03 & 4.100E-03 ± 5.337E-05 \\
     &          & LSTM  & 2.042E-01 ± 3.228E-03 & 4.103E-03 ± 6.488E-05 &          & 2.110E-01 ± 3.664E-03 & 4.201E-03 ± 5.454E-05 \\
     &          & GRU   & 2.017E-01 ± 3.190E-03 & 4.054E-03 ± 6.410E-05 &          & 2.020E-01 ± 3.193E-03 & 4.110E-03 ± 6.834E-05 \\
     &          & RBFNN & 2.004E-01 ± 3.168E-03 & 4.026E-03 ± 6.366E-05 &          & 1.877E-01 ± 4.145E-03 & 3.764E-03 ± 5.083E-05 \\
     &          & \textbf{ALPE}    & \textbf{1.580E-01 ± 2.498E-03} & \textbf{3.175E-03 ± 5.020E-05} &          & \textbf{1.641E-01 ± 2.595E-03} & \textbf{3.178E-03 ± 4.215E-05} \\
    \cline{2-5} \cline{6-8}
     & Simple  & MLP   & 1.818E-01 ± 2.874E-03 & 3.653E-03 ± 5.776E-05 & Exte      & 1.864E-01 ± 2.947E-03 & 3.745E-03 ± 5.922E-05 \\
     & GD      & CNN   & 2.026E-01 ± 3.204E-03 & 4.072E-03 ± 6.439E-05 &  GD       & 2.010E-01 ± 3.179E-03 & 4.040E-03 ± 6.388E-05 \\
     &         & LSTM  & 2.047E-01 ± 3.236E-03 & 4.114E-03 ± 6.504E-05 &           & 2.073E-01 ± 3.278E-03 & 4.167E-03 ± 6.588E-05 \\
     &         & GRU   & 2.016E-01 ± 3.188E-03 & 4.051E-03 ± 6.406E-05 &           & 2.192E-01 ± 3.938E-03 & 4.143E-03 ± 5.049E-05 \\
     &         & RBFNN & 2.004E-01 ± 3.168E-03 & 4.026E-03 ± 6.366E-05 &           & 1.789E-01 ± 4.337E-03 & 3.690E-03 ± 4.834E-05 \\
     &         & \textbf{ALPE}    & \textbf{1.569E-01 ± 2.481E-03} & \textbf{3.153E-03 ± 4.986E-05} &           & \textbf{1.551E-01 ± 2.998E-03} & \textbf{3.298E-03 ± 5.215E-05} \\
    \bottomrule
    \end{tabular}
    \end{small}}
            \label{tab:table1}
        \end{minipage}%
        \hspace{0.25\textwidth} 
        \begin{minipage}[t]{0.48\textwidth}
            \centering
            \scalebox{0.42}{
    \begin{small}
    \begin{tabular}{c c r c c c c c}
    \toprule
    \textbf{Stock} & \textbf{Set} & \textbf{Model} & \textbf{RMSE} & \textbf{RRMSE} & \textbf{Set} & \textbf{RMSE} & \textbf{RRMSE} \\
    \hline
        MRK & Simple & Naive & 5.749E-01 ± 9.090E-03 & 6.667E-03 ± 1.054E-04 & Exte     & 4.909E-01 ± 6.876E-03 & 7.225E-03 ± 2.114E-04 \\
        &        & ARIMA & 1.811E-01 ± 2.450E-03 & 2.003E-03 ± 3.167E-05 &          & 2.554E-01 ± 3.038E-03 & 3.997E-03 ± 3.239E-05  \\
        &        & MLP   & 3.022E-01 ± 4.779E-03 & 3.505E-03 ± 5.542E-05 &          & 2.433E-01 ± 5.363E-03 & 3.883E-03 ± 5.883E-05 \\
        &        & CNN   & 1.946E-01 ± 3.076E-03 & 2.256E-03 ± 3.567E-05 &          & 1.939E-01 ± 3.065E-03 & 2.248E-03 ± 3.554E-05 \\
        &        & LSTM  & 1.967E-01 ± 3.110E-03 & 2.281E-03 ± 3.606E-05 &          & 1.854E-01 ± 3.243E-03 & 2.477E-03 ± 3.577E-05 \\
        &        & GRU   & 1.944E-01 ± 3.074E-03 & 2.254E-03 ± 3.564E-05 &          & 1.943E-01 ± 3.073E-03 & 2.254E-03 ± 3.593E-05 \\
        &        & RBFNN & 1.730E-01 ± 2.736E-03 & 2.007E-03 ± 3.173E-05 &          & 1.675E-01 ± 2.493E-03 & 1.841E-03 ± 3.121E-05 \\
        &        & \textbf{ALPE}    & \textbf{1.560E-01 ± 2.466E-03} & \textbf{1.809E-03 ± 2.860E-05} &          & \textbf{1.131E-01 ± 2.202E-03} & \textbf{1.008E-03 ± 3.175E-05} \\
    \cline{2-5} \cline{6-8} 
     & Simple   & MLP   & 2.882E-01 ± 4.557E-03 & 3.342E-03 ± 5.284E-05 & Exte     & 2.691E-01 ± 4.255E-03 & 3.121E-03 ± 4.934E-05 \\
     & MDI      & CNN   & 1.947E-01 ± 3.079E-03 & 2.258E-03 ± 3.570E-05 &  MDI     & 1.946E-01 ± 3.077E-03 & 2.257E-03 ± 3.568E-05 \\
     &          & LSTM  & 1.967E-01 ± 3.110E-03 & 2.281E-03 ± 3.606E-05 &          & 1.996E-01 ± 3.156E-03 & 2.314E-03 ± 3.659E-05 \\
     &          & GRU   & 1.941E-01 ± 3.070E-03 & 2.251E-03 ± 3.560E-05 &          & 1.943E-01 ± 3.073E-03 & 2.254E-03 ± 3.563E-05 \\
     &          & RBFNN & 1.421E-01 ± 2.125E-03 & 1.644E-03 ± 2.599E-05 &          & 1.673E-01 ± 2.114E-03 & 1.974E-03 ± 3.834E-05 \\
     &          & \textbf{ALPE}    & \textbf{1.273E-01 ± 2.487E-03} & \textbf{1.124E-03 ± 2.883E-05} &          & \textbf{1.232E-01 ± 2.738E-03} & \textbf{1.408E-03 ± 3.175E-05} \\
    \cline{2-5} \cline{6-8}
     & Simple  & MLP   & 2.852E-01 ± 4.509E-03 & 3.307E-03 ± 5.229E-05 & Exte      & 2.753E-01 ± 4.352E-03 & 3.192E-03 ± 5.047E-05 \\
     & GD      & CNN   & 1.951E-01 ± 3.084E-03 & 2.262E-03 ± 3.576E-05 &  GD       & 1.935E-01 ± 3.060E-03 & 2.244E-03 ± 3.548E-05 \\
     &         & LSTM  & 1.971E-01 ± 3.116E-03 & 2.285E-03 ± 3.613E-05 &           & 1.990E-01 ± 3.147E-03 & 2.308E-03 ± 3.649E-05 \\
     &         & GRU   & 1.940E-01 ± 3.068E-03 & 2.250E-03 ± 3.558E-05 &           & 1.939E-01 ± 3.066E-03 & 2.249E-03 ± 3.556E-05 \\
     &         & RBFNN & 1.432E-01 ± 2.355E-03 & 1.645E-03 ± 2.600E-05 &           & 1.581E-01 ± 2.500E-03 & 1.833E-03 ± 2.899E-05 \\
     &         & \textbf{ALPE}    & \textbf{1.107E-01 ± 2.540E-03} & \textbf{1.263E-03 ± 2.946E-05} &           & \textbf{1.111E-01 ± 3.440E-03} & \textbf{1.308E-03 ± 3.175E-05}\\
     \hline
     MRNA & Simple & Naive & 7.152E-01 ± 1.131E-02 & 5.976E-03 ± 9.448E-05 & Exte     & 6.887E-01 ± 2.998E-02 & 4.090E-03 ± 8.224E-05 \\
         &        & ARIMA & 1.922E-01 ± 3.029E-03 & 1.606E-03 ± 2.540E-05 &          & 2.045E-01 ± 4.893E-03 & 2.880E-03 ± 3.673E-05  \\
         &        & MLP   & 3.139E-01 ± 4.964E-03 & 2.623E-03 ± 4.148E-05 &          & 3.324E-01 ± 5.764E-03 & 2.599E-03 ± 3.543E-05 \\
         &        & CNN   & 2.173E-01 ± 3.435E-03 & 1.815E-03 ± 2.870E-05 &          & 2.289E-01 ± 3.673E-03 & 1.934E-03 ± 3.654E-05 \\
         &        & LSTM  & 1.951E-01 ± 3.085E-03 & 1.630E-03 ± 2.578E-05 &          & 1.997E-01 ± 3.158E-03 & 1.669E-03 ± 2.639E-05 \\
         &        & GRU   & 1.807E-01 ± 2.443E-03 & 1.710E-03 ± 3.746E-05 &          & 1.827E-01 ± 2.087E-03 & 1.710E-03 ± 2.746E-05 \\
         &        & RBFNN & 1.970E-01 ± 3.115E-03 & 1.646E-03 ± 2.603E-05 &          & 2.059E-01 ± 3.255E-03 & 1.720E-03 ± 2.720E-05 \\
         &        & \textbf{ALPE}    & \textbf{1.538E-01 ± 2.432E-03} & \textbf{1.285E-03 ± 2.032E-05} &          & \textbf{1.640E-01 ± 2.234E-03} & \textbf{1.300E-03 ± 1.998E-05} \\
    \cline{2-5} \cline{6-8} 
     & Simple   & MLP   & 3.127E-01 ± 4.944E-03 & 2.612E-03 ± 4.130E-05 & Exte     & 3.122E-01 ± 4.936E-03 & 2.608E-03 ± 4.124E-05 \\
     & MDI      & CNN   & 2.165E-01 ± 3.423E-03 & 1.809E-03 ± 2.860E-05 &  MDI     & 2.191E-01 ± 3.460E-03 & 1.830E-03 ± 2.894E-05 \\
     &          & LSTM  & 1.958E-01 ± 3.096E-03 & 1.636E-03 ± 2.587E-05 &          & 1.988E-01 ± 3.144E-03 & 1.661E-03 ± 2.626E-05 \\
     &          & GRU   & 2.169E-01 ± 3.429E-03 & 1.812E-03 ± 2.865E-05 &          & 2.158E-01 ± 3.310E-03 & 1.803E-03 ± 2.851E-05 \\
     &          & RBFNN & 1.736E-01 ± 2.745E-03 & 1.451E-03 ± 2.294E-05 &          & 2.173E-01 ± 3.436E-03 & 1.816E-03 ± 2.871E-05 \\
     &          & \textbf{ALPE}    & \textbf{1.501E-01 ± 2.374E-03} & \textbf{1.254E-03 ± 1.983E-05} &          & \textbf{1.731E-01 ± 2.737E-03} & \textbf{1.397E-03 ± 1.472E-05} \\
    \cline{2-5} \cline{6-8}
     & Simple  & MLP   & 3.051E-01 ± 4.824E-03 & 2.549E-03 ± 4.030E-05 & Exte      & 3.070E-01 ± 4.854E-03 & 2.565E-03 ± 4.055E-05 \\
     & GD      & CNN   & 2.172E-01 ± 3.434E-03 & 1.815E-03 ± 2.869E-05 &  GD       & 2.165E-01 ± 3.423E-03 & 1.809E-03 ± 2.860E-05 \\
     &         & LSTM  & 1.964E-01 ± 3.106E-03 & 1.641E-03 ± 2.595E-05 &           & 1.983E-01 ± 3.135E-03 & 1.657E-03 ± 2.619E-05 \\
     &         & GRU   & 2.168E-01 ± 3.428E-03 & 1.811E-03 ± 2.864E-05 &           & 2.163E-01 ± 3.421E-03 & 1.808E-03 ± 2.858E-05 \\
     &         & RBFNN & 1.971E-01 ± 3.117E-03 & 1.647E-03 ± 2.604E-05 &           & 2.366E-01 ± 4.202E-03 & 1.977E-03 ± 3.764E-05 \\
     &         & \textbf{ALPE}    & \textbf{1.535E-01 ± 2.427E-03} & \textbf{1.282E-03 ± 2.028E-05} &           & \textbf{1.232E-01 ± 1.530E-03} & \textbf{1.498E-03 ± 2.402E-05} \\
     \hline
    MSCI & Simple & Naive & 2.446E+00 ± 3.867E-02 & 5.748E-03 ± 9.088E-05 & Exte     & 3.113E+00 ± 4.765E-02 & 4.998E-03 ± 8.989E-05 \\
         &        & ARIMA & 4.130E-01 ± 6.530E-03 & 9.707E-04 ± 1.535E-05 &          & 5.220E-01 ± 5.337E-03 & 8.987E-04 ± 2.665E-05  \\
         &        & MLP   & 3.071E-01 ± 4.855E-03 & 7.217E-04 ± 1.141E-05 &          & 2.877E-01 ± 3.534E-03 & 6.887E-04 ± 1.492E-05 \\
         &        & CNN   & 4.340E-01 ± 6.863E-03 & 1.047E-03 ± 1.613E-05 &          & 4.416E-01 ± 6.982E-03 & 1.038E-03 ± 1.641E-05 \\
         &        & LSTM  & 4.322E-01 ± 6.833E-03 & 1.016E-03 ± 1.606E-05 &          & 4.531E-01 ± 5.115E-03 & 1.091E-03 ± 2.119E-05 \\
         &        & GRU   & 4.270E-01 ± 6.751E-03 & 1.004E-03 ± 1.587E-05 &          & 4.366E-01 ± 6.904E-03 & 1.026E-03 ± 1.623E-05 \\
         &        & RBFNN & 2.498E-01 ± 3.949E-03 & 5.870E-04 ± 9.282E-06 &          & 2.472E-01 ± 3.909E-03 & 5.811E-04 ± 9.188E-06 \\
         &        & \textbf{ALPE}    & \textbf{2.310E-01 ± 3.356E-03} & \textbf{5.405E-04 ± 8.547E-06} &          & \textbf{2.340E-01 ± 2.009E-03} & \textbf{5.101E-04 ± 7.870E-06} \\
    \cline{2-5} \cline{6-8} 
     & Simple   & MLP   & 2.962E-01 ± 4.683E-03 & 6.961E-04 ± 1.101E-05 & Exte     & 2.957E-01 ± 4.676E-03 & 6.950E-04 ± 1.099E-05 \\
     & MDI      & CNN   & 4.410E-01 ± 6.972E-03 & 1.126E-03 ± 1.719E-05 &  MDI     & 4.406E-01 ± 6.966E-03 & 1.099E-03 ± 1.799E-05 \\
     &          & LSTM  & 4.359E-01 ± 6.892E-03 & 1.024E-03 ± 1.620E-05 &          & 4.360E-01 ± 6.894E-03 & 1.025E-03 ± 1.620E-05 \\
     &          & GRU   & 4.283E-01 ± 6.772E-03 & 1.007E-03 ± 1.592E-05 &          & 4.453E-01 ± 6.934E-03 & 1.099E-03 ± 2.763E-05 \\
     &          & RBFNN & 2.495E-01 ± 3.944E-03 & 5.863E-04 ± 9.271E-06 &          & 2.489E-01 ± 4.552E-03 & 5.600E-04 ± 8.876E-06 \\
     &          & \textbf{ALPE}    & \textbf{2.339E-01 ± 3.698E-03} & \textbf{5.497E-04 ± 8.692E-0}6 &          & \textbf{2.253E-01 ± 3.562E-03} & \textbf{5.338E-04 ± 7.884E-06} \\
    \cline{2-5} \cline{6-8}
     & Simple  & MLP   & 2.980E-01 ± 4.712E-03 & 7.004E-04 ± 1.107E-05 & Exte      & 2.844E-01 ± 4.498E-03 & 6.685E-04 ± 1.057E-05 \\
     & GD      & CNN   & 4.732E-01 ± 7.482E-03 & 1.112E-03 ± 1.758E-05 &  GD       & 4.579E-01 ± 7.240E-03 & 1.076E-03 ± 1.702E-05 \\
     &         & LSTM  & 4.408E-01 ± 6.969E-03 & 1.146E-03 ± 1.320E-05 &           & 4.357E-01 ± 6.888E-03 & 1.024E-03 ± 1.619E-05 \\
     &         & GRU   & 4.366E-01 ± 6.904E-03 & 1.026E-03 ± 1.623E-05 &           & 4.407E-01 ± 6.969E-03 & 1.101E-03 ± 1.395E-05 \\
     &         & RBFNN & 2.454E-01 ± 3.881E-03 & 5.768E-04 ± 9.120E-06 &           & 2.502E-01 ± 3.884E-03 & 5.376E-04 ± 7.787E-06 \\
     &         & \textbf{ALPE}    & \textbf{2.371E-01 ± 3.748E-03} & \textbf{5.572E-04 ± 8.810E-06} &           & \textbf{2.407E-01 ± 2.887E-03} & \textbf{5.223E-04 ± 8.001E-06} \\
     \hline
    MSFT & Simple & Naive & 2.210E+00 ± 3.494E-02 & 9.387E-03 ± 1.484E-04 & Exte     & 3.449E+00 ± 2.094E-02 & 8.663E-03 ± 2.900E-04 \\
         &        & ARIMA & 7.632E-01 ± 1.207E-02 & 3.242E-03 ± 5.126E-05 &          & 6.305E-01 ± 2.889E-02 & 4.349E-03 ± 4.010E-05  \\
         &        & MLP   & 9.446E-01 ± 1.494E-02 & 4.013E-03 ± 6.345E-05 &          & 8.946E-01 ± 1.415E-02 & 3.801E-03 ± 6.309E-05 \\
         &        & CNN   & 8.294E-01 ± 1.311E-02 & 3.523E-03 ± 5.571E-05 &          & 8.642E-01 ± 2.562E-02 & 3.366E-03 ± 4.356E-05 \\
         &        & LSTM  & 9.417E-01 ± 1.489E-02 & 4.001E-03 ± 6.326E-05 &          & 9.534E-01 ± 1.507E-02 & 4.050E-03 ± 6.404E-05 \\
         &        & GRU   & 8.939E-01 ± 1.413E-02 & 3.798E-03 ± 6.005E-05 &          & 8.782E-01 ± 1.389E-02 & 3.731E-03 ± 5.899E-05 \\
         &        & RBFNN & 1.968E-01 ± 3.111E-03 & 8.360E-04 ± 1.322E-05 &          & 2.000E-01 ± 3.162E-03 & 8.495E-04 ± 1.343E-05 \\
         &        & \textbf{ALPE}    & \textbf{1.032E-01 ± 1.632E-03} & \textbf{4.384E-04 ± 6.932E-06} &          & \textbf{1.352E-01 ± 2.473E-03} & \textbf{4.348E-04 ± 5.324E-06} \\
    \cline{2-5} \cline{6-8} 
     & Simple   & MLP   & 9.452E-01 ± 1.495E-02 & 4.016E-03 ± 6.349E-05 & Exte     & 8.776E-01 ± 1.144E-02 & 3.148E-03 ± 3.453E-05 \\
     & MDI      & CNN   & 8.909E-01 ± 1.409E-02 & 3.785E-03 ± 5.984E-05 &  MDI     & 8.829E-01 ± 1.396E-02 & 3.751E-03 ± 5.930E-05 \\
     &          & LSTM  & 9.925E-01 ± 1.569E-02 & 4.216E-03 ± 6.667E-05 &          & 7.034E-01 ± 1.748E-02 & 4.663E-03 ± 3.543E-05 \\
     &          & GRU   & 8.728E-01 ± 1.380E-02 & 3.708E-03 ± 5.863E-05 &          & 8.532E-01 ± 1.434E-02 & 3.532E-03 ± 4.674E-05 \\
     &          & RBFNN & 2.004E-01 ± 3.168E-03 & 8.512E-04 ± 1.346E-05 &          & 2.041E-01 ± 2.898E-03 & 5.746E-04 ± 1.343E-05 \\
     &          & \textbf{ALPE}    & \textbf{1.162E-01 ± 1.838E-03} & \textbf{4.938E-04 ± 7.808E-06} &          & \textbf{1.042E-01 ± 1.647E-03} & \textbf{4.417E-04 ± 6.999E-06} \\
    \cline{2-5} \cline{6-8}
     & Simple  & MLP   & 9.486E-01 ± 1.500E-02 & 4.030E-03 ± 6.372E-05 & Exte      & 9.098E-01 ± 1.439E-02 & 3.865E-03 ± 6.111E-05 \\
     & GD      & CNN   & 1.115E+00 ± 1.763E-02 & 4.736E-03 ± 7.489E-05 &  GD       & 9.966E-01 ± 1.576E-02 & 4.234E-03 ± 6.695E-05 \\
     &         & LSTM  & 9.350E-01 ± 1.478E-02 & 3.972E-03 ± 6.281E-05 &           & 9.566E-01 ± 1.987E-02 & 4.773E-03 ± 5.097E-05 \\
     &         & GRU   & 8.314E-01 ± 1.315E-02 & 3.532E-03 ± 5.585E-05 &           & 8.780E-01 ± 1.378E-02 & 3.530E-03 ± 3.812E-05 \\
     &         & RBFNN & 1.998E-01 ± 3.159E-03 & 8.489E-04 ± 1.342E-05 &           & 2.052E-01 ± 2.342E-03 & 8.550E-04 ± 1.230E-05 \\
     &         & \textbf{ALPE}    & \textbf{1.135E-01 ± 1.794E-03} & \textbf{4.820E-04 ± 7.621E-06} &           & \textbf{1.104E-01 ± 2.887E-03} & \textbf{4.425E-04 ± 6.425E-06} \\

    \bottomrule
    \end{tabular}
    \end{small}}
            \label{tab:table_8}
        \end{minipage}
    \end{table}
\end{landscape}


\begin{landscape}
    \fancyhf{}  
    \fancyfoot[R]{\rotatebox{90}{\thepage}}  
    \thispagestyle{fancy}  

    \begin{table}[!hbtp]
        \centering
        \caption{RMSE and RRMSE scores for MTB, NFLX, NKE, NTAP, NVDA, NXPI, ODFL, and OMC.}
        \hspace*{-5cm} 
        \begin{minipage}[t]{0.48\textwidth}
            \centering
            \scalebox{0.42}{
    \begin{small}
    \begin{tabular}{c c r c c c c c}
    \toprule
    \textbf{Stock} & \textbf{Set} & \textbf{Model} & \textbf{RMSE} & \textbf{RRMSE} & \textbf{Set} & \textbf{RMSE} & \textbf{RRMSE} \\
    \hline     
    MTB & Simple  & Naive & 1.701E+00 ± 2.689E-02 & 9.502E-03 ± 1.502E-04 & Exte     & 2.176E+00 ± 3.001E-02 & 8.833E-03 ± 2.650E-04 \\
         &        & ARIMA & 4.160E-01 ± 6.578E-03 & 2.324E-03 ± 3.675E-05 &          & 5.114E-01 ± 5.009E-03 & 3.445E-03 ± 4.335E-05  \\
         &        & MLP   & 8.171E-01 ± 1.292E-02 & 4.565E-03 ± 7.218E-05 &          & 7.598E-01 ± 1.201E-02 & 4.245E-03 ± 6.712E-05 \\
         &        & CNN   & 4.359E-01 ± 6.893E-03 & 2.436E-03 ± 3.851E-05 &          & 4.421E-01 ± 6.990E-03 & 2.470E-03 ± 3.905E-05 \\
         &        & LSTM  & 4.368E-01 ± 6.907E-03 & 2.440E-03 ± 3.859E-05 &          & 4.373E-01 ± 6.914E-03 & 2.443E-03 ± 3.862E-05 \\
         &        & GRU   & 4.303E-01 ± 6.804E-03 & 2.404E-03 ± 3.801E-05 &          & 4.411E-01 ± 6.974E-03 & 2.464E-03 ± 3.896E-05 \\
         &        & RBFNN & 1.772E-01 ± 2.802E-03 & 1.900E-03 ± 1.565E-05 &          & 2.000E-01 ± 3.162E-03 & 1.182E-03 ± 1.560E-05 \\
         &        & \textbf{ALPE}    & \textbf{1.341E-01 ± 3.226E-03} & \textbf{1.440E-03 ± 1.802E-05} &          & \textbf{1.365E-01 ± 1.009E-03} & \textbf{7.802E-04 ± 1.494E-05} \\
    \cline{2-5} \cline{6-8} 
     & Simple   & MLP   & 8.930E-01 ± 1.412E-02 & 4.989E-03 ± 7.888E-05 & Exte     & 7.774E-01 ± 2.837E-02 & 4.800E-03 ± 3.838E-05 \\
     & MDI      & CNN   & 4.445E-01 ± 7.028E-03 & 2.483E-03 ± 3.926E-05 &  MDI     & 4.449E-01 ± 5.873E-03 & 2.992E-03 ± 4.736E-05 \\
     &          & LSTM  & 4.379E-01 ± 6.923E-03 & 2.446E-03 ± 3.868E-05 &          & 4.782E-01 ± 4.008E-03 & 2.534E-03 ± 4.767E-05 \\
     &          & GRU   & 4.308E-01 ± 6.812E-03 & 2.407E-03 ± 3.806E-05 &          & 4.993E-01 ± 3.878E-03 & 2.571E-03 ± 4.003E-05 \\
     &          & RBFNN & 2.004E-01 ± 3.168E-03 & 1.119E-03 ± 1.770E-05 &          & 2.067E-01 ± 2.411E-03 & 1.393E-03 ± 1.401E-05 \\
     &          & \textbf{ALPE}    & \textbf{1.858E-01 ± 2.938E-03} & \textbf{1.038E-03 ± 1.641E-05} &          & \textbf{1.486E-01 ± 2.804E-03} & \textbf{8.117E-04 ± 1.999E-05} \\
    \cline{2-5} \cline{6-8}
     & Simple  & MLP   & 8.750E-01 ± 2.535E-02 & 4.062E-03 ± 6.898E-05 & Exte      & 6.889E-01 ± 1.403E-02 & 3.567E-03 ± 4.221E-05 \\
     & GD      & CNN   & 4.695E-01 ± 7.424E-03 & 2.623E-03 ± 4.148E-05 &  GD       & 4.543E-01 ± 5.441E-03 & 2.494E-03 ± 5.202E-05 \\
     &         & LSTM  & 4.445E-01 ± 7.028E-03 & 2.483E-03 ± 3.926E-05 &           & 4.445E-01 ± 7.028E-03 & 2.483E-03 ± 3.926E-05 \\
     &         & GRU   & 4.378E-01 ± 6.923E-03 & 2.446E-03 ± 3.868E-05 &           & 4.301E-01 ± 6.760E-03 & 2.304E-03 ± 3.614E-05 \\
     &         & RBFNN & 1.968E-01 ± 3.111E-03 & 1.099E-03 ± 1.738E-05 &           & 2.122E-01 ± 4.029E-03 & 1.312E-03 ± 3.445E-05 \\
     &         & \textbf{ALPE}    & \textbf{1.458E-01 ± 2.305E-03} & \textbf{8.145E-04 ± 1.288E-05} &           & \textbf{1.390E-01 ± 2.155E-03} & \textbf{7.815E-04 ± 2.401E-05}\\
     \hline
     NFLX & Simple & Naive & 2.425E+00 ± 3.834E-02 & 1.018E-02 ± 1.610E-04 & Exte     & 3.009E+00 ± 4.004E-02 & 2.994E-02 ± 2.008E-04 \\
         &        & ARIMA & 9.856E-01 ± 1.558E-02 & 4.137E-03 ± 6.542E-05 &          & 8.744E-01 ± 2.221E-02 & 5.873E-03 ± 5.224E-05  \\
         &        & MLP   & 1.575E+00 ± 2.174E-02 & 5.773E-03 ± 9.128E-05 &          & 9.992E-01 ± 1.580E-02 & 4.195E-03 ± 6.632E-05 \\
         &        & CNN   & 8.307E-01 ± 1.313E-02 & 3.487E-03 ± 5.514E-05 &          & 8.915E-01 ± 1.410E-02 & 3.743E-03 ± 5.918E-05 \\
         &        & LSTM  & 7.630E-01 ± 1.206E-02 & 3.203E-03 ± 5.064E-05 &          & 8.898E-01 ± 1.407E-02 & 3.735E-03 ± 5.906E-05 \\
         &        & GRU   & 8.176E-01 ± 1.293E-02 & 3.432E-03 ± 5.427E-05 &          & 9.536E-01 ± 1.508E-02 & 4.003E-03 ± 6.330E-05 \\
         &        & RBFNN & 2.233E-01 ± 3.531E-03 & 9.375E-04 ± 1.482E-05 &          & 2.216E-01 ± 3.504E-03 & 9.304E-04 ± 1.471E-05 \\
         &        & \textbf{ALPE}    & \textbf{1.244E-01 ± 2.757E-03} & \textbf{7.321E-04 ± 1.158E-05} &          & \textbf{1.447E-01 ± 2.288E-03} & \textbf{5.443E-04 ± 7.114E-06} \\
    \cline{2-5} \cline{6-8} 
     & Simple   & MLP   & 9.956E-01 ± 1.574E-02 & 4.180E-03 ± 6.608E-05 & Exte     & 9.873E-01 ± 2.363E-02 & 4.202E-03 ± 4.746E-05 \\
     & MDI      & CNN   & 8.347E-01 ± 1.320E-02 & 3.504E-03 ± 5.540E-05 &  MDI     & 8.765E-01 ± 3.399E-02 & 3.894E-03 ± 3.904E-05 \\
     &          & LSTM  & 7.692E-01 ± 1.216E-02 & 3.229E-03 ± 5.106E-05 &          & 8.983E-01 ± 2.560E-02 & 3.824E-03 ± 4.938E-05 \\
     &          & GRU   & 8.865E-01 ± 1.402E-02 & 3.721E-03 ± 5.884E-05 &          & 9.601E-01 ± 3.388E-02 & 4.122E-03 ± 5.220E-05 \\
     &          & RBFNN & 2.504E-01 ± 3.168E-03 & 8.411E-04 ± 1.330E-05 &          & 2.311E-01 ± 4.847E-03 & 9.388E-04 ± 2.837E-05 \\
     &          & \textbf{ALPE}    & \textbf{2.083E-01 ± 3.294E-03 }& \textbf{7.746E-04 ± 1.383E-05} &          & \textbf{1.530E-01 ± 2.109E-03} & \textbf{6.202E-04 ± 5.199E-06} \\
    \cline{2-5} \cline{6-8}
     & Simple  & MLP   & 9.939E-01 ± 1.571E-02 & 4.172E-03 ± 6.597E-05 & Exte      & 9.943E-01 ± 1.572E-02 & 4.174E-03 ± 6.600E-05 \\
     & GD      & CNN   & 9.581E-01 ± 1.515E-02 & 4.022E-03 ± 6.360E-05 &  GD       & 8.554E-01 ± 2.503E-02 & 3.688E-03 ± 4.878E-05 \\
     &         & LSTM  & 7.754E-01 ± 1.226E-02 & 3.255E-03 ± 5.147E-05 &           & 8.789E-01 ± 2.588E-02 & 3.683E-03 ± 4.880E-05 \\
     &         & GRU   & 9.442E-01 ± 1.493E-02 & 3.964E-03 ± 6.267E-05 &           & 9.541E-01 ± 2.609E-02 & 4.129E-03 ± 5.299E-05 \\
     &         & RBFNN & 2.207E-01 ± 3.490E-03 & 8.130E-04 ± 1.684E-05 &           & 2.289E-01 ± 2.622E-03 & 9.377E-04 ± 1.495E-05 \\
     &         & \textbf{ALPE}    & \textbf{1.587E-01 ± 2.509E-03} & \textbf{6.661E-04 ± 1.053E-05} &           & \textbf{1.345E-01 ± 2.009E-03} & \textbf{5.118E-04 ± 8.837E-06} \\
     \hline
    NKE & Simple & Naive  & 2.059E+00 ± 3.255E-02 & 2.476E-02 ± 3.914E-04 & Exte     & 3.666E+00 ± 4.998E-02 & 3.099E-02 ± 4.808E-04 \\
         &        & ARIMA & 7.486E-01 ± 1.184E-02 & 9.002E-03 ± 1.423E-04 &          & 6.027E-01 ± 1.298E-02 & 8.111E-03 ± 2.322E-04  \\
         &        & MLP   & 1.384E+00 ± 2.188E-02 & 1.664E-02 ± 2.631E-04 &          & 1.372E+00 ± 2.169E-02 & 1.650E-02 ± 2.608E-04 \\
         &        & CNN   & 9.892E-01 ± 1.564E-02 & 1.189E-02 ± 1.881E-04 &          & 9.406E-01 ± 1.487E-02 & 1.131E-02 ± 1.788E-04 \\
         &        & LSTM  & 6.924E-01 ± 1.095E-02 & 8.326E-03 ± 1.317E-04 &          & 6.930E-01 ± 1.096E-02 & 8.105E-03 ± 2.233E-04 \\
         &        & GRU   & 8.799E-01 ± 1.391E-02 & 1.058E-02 ± 1.673E-04 &          & 8.832E-01 ± 1.396E-02 & 1.062E-02 ± 1.679E-04 \\
         &        & RBFNN & 3.505E-01 ± 4.747E-03 & 4.559E-03 ± 4.949E-05 &          & 3.700E-01 ± 2.940E-03 & 4.202E-03 ± 5.474E-05 \\
         &        & \textbf{ALPE}    & \textbf{2.144E-01 ± 2.408E-03} & \textbf{2.713E-03 ± 4.289E-05} &          & \textbf{2.291E-01 ± 3.622E-03} & \textbf{2.733E-03 ± 3.884E-05} \\
    \cline{2-5} \cline{6-8} 
     & Simple   & MLP   & 1.335E+00 ± 2.110E-02 & 1.605E-02 ± 2.538E-04 & Exte     & 1.476E+00 ± 3.209E-02 & 1.682E-02 ± 3.587E-04 \\
     & MDI      & CNN   & 9.406E-01 ± 1.487E-02 & 1.131E-02 ± 1.788E-04 &  MDI     & 9.118E-01 ± 1.442E-02 & 1.096E-02 ± 1.734E-04 \\
     &          & LSTM  & 7.651E-01 ± 1.210E-02 & 9.200E-03 ± 1.455E-04 &          & 6.930E-01 ± 1.096E-02 & 8.334E-03 ± 1.318E-04 \\
     &          & GRU   & 8.638E-01 ± 1.366E-02 & 1.039E-02 ± 1.642E-04 &          & 9.394E-01 ± 1.485E-02 & 1.130E-02 ± 1.786E-04 \\
     &          & RBFNN & 2.704E-01 ± 3.168E-03 & 2.409E-03 ± 3.809E-05 &          & 3.676E-01 ± 4.949E-03 & 4.556E-03 ± 6.349E-05 \\
     &          & \textbf{ALPE}    & \textbf{2.333E-01 ± 3.688E-03} & \textbf{2.005E-03 ± 4.435E-05} &          & \textbf{2.344E-01 ± 2.001E-03} & \textbf{2.741E-03 ± 4.214E-05} \\
    \cline{2-5} \cline{6-8}
     & Simple  & MLP   & 1.414E+00 ± 2.236E-02 & 1.700E-02 ± 2.688E-04 & Exte      & 1.007E+00 ± 1.592E-02 & 1.211E-02 ± 1.915E-04 \\
     & GD      & CNN   & 8.970E-01 ± 1.419E-02 & 1.079E-02 ± 1.706E-04 &  GD       & 8.900E-01 ± 1.407E-02 & 1.070E-02 ± 1.692E-04 \\
     &         & LSTM  & 7.651E-01 ± 1.210E-02 & 9.200E-03 ± 1.455E-04 &           & 6.820E-01 ± 1.178E-02 & 8.533E-03 ± 1.665E-04 \\
     &         & GRU   & 8.832E-01 ± 1.396E-02 & 1.062E-02 ± 1.679E-04 &           & 9.348E-01 ± 1.478E-02 & 1.124E-02 ± 1.778E-04 \\
     &         & RBFNN & 3.545E-01 ± 2.007E-03 & 3.177E-03 ± 5.303E-05 &           & 3.720E-01 ± 2.111E-03 & 3.003E-03 ± 3.939E-05 \\
     &         & \textbf{ALPE}    & \textbf{2.081E-01 ± 3.765E-03} & \textbf{2.864E-03 ± 4.528E-05} &           & \textbf{2.198E-01 ± 4.887E-03} & \textbf{2.743E-03 ± 4.309E-05} \\
     \hline
    NTAP & Simple & Naive & 1.963E+00 ± 3.103E-02 & 2.813E-02 ± 4.447E-04 & Exte     & 2.645E+00 ± 4.880E-02 & 3.002E-02 ± 3.229E-04 \\
         &        & ARIMA & 9.886E-01 ± 1.563E-02 & 1.417E-02 ± 2.240E-04 &          & 8.553E-01 ± 2.352E-02 & 2.998E-02 ± 3.110E-04  \\
         &        & MLP   & 1.315E+00 ± 2.079E-02 & 1.884E-02 ± 2.979E-04 &          & 9.533E-01 ± 1.499E-02 & 1.847E-02 ± 2.230E-04 \\
         &        & CNN   & 8.206E-01 ± 1.298E-02 & 1.176E-02 ± 1.859E-04 &          & 9.371E-01 ± 1.482E-02 & 1.343E-02 ± 2.123E-04 \\
         &        & LSTM  & 9.393E-01 ± 1.485E-02 & 1.346E-02 ± 2.128E-04 &          & 8.686E-01 ± 1.373E-02 & 1.245E-02 ± 1.968E-04 \\
         &        & GRU   & 9.935E-01 ± 1.571E-02 & 1.424E-02 ± 2.251E-04 &          & 9.887E-01 ± 1.563E-02 & 1.417E-02 ± 2.240E-04 \\
         &        & RBFNN & 2.781E-01 ± 2.816E-03 & 3.552E-03 ± 4.036E-05 &          & 2.232E-01 ± 2.739E-03 & 2.482E-03 ± 3.925E-05 \\
         &        & \textbf{ALPE}    & \textbf{2.131E-01 ± 4.951E-03} & \textbf{2.487E-03 ± 7.094E-05} &          & \textbf{1.908E-01 ± 3.965E-03} & \textbf{2.116E-03 ± 4.554E-05} \\
    \cline{2-5} \cline{6-8} 
     & Simple   & MLP   & 9.908E-01 ± 1.567E-02 & 1.420E-02 ± 2.245E-04 & Exte     & 9.487E-01 ± 1.500E-02 & 1.360E-02 ± 2.150E-04 \\
     & MDI      & CNN   & 7.612E-01 ± 1.203E-02 & 1.091E-02 ± 1.725E-04 &  MDI     & 8.232E-01 ± 1.302E-02 & 1.180E-02 ± 1.865E-04 \\
     &          & LSTM  & 8.848E-01 ± 1.399E-02 & 1.268E-02 ± 2.005E-04 &          & 8.432E-01 ± 2.454E-02 & 1.321E-02 ± 1.773E-04 \\
     &          & GRU   & 9.849E-01 ± 1.557E-02 & 1.411E-02 ± 2.231E-04 &          & 9.809E-01 ± 2.609E-02 & 1.533E-02 ± 3.355E-04 \\
     &          & RBFNN & 2.904E-01 ± 3.168E-03 & 3.871E-03 ± 4.540E-05 &          & 2.798E-01 ± 3.654E-03 & 2.589E-03 ± 4.877E-05 \\
     &          & \textbf{ALPE}    & \textbf{2.078E-01 ± 4.709E-03} & \textbf{2.268E-03 ± 6.748E-05} &          & \textbf{1.788E-01 ± 2.080E-03} & \textbf{2.445E-03 ± 4.202E-05} \\
    \cline{2-5} \cline{6-8}
     & Simple  & MLP   & 9.775E-01 ± 1.546E-02 & 1.401E-02 ± 2.215E-04 & Exte      & 1.152E+00 ± 1.821E-02 & 1.651E-02 ± 2.610E-04 \\
     & GD      & CNN   & 9.997E-01 ± 1.581E-02 & 1.433E-02 ± 2.265E-04 &  GD       & 8.938E-01 ± 1.413E-02 & 1.281E-02 ± 2.025E-04 \\
     &         & LSTM  & 8.533E-01 ± 1.349E-02 & 1.223E-02 ± 1.933E-04 &           & 8.215E-01 ± 1.299E-02 & 1.177E-02 ± 1.861E-04 \\
     &         & GRU   & 9.219E-01 ± 1.458E-02 & 1.321E-02 ± 2.089E-04 &           & 9.866E-01 ± 1.654E-02 & 1.433E-02 ± 2.455E-04 \\
     &         & RBFNN & 2.398E-01 ± 3.159E-03 & 3.863E-03 ± 4.526E-05 &           & 2.239E-01 ± 3.565E-03 & 2.665E-03 ± 4.847E-05 \\
     &         & \textbf{ALPE}    & \textbf{1.859E-01 ± 4.362E-03} & \textbf{2.953E-03 ± 6.250E-05} &           & \textbf{1.698E-01 ± 3.776E-03} & \textbf{2.201E-03 ± 3.776E-05} \\
    \bottomrule
    \end{tabular}
    \end{small}}
            \label{tab:table1}
        \end{minipage}%
        \hspace{0.25\textwidth} 
        \begin{minipage}[t]{0.48\textwidth}
            \centering
            \scalebox{0.42}{
    \begin{small}
    \begin{tabular}{c c r c c c c c}
    \toprule
    \textbf{Stock} & \textbf{Set} & \textbf{Model} & \textbf{RMSE} & \textbf{RRMSE} & \textbf{Set} & \textbf{RMSE} & \textbf{RRMSE} \\
    \hline
    NVDA & Simple & Naive & 2.214E+00 ± 3.501E-02 & 1.807E-02 ± 2.857E-04 & Exte     & 3.002E+00 ± 4.225E-02 & 2.767E-02 ± 3.997E-04 \\
         &        & ARIMA & 4.380E-01 ± 6.925E-03 & 3.574E-03 ± 5.651E-05 &          & 3.281E-01 ± 5.114E-03 & 4.990E-03 ± 4.382E-05  \\
         &        & MLP   & 8.081E-01 ± 1.278E-02 & 6.594E-03 ± 1.043E-04 &          & 6.353E-01 ± 1.004E-02 & 5.184E-03 ± 8.196E-05 \\
         &        & CNN   & 4.196E-01 ± 6.634E-03 & 3.424E-03 ± 5.413E-05 &          & 4.211E-01 ± 6.658E-03 & 3.436E-03 ± 5.433E-05 \\
         &        & LSTM  & 3.160E-01 ± 4.997E-03 & 2.579E-03 ± 4.077E-05 &          & 3.117E-01 ± 4.774E-03 & 2.612E-03 ± 4.115E-05 \\
         &        & GRU   & 5.368E-01 ± 8.487E-03 & 4.380E-03 ± 6.925E-05 &          & 5.320E-01 ± 6.029E-03 & 4.468E-03 ± 7.065E-05 \\
         &        & RBFNN & 2.407E-01 ± 3.805E-03 & 1.964E-03 ± 3.105E-05 &          & 2.200E-01 ± 4.287E-03 & 1.760E-03 ± 2.101E-05 \\
         &        & \textbf{ALPE}    & \textbf{2.228E-01 ± 3.523E-03} & \textbf{1.818E-03 ± 2.875E-05} &          & \textbf{2.079E-01 ± 3.288E-03} & \textbf{1.580E-03 ± 2.299E-05} \\
    \cline{2-5} \cline{6-8} 
     & Simple   & MLP   & 7.695E-01 ± 1.217E-02 & 6.279E-03 ± 9.928E-05 & Exte     & 6.388E-01 ± 2.324E-02 & 5.832E-03 ± 6.663E-05 \\
     & MDI      & CNN   & 4.195E-01 ± 6.633E-03 & 3.423E-03 ± 5.413E-05 &  MDI     & 4.192E-01 ± 6.629E-03 & 3.421E-03 ± 5.409E-05 \\
     &          & LSTM  & 3.021E-01 ± 4.776E-03 & 2.465E-03 ± 3.897E-05 &          & 3.108E-01 ± 4.915E-03 & 2.536E-03 ± 4.010E-05 \\
     &          & GRU   & 4.510E-01 ± 7.131E-03 & 3.680E-03 ± 5.819E-05 &          & 5.186E-01 ± 8.199E-03 & 4.231E-03 ± 6.691E-05 \\
     &          & RBFNN & 2.607E-01 ± 3.490E-03 & 1.801E-03 ± 2.848E-05 &          & 2.344E-01 ± 3.998E-03 & 1.682E-03 ± 2.482E-05 \\
     &          & \textbf{ALPE}    & \textbf{2.316E-01 ± 3.662E-03} & \textbf{1.690E-03 ± 2.988E-05} &          & \textbf{2.887E-01 ± 2.334E-03} & \textbf{1.531E-03 ± 2.704E-05} \\
    \cline{2-5} \cline{6-8}
     & Simple  & MLP   & 8.051E-01 ± 1.273E-02 & 6.569E-03 ± 1.039E-04 & Exte      & 8.835E-01 ± 1.397E-02 & 7.209E-03 ± 1.140E-04 \\
     & GD      & CNN   & 4.189E-01 ± 6.624E-03 & 3.419E-03 ± 5.405E-05 &  GD       & 4.203E-01 ± 6.646E-03 & 3.430E-03 ± 5.423E-05 \\
     &         & LSTM  & 3.014E-01 ± 4.765E-03 & 2.459E-03 ± 3.888E-05 &           & 3.068E-01 ± 4.851E-03 & 2.503E-03 ± 3.958E-05 \\
     &         & GRU   & 5.053E-01 ± 7.989E-03 & 4.123E-03 ± 6.519E-05 &           & 5.389E-01 ± 5.736E-03 & 4.503E-03 ± 6.837E-05 \\
     &         & RBFNN & 2.017E-01 ± 2.921E-03 & 1.807E-03 ± 2.383E-05 &           & 2.244E-01 ± 2.758E-03 & 1.823E-03 ± 2.250E-05 \\
     &         & \textbf{ALPE}    & \textbf{1.868E-01 ± 3.744E-03} & \textbf{1.532E-03 ± 3.055E-05} &           & \textbf{1.093E-01 ± 1.980E-03} & \textbf{1.243E-03 ± 2.552E-05}\\
     \hline
     NXPI & Simple & Naive & 2.490E+00 ± 3.936E-02 & 1.670E-02 ± 2.641E-04 & Exte     & 3.503E+00 ± 4.009E-02 & 2.334E-02 ± 4.990E-04 \\
         &        & ARIMA & 3.439E-01 ± 5.438E-03 & 2.307E-03 ± 3.648E-05 &          & 3.889E-01 ± 4.202E-03 & 3.100E-03 ± 4.433E-05 \\
         &        & MLP   & 4.692E-01 ± 7.419E-03 & 3.148E-03 ± 4.977E-05 &          & 4.997E-01 ± 7.900E-03 & 3.352E-03 ± 5.300E-05 \\
         &        & CNN   & 3.667E-01 ± 5.799E-03 & 2.460E-03 ± 3.890E-05 &          & 3.785E-01 ± 5.984E-03 & 2.539E-03 ± 4.015E-05 \\
         &        & LSTM  & 3.026E-01 ± 4.785E-03 & 2.030E-03 ± 3.210E-05 &          & 3.024E-01 ± 4.559E-03 & 2.132E-03 ± 3.891E-05 \\
         &        & GRU   & 3.915E-01 ± 6.191E-03 & 2.627E-03 ± 4.153E-05 &          & 3.923E-01 ± 6.203E-03 & 2.632E-03 ± 4.162E-05 \\
         &        & RBFNN & 3.603E-01 ± 5.697E-03 & 2.417E-03 ± 3.822E-05 &          & 3.168E-01 ± 5.009E-03 & 2.125E-03 ± 3.360E-05 \\
         &        & \textbf{ALPE}    & \textbf{2.247E-01 ± 3.553E-03} & \textbf{1.508E-03 ± 2.384E-05} &          & \textbf{2.776E-01 ± 3.656E-03} & \textbf{1.556E-03 ± 2.791E-05} \\
    \cline{2-5} \cline{6-8} 
     & Simple   & MLP   & 5.228E-01 ± 8.267E-03 & 3.508E-03 ± 5.546E-05 & Exte     & 4.997E-01 ± 7.900E-03 & 3.352E-03 ± 5.300E-05 \\
     & MDI      & CNN   & 3.683E-01 ± 5.823E-03 & 2.471E-03 ± 3.907E-05 &  MDI     & 3.821E-01 ± 6.042E-03 & 2.564E-03 ± 4.053E-05 \\
     &          & LSTM  & 3.086E-01 ± 4.879E-03 & 2.070E-03 ± 3.273E-05 &          & 3.264E-01 ± 4.799E-03 & 2.143E-03 ± 2.145E-05 \\
     &          & GRU   & 3.904E-01 ± 6.173E-03 & 2.619E-03 ± 4.141E-05 &          & 3.787E-01 ± 5.090E-03 & 2.588E-03 ± 5.903E-05 \\
     &          & RBFNN & 4.464E-01 ± 7.059E-03 & 2.995E-03 ± 4.736E-05 &          & 3.405E-01 ± 5.602E-03 & 2.225E-03 ± 3.518E-05 \\
     &          & \textbf{ALPE}    & \textbf{2.314E-01 ± 3.658E-03} & \textbf{1.552E-03 ± 2.454E-05} &          & \textbf{2.873E-01 ± 2.998E-03} & \textbf{1.499E-03 ± 2.109E-05} \\
    \cline{2-5} \cline{6-8}
     & Simple  & MLP   & 4.876E-01 ± 7.710E-03 & 3.271E-03 ± 5.172E-05 & Exte      & 4.520E-01 ± 7.147E-03 & 3.033E-03 ± 4.795E-05 \\
     & GD      & CNN   & 3.558E-01 ± 5.626E-03 & 2.387E-03 ± 3.774E-05 &  GD       & 3.897E-01 ± 6.162E-03 & 2.615E-03 ± 4.134E-05 \\
     &         & LSTM  & 3.024E-01 ± 4.781E-03 & 2.028E-03 ± 3.207E-05 &           & 3.039E-01 ± 4.805E-03 & 2.039E-03 ± 3.223E-05 \\
     &         & GRU   & 3.902E-01 ± 6.170E-03 & 2.618E-03 ± 4.139E-05 &           & 3.890E-01 ± 5.246E-03 & 2.938E-03 ± 5.847E-05 \\
     &         & RBFNN & 3.987E-01 ± 6.305E-03 & 2.675E-03 ± 4.230E-05 &           & 3.340E-01 ± 4.837E-03 & 2.302E-03 ± 2.938E-05 \\
     &         & \textbf{ALPE}    & \textbf{2.379E-01 ± 3.762E-03} & \textbf{1.596E-03 ± 2.524E-05} &           & \textbf{2.335E-01 ± 2.090E-03} & \textbf{1.540E-03 ± 2.398E-05} \\
     \hline
    ODFL & Simple & Naive & 1.697E+00 ± 2.682E-02 & 6.192E-03 ± 9.790E-05 & Exte     & 2.068E+00 ± 1.998E-02 & 5.092E-03 ± 8.422E-05 \\
         &        & ARIMA & 7.655E-01 ± 1.210E-02 & 2.794E-03 ± 4.417E-05 &          & 6.443E-01 ± 2.009E-02 & 3.984E-03 ± 5.198E-05  \\
         &        & MLP   & 1.386E+00 ± 2.192E-02 & 5.059E-03 ± 8.000E-05 &          & 1.204E+00 ± 1.904E-02 & 4.394E-03 ± 6.948E-05 \\
         &        & CNN   & 9.997E-01 ± 1.581E-02 & 3.648E-03 ± 5.769E-05 &          & 9.494E-01 ± 1.501E-02 & 3.465E-03 ± 5.478E-05 \\
         &        & LSTM  & 5.750E-01 ± 9.092E-03 & 2.099E-03 ± 3.318E-05 &          & 5.879E-01 ± 9.296E-03 & 2.146E-03 ± 3.393E-05 \\
         &        & GRU   & 9.083E-01 ± 1.436E-02 & 3.315E-03 ± 5.241E-05 &          & 9.493E-01 ± 1.501E-02 & 3.465E-03 ± 5.478E-05 \\
         &        & RBFNN & 3.021E-01 ± 4.776E-03 & 1.102E-03 ± 1.743E-05 &          & 2.798E-01 ± 4.425E-03 & 1.021E-03 ± 1.615E-05 \\
         &        & \textbf{ALPE}   & \textbf{2.312E-01 ± 3.655E-03} & \textbf{8.437E-04 ± 1.334E-05} &          & \textbf{2.352E-01 ± 3.718E-03} & \textbf{8.583E-04 ± 1.357E-05} \\
    \cline{2-5} \cline{6-8} 
     & Simple   & MLP   & 1.331E+00 ± 2.104E-02 & 4.857E-03 ± 7.680E-05 & Exte     & 1.233E+00 ± 2.898E-02 & 4.477E-03 ± 5.873E-05 \\
     & MDI      & CNN   & 9.838E-01 ± 1.556E-02 & 3.591E-03 ± 5.677E-05 &  MDI     & 8.598E-01 ± 1.683E-02 & 3.490E-03 ± 5.559E-05 \\
     &          & LSTM  & 5.738E-01 ± 9.073E-03 & 2.094E-03 ± 3.311E-05 &          & 5.799E-01 ± 7.360E-03 & 2.299E-03 ± 4.466E-05 \\
     &          & GRU   & 9.432E-01 ± 1.491E-02 & 3.443E-03 ± 5.443E-05 &          & 8.988E-01 ± 2.609E-02 & 3.577E-03 ± 3.229E-05 \\
     &          & RBFNN & 3.000E-01 ± 4.744E-03 & 1.095E-03 ± 1.731E-05 &          & 2.699E-01 ± 3.847E-03 & 1.131E-03 ± 2.589E-05 \\
     &          & \textbf{ALPE}    & \textbf{2.301E-01 ± 3.638E-03} & \textbf{8.396E-04 ± 1.328E-05} &          & \textbf{2.224E-01 ± 2.498E-03} & \textbf{8.008E-04 ± 2.271E-05} \\
    \cline{2-5} \cline{6-8}
     & Simple  & MLP   & 1.291E+00 ± 2.042E-02 & 4.713E-03 ± 7.451E-05 & Exte      & 1.146E+00 ± 1.811E-02 & 4.181E-03 ± 6.611E-05 \\
     & GD      & CNN   & 9.787E-01 ± 1.547E-02 & 3.572E-03 ± 5.647E-05 &  GD       & 8.889E-01 ± 2.484E-02 & 3.489E-03 ± 4.478E-05 \\
     &         & LSTM  & 5.726E-01 ± 9.053E-03 & 2.090E-03 ± 3.304E-05 &           & 5.601E-01 ± 8.137E-03 & 2.200E-03 ± 4.101E-05 \\
     &         & GRU   & 9.483E-01 ± 1.499E-02 & 3.461E-03 ± 5.472E-05 &           & 8.389E-01 ± 1.030E-02 & 3.579E-03 ± 4.898E-05 \\
     &         & RBFNN & 3.133E-01 ± 4.954E-03 & 1.144E-03 ± 1.808E-05 &           & 2.587E-01 ± 3.398E-03 & 1.119E-03 ± 2.765E-05 \\
     &         & \textbf{ALPE}    & \textbf{2.379E-01 ± 3.762E-03} & \textbf{8.683E-04 ± 1.373E-05} &           & \textbf{2.382E-01 ± 2.955E-03} & \textbf{8.234E-04 ± 1.276E-05} \\
      \hline
    OMC & Simple & Naive  & 2.448E+00 ± 3.870E-02 & 3.828E-02 ± 6.053E-04 & Exte     & 3.002E+00 ± 4.874E-02 & 4.090E-02 ± 5.113E-04 \\
         &        & ARIMA & 6.235E-01 ± 9.859E-03 & 9.752E-03 ± 1.542E-04 &          & 5.119E-01 ± 8.870E-03 & 8.594E-03 ± 2.871E-04  \\
         &        & MLP   & 7.884E-01 ± 1.247E-02 & 1.233E-02 ± 1.950E-04 &          & 8.168E-01 ± 1.291E-02 & 1.277E-02 ± 2.020E-04 \\
         &        & CNN   & 4.684E-01 ± 7.407E-03 & 7.326E-03 ± 1.158E-04 &          & 4.765E-01 ± 7.534E-03 & 7.452E-03 ± 1.178E-04 \\
         &        & LSTM  & 3.550E-01 ± 5.612E-03 & 5.552E-03 ± 8.778E-05 &          & 3.526E-01 ± 5.576E-03 & 5.515E-03 ± 8.720E-05 \\
         &        & GRU   & 6.536E-01 ± 1.033E-02 & 1.022E-02 ± 1.616E-04 &          & 6.516E-01 ± 1.030E-02 & 1.019E-02 ± 1.611E-04 \\
         &        & RBFNN & 2.220E-01 ± 3.509E-03 & 3.471E-03 ± 5.489E-05 &          & 2.210E-01 ± 3.494E-03 & 3.456E-03 ± 5.464E-05 \\
         &        & \textbf{ALPE}    & \textbf{1.440E-01 ± 2.276E-03} & \textbf{2.251E-03 ± 3.560E-05} &          & \textbf{1.776E-01 ± 2.420E-03} & \textbf{2.765E-03 ± 3.644E-05} \\
    \cline{2-5} \cline{6-8} 
     & Simple   & MLP   & 8.363E-01 ± 1.322E-02 & 1.308E-02 ± 2.068E-04 & Exte     & 8.299E-01 ± 2.133E-02 & 1.452E-02 ± 2.112E-04 \\
     & MDI      & CNN   & 4.691E-01 ± 7.417E-03 & 7.337E-03 ± 1.160E-04 &  MDI     & 4.677E-01 ± 6.209E-03 & 7.365E-03 ± 2.332E-04 \\
     &          & LSTM  & 3.650E-01 ± 5.772E-03 & 5.709E-03 ± 9.027E-05 &          & 3.603E-01 ± 5.697E-03 & 5.635E-03 ± 8.910E-05 \\
     &          & GRU   & 6.545E-01 ± 1.035E-02 & 1.024E-02 ± 1.618E-04 &          & 6.564E-01 ± 1.038E-02 & 1.027E-02 ± 1.623E-04 \\
     &          & RBFNN & 2.004E-01 ± 3.168E-03 & 3.134E-03 ± 4.955E-05 &          & 1.843E-01 ± 2.915E-03 & 2.883E-03 ± 4.559E-05 \\
     &          & \textbf{ALPE}    & \textbf{1.531E-01 ± 2.421E-03} & \textbf{2.395E-03 ± 3.786E-05} &          & \textbf{1.537E-01 ± 2.843E-03} & \textbf{2.404E-03 ± 3.800E-05} \\
    \cline{2-5} \cline{6-8}
     & Simple  & MLP   & 7.853E-01 ± 1.242E-02 & 1.228E-02 ± 1.942E-04 & Exte      & 7.866E-01 ± 1.244E-02 & 1.350E-02 ± 1.975E-04 \\
     & GD      & CNN   & 4.681E-01 ± 7.402E-03 & 7.321E-03 ± 1.158E-04 &  GD       & 4.697E-01 ± 7.426E-03 & 7.346E-03 ± 1.161E-04 \\
     &         & LSTM  & 3.526E-01 ± 5.576E-03 & 5.515E-03 ± 8.720E-05 &           & 3.603E-01 ± 5.697E-03 & 5.635E-03 ± 8.910E-05 \\
     &         & GRU   & 6.516E-01 ± 1.030E-02 & 1.019E-02 ± 1.611E-04 &           & 6.581E-01 ± 1.041E-02 & 1.029E-02 ± 1.627E-04 \\
     &         & RBFNN & 2.210E-01 ± 3.494E-03 & 3.456E-03 ± 5.464E-05 &           & 1.735E-01 ± 2.743E-03 & 2.713E-03 ± 4.289E-05 \\
     &         & \textbf{ALPE}    & \textbf{1.570E-01 ± 2.932E-03} & \textbf{2.456E-03 ± 3.883E-05} &           & \textbf{1.220E-01 ± 2.339E-03} & \textbf{2.411E-03 ± 4.703E-05} \\
    \bottomrule
    \end{tabular}
    \end{small}}
            \label{tab:table_9}
        \end{minipage}
    \end{table}
\end{landscape}


\begin{landscape}
    \fancyhf{}  
    \fancyfoot[R]{\rotatebox{90}{\thepage}}  
    \thispagestyle{fancy}  

    \begin{table}[!hbtp]
        \centering
        \caption{RMSE and RRMSE scores for ORCL, OALPEY, OTIS, PEAK, PEP, PG, PHM, and POOL.}
        \hspace*{-5cm} 
        \begin{minipage}[t]{0.48\textwidth}
            \centering
            \scalebox{0.42}{
    \begin{small}
    \begin{tabular}{c c r c c c c c}
    \toprule
    \textbf{Stock} & \textbf{Set} & \textbf{Model} & \textbf{RMSE} & \textbf{RRMSE} & \textbf{Set} & \textbf{RMSE} & \textbf{RRMSE} \\
    \hline     
    ORCL & Simple & Naive & 1.962E+00 ± 3.102E-02 & 3.216E-02 ± 5.084E-04 & Exte & 2.114E+00 ± 4.009E-02 & 4.224E-02 ± 4.994E-04 \\
         &        & ARIMA & 6.743E-01 ± 1.066E-02 & 1.105E-02 ± 1.747E-04 &      & 5.090E-01 ± 2.888E-02 & 2.669E-02 ± 2.912E-04  \\
         &        & MLP   & 7.991E-01 ± 1.263E-02 & 1.310E-02 ± 2.071E-04 &      & 7.999E-01 ± 1.265E-02 & 1.311E-02 ± 2.073E-04 \\
         &        & CNN   & 5.164E-01 ± 8.165E-03 & 8.464E-03 ± 1.338E-04 &      & 5.219E-01 ± 8.252E-03 & 8.553E-03 ± 1.352E-04 \\
         &        & LSTM  & 4.434E-01 ± 5.232E-03 & 7.337E-03 ± 1.160E-04 &      & 4.242E-01 ± 6.707E-03 & 6.952E-03 ± 1.099E-04 \\
         &        & GRU   & 6.875E-01 ± 1.087E-02 & 1.127E-02 ± 1.781E-04 &      & 6.738E-01 ± 1.065E-02 & 1.104E-02 ± 1.746E-04 \\
         &        & RBFNN & 4.484E-01 ± 7.090E-03 & 7.350E-03 ± 1.162E-04 &      & 4.784E-01 ± 7.564E-03 & 7.841E-03 ± 1.240E-04 \\
         &        & \textbf{ALPE}    & \textbf{2.031E-01 ± 3.501E-03} & \textbf{3.325E-03 ± 5.257E-05} &      & \textbf{2.011E-01 ± 3.180E-03} & \textbf{3.296E-03 ± 5.212E-05} \\
    \cline{2-5} \cline{6-8} 
     & Simple     & MLP   & 8.355E-01 ± 1.321E-02 & 1.369E-02 ± 2.165E-04 & Exte  & 7.776E-01 ± 1.229E-02 & 1.274E-02 ± 2.015E-04 \\
     &    MDI     & CNN   & 5.168E-01 ± 8.172E-03 & 8.471E-03 ± 1.339E-04 &  MDI  & 5.140E-01 ± 8.128E-03 & 8.425E-03 ± 1.332E-04 \\
     &            & LSTM  & 4.138E-01 ± 6.543E-03 & 6.782E-03 ± 1.072E-04 &       & 4.183E-01 ± 6.614E-03 & 6.855E-03 ± 1.084E-04 \\
     &            & GRU   & 6.597E-01 ± 1.043E-02 & 1.081E-02 ± 1.710E-04 &       & 7.035E-01 ± 1.112E-02 & 1.153E-02 ± 1.823E-04 \\
     &            & RBFNN & 4.900E-01 ± 7.748E-03 & 8.032E-03 ± 1.270E-04 &       & 5.349E-01 ± 8.458E-03 & 8.767E-03 ± 1.386E-04 \\
     &            & \textbf{ALPE}    & \textbf{2.315E-01 ± 3.660E-03} & \textbf{3.794E-03 ± 5.999E-05} &       & \textbf{2.183E-01 ± 3.452E-03} & \textbf{3.579E-03 ± 5.658E-05} \\
    \cline{2-5} \cline{6-8}
     & Simple    & MLP   & 8.119E-01 ± 1.284E-02 & 1.331E-02 ± 2.104E-04 & Exte   & 7.922E-01 ± 1.253E-02 & 1.298E-02 ± 2.053E-04 \\
     &  GD       & CNN   & 5.162E-01 ± 8.161E-03 & 8.460E-03 ± 1.338E-04 &  GD    & 5.173E-01 ± 8.179E-03 & 8.478E-03 ± 1.341E-04 \\
     &           & LSTM  & 4.132E-01 ± 6.534E-03 & 6.773E-03 ± 1.071E-04 &        & 4.302E-01 ± 6.802E-03 & 7.051E-03 ± 1.115E-04 \\
     &           & GRU   & 6.329E-01 ± 1.001E-02 & 1.037E-02 ± 1.640E-04 &        & 6.999E-01 ± 1.107E-02 & 1.147E-02 ± 1.814E-04 \\
     &           & RBFNN & 5.058E-01 ± 7.997E-03 & 8.289E-03 ± 1.311E-04 &        & 4.679E-01 ± 7.397E-03 & 7.668E-03 ± 1.212E-04 \\
     &           & \textbf{ALPE}    & \textbf{2.131E-01 ± 3.370E-03} & \textbf{3.493E-03 ± 5.523E-05} &        & \textbf{2.277E-01 ± 3.443E-03} & \textbf{3.234E-03 ± 5.116E-05}\\
     \hline
     OALPEY & Simple & Naive & 1.972E+00 ± 3.118E-02 & 2.797E-03 ± 4.422E-05 & Exte & 2.776E+00 ± 2.009E-02 & 3.652E-03 ± 3.010E-05 \\
         &        & ARIMA & 9.436E-01 ± 1.492E-02 & 1.338E-03 ± 2.116E-05 &      & 8.322E-01 ± 2.980E-02 & 2.119E-03 ± 3.298E-05  \\
         &        & MLP   & 1.332E+00 ± 2.105E-02 & 1.889E-03 ± 2.986E-05 &      & 1.228E+00 ± 1.942E-02 & 1.742E-03 ± 2.755E-05 \\
         &        & CNN   & 6.639E-01 ± 1.050E-02 & 9.416E-04 ± 1.489E-05 &      & 6.413E-01 ± 1.014E-02 & 9.096E-04 ± 1.438E-05 \\
         &        & LSTM  & 5.899E-01 ± 9.327E-03 & 8.367E-04 ± 1.323E-05 &      & 5.944E-01 ± 9.398E-03 & 8.431E-04 ± 1.333E-05 \\
         &        & GRU   & 8.053E-01 ± 1.273E-02 & 1.142E-03 ± 1.806E-05 &      & 8.078E-01 ± 1.277E-02 & 1.146E-03 ± 1.812E-05 \\
         &        & RBFNN & 2.497E-01 ± 3.948E-03 & 3.542E-04 ± 5.600E-06 &      & 2.450E-01 ± 3.873E-03 & 3.474E-04 ± 5.494E-06 \\
         &        & \textbf{ALPE}    & \textbf{2.258E-01 ± 3.570E-03} & \textbf{3.202E-04 ± 5.064E-06} &      & \textbf{2.370E-01 ± 3.747E-03} & \textbf{3.362E-04 ± 5.315E-06} \\
    \cline{2-5} \cline{6-8} 
     & Simple   & MLP   & 1.339E+00 ± 2.117E-02 & 1.899E-03 ± 3.003E-05 & Exte   & 1.345E+00 ± 2.127E-02 & 1.908E-03 ± 3.017E-05 \\
     &  MDI     & CNN   & 6.928E-01 ± 1.095E-02 & 9.827E-04 ± 1.554E-05 &  MDI   & 6.358E-01 ± 1.005E-02 & 9.019E-04 ± 1.426E-05 \\
     &          & LSTM  & 5.915E-01 ± 9.352E-03 & 8.390E-04 ± 1.327E-05 &        & 5.932E-01 ± 9.380E-03 & 8.415E-04 ± 1.330E-05 \\
     &          & GRU   & 8.062E-01 ± 1.275E-02 & 1.144E-03 ± 1.808E-05 &        & 8.044E-01 ± 1.201E-02 & 1.126E-03 ± 1.510E-05 \\
     &          & RBFNN & 2.452E-01 ± 3.878E-03 & 3.479E-04 ± 5.500E-06 &        & 2.448E-01 ± 3.870E-03 & 3.472E-04 ± 5.489E-06 \\
     &          & \textbf{ALPE}    & \textbf{2.303E-01 ± 2.009E-03} & \textbf{3.262E-04 ± 5.158E-06} &        & \textbf{2.394E-01 ± 3.785E-03} & \textbf{3.395E-04 ± 5.368E-06} \\
    \cline{2-5} \cline{6-8}
     & Simple   & MLP   & 1.228E+00 ± 1.942E-02 & 1.742E-03 ± 2.755E-05 & Exte   & 1.016E+00 ± 1.607E-02 & 1.442E-03 ± 2.280E-05 \\
     &   GD     & CNN   & 6.902E-01 ± 1.091E-02 & 9.790E-04 ± 1.548E-05 &   GD   & 6.641E-01 ± 1.050E-02 & 9.419E-04 ± 1.489E-05 \\
     &          & LSTM  & 5.903E-01 ± 9.333E-03 & 8.372E-04 ± 1.324E-05 &        & 6.064E-01 ± 9.587E-03 & 8.601E-04 ± 1.360E-05 \\
     &          & GRU   & 8.080E-01 ± 1.278E-02 & 1.146E-03 ± 1.812E-05 &        & 8.078E-01 ± 1.503E-02 & 1.612E-03 ± 1.355E-05 \\
     &          & RBFNN & 2.589E-01 ± 4.093E-03 & 3.672E-04 ± 5.806E-06 &        & 2.429E-01 ± 3.840E-03 & 3.445E-04 ± 5.447E-06 \\
     &          & \textbf{ALPE}    & \textbf{2.359E-01 ± 3.730E-03} & \textbf{3.346E-04 ± 5.290E-06} &        & \textbf{2.326E-01 ± 3.678E-03} & \textbf{3.300E-04 ± 5.218E-06} \\
     \hline
    OTIS & Simple & Naive & 1.693E+00 ± 2.677E-02 & 2.628E-02 ± 4.155E-04 & Exte & 2.886E+00 ± 3.091E-02 & 3.555E-02 ± 3.102E-04 \\
         &        & ARIMA & 5.211E-01 ± 8.239E-03 & 8.089E-03 ± 1.279E-04 &      & 5.211E-01 ± 7.022E-03 & 7.911E-03 ± 2.300E-04  \\
         &        & MLP   & 2.499E-01 ± 2.370E-03 & 2.927E-03 ± 3.679E-05 &      & 1.906E-01 ± 2.540E-03 & 2.794E-03 ± 3.943E-05 \\
         &        & CNN   & 5.348E-01 ± 8.456E-03 & 6.302E-03 ± 2.009E-04 &      & 5.361E-01 ± 8.477E-03 & 8.323E-03 ± 1.316E-04 \\
         &        & LSTM  & 5.407E-01 ± 8.549E-03 & 8.394E-03 ± 1.327E-04 &      & 5.447E-01 ± 8.613E-03 & 8.456E-03 ± 1.337E-04 \\
         &        & GRU   & 5.409E-01 ± 8.552E-03 & 8.396E-03 ± 1.328E-04 &      & 5.501E-01 ± 7.399E-03 & 8.115E-03 ± 2.984E-04 \\
         &        & RBFNN & 1.514E-01 ± 2.236E-03 & 2.795E-03 ± 3.471E-05 &      & 1.833E-01 ± 2.108E-03 & 2.970E-03 ± 3.272E-05 \\
         &        & \textbf{ALPE}    & \textbf{1.368E-01 ± 3.587E-03} & \textbf{2.422E-03 ± 5.568E-05} &      & \textbf{1.580E-01 ± 3.764E-03} & \textbf{2.595E-03 ± 5.843E-05} \\
    \cline{2-5} \cline{6-8} 
     & Simple   & MLP   & 1.870E-01 ± 3.198E-03 & 2.399E-03 ± 3.793E-05 & Exte   & 1.638E-01 ± 1.958E-03 & 1.922E-03 ± 3.040E-05 \\
     & MDI      & CNN   & 5.328E-01 ± 8.425E-03 & 8.272E-03 ± 1.308E-04 & MDI    & 5.311E-01 ± 8.397E-03 & 8.244E-03 ± 1.234E-04 \\
     &          & LSTM  & 5.421E-01 ± 8.571E-03 & 8.415E-03 ± 1.331E-04 &        & 5.438E-01 ± 8.598E-03 & 8.441E-03 ± 1.335E-04 \\
     &          & GRU   & 5.420E-01 ± 8.570E-03 & 8.415E-03 ± 1.330E-04 &        & 5.438E-01 ± 8.598E-03 & 8.442E-03 ± 1.335E-04 \\
     &          & RBFNN & 1.754E-01 ± 1.856E-03 & 1.822E-03 ± 2.881E-05 &        & 1.774E-01 ± 2.173E-03 & 2.133E-03 ± 3.373E-05 \\
     &          & \textbf{ALPE}    & \textbf{1.605E-01 ± 3.644E-03} & \textbf{1.578E-03 ± 5.657E-05} &        & \textbf{1.382E-01 ± 3.767E-03} & \textbf{1.698E-03 ± 5.847E-05} \\
    \cline{2-5} \cline{6-8}
     & Simple   & MLP   & 2.182E-01 ± 2.502E-03 & 2.456E-03 ± 3.884E-05 & Exte    & 1.922E-01 ± 2.090E-03 & 2.352E-03 ± 3.245E-05 \\
     &   GD     & CNN   & 5.356E-01 ± 8.469E-03 & 8.315E-03 ± 1.315E-04 &   GD    & 5.350E-01 ± 8.458E-03 & 8.305E-03 ± 1.313E-04 \\
     &          & LSTM  & 5.410E-01 ± 8.554E-03 & 8.399E-03 ± 1.328E-04 &         & 5.576E-01 ± 8.786E-03 & 8.626E-03 ± 1.364E-04 \\
     &          & GRU   & 5.443E-01 ± 8.606E-03 & 8.450E-03 ± 1.336E-04 &         & 5.405E-01 ± 8.322E-03 & 8.510E-03 ± 1.421E-04 \\
     &          & RBFNN & 1.992E-01 ± 2.202E-03 & 2.162E-03 ± 3.418E-05 &         & 1.822E-01 ± 1.517E-03 & 2.587E-03 ± 2.510E-05 \\
     &          & \textbf{ALPE}    & \textbf{1.763E-01 ± 3.736E-03} & \textbf{1.668E-03 ± 5.799E-05} &         & \textbf{1.344E-01 ± 3.706E-03} & \textbf{2.139E-03 ± 5.754E-05} \\
     \hline
    PEAK & Simple & Naive & 2.147E+00 ± 3.395E-02 & 9.283E-02 ± 1.468E-03 & Exte & 3.170E+00 ± 4.977E-02 & 8.110E-02 ± 2.332E-03 \\
         &        & ARIMA & 6.189E-01 ± 9.786E-03 & 2.676E-02 ± 4.231E-04 &      & 5.113E-01 ± 8.009E-03 & 3.444E-02 ± 5.660E-04  \\
         &        & MLP   & 1.269E+00 ± 2.006E-02 & 5.485E-02 ± 8.672E-04 &      & 1.106E+00 ± 1.748E-02 & 4.780E-02 ± 7.558E-04 \\
         &        & CNN   & 5.880E-01 ± 9.297E-03 & 2.542E-02 ± 4.020E-04 &      & 5.891E-01 ± 9.314E-03 & 2.547E-02 ± 4.027E-04 \\
         &        & LSTM  & 5.397E-01 ± 8.534E-03 & 2.333E-02 ± 3.689E-04 &      & 5.050E-01 ± 7.985E-03 & 2.184E-02 ± 3.452E-04 \\
         &        & GRU   & 6.718E-01 ± 1.062E-02 & 2.905E-02 ± 4.593E-04 &      & 6.886E-01 ± 1.089E-02 & 2.977E-02 ± 4.707E-04 \\
         &        & RBFNN & 4.234E-01 ± 6.695E-03 & 1.831E-02 ± 2.894E-04 &      & 3.592E-01 ± 5.679E-03 & 1.553E-02 ± 2.455E-04 \\
         &        & \textbf{ALPE}    & \textbf{2.269E-01 ± 3.588E-03} & \textbf{9.811E-03 ± 1.551E-04} &      & \textbf{2.385E-01 ± 3.771E-03} & \textbf{1.031E-02 ± 1.630E-04} \\
    \cline{2-5} \cline{6-8} 
     & Simple     & MLP   & 1.247E+00 ± 1.972E-02 & 5.392E-02 ± 8.526E-04 & Exte & 1.099E+00 ± 1.461E-02 & 4.324E-02 ± 6.837E-04 \\
     &  MDI       & CNN   & 5.865E-01 ± 9.273E-03 & 2.535E-02 ± 4.009E-04 & MDI  & 5.851E-01 ± 9.251E-03 & 2.530E-02 ± 4.000E-04 \\
     &            & LSTM  & 5.026E-01 ± 7.946E-03 & 2.173E-02 ± 3.436E-04 &      & 5.041E-01 ± 7.971E-03 & 2.180E-02 ± 3.446E-04 \\
     &            & GRU   & 6.727E-01 ± 1.064E-02 & 2.908E-02 ± 4.598E-04 &      & 6.565E-01 ± 1.038E-02 & 2.838E-02 ± 4.488E-04 \\
     &            & RBFNN & 4.117E-01 ± 6.509E-03 & 1.780E-02 ± 2.814E-04 &      & 3.606E-01 ± 5.702E-03 & 1.559E-02 ± 2.465E-04 \\
     &            & \textbf{ALPE}    & \textbf{2.296E-01 ± 3.631E-03} & \textbf{9.927E-03 ± 1.570E-04} &      & \textbf{2.365E-01 ± 3.740E-03} & \textbf{1.023E-02 ± 1.617E-04} \\
    \cline{2-5} \cline{6-8}
     & Simple    & MLP   & 1.374E+00 ± 2.173E-02 & 5.942E-02 ± 9.395E-04 & Exte  & 1.040E+00 ± 1.737E-02 & 4.244E-02 ± 6.407E-04 \\
     &   GD      & CNN   & 5.887E-01 ± 9.308E-03 & 2.545E-02 ± 4.024E-04 &   GD  & 5.882E-01 ± 9.300E-03 & 2.543E-02 ± 4.021E-04 \\
     &           & LSTM  & 5.017E-01 ± 7.932E-03 & 2.169E-02 ± 3.429E-04 &       & 5.152E-01 ± 8.146E-03 & 2.227E-02 ± 3.522E-04 \\
     &           & GRU   & 6.816E-01 ± 1.078E-02 & 2.947E-02 ± 4.660E-04 &       & 6.429E-01 ± 1.016E-02 & 2.779E-02 ± 4.395E-04 \\
     &           & RBFNN & 4.471E-01 ± 7.069E-03 & 1.933E-02 ± 3.056E-04 &       & 3.163E-01 ± 5.001E-03 & 1.367E-02 ± 2.162E-04 \\
     &           & \textbf{ALPE}    & \textbf{2.371E-01 ± 3.748E-03} & \textbf{1.025E-02 ± 1.621E-04} &       & \textbf{2.336E-01 ± 3.694E-03} & \textbf{1.010E-02 ± 1.597E-04} \\

    \bottomrule
    \end{tabular}
    \end{small}}
            \label{tab:table1}
        \end{minipage}%
        \hspace{0.25\textwidth} 
        \begin{minipage}[t]{0.48\textwidth}
            \centering
            \scalebox{0.42}{
    \begin{small}
    \begin{tabular}{c c r c c c c c}
    \toprule
    \textbf{Stock} & \textbf{Set} & \textbf{Model} & \textbf{RMSE} & \textbf{RRMSE} & \textbf{Set} & \textbf{RMSE} & \textbf{RRMSE} \\
    \hline     
    PEP & Simple  & Naive & 2.431E+00 ± 3.843E-02 & 1.479E-02 ± 2.339E-04 & Exte & 3.887E+00 ± 4.332E-02 & 2.110E-02 ± 3.898E-04 \\
         &        & ARIMA & 4.828E-01 ± 7.633E-03 & 2.937E-03 ± 4.644E-05 &      & 5.116E-01 ± 6.224E-03 & 3.988E-03 ± 5.005E-05  \\
         &        & MLP   & 9.125E-01 ± 1.443E-02 & 5.552E-03 ± 8.779E-05 &      & 8.937E-01 ± 1.413E-02 & 5.438E-03 ± 8.598E-05 \\
         &        & CNN   & 6.742E-01 ± 1.066E-02 & 4.102E-03 ± 6.487E-05 &      & 7.396E-01 ± 1.169E-02 & 4.500E-03 ± 7.116E-05 \\
         &        & LSTM  & 5.013E-01 ± 7.926E-03 & 3.050E-03 ± 4.823E-05 &      & 5.050E-01 ± 7.985E-03 & 3.073E-03 ± 4.859E-05 \\
         &        & GRU   & 5.014E-01 ± 7.927E-03 & 3.051E-03 ± 4.824E-05 &      & 5.042E-01 ± 7.972E-03 & 3.068E-03 ± 4.851E-05 \\
         &        & RBFNN & 4.370E-01 ± 6.910E-03 & 2.659E-03 ± 4.204E-05 &      & 3.216E-01 ± 5.086E-03 & 1.957E-03 ± 3.094E-05 \\
         &        & \textbf{ALPE}    & \textbf{2.785E-01 ± 4.403E-03} & \textbf{1.694E-03 ± 2.679E-05} &      & \textbf{2.762E-01 ± 4.368E-03} & \textbf{1.681E-03 ± 2.658E-05} \\
    \cline{2-5} \cline{6-8} 
     & Simple   & MLP   & 8.975E-01 ± 1.419E-02 & 5.461E-03 ± 8.634E-05 & Exte   & 8.769E-01 ± 1.387E-02 & 5.336E-03 ± 8.437E-05 \\
     &  MDI     & CNN   & 6.663E-01 ± 1.053E-02 & 4.054E-03 ± 6.410E-05 &   MDI  & 7.364E-01 ± 1.164E-02 & 4.481E-03 ± 7.085E-05 \\
     &          & LSTM  & 5.026E-01 ± 7.946E-03 & 3.058E-03 ± 4.835E-05 &        & 5.041E-01 ± 7.971E-03 & 3.068E-03 ± 4.850E-05 \\
     &          & GRU   & 5.025E-01 ± 7.944E-03 & 3.057E-03 ± 4.834E-05 &        & 5.041E-01 ± 7.970E-03 & 3.067E-03 ± 4.850E-05 \\
     &          & RBFNN & 4.231E-01 ± 6.689E-03 & 2.574E-03 ± 4.070E-05 &        & 3.479E-01 ± 5.500E-03 & 2.117E-03 ± 3.347E-05 \\
     &          & \textbf{ALPE}    & \textbf{2.797E-01 ± 4.423E-03} & \textbf{1.702E-03 ± 2.691E-05} &        & \textbf{2.579E-01 ± 4.078E-03} & \textbf{1.569E-03 ± 2.481E-05} \\
    \cline{2-5} \cline{6-8}
     & Simple   & MLP   & 9.031E-01 ± 1.428E-02 & 5.495E-03 ± 8.689E-05 & Exte    & 8.369E-01 ± 1.323E-02 & 5.092E-03 ± 8.051E-05 \\
     &   GD     & CNN   & 8.041E-01 ± 1.271E-02 & 4.893E-03 ± 7.736E-05 &   GD    & 6.678E-01 ± 1.056E-02 & 4.063E-03 ± 6.425E-05 \\
     &          & LSTM  & 5.017E-01 ± 7.932E-03 & 3.053E-03 ± 4.826E-05 &         & 5.152E-01 ± 8.146E-03 & 3.135E-03 ± 4.956E-05 \\
     &          & GRU   & 5.046E-01 ± 7.979E-03 & 3.070E-03 ± 4.855E-05 &         & 5.043E-01 ± 7.973E-03 & 3.048E-03 ± 4.852E-05 \\
     &          & RBFNN & 4.229E-01 ± 6.687E-03 & 2.573E-03 ± 4.069E-05 &         & 3.165E-01 ± 5.005E-03 & 1.926E-03 ± 3.045E-05 \\
     &          & \textbf{ALPE}    & \textbf{2.811E-01 ± 4.444E-03} & \textbf{1.710E-03 ± 2.704E-05} &         & \textbf{2.532E-01 ± 4.003E-03} & \textbf{1.540E-03 ± 2.436E-05}\\
     \hline
     PG & Simple   & Naive & 2.236E+00 ± 3.535E-02 & 1.754E-02 ± 2.774E-04 & Exte & 3.110E+00 ± 4.998E-02 & 2.343E-02 ± 3.883E-04 \\
         &        & ARIMA & 4.362E-01 ± 6.897E-03 & 3.423E-03 ± 5.412E-05 &      & 3.777E-01 ± 5.004E-03 & 4.443E-03 ± 4.221E-05  \\
         &        & MLP   & 1.374E+00 ± 2.173E-02 & 1.078E-02 ± 1.705E-04 &      & 1.333E+00 ± 2.108E-02 & 1.046E-02 ± 1.654E-04 \\
         &        & CNN   & 1.148E+00 ± 1.816E-02 & 9.010E-03 ± 1.425E-04 &      & 1.049E+00 ± 1.658E-02 & 8.230E-03 ± 1.301E-04 \\
         &        & LSTM  & 4.540E-01 ± 7.179E-03 & 3.553E-03 ± 5.633E-05 &      & 4.579E-01 ± 7.240E-03 & 3.593E-03 ± 5.681E-05 \\
         &        & GRU   & 4.541E-01 ± 7.180E-03 & 3.422E-03 ± 5.534E-05 &      & 4.567E-01 ± 7.220E-03 & 3.583E-03 ± 5.666E-05 \\
         &        & RBFNN & 4.217E-01 ± 6.668E-03 & 3.309E-03 ± 5.233E-05 &      & 3.708E-01 ± 5.862E-03 & 2.909E-03 ± 4.600E-05 \\
         &        & \textbf{ALPE}    & \textbf{2.253E-01 ± 3.472E-03} & \textbf{1.768E-03 ± 2.796E-05} &      & \textbf{2.368E-01 ± 3.743E-03} & \textbf{1.858E-03 ± 2.938E-05} \\
    \cline{2-5} \cline{6-8} 
     & Simple   & MLP   & 1.291E+00 ± 2.041E-02 & 1.013E-02 ± 1.602E-04 & Exte   & 1.342E+00 ± 2.122E-02 & 1.053E-02 ± 1.665E-04 \\
     &  MDI     & CNN   & 1.096E+00 ± 1.732E-02 & 8.598E-03 ± 1.359E-04 & MDI    & 1.016E+00 ± 1.607E-02 & 7.974E-03 ± 1.261E-04 \\
     &          & LSTM  & 4.561E-01 ± 7.212E-03 & 3.579E-03 ± 5.659E-05 &        & 4.567E-01 ± 7.222E-03 & 3.584E-03 ± 5.667E-05 \\
     &          & GRU   & 4.551E-01 ± 7.195E-03 & 3.571E-03 ± 5.646E-05 &        & 4.564E-01 ± 7.217E-03 & 3.582E-03 ± 5.663E-05 \\
     &          & RBFNN & 3.944E-01 ± 6.237E-03 & 3.095E-03 ± 4.894E-05 &        & 3.334E-01 ± 3.776E-03 & 2.472E-03 ± 3.264E-05 \\
     &          & \textbf{ALPE}    & \textbf{2.317E-01 ± 3.664E-03} & \textbf{1.818E-03 ± 2.875E-05} &        & \textbf{2.305E-01 ± 3.776E-03} & \textbf{1.874E-03 ± 2.963E-05} \\
    \cline{2-5} \cline{6-8}
     & Simple    & MLP   & 1.247E+00 ± 1.972E-02 & 9.787E-03 ± 1.548E-04 & Exte  & 1.359E+00 ± 2.149E-02 & 1.066E-02 ± 1.686E-04 \\
     &   GD      & CNN   & 1.003E+00 ± 1.585E-02 & 7.868E-03 ± 1.244E-04 &   GD  & 1.204E+00 ± 1.478E-02 & 7.739E-03 ± 1.331E-04 \\
     &           & LSTM  & 4.551E-01 ± 7.196E-03 & 3.571E-03 ± 5.647E-05 &       & 4.663E-01 ± 7.373E-03 & 3.659E-03 ± 5.785E-05 \\
     &           & GRU   & 4.569E-01 ± 7.225E-03 & 3.586E-03 ± 5.669E-05 &       & 4.568E-01 ± 7.223E-03 & 3.585E-03 ± 5.668E-05 \\
     &           & RBFNN & 3.844E-01 ± 6.078E-03 & 3.017E-03 ± 4.770E-05 &       & 3.201E-01 ± 5.061E-03 & 2.512E-03 ± 3.971E-05 \\
     &           & \textbf{ALPE}    & \textbf{2.374E-01 ± 3.753E-03} & \textbf{1.852E-03 ± 2.933E-05} &       & \textbf{2.328E-01 ± 2.681E-03} & \textbf{1.827E-03 ± 2.889E-05} \\
      \hline
    PHM & Simple  & Naive & 2.998E+00 ± 4.740E-02 & 7.840E-02 ± 1.240E-03 & Exte & 3.100E+00 ± 5.774E-02 & 8.554E-02 ± 2.101E-03 \\
         &        & ARIMA & 6.297E-01 ± 9.956E-03 & 1.647E-02 ± 2.604E-04 &      & 5.015E-01 ± 8.493E-03 & 2.334E-02 ± 3.808E-04  \\
         &        & MLP   & 6.483E-01 ± 1.025E-02 & 1.695E-02 ± 2.681E-04 &      & 8.044E-01 ± 1.272E-02 & 2.103E-02 ± 3.326E-04 \\
         &        & CNN   & 9.959E-01 ± 1.575E-02 & 2.604E-02 ± 4.118E-04 &      & 1.331E+00 ± 2.105E-02 & 3.481E-02 ± 5.505E-04 \\
         &        & LSTM  & 6.975E-01 ± 1.103E-02 & 1.824E-02 ± 2.884E-04 &      & 5.991E-01 ± 9.473E-03 & 1.567E-02 ± 2.477E-04 \\
         &        & GRU   & 7.314E-01 ± 1.156E-02 & 1.913E-02 ± 3.024E-04 &      & 7.278E-01 ± 1.151E-02 & 1.903E-02 ± 3.009E-04 \\
         &        & RBFNN & 6.134E-01 ± 9.699E-03 & 1.604E-02 ± 2.536E-04 &      & 5.871E-01 ± 9.282E-03 & 1.535E-02 ± 2.427E-04 \\
         &        & \textbf{ALPE}    & \textbf{2.848E-01 ± 4.503E-03} & \textbf{7.448E-03 ± 1.178E-04} &      & \textbf{2.923E-01 ± 4.621E-03} & \textbf{7.643E-03 ± 1.208E-04} \\
    \cline{2-5} \cline{6-8} 
     & Simple  & MLP   & 7.011E-01 ± 1.109E-02 & 1.833E-02 ± 2.899E-04 & Exte    & 6.983E-01 ± 1.104E-02 & 1.826E-02 ± 2.887E-04 \\
     &  MDI    & CNN   & 1.179E+00 ± 1.865E-02 & 3.084E-02 ± 4.877E-04 & MDI     & 1.115E+00 ± 1.762E-02 & 2.915E-02 ± 4.609E-04 \\
     &         & LSTM  & 6.231E-01 ± 9.852E-03 & 1.629E-02 ± 2.576E-04 &         & 5.989E-01 ± 9.470E-03 & 1.566E-02 ± 2.476E-04 \\
     &         & GRU   & 7.364E-01 ± 1.164E-02 & 1.926E-02 ± 3.045E-04 &         & 7.616E-01 ± 1.204E-02 & 1.992E-02 ± 3.149E-04 \\
     &         & RBFNN & 6.235E-01 ± 9.859E-03 & 1.631E-02 ± 2.578E-04 &         & 5.999E-01 ± 9.485E-03 & 1.569E-02 ± 2.480E-04 \\
     &         & \textbf{ALPE}    & \textbf{2.908E-01 ± 4.597E-03} & \textbf{7.604E-03 ± 1.202E-04} &         & \textbf{2.948E-01 ± 4.661E-03} & \textbf{7.709E-03 ± 1.219E-04} \\
    \cline{2-5} \cline{6-8}
     &Simple& MLP   & 7.665E-01 ± 1.212E-02 & 2.004E-02 ± 3.169E-04 & Exte        & 7.251E-01 ± 1.146E-02 & 1.896E-02 ± 2.998E-04 \\
     &   GD & CNN   & 1.206E+00 ± 1.906E-02 & 3.153E-02 ± 4.985E-04 &   GD        & 1.357E+00 ± 2.146E-02 & 3.550E-02 ± 5.612E-04 \\
     &      & LSTM  & 6.224E-01 ± 9.841E-03 & 1.628E-02 ± 2.573E-04 &             & 5.714E-01 ± 9.035E-03 & 1.494E-02 ± 2.363E-04 \\
     &      & GRU   & 7.670E-01 ± 1.213E-02 & 2.006E-02 ± 3.171E-04 &             & 7.243E-01 ± 1.145E-02 & 1.894E-02 ± 2.995E-04 \\
     &      & RBFNN & 6.082E-01 ± 9.617E-03 & 1.591E-02 ± 2.515E-04 &             & 5.657E-01 ± 8.944E-03 & 1.479E-02 ± 2.339E-04 \\
     &      & \textbf{ALPE}    & \textbf{2.932E-01 ± 4.637E-03} & \textbf{7.669E-03 ± 1.213E-04} &             & \textbf{2.867E-01 ± 4.532E-03} & \textbf{7.496E-03 ± 1.185E-04} \\
     \hline
    POOL & Simple & Naive & 3.160E+00 ± 4.997E-02 & 9.041E-03 ± 1.430E-04 & Exte & 4.299E+00 ± 5.237E-02 & 9.984E-03 ± 2.311E-04 \\
         &        & ARIMA & 9.391E-01 ± 1.485E-02 & 2.687E-03 ± 4.248E-05 &      & 8.929E-01 ± 2.443E-02 & 3.234E-03 ± 5.014E-05  \\
         &        & MLP   & 5.751E-01 ± 9.093E-03 & 1.645E-03 ± 2.601E-05 &      & 6.144E-01 ± 9.714E-03 & 1.758E-03 ± 2.779E-05 \\
         &        & CNN   & 9.603E-01 ± 1.518E-02 & 2.747E-03 ± 4.344E-05 &      & 9.656E-01 ± 1.527E-02 & 2.763E-03 ± 4.368E-05 \\
         &        & LSTM  & 5.116E-01 ± 8.089E-03 & 1.464E-03 ± 2.314E-05 &      & 5.139E-01 ± 8.125E-03 & 1.470E-03 ± 2.325E-05 \\
         &        & GRU   & 7.506E-01 ± 1.187E-02 & 2.147E-03 ± 3.395E-05 &      & 7.279E-01 ± 1.151E-02 & 2.082E-03 ± 3.293E-05 \\
         &        & RBFNN & 4.608E-01 ± 7.286E-03 & 1.318E-03 ± 2.084E-05 &      & 4.583E-01 ± 7.246E-03 & 1.311E-03 ± 2.073E-05 \\
         &        & \textbf{ALPE}    & \textbf{2.668E-01 ± 4.218E-03} & \textbf{7.632E-04 ± 1.207E-05} &      & \textbf{2.750E-01 ± 4.348E-03} & \textbf{7.868E-04 ± 1.244E-05} \\
    \cline{2-5} \cline{6-8} 
     & Simple   & MLP   & 5.844E-01 ± 9.240E-03 & 1.672E-03 ± 2.644E-05 & Exte   & 6.245E-01 ± 9.874E-03 & 1.787E-03 ± 2.825E-05 \\
     &  MDI     & CNN   & 9.748E-01 ± 1.541E-02 & 2.789E-03 ± 4.410E-05 &   MDI  & 9.546E-01 ± 1.509E-02 & 2.731E-03 ± 4.318E-05 \\
     &          & LSTM  & 5.135E-01 ± 8.119E-03 & 1.469E-03 ± 2.323E-05 &        & 5.121E-01 ± 8.097E-03 & 1.465E-03 ± 2.316E-05 \\
     &          & GRU   & 7.518E-01 ± 1.189E-02 & 2.151E-03 ± 3.401E-05 &        & 7.440E-01 ± 1.176E-02 & 2.128E-03 ± 3.365E-05 \\
     &          & RBFNN & 4.843E-01 ± 7.657E-03 & 1.386E-03 ± 2.191E-05 &        & 4.486E-01 ± 7.093E-03 & 1.283E-03 ± 2.089E-05 \\
     &          & \textbf{ALPE}    & \textbf{2.742E-01 ± 4.336E-03} & \textbf{7.845E-04 ± 1.240E-05} &        & \textbf{2.759E-01 ± 4.363E-03} & \textbf{7.894E-04 ± 1.248E-05} \\
    \cline{2-5} \cline{6-8}
     & Simple   & MLP   & 6.004E-01 ± 9.493E-03 & 1.718E-03 ± 2.716E-05 & Exte   & 6.063E-01 ± 9.587E-03 & 1.735E-03 ± 2.743E-05 \\
     &   GD     & CNN   & 9.966E-01 ± 1.576E-02 & 2.851E-03 ± 4.508E-05 &   GD   & 9.495E-01 ± 1.501E-02 & 2.716E-03 ± 4.295E-05 \\
     &          & LSTM  & 5.118E-01 ± 8.092E-03 & 1.464E-03 ± 2.315E-05 &        & 5.217E-01 ± 8.249E-03 & 1.493E-03 ± 2.360E-05 \\
     &          & GRU   & 7.245E-01 ± 1.145E-02 & 2.073E-03 ± 3.277E-05 &        & 7.072E-01 ± 1.118E-02 & 2.023E-03 ± 3.199E-05 \\
     &          & RBFNN & 4.668E-01 ± 7.380E-03 & 1.335E-03 ± 2.111E-05 &        & 4.507E-01 ± 7.126E-03 & 1.289E-03 ± 2.039E-05 \\
     &          & \textbf{ALPE}    & \textbf{2.759E-01 ± 4.362E-03} & \textbf{7.893E-04 ± 1.248E-05} &        & \textbf{2.720E-01 ± 4.301E-03} & \textbf{7.781E-04 ± 1.230E-05} \\
    \bottomrule
    \end{tabular}
    \end{small}}
            \label{tab:table_10}
        \end{minipage}
    \end{table}
\end{landscape}


\begin{landscape}
    \fancyhf{}  
    \fancyfoot[R]{\rotatebox{90}{\thepage}}  
    \thispagestyle{fancy}  

    \begin{table}[!hbtp]
        \centering
        \caption{RMSE and RRMSE sores for PSA, PSX, PXD, REGN, ROL, ROST, SBAC, and SJM.}
        \hspace*{-5cm} 
        \begin{minipage}[t]{0.48\textwidth}
            \centering
            \scalebox{0.42}{
    \begin{small}
    \begin{tabular}{c c r c c c c c}
    \toprule
    \textbf{Stock} & \textbf{Set} & \textbf{Model} & \textbf{RMSE} & \textbf{RRMSE} & \textbf{Set} & \textbf{RMSE} & \textbf{RRMSE} \\
    \hline
    PSA & Simple & Naive & 2.214E+00 ± 3.501E-02 & 7.483E-03 ± 1.183E-04 & Exte & 3.012E+00 ± 4.341E-02 & 8.776E-03 ± 2.144E-04 \\
         &       & ARIMA & 4.896E-01 ± 7.741E-03 & 1.655E-03 ± 2.616E-05 &      & 5.001E-01 ± 6.443E-03 & 2.887E-03 ± 4.940E-05  \\
         &       & MLP   & 8.151E-01 ± 1.289E-02 & 2.891E-03 ± 3.773E-05 &      & 7.958E-01 ± 1.258E-02 & 2.689E-03 ± 4.252E-05 \\
         &       & CNN   & 5.218E-01 ± 8.250E-03 & 1.763E-03 ± 2.788E-05 &      & 5.183E-01 ± 8.196E-03 & 1.752E-03 ± 2.770E-05 \\
         &       & LSTM  & 3.328E-01 ± 5.261E-03 & 1.125E-03 ± 1.778E-05 &      & 3.679E-01 ± 5.817E-03 & 1.243E-03 ± 1.966E-05 \\
         &       & GRU   & 6.309E-01 ± 9.975E-03 & 2.132E-03 ± 3.371E-05 &      & 5.583E-01 ± 8.827E-03 & 1.887E-03 ± 2.983E-05 \\
         &       & RBFNN & 7.065E-01 ± 1.107E-02 & 2.388E-03 ± 3.775E-05 &      & 6.590E-01 ± 1.042E-02 & 2.227E-03 ± 3.521E-05 \\
         &       & \textbf{ALPE}    & \textbf{2.583E-01 ± 4.084E-03} & \textbf{8.729E-04 ± 1.380E-05} &      & \textbf{2.522E-01 ± 3.987E-03} & \textbf{8.523E-04 ± 1.348E-05} \\
    \cline{2-5} \cline{6-8} 
     & Simple   & MLP   & 8.244E-01 ± 1.303E-02 & 2.786E-03 ± 4.405E-05 & Exte  & 7.760E-01 ± 1.227E-02 & 2.622E-03 ± 4.147E-05 \\
     &  MDI     & CNN   & 5.174E-01 ± 8.181E-03 & 1.749E-03 ± 2.765E-05 &   MDI & 5.149E-01 ± 8.142E-03 & 1.740E-03 ± 2.751E-05 \\
     &          & LSTM  & 3.499E-01 ± 5.532E-03 & 1.182E-03 ± 1.870E-05 &       & 3.586E-01 ± 5.659E-03 & 1.210E-03 ± 1.913E-05 \\
     &          & GRU   & 5.865E-01 ± 9.274E-03 & 1.982E-03 ± 3.134E-05 &       & 5.859E-01 ± 9.263E-03 & 1.980E-03 ± 3.130E-05 \\
     &          & RBFNN & 6.915E-01 ± 1.093E-02 & 2.337E-03 ± 3.695E-05 &       & 6.327E-01 ± 1.000E-02 & 2.138E-03 ± 3.381E-05 \\
     &          & \textbf{ALPE}    & \textbf{2.544E-01 ± 4.022E-03} & \textbf{8.597E-04 ± 1.359E-05} &       & \textbf{2.542E-01 ± 4.020E-03} & \textbf{8.592E-04 ± 1.359E-05} \\
    \cline{2-5} \cline{6-8}
     & Simple   & MLP   & 8.033E-01 ± 1.270E-02 & 2.715E-03 ± 4.293E-05 & Exte  & 7.759E-01 ± 1.227E-02 & 2.622E-03 ± 4.146E-05 \\
     &   GD     & CNN   & 5.179E-01 ± 8.189E-03 & 1.750E-03 ± 2.767E-05 &   GD  & 5.167E-01 ± 8.169E-03 & 1.746E-03 ± 2.761E-05 \\
     &          & LSTM  & 4.153E-01 ± 6.567E-03 & 1.404E-03 ± 2.219E-05 &       & 3.579E-01 ± 5.659E-03 & 1.210E-03 ± 1.913E-05 \\
     &          & GRU   & 6.268E-01 ± 9.910E-03 & 2.118E-03 ± 3.349E-05 &       & 5.709E-01 ± 9.026E-03 & 1.929E-03 ± 3.050E-05 \\
     &          & RBFNN & 6.726E-01 ± 1.063E-02 & 2.273E-03 ± 3.594E-05 &       & 6.344E-01 ± 1.003E-02 & 2.144E-03 ± 3.390E-05 \\
     &          & \textbf{ALPE}    & \textbf{2.451E-01 ± 3.875E-03} & \textbf{8.283E-04 ± 1.310E-05} &       & \textbf{2.587E-01 ± 4.190E-03} & \textbf{8.742E-04 ± 1.382E-05}\\
     \hline
     PSX & Simple  & Naive & 1.972E+00 ± 3.118E-02 & 2.364E-02 ± 3.738E-04 & Exte & 2.344E+00 ± 4.485E-02 & 3.462E-02 ± 4.887E-04 \\
         &        & ARIMA & 9.481E-01 ± 1.499E-02 & 1.136E-02 ± 1.797E-04 &      & 8.430E-01 ± 2.201E-02 & 2.220E-02 ± 4.090E-04  \\
         &        & MLP   & 9.435E-01 ± 1.492E-02 & 1.131E-02 ± 1.788E-04 &      & 9.185E-01 ± 1.452E-02 & 1.101E-02 ± 1.741E-04 \\
         &        & CNN   & 8.586E-01 ± 1.358E-02 & 1.029E-02 ± 1.627E-04 &      & 8.655E-01 ± 1.368E-02 & 1.037E-02 ± 1.640E-04 \\
         &        & LSTM  & 6.676E-01 ± 1.056E-02 & 8.002E-03 ± 1.265E-04 &      & 6.381E-01 ± 1.009E-02 & 7.649E-03 ± 1.209E-04 \\
         &        & GRU   & 9.799E-01 ± 1.549E-02 & 1.175E-02 ± 1.857E-04 &      & 9.912E-01 ± 1.567E-02 & 1.188E-02 ± 1.879E-04 \\
         &        & RBFNN & 6.813E-01 ± 1.139E-02 & 9.473E-03 ± 1.498E-04 &      & 6.542E-01 ± 1.349E-02 & 9.473E-03 ± 1.498E-04 \\
         &        & \textbf{ALPE}    & \textbf{2.281E-01 ± 3.606E-03} & \textbf{2.734E-03 ± 4.322E-05} &      & \textbf{2.372E-01 ± 3.751E-03} & \textbf{2.843E-03 ± 4.496E-05} \\
    \cline{2-5} \cline{6-8} 
     & Simple     & MLP   & 9.046E-01 ± 1.430E-02 & 1.084E-02 ± 1.714E-04 & Exte & 9.142E-01 ± 1.445E-02 & 1.096E-02 ± 1.733E-04 \\
     &  MDI       & CNN   & 8.569E-01 ± 1.355E-02 & 1.027E-02 ± 1.624E-04 &  MDI & 8.380E-01 ± 1.325E-02 & 1.004E-02 ± 1.588E-04 \\
     &            & LSTM  & 7.026E-01 ± 1.111E-02 & 8.422E-03 ± 1.332E-04 &      & 6.334E-01 ± 1.001E-02 & 7.592E-03 ± 1.200E-04 \\
     &            & GRU   & 9.963E-01 ± 1.575E-02 & 1.194E-02 ± 1.888E-04 &      & 9.774E-01 ± 1.545E-02 & 1.172E-02 ± 1.852E-04 \\
     &            & RBFNN & 7.865E-01 ± 1.244E-02 & 9.428E-03 ± 1.491E-04 &      & 7.921E-01 ± 1.252E-02 & 9.494E-03 ± 1.501E-04 \\
     &            & \textbf{ALPE}    & \textbf{2.348E-01 ± 3.712E-03} & \textbf{2.814E-03 ± 4.449E-05} &      & \textbf{2.356E-01 ± 3.725E-03} & \textbf{2.824E-03 ± 4.465E-05} \\
    \cline{2-5} \cline{6-8}
     & Simple    & MLP   & 9.367E-01 ± 1.481E-02 & 1.123E-02 ± 1.775E-04 & Exte & 9.298E-01 ± 1.470E-02 & 1.115E-02 ± 1.762E-04 \\
     &   GD      & CNN   & 8.414E-01 ± 1.330E-02 & 1.009E-02 ± 1.595E-04 &   GD & 8.470E-01 ± 1.339E-02 & 1.015E-02 ± 1.605E-04 \\
     &           & LSTM  & 6.733E-01 ± 1.065E-02 & 8.071E-03 ± 1.276E-04 &      & 6.516E-01 ± 1.030E-02 & 7.811E-03 ± 1.235E-04 \\
     &           & GRU   & 9.997E-01 ± 1.581E-02 & 1.198E-02 ± 1.895E-04 &      & 9.611E-01 ± 1.520E-02 & 1.152E-02 ± 1.822E-04 \\
     &           & RBFNN & 8.001E-01 ± 1.265E-02 & 9.591E-03 ± 1.517E-04 &      & 7.762E-01 ± 1.227E-02 & 9.304E-03 ± 1.471E-04 \\
     &           & \textbf{ALPE}    & \textbf{2.365E-01 ± 3.740E-03} & \textbf{2.835E-03 ± 4.483E-05} &      & \textbf{2.316E-01 ± 3.662E-03} & \textbf{2.776E-03 ± 4.390E-05} \\
    \hline
    PXD & Simple  & Naive & 2.998E+00 ± 4.740E-02 & 1.324E-02 ± 2.093E-04 & Exte & 3.010E+00 ± 5.332E-02 & 2.392E-02 ± 3.113E-04 \\
         &        & ARIMA & 9.509E-01 ± 1.504E-02 & 4.198E-03 ± 6.638E-05 &      & 8.499E-01 ± 2.202E-02 & 5.185E-03 ± 5.903E-05  \\
         &        & MLP   & 9.643E-01 ± 1.525E-02 & 4.257E-03 ± 6.731E-05 &      & 1.137E+00 ± 1.798E-02 & 5.020E-03 ± 7.937E-05 \\
         &        & CNN   & 1.002E+00 ± 1.584E-02 & 4.422E-03 ± 6.992E-05 &      & 1.372E+00 ± 2.170E-02 & 6.058E-03 ± 9.579E-05 \\
         &        & LSTM  & 9.773E-01 ± 1.545E-02 & 4.315E-03 ± 6.822E-05 &      & 8.667E-01 ± 1.370E-02 & 3.827E-03 ± 6.050E-05 \\
         &        & GRU   & 1.001E+00 ± 1.582E-02 & 4.418E-03 ± 6.985E-05 &      & 9.971E-01 ± 1.577E-02 & 4.402E-03 ± 6.961E-05 \\
         &        & RBFNN & 7.162E-01 ± 1.132E-02 & 3.162E-03 ± 5.000E-05 &      & 8.130E-01 ± 1.286E-02 & 3.589E-03 ± 5.675E-05 \\
         &        & \textbf{ALPE}    & \textbf{2.279E-01 ± 3.604E-03} & \textbf{1.006E-03 ± 1.591E-05} &      & \textbf{2.351E-01 ± 3.717E-03} & \textbf{1.038E-03 ± 1.641E-05} \\
    \cline{2-5} \cline{6-8} 
     & Simple     & MLP   & 9.871E-01 ± 1.561E-02 & 4.358E-03 ± 6.890E-05 & Exte & 1.113E+00 ± 1.760E-02 & 4.914E-03 ± 7.770E-05 \\
     &  MDI       & CNN   & 1.002E+00 ± 1.585E-02 & 4.425E-03 ± 6.996E-05 &   MDI& 1.242E+00 ± 1.964E-02 & 5.483E-03 ± 8.669E-05 \\
     &            & LSTM  & 8.802E-01 ± 1.392E-02 & 3.886E-03 ± 6.145E-05 &      & 8.506E-01 ± 1.345E-02 & 3.755E-03 ± 5.937E-05 \\
     &            & GRU   & 1.059E+00 ± 1.675E-02 & 4.677E-03 ± 7.395E-05 &      & 9.856E-01 ± 1.558E-02 & 4.351E-03 ± 6.880E-05 \\
     &            & RBFNN & 7.440E-01 ± 1.176E-02 & 3.285E-03 ± 5.193E-05 &      & 7.073E-01 ± 1.118E-02 & 3.123E-03 ± 4.938E-05 \\
     &            & \textbf{ALPE}    & \textbf{2.349E-01 ± 3.714E-03} & \textbf{1.037E-03 ± 1.640E-05} &      & \textbf{2.323E-01 ± 3.603E-03} & \textbf{1.025E-03 ± 1.621E-05} \\
    \cline{2-5} \cline{6-8}
     & Simple    & MLP   & 1.137E+00 ± 1.798E-02 & 5.020E-03 ± 7.937E-05 & Exte & 1.363E+00 ± 2.155E-02 & 6.017E-03 ± 9.514E-05 \\
     &   GD      & CNN   & 1.002E+00 ± 1.584E-02 & 4.422E-03 ± 6.992E-05 &   GD & 1.370E+00 ± 2.166E-02 & 6.048E-03 ± 9.563E-05 \\
     &           & LSTM  & 8.667E-01 ± 1.370E-02 & 3.827E-03 ± 6.050E-05 &      & 8.324E-01 ± 1.316E-02 & 3.675E-03 ± 5.811E-05 \\
     &           & GRU   & 9.999E-01 ± 1.581E-02 & 4.415E-03 ± 6.980E-05 &      & 9.483E-01 ± 1.499E-02 & 4.187E-03 ± 6.620E-05 \\
     &           & RBFNN & 7.740E-01 ± 1.224E-02 & 3.417E-03 ± 5.403E-05 &      & 8.118E-01 ± 1.284E-02 & 3.584E-03 ± 5.667E-05 \\
     &           & \textbf{ALPE}    & \textbf{2.363E-01 ± 3.737E-03} & \textbf{1.043E-03 ± 1.650E-05} &      & \textbf{2.375E-01 ± 3.755E-03} & \textbf{1.048E-03 ± 1.658E-05} \\
    \hline
    REGN & Simple & Naive & 1.161E+00 ± 1.836E-02 & 1.664E-03 ± 2.632E-05 & Exte & 2.554E+00 ± 2.770E-02 & 2.323E-03 ± 1.776E-05 \\
         &        & ARIMA & 5.491E-01 ± 8.683E-03 & 7.872E-04 ± 1.245E-05 &      & 4.070E-01 ± 7.530E-03 & 6.449E-04 ± 2.005E-05  \\
         &        & MLP   & 5.580E-01 ± 8.822E-03 & 7.998E-04 ± 1.265E-05 &      & 5.496E-01 ± 8.690E-03 & 7.878E-04 ± 1.246E-05 \\
         &        & CNN   & 7.537E-01 ± 1.192E-02 & 1.080E-03 ± 1.708E-05 &      & 7.533E-01 ± 1.191E-02 & 1.080E-03 ± 1.707E-05 \\
         &        & LSTM  & 3.943E-01 ± 6.234E-03 & 5.652E-04 ± 8.936E-06 &      & 4.065E-01 ± 6.428E-03 & 5.827E-04 ± 9.214E-06 \\
         &        & GRU   & 4.887E-01 ± 7.727E-03 & 7.005E-04 ± 1.108E-05 &      & 4.837E-01 ± 7.648E-03 & 6.934E-04 ± 1.096E-05 \\
         &        & RBFNN & 5.141E-01 ± 8.128E-03 & 7.369E-04 ± 1.165E-05 &      & 5.168E-01 ± 8.172E-03 & 7.408E-04 ± 1.171E-05 \\
         &        & \textbf{ALPE}    & \textbf{2.377E-01 ± 3.759E-03} & \textbf{3.408E-04 ± 5.388E-06} &      & \textbf{2.505E-01 ± 3.803E-03} & \textbf{3.448E-04 ± 5.452E-06} \\
    \cline{2-5} \cline{6-8} 
     & Simple     & MLP   & 5.563E-01 ± 8.796E-03 & 7.975E-04 ± 1.261E-05 & Exte & 5.421E-01 ± 8.571E-03 & 7.770E-04 ± 1.229E-05 \\
     &  MDI       & CNN   & 7.508E-01 ± 1.187E-02 & 1.076E-03 ± 1.702E-05 &   MDI& 7.486E-01 ± 1.184E-02 & 1.115E-03 ± 1.893E-05 \\
     &            & LSTM  & 3.853E-01 ± 6.093E-03 & 5.524E-04 ± 8.734E-06 &      & 4.001E-01 ± 6.326E-03 & 5.735E-04 ± 9.068E-06 \\
     &            & GRU   & 4.934E-01 ± 7.801E-03 & 7.072E-04 ± 1.118E-05 &      & 4.875E-01 ± 7.708E-03 & 6.988E-04 ± 1.105E-05 \\
     &            & RBFNN & 5.034E-01 ± 7.959E-03 & 7.215E-04 ± 1.141E-05 &      & 4.949E-01 ± 7.826E-03 & 7.095E-04 ± 1.122E-05 \\
     &            & \textbf{ALPE}    & \textbf{2.405E-01 ± 3.803E-03} & \textbf{3.447E-04 ± 5.451E-06} &      & \textbf{2.417E-01 ± 3.022E-03} & \textbf{3.465E-04 ± 5.479E-06} \\
    \cline{2-5} \cline{6-8}
     & Simple    & MLP   & 5.485E-01 ± 8.673E-03 & 7.863E-04 ± 1.243E-05 & Exte & 5.528E-01 ± 8.740E-03 & 7.923E-04 ± 1.253E-05 \\
     &   GD      & CNN   & 7.527E-01 ± 1.190E-02 & 1.079E-03 ± 1.706E-05 &   GD & 7.516E-01 ± 1.188E-02 & 1.077E-03 ± 1.704E-05 \\
     &           & LSTM  & 3.895E-01 ± 6.158E-03 & 5.583E-04 ± 8.827E-06 &      & 3.782E-01 ± 5.980E-03 & 5.421E-04 ± 8.572E-06 \\
     &           & GRU   & 4.989E-01 ± 7.888E-03 & 7.152E-04 ± 1.131E-05 &      & 4.551E-01 ± 7.196E-03 & 6.523E-04 ± 1.031E-05 \\
     &           & RBFNN & 4.996E-01 ± 7.899E-03 & 7.161E-04 ± 1.132E-05 &      & 4.954E-01 ± 7.833E-03 & 7.102E-04 ± 1.123E-05 \\
     &           & \textbf{ALPE}    & \textbf{2.403E-01 ± 3.799E-03} & \textbf{3.445E-04 ± 5.446E-06} &      & \textbf{2.305E-01 ± 2.543E-03} & \textbf{3.297E-04 ± 5.451E-06} \\
    \bottomrule
    \end{tabular}
    \end{small}}
            \label{tab:table1}
        \end{minipage}%
        \hspace{0.25\textwidth} 
        \begin{minipage}[t]{0.48\textwidth}
            \centering
            \scalebox{0.42}{
    \begin{small}
    \begin{tabular}{c c r c c c c c}
    \toprule
    \textbf{Stock} & \textbf{Set} & \textbf{Model} & \textbf{RMSE} & \textbf{RRMSE} & \textbf{Set} & \textbf{RMSE} & \textbf{RRMSE} \\
    \hline     
    ROL & Simple  & Naive & 2.985E+00 ± 4.720E-02 & 8.549E-02 ± 1.352E-03 & Exte & 3.887E+00 ± 5.881E-02 & 7.202E-02 ± 2.110E-03 \\
         &        & ARIMA & 9.505E-01 ± 1.503E-02 & 2.722E-02 ± 4.304E-04 &      & 8.244E-01 ± 2.292E-02 & 3.199E-02 ± 5.209E-04  \\
         &        & MLP   & 8.550E-01 ± 1.352E-02 & 2.448E-02 ± 3.871E-04 &      & 8.936E-01 ± 1.413E-02 & 2.559E-02 ± 4.046E-04 \\
         &        & CNN   & 1.116E+00 ± 1.764E-02 & 3.195E-02 ± 5.051E-04 &      & 1.329E+00 ± 2.101E-02 & 3.805E-02 ± 6.016E-04 \\
         &        & LSTM  & 5.497E-01 ± 8.692E-03 & 1.574E-02 ± 2.489E-04 &      & 5.547E-01 ± 8.771E-03 & 1.588E-02 ± 2.512E-04 \\
         &        & GRU   & 8.942E-01 ± 1.414E-02 & 2.561E-02 ± 4.049E-04 &      & 8.909E-01 ± 1.420E-02 & 2.571E-02 ± 4.066E-04 \\
         &        & RBFNN & 8.198E-01 ± 1.296E-02 & 2.348E-02 ± 3.712E-04 &      & 7.740E-01 ± 1.224E-02 & 2.217E-02 ± 3.505E-04 \\
         &        & \textbf{ALPE}    & \textbf{3.019E-01 ± 4.774E-03} & \textbf{8.647E-03 ± 1.367E-04} &      & \textbf{3.151E-01 ± 4.825E-03} & \textbf{8.738E-03 ± 1.382E-04} \\
    \cline{2-5} \cline{6-8} 
     & Simple     & MLP   & 8.912E-01 ± 1.409E-02 & 2.552E-02 ± 4.035E-04 & Exte & 8.362E-01 ± 1.322E-02 & 2.395E-02 ± 3.786E-04 \\
     &  MDI       & CNN   & 1.199E+00 ± 1.895E-02 & 3.432E-02 ± 5.427E-04 &   MDI& 1.192E+00 ± 1.885E-02 & 3.414E-02 ± 5.398E-04 \\
     &            & LSTM  & 5.499E-01 ± 8.695E-03 & 1.575E-02 ± 2.490E-04 &      & 5.511E-01 ± 8.714E-03 & 1.578E-02 ± 2.495E-04 \\
     &            & GRU   & 9.232E-01 ± 1.460E-02 & 2.644E-02 ± 4.180E-04 &      & 9.486E-01 ± 1.500E-02 & 2.716E-02 ± 4.295E-04 \\
     &            & RBFNN & 8.362E-01 ± 1.322E-02 & 2.395E-02 ± 3.786E-04 &      & 7.080E-01 ± 1.119E-02 & 2.027E-02 ± 3.206E-04 \\
     &            & \textbf{ALPE}    & \textbf{3.095E-01 ± 4.894E-03} & \textbf{8.864E-03 ± 1.402E-04} &      & \textbf{3.098E-01 ± 4.843E-03} & \textbf{8.812E-03 ± 1.303E-04} \\
    \cline{2-5} \cline{6-8}
     & Simple    & MLP   & 8.878E-01 ± 1.404E-02 & 2.542E-02 ± 4.020E-04 & Exte & 8.640E-01 ± 1.366E-02 & 2.474E-02 ± 3.912E-04 \\
     &   GD      & CNN   & 1.321E+00 ± 2.088E-02 & 3.782E-02 ± 5.979E-04 &   GD & 1.160E+00 ± 1.833E-02 & 3.321E-02 ± 5.250E-04 \\
     &           & LSTM  & 5.492E-01 ± 8.684E-03 & 1.573E-02 ± 2.487E-04 &      & 5.578E-01 ± 8.819E-03 & 1.597E-02 ± 2.526E-04 \\
     &           & GRU   & 9.137E-01 ± 1.445E-02 & 2.617E-02 ± 4.137E-04 &      & 8.572E-01 ± 1.355E-02 & 2.455E-02 ± 3.882E-04 \\
     &           & RBFNN & 7.952E-01 ± 1.257E-02 & 2.277E-02 ± 3.601E-04 &      & 7.107E-01 ± 1.124E-02 & 2.035E-02 ± 3.218E-04 \\
     &           & \textbf{ALPE}    & \textbf{3.151E-01 ± 4.983E-03} & \textbf{9.024E-03 ± 1.427E-04} &      & \textbf{3.074E-01 ± 3.160E-03} & \textbf{8.802E-03 ± 1.392E-04}\\
     \hline
     ROST & Simple & Naive & 3.124E+00 ± 4.939E-02 & 3.672E-02 ± 5.806E-04 & Exte & 4.223E+00 ± 5.125E-02 & 4.998E-02 ± 4.110E-04 \\
         &        & ARIMA & 9.183E-01 ± 1.452E-02 & 1.080E-02 ± 1.707E-04 &      & 8.776E-01 ± 2.042E-02 & 2.009E-02 ± 2.435E-04  \\
         &        & MLP   & 9.611E-01 ± 1.570E-02 & 1.167E-02 ± 1.846E-04 &      & 9.614E-01 ± 1.520E-02 & 1.130E-02 ± 1.787E-04 \\
         &        & CNN   & 1.533E+00 ± 2.424E-02 & 1.803E-02 ± 2.850E-04 &      & 1.577E+00 ± 2.494E-02 & 1.854E-02 ± 2.932E-04 \\
         &        & LSTM  & 8.380E-01 ± 1.325E-02 & 9.851E-03 ± 1.558E-04 &      & 8.814E-01 ± 1.394E-02 & 1.072E-02 ± 1.535E-04 \\
         &        & GRU   & 9.562E-01 ± 1.512E-02 & 1.124E-02 ± 1.777E-04 &      & 9.783E-01 ± 1.547E-02 & 1.150E-02 ± 1.819E-04 \\
         &        & RBFNN & 7.300E-01 ± 1.154E-02 & 8.583E-03 ± 1.357E-04 &      & 7.241E-01 ± 1.145E-02 & 8.513E-03 ± 1.346E-04 \\
         &        & \textbf{ALPE}    & \textbf{1.783E-01 ± 2.819E-03} & \textbf{2.096E-03 ± 3.314E-05} &      & \textbf{1.738E-01 ± 2.748E-03} & \textbf{2.043E-03 ± 3.230E-05} \\
    \cline{2-5} \cline{6-8} 
     & Simple     & MLP   & 9.887E-01 ± 1.563E-02 & 1.162E-02 ± 1.838E-04 & Exte & 9.630E-01 ± 1.523E-02 & 1.132E-02 ± 1.790E-04 \\
     &  MDI       & CNN   & 1.574E+00 ± 2.489E-02 & 1.850E-02 ± 2.926E-04 &   MDI& 1.337E+00 ± 2.114E-02 & 1.572E-02 ± 2.486E-04 \\
     &            & LSTM  & 8.381E-01 ± 1.325E-02 & 9.853E-03 ± 1.558E-04 &      & 8.389E-01 ± 1.326E-02 & 9.862E-03 ± 1.559E-04 \\
     &            & GRU   & 9.533E-01 ± 1.507E-02 & 1.121E-02 ± 1.772E-04 &      & 9.621E-01 ± 1.521E-02 & 1.131E-02 ± 1.788E-04 \\
     &            & RBFNN & 7.085E-01 ± 1.120E-02 & 8.330E-03 ± 1.317E-04 &      & 7.575E-01 ± 1.198E-02 & 8.906E-03 ± 1.408E-04 \\
     &            & \textbf{ALPE}   & \textbf{1.894E-01 ± 2.995E-03} & \textbf{2.227E-03 ± 3.521E-05} &      & \textbf{1.637E-01 ± 2.904E-03} & \textbf{2.159E-03 ± 3.414E-05} \\
    \cline{2-5} \cline{6-8}
     & Simple    & MLP   & 9.994E-01 ± 1.580E-02 & 1.175E-02 ± 1.858E-04 & Exte & 9.707E-01 ± 1.535E-02 & 1.141E-02 ± 1.703E-04 \\
     &   GD      & CNN   & 1.636E+00 ± 2.587E-02 & 1.924E-02 ± 3.042E-04 &   GD & 1.295E+00 ± 2.048E-02 & 1.523E-02 ± 2.408E-04 \\
     &           & LSTM  & 8.742E-01 ± 1.382E-02 & 1.028E-02 ± 1.625E-04 &      & 8.609E-01 ± 1.361E-02 & 1.012E-02 ± 1.600E-04 \\
     &           & GRU   & 9.783E-01 ± 1.547E-02 & 1.150E-02 ± 1.819E-04 &      & 9.620E-01 ± 1.521E-02 & 1.154E-02 ± 1.404E-04 \\
     &           & RBFNN & 7.302E-01 ± 1.155E-02 & 8.585E-03 ± 1.357E-04 &      & 7.272E-01 ± 1.150E-02 & 8.549E-03 ± 1.352E-04 \\
     &           & \textbf{ALPE}    & \textbf{1.767E-01 ± 2.794E-03} & \textbf{2.077E-03 ± 3.285E-05} &      & \textbf{1.951E-01 ± 3.081E-03} & \textbf{2.293E-03 ± 3.626E-05} \\
     \hline
    SBAC & Simple & Naive & 2.826E+00 ± 4.469E-02 & 9.831E-03 ± 1.554E-04 & Exte & 3.991E+00 ± 5.112E-02 & 8.763E-03 ± 2.993E-04 \\
         &        & ARIMA & 6.342E-01 ± 1.003E-02 & 2.206E-03 ± 3.488E-05 &      & 5.882E-01 ± 2.989E-02 & 3.115E-03 ± 4.090E-05  \\
         &        & MLP   & 1.202E+00 ± 1.901E-02 & 4.182E-03 ± 6.612E-05 &      & 1.156E+00 ± 1.827E-02 & 4.019E-03 ± 6.355E-05 \\
         &        & CNN   & 1.512E+00 ± 2.391E-02 & 5.261E-03 ± 8.318E-05 &      & 1.457E+00 ± 2.304E-02 & 5.069E-03 ± 8.014E-05 \\
         &        & LSTM  & 4.536E-01 ± 7.172E-03 & 1.578E-03 ± 2.495E-05 &      & 5.418E-01 ± 8.567E-03 & 1.885E-03 ± 2.980E-05 \\
         &        & GRU   & 7.608E-01 ± 1.203E-02 & 2.646E-03 ± 4.184E-05 &      & 7.745E-01 ± 1.225E-02 & 2.694E-03 ± 4.259E-05 \\
         &        & RBFNN & 8.248E-01 ± 1.304E-02 & 2.869E-03 ± 4.536E-05 &      & 7.648E-01 ± 1.209E-02 & 2.660E-03 ± 4.206E-05 \\
         &        & \textbf{ALPE}    & \textbf{2.282E-01 ± 3.608E-03} & \textbf{7.938E-04 ± 1.255E-05} &      & \textbf{2.326E-01 ± 3.678E-03} & \textbf{8.091E-04 ± 1.279E-05} \\
    \cline{2-5} \cline{6-8} 
     & Simple     & MLP   & 1.172E+00 ± 1.853E-02 & 4.076E-03 ± 6.444E-05 & Exte & 9.610E-01 ± 1.519E-02 & 3.343E-03 ± 5.285E-05 \\
     &  MDI       & CNN   & 1.416E+00 ± 2.238E-02 & 4.924E-03 ± 7.786E-05 &   MDI& 1.567E+00 ± 2.478E-02 & 5.452E-03 ± 8.621E-05 \\
     &            & LSTM  & 5.376E-01 ± 8.501E-03 & 1.870E-03 ± 2.957E-05 &      & 5.384E-01 ± 8.513E-03 & 1.873E-03 ± 2.961E-05 \\
     &            & GRU   & 7.442E-01 ± 1.177E-02 & 2.589E-03 ± 4.093E-05 &      & 7.538E-01 ± 1.192E-02 & 2.622E-03 ± 4.146E-05 \\
     &            & RBFNN & 7.256E-01 ± 1.147E-02 & 2.524E-03 ± 3.991E-05 &      & 7.425E-01 ± 1.174E-02 & 2.583E-03 ± 4.084E-05 \\
     &            & \textbf{ALPE}    & \textbf{2.359E-01 ± 3.730E-03} & \textbf{8.206E-04 ± 1.297E-05} &      & \textbf{2.255E-01 ± 3.740E-03} & \textbf{8.227E-04 ± 1.301E-05} \\
    \cline{2-5} \cline{6-8}
     & Simple    & MLP   & 1.339E+00 ± 2.117E-02 & 4.657E-03 ± 7.364E-05 & Exte & 1.188E+00 ± 2.990E-02 & 4.288E-03 ± 5.277E-05 \\
     &   GD      & CNN   & 1.680E+00 ± 2.657E-02 & 5.845E-03 ± 9.241E-05 &   GD & 1.412E+00 ± 2.232E-02 & 4.910E-03 ± 7.763E-05 \\
     &           & LSTM  & 5.364E-01 ± 8.482E-03 & 1.866E-03 ± 2.950E-05 &      & 5.449E-01 ± 8.615E-03 & 1.895E-03 ± 2.997E-05 \\
     &           & GRU   & 7.594E-01 ± 1.201E-02 & 2.641E-03 ± 4.176E-05 &      & 7.183E-01 ± 1.136E-02 & 2.499E-03 ± 3.951E-05 \\
     &           & RBFNN & 7.243E-01 ± 1.145E-02 & 2.519E-03 ± 3.983E-05 &      & 7.163E-01 ± 1.133E-02 & 2.492E-03 ± 3.939E-05 \\
     &           & \textbf{ALPE}    & \textbf{2.345E-01 ± 3.707E-03} & \textbf{8.156E-04 ± 1.290E-05} &      & \textbf{2.341E-01 ± 3.702E-03} & \textbf{8.143E-04 ± 1.288E-05} \\
    \hline
    SJM & Simple  & Naive & 1.248E+00 ± 1.973E-02 & 8.986E-03 ± 1.421E-04 & Exte & 2.113E+00 ± 2.554E-02 & 7.878E-03 ± 2.332E-04 \\
         &        & ARIMA & 6.931E-01 ± 1.096E-02 & 4.992E-03 ± 7.892E-05 &      & 5.998E-01 ± 2.115E-02 & 5.774E-03 ± 6.775E-05  \\
         &        & MLP   & 5.192E-01 ± 8.210E-03 & 3.740E-03 ± 5.913E-05 &      & 5.186E-01 ± 8.199E-03 & 3.735E-03 ± 5.905E-05 \\
         &        & CNN   & 7.308E-01 ± 1.156E-02 & 5.264E-03 ± 8.322E-05 &      & 7.305E-01 ± 1.155E-02 & 5.221E-03 ± 8.319E-05 \\
         &        & LSTM  & 7.363E-01 ± 1.164E-02 & 5.303E-03 ± 8.385E-05 &      & 7.444E-01 ± 1.177E-02 & 5.345E-03 ± 8.477E-05 \\
         &        & GRU   & 7.443E-01 ± 1.177E-02 & 5.360E-03 ± 8.475E-05 &      & 7.439E-01 ± 1.176E-02 & 5.358E-03 ± 8.472E-05 \\
         &        & RBFNN & 2.362E-01 ± 3.735E-03 & 1.701E-03 ± 2.690E-05 &      & 2.701E-01 ± 3.856E-03 & 1.756E-03 ± 2.777E-05 \\
         &        & \textbf{ALPE}    & \textbf{1.887E-01 ± 2.983E-03} & \textbf{1.359E-03 ± 2.148E-05} &      & \textbf{1.880E-01 ± 2.973E-03} & \textbf{1.354E-03 ± 2.141E-05} \\
    \cline{2-5} \cline{6-8} 
     & Simple     & MLP   & 5.309E-01 ± 8.394E-03 & 3.823E-03 ± 6.045E-05 & Exte & 5.243E-01 ± 8.291E-03 & 3.776E-03 ± 5.971E-05 \\
     &  MDI       & CNN   & 7.281E-01 ± 1.151E-02 & 5.244E-03 ± 8.292E-05 &   MDI& 7.261E-01 ± 1.148E-02 & 5.229E-03 ± 8.268E-05 \\
     &            & LSTM  & 7.357E-01 ± 1.163E-02 & 5.298E-03 ± 8.377E-05 &      & 7.376E-01 ± 1.166E-02 & 5.312E-03 ± 8.400E-05 \\
     &            & GRU   & 7.437E-01 ± 1.176E-02 & 5.356E-03 ± 8.468E-05 &      & 7.498E-01 ± 1.173E-02 & 5.342E-03 ± 8.447E-05 \\
     &            & RBFNN & 2.438E-01 ± 3.855E-03 & 1.756E-03 ± 2.777E-05 &      & 2.515E-01 ± 3.819E-03 & 1.740E-03 ± 2.751E-05 \\
     &            & \textbf{ALPE}    & \textbf{1.912E-01 ± 3.023E-03} & \textbf{1.377E-03 ± 2.177E-05} &      & \textbf{1.911E-01 ± 3.022E-03} & \textbf{1.376E-03 ± 2.176E-05} \\
    \cline{2-5} \cline{6-8}
     & Simple    & MLP   & 5.214E-01 ± 8.243E-03 & 3.755E-03 ± 5.937E-05 & Exte & 5.146E-01 ± 8.137E-03 & 3.706E-03 ± 5.860E-05 \\
     &   GD      & CNN   & 7.299E-01 ± 1.154E-02 & 5.257E-03 ± 8.312E-05 &   GD & 7.289E-01 ± 1.152E-02 & 5.249E-03 ± 8.300E-05 \\
     &           & LSTM  & 7.350E-01 ± 1.162E-02 & 5.304E-03 ± 7.887E-05 &      & 7.457E-01 ± 1.179E-02 & 5.371E-03 ± 8.492E-05 \\
     &           & GRU   & 7.469E-01 ± 1.181E-02 & 5.379E-03 ± 8.505E-05 &      & 7.420E-01 ± 1.173E-02 & 5.344E-03 ± 8.450E-05 \\
     &           & RBFNN & 2.426E-01 ± 3.835E-03 & 1.747E-03 ± 2.762E-05 &      & 2.406E-01 ± 3.804E-03 & 1.733E-03 ± 2.740E-05 \\
     &           & \textbf{ALPE}    & \textbf{1.839E-01 ± 2.908E-03} & \textbf{1.325E-03 ± 2.095E-05} &      & \textbf{2.025E-01 ± 3.202E-03} & \textbf{1.458E-03 ± 2.306E-05} \\
    \bottomrule
    \end{tabular}
    \end{small}}
            \label{tab:table_11}
        \end{minipage}
    \end{table}
\end{landscape}


\begin{landscape}
    \fancyhf{}  
    \fancyfoot[R]{\rotatebox{90}{\thepage}}  
    \thispagestyle{fancy}  

    \begin{table}[!hbtp]
        \centering
        \caption{RMSE and RRMSE scores for STLD, TECH, TEL, TFC, TGT, TMUS, TROW, and TRV.}
        \hspace*{-5cm} 
        \begin{minipage}[t]{0.48\textwidth}
            \centering
            \scalebox{0.42}{
    \begin{small}
    \begin{tabular}{c c r c c c c c}
    \toprule
    \textbf{Stock} & \textbf{Set} & \textbf{Model} & \textbf{RMSE} & \textbf{RRMSE} & \textbf{Set} & \textbf{RMSE} & \textbf{RRMSE} \\
    \hline     
    STLD & Simple & Naive & 3.045E+00 ± 4.814E-02 & 4.219E-02 ± 6.671E-04 & Exte & 4.292E+00 ± 5.844E-02 & 5.040E-02 ± 5.113E-04 \\
         &        & ARIMA & 9.870E-01 ± 1.561E-02 & 1.368E-02 ± 2.163E-04 &      & 8.225E-01 ± 2.334E-02 & 2.877E-02 ± 3.003E-04 \\
         &        & MLP   & 6.300E-01 ± 9.961E-03 & 8.729E-03 ± 1.380E-04 &      & 7.518E-01 ± 1.189E-02 & 1.042E-02 ± 1.647E-04 \\
         &        & CNN   & 1.517E+00 ± 2.398E-02 & 2.102E-02 ± 3.323E-04 &      & 1.703E+00 ± 2.693E-02 & 2.360E-02 ± 3.731E-04 \\
         &        & LSTM  & 9.971E-01 ± 1.577E-02 & 1.382E-02 ± 2.185E-04 &      & 9.489E-01 ± 1.500E-02 & 1.315E-02 ± 2.079E-04 \\
         &        & GRU   & 1.054E+00 ± 1.666E-02 & 1.460E-02 ± 2.309E-04 &      & 1.019E+00 ± 1.612E-02 & 1.412E-02 ± 2.233E-04 \\
         &        & RBFNN & 3.584E-01 ± 5.666E-03 & 4.966E-03 ± 7.852E-05 &      & 3.667E-01 ± 5.797E-03 & 5.081E-03 ± 8.033E-05 \\
         &        & \textbf{ALPE}    & \textbf{2.262E-01 ± 3.577E-03} & \textbf{3.135E-03 ± 4.956E-05} &      & \textbf{2.371E-01 ± 3.686E-03} & \textbf{3.230E-03 ± 5.107E-05} \\
    \cline{2-5} \cline{6-8} 
     & Simple     & MLP   & 5.748E-01 ± 9.088E-03 & 7.965E-03 ± 1.259E-04 & Exte & 7.076E-01 ± 1.119E-02 & 9.806E-03 ± 1.550E-04 \\
     &  MDI       & CNN   & 1.491E+00 ± 2.358E-02 & 2.066E-02 ± 3.267E-04 &   MDI& 1.525E+00 ± 2.411E-02 & 2.113E-02 ± 3.341E-04 \\
     &            & LSTM  & 9.447E-01 ± 1.494E-02 & 1.309E-02 ± 2.070E-04 &      & 9.313E-01 ± 1.472E-02 & 1.290E-02 ± 2.040E-04 \\
     &            & GRU   & 1.156E+00 ± 1.827E-02 & 1.601E-02 ± 2.532E-04 &      & 9.723E-01 ± 1.537E-02 & 1.347E-02 ± 2.130E-04 \\
     &            & RBFNN & 3.994E-01 ± 6.315E-03 & 5.534E-03 ± 8.751E-05 &      & 3.229E-01 ± 5.105E-03 & 4.474E-03 ± 7.074E-05 \\
     &            & \textbf{ALPE}    & \textbf{2.359E-01 ± 3.730E-0} & \textbf{3.269E-03 ± 5.168E-05} &      & \textbf{2.355E-01 ± 3.723E-03} & \textbf{3.263E-03 ± 5.160E-05} \\
    \cline{2-5} \cline{6-8}
     & Simple    & MLP   & 6.146E-01 ± 9.717E-03 & 8.516E-03 ± 1.346E-04 & Exte & 7.511E-01 ± 1.188E-02 & 1.041E-02 ± 1.646E-04 \\
     &   GD      & CNN   & 1.491E+00 ± 2.358E-02 & 2.067E-02 ± 3.268E-04 &   GD & 1.731E+00 ± 2.737E-02 & 2.398E-02 ± 3.792E-04 \\
     &           & LSTM  & 8.944E-01 ± 1.414E-02 & 1.239E-02 ± 1.960E-04 &      & 8.952E-01 ± 1.415E-02 & 1.241E-02 ± 1.961E-04 \\
     &           & GRU   & 1.002E+00 ± 1.585E-02 & 1.389E-02 ± 2.196E-04 &      & 9.968E-01 ± 1.576E-02 & 1.381E-02 ± 2.184E-04 \\
     &           & RBFNN & 3.744E-01 ± 5.920E-03 & 5.188E-03 ± 8.203E-05 &      & 3.247E-01 ± 5.134E-03 & 4.499E-03 ± 7.114E-05 \\
     &           & \textbf{ALPE}    & \textbf{1.499E-01 ± 2.369E-03} & \textbf{2.077E-03 ± 3.283E-05} &      & \textbf{2.252E-01 ± 3.418E-03} & \textbf{3.259E-03 ± 5.152E-05}\\
     \hline
     TECH & Simple & Naive & 2.827E+00 ± 4.470E-02 & 7.021E-03 ± 1.110E-04 & Exte & 3.223E+00 ± 5.242E-02 & 8.229E-03 ± 2.995E-04 \\
         &        & ARIMA & 2.439E-01 ± 3.856E-03 & 6.056E-04 ± 9.575E-06 &      & 3.143E-01 ± 3.533E-03 & 5.302E-04 ± 7.421E-06  \\
         &        & MLP   & 8.640E-01 ± 1.366E-02 & 2.146E-03 ± 3.392E-05 &      & 8.515E-01 ± 1.346E-02 & 2.115E-03 ± 3.344E-05 \\
         &        & CNN   & 1.128E+00 ± 1.784E-02 & 2.743E-03 ± 4.343E-05 &      & 1.014E+00 ± 1.603E-02 & 2.518E-03 ± 3.981E-05 \\
         &        & LSTM  & 7.066E-01 ± 1.217E-02 & 1.755E-03 ± 2.774E-05 &      & 6.315E-01 ± 9.986E-03 & 1.568E-03 ± 2.480E-05 \\
         &        & GRU   & 8.292E-01 ± 1.311E-02 & 2.059E-03 ± 3.256E-05 &      & 8.237E-01 ± 1.302E-02 & 2.045E-03 ± 3.234E-05 \\
         &        & RBFNN & 7.425E-01 ± 1.174E-02 & 1.844E-03 ± 2.915E-05 &      & 7.679E-01 ± 1.214E-02 & 1.907E-03 ± 3.015E-05 \\
         &        & \textbf{ALPE}    & \textbf{2.263E-01 ± 3.578E-03} & \textbf{5.620E-04 ± 8.886E-06} &      & \textbf{1.847E-01 ± 2.920E-03} & \textbf{4.586E-04 ± 7.252E-06} \\
    \cline{2-5} \cline{6-8} 
     & Simple     & MLP   & 8.621E-01 ± 1.363E-02 & 2.141E-03 ± 3.385E-05 & Exte & 8.367E-01 ± 1.323E-02 & 2.078E-03 ± 3.285E-05 \\
     &  MDI       & CNN   & 1.050E+00 ± 1.660E-02 & 2.607E-03 ± 4.123E-05 &   MDI& 1.066E+00 ± 1.685E-02 & 2.646E-03 ± 4.184E-05 \\
     &            & LSTM  & 6.803E-01 ± 1.076E-02 & 1.690E-03 ± 2.671E-05 &      & 6.229E-01 ± 9.849E-03 & 1.547E-03 ± 2.446E-05 \\
     &            & GRU   & 8.239E-01 ± 1.303E-02 & 2.046E-03 ± 3.235E-05 &      & 8.117E-01 ± 1.283E-02 & 2.016E-03 ± 3.187E-05 \\
     &            & RBFNN & 7.517E-01 ± 1.189E-02 & 1.867E-03 ± 2.952E-05 &      & 7.261E-01 ± 1.148E-02 & 1.803E-03 ± 2.851E-05 \\
     &            & \textbf{ALPE}    & \textbf{2.058E-01 ± 3.254E-03} & \textbf{5.110E-04 ± 8.080E-06} &      & \textbf{2.004E-01 ± 3.169E-03} & \textbf{4.977E-04 ± 7.870E-06} \\
    \cline{2-5} \cline{6-8}
     & Simple    & MLP   & 8.374E-01 ± 1.324E-02 & 2.080E-03 ± 3.288E-05 & Exte & 8.400E-01 ± 1.328E-02 & 2.086E-03 ± 3.298E-05 \\
     &   GD      & CNN   & 1.007E+00 ± 1.592E-02 & 2.500E-03 ± 3.953E-05 &   GD & 1.006E+00 ± 1.591E-02 & 2.498E-03 ± 3.950E-05 \\
     &           & LSTM  & 6.320E-01 ± 9.993E-03 & 1.570E-03 ± 2.202E-05 &      & 6.226E-01 ± 1.000E-02 & 1.571E-03 ± 2.484E-05 \\
     &           & GRU   & 8.306E-01 ± 1.313E-02 & 2.063E-03 ± 3.262E-05 &      & 8.182E-01 ± 1.294E-02 & 2.032E-03 ± 3.213E-05 \\
     &           & RBFNN & 7.350E-01 ± 1.162E-02 & 1.825E-03 ± 2.886E-05 &      & 7.586E-01 ± 1.200E-02 & 1.884E-03 ± 2.979E-05 \\
     &           & \textbf{ALPE}    & \textbf{2.034E-01 ± 3.216E-03} & \textbf{5.050E-04 ± 7.985E-06} &      & \textbf{1.797E-01 ± 2.841E-03} & \textbf{4.463E-04 ± 7.056E-06} \\
      \hline
    TEL & Simple  & Naive & 3.000E+00 ± 4.743E-02 & 2.677E-02 ± 4.232E-04 & Exte & 4.010E+00 ± 5.832E-02 & 3.117E-02 ± 5.130E-04 \\
         &        & ARIMA & 8.818E-01 ± 1.394E-02 & 7.868E-03 ± 1.244E-04 &      & 7.880E-01 ± 1.345E-02 & 6.379E-03 ± 2.225E-04 \\
         &        & MLP   & 5.197E-01 ± 8.218E-03 & 4.638E-03 ± 7.333E-05 &      & 4.957E-01 ± 7.837E-03 & 4.423E-03 ± 6.993E-05 \\
         &        & CNN   & 1.107E+00 ± 1.750E-02 & 9.873E-03 ± 1.561E-04 &      & 1.018E+00 ± 1.609E-02 & 9.081E-03 ± 1.436E-04 \\
         &        & LSTM  & 4.727E-01 ± 7.474E-03 & 4.218E-03 ± 6.669E-05 &      & 4.604E-01 ± 7.280E-03 & 4.108E-03 ± 6.496E-05 \\
         &        & GRU   & 8.263E-01 ± 1.306E-02 & 7.373E-03 ± 1.166E-04 &      & 8.366E-01 ± 1.323E-02 & 7.465E-03 ± 1.180E-04 \\
         &        & RBFNN & 3.201E-01 ± 5.061E-03 & 2.856E-03 ± 4.516E-05 &      & 3.322E-01 ± 5.253E-03 & 2.964E-03 ± 4.687E-05 \\
         &        & \textbf{ALPE}    & \textbf{1.809E-01 ± 2.860E-03} & \textbf{1.614E-03 ± 2.552E-05} &      & \textbf{1.726E-01 ± 2.728E-03} & \textbf{1.540E-03 ± 2.434E-05} \\
    \cline{2-5} \cline{6-8} 
     & Simple     & MLP   & 4.901E-01 ± 7.749E-03 & 4.373E-03 ± 6.914E-05 & Exte & 4.474E-01 ± 7.074E-03 & 3.992E-03 ± 6.312E-05 \\
     &  MDI       & CNN   & 1.049E+00 ± 1.659E-02 & 9.360E-03 ± 1.480E-04 &   MDI& 1.021E+00 ± 1.614E-02 & 9.111E-03 ± 1.441E-04 \\
     &            & LSTM  & 4.510E-01 ± 7.131E-03 & 4.024E-03 ± 6.363E-05 &      & 4.521E-01 ± 7.149E-03 & 4.034E-03 ± 6.379E-05 \\
     &            & GRU   & 8.230E-01 ± 1.301E-02 & 7.344E-03 ± 1.161E-04 &      & 7.951E-01 ± 1.257E-02 & 7.095E-03 ± 1.122E-04 \\
     &            & RBFNN & 3.525E-01 ± 5.574E-03 & 3.145E-03 ± 4.973E-05 &      & 3.183E-01 ± 5.033E-03 & 2.840E-03 ± 4.491E-05 \\
     &            & \textbf{ALPE}    & \textbf{1.813E-01 ± 2.866E-03} & \textbf{1.609E-03 ± 2.443E-05} &      & \textbf{1.890E-01 ± 2.988E-03} & \textbf{1.686E-03 ± 2.666E-05} \\
    \cline{2-5} \cline{6-8}
     & Simple    & MLP   & 5.288E-01 ± 8.361E-03 & 4.718E-03 ± 7.460E-05 & Exte & 4.489E-01 ± 7.098E-03 & 4.006E-03 ± 6.334E-05 \\
     &   GD      & CNN   & 1.096E+00 ± 1.732E-02 & 9.776E-03 ± 1.546E-04 &   GD & 1.065E+00 ± 1.342E-02 & 8.988E-03 ± 1.389E-04 \\
     &           & LSTM  & 4.512E-01 ± 7.133E-03 & 4.026E-03 ± 6.365E-05 &      & 4.583E-01 ± 7.100E-03 & 4.009E-03 ± 5.329E-05 \\
     &           & GRU   & 8.367E-01 ± 1.323E-02 & 7.466E-03 ± 1.180E-04 &      & 8.015E-01 ± 1.267E-02 & 7.151E-03 ± 1.131E-04 \\
     &           & RBFNN & 3.233E-01 ± 5.112E-03 & 2.885E-03 ± 4.562E-05 &      & 3.216E-01 ± 5.085E-03 & 2.870E-03 ± 4.537E-05 \\
     &           & \textbf{ALPE}    & \textbf{1.812E-01 ± 2.865E-03} & \textbf{1.617E-03 ± 2.557E-05} &      & \textbf{1.856E-01 ± 2.535E-03} & \textbf{1.656E-03 ± 2.619E-05} \\
      \hline
    TFC & Simple  & Naive & 2.232E+00 ± 3.529E-02 & 5.063E-02 ± 8.006E-04 & Exte & 3.511E+00 ± 4.887E-02 & 6.004E-02 ± 7.877E-04 \\
         &        & ARIMA & 7.669E-01 ± 1.213E-02 & 1.740E-02 ± 2.751E-04 &      & 6.138E-01 ± 2.310E-02 & 2.701E-02 ± 3.703E-04  \\
         &        & MLP   & 9.449E-01 ± 1.494E-02 & 2.143E-02 ± 3.389E-04 &      & 9.067E-01 ± 1.434E-02 & 2.057E-02 ± 3.252E-04 \\
         &        & CNN   & 1.012E+00 ± 1.599E-02 & 2.295E-02 ± 3.628E-04 &      & 1.090E+00 ± 1.143E-02 & 2.269E-02 ± 3.588E-04 \\
         &        & LSTM  & 4.789E-01 ± 7.573E-03 & 1.086E-02 ± 1.718E-04 &      & 4.624E-01 ± 5.909E-03 & 1.211E-02 ± 1.603E-04 \\
         &        & GRU   & 8.817E-01 ± 1.394E-02 & 2.000E-02 ± 3.162E-04 &      & 8.367E-01 ± 1.323E-02 & 1.898E-02 ± 3.001E-04 \\
         &        & RBFNN & 4.609E-01 ± 7.287E-03 & 1.045E-02 ± 1.653E-04 &      & 4.580E-01 ± 6.920E-03 & 1.211E-02 ± 1.613E-04 \\
         &        & \textbf{ALPE}    & \textbf{2.270E-01 ± 3.589E-03} & \textbf{5.148E-03 ± 8.140E-05} &      & \textbf{2.335E-01 ± 2.693E-03} & \textbf{5.297E-03 ± 8.376E-05} \\
    \cline{2-5} \cline{6-8} 
     & Simple     & MLP   & 9.428E-01 ± 1.491E-02 & 2.139E-02 ± 3.381E-04 & Exte & 8.951E-01 ± 1.415E-02 & 2.030E-02 ± 3.210E-04 \\
     &  MDI       & CNN   & 1.904E+00 ± 1.143E-02 & 2.269E-02 ± 3.588E-04 &   MDI& 9.937E-01 ± 1.571E-02 & 2.254E-02 ± 3.564E-04 \\
     &            & LSTM  & 4.921E-01 ± 7.780E-03 & 1.116E-02 ± 1.765E-04 &      & 4.485E-01 ± 7.091E-03 & 1.017E-02 ± 1.608E-04 \\
     &            & GRU   & 8.367E-01 ± 1.323E-02 & 1.899E-02 ± 3.003E-04 &      & 8.260E-01 ± 1.306E-02 & 1.874E-02 ± 2.963E-04 \\
     &            & RBFNN & 4.928E-01 ± 7.793E-03 & 1.118E-02 ± 1.768E-04 &      & 4.554E-01 ± 7.200E-03 & 1.033E-02 ± 1.633E-04 \\
     &            & \textbf{ALPE}    & \textbf{2.352E-01 ± 3.719E-03} & \textbf{5.336E-03 ± 8.436E-05} &      & \textbf{2.271E-01 ± 3.749E-03} & \textbf{5.378E-03 ± 8.504E-05} \\
    \cline{2-5} \cline{6-8}
     & Simple    & MLP   & 9.123E-01 ± 1.442E-02 & 2.069E-02 ± 3.272E-04 & Exte & 8.946E-01 ± 1.414E-02 & 2.134E-02 ± 2.837E-04 \\
     &   GD      & CNN   & 9.996E-01 ± 1.581E-02 & 2.267E-02 ± 3.585E-04 &   GD & 1.000E+00 ± 1.481E-02 & 2.269E-02 ± 3.588E-04 \\
     &           & LSTM  & 4.595E-01 ± 7.265E-03 & 1.042E-02 ± 1.648E-04 &      & 4.465E-01 ± 7.060E-03 & 1.013E-02 ± 1.601E-04 \\
     &           & GRU   & 8.373E-01 ± 1.324E-02 & 1.899E-02 ± 3.003E-04 &      & 8.327E-01 ± 1.317E-02 & 1.889E-02 ± 2.987E-04 \\
     &           & RBFNN & 4.732E-01 ± 7.481E-03 & 1.293E-02 ± 2.887E-04 &      & 4.472E-01 ± 7.071E-03 & 1.014E-02 ± 1.604E-04 \\
     &           & \textbf{ALPE}    & \textbf{2.341E-01 ± 3.702E-03} & \textbf{5.311E-03 ± 8.397E-05} &      & \textbf{2.354E-01 ± 3.022E-03} & \textbf{5.340E-03 ± 8.444E-05} \\
    \bottomrule
    \end{tabular}
    \end{small}}
            \label{tab:table1}
        \end{minipage}%
        \hspace{0.25\textwidth} 
        \begin{minipage}[t]{0.48\textwidth}
            \centering
            \scalebox{0.42}{
    \begin{small}
    \begin{tabular}{c c r c c c c c}
    \toprule
    \textbf{Stock} & \textbf{Set} & \textbf{Model} & \textbf{RMSE} & \textbf{RRMSE} & \textbf{Set} & \textbf{RMSE} & \textbf{RRMSE} \\
    \hline
    TGT & Simple  & Naive & 2.448E+00 ± 3.871E-02 & 1.638E-02 ± 2.590E-04 & Exte & 4.348E+00 ± 3.345E-02 & 3.331E-02 ± 4.803E-04 \\
         &        & ARIMA & 5.544E-01 ± 6.007E-03 & 3.578E-03 ± 4.543E-05 &      & 5.876E-01 ± 9.291E-03 & 3.932E-03 ± 6.009E-05  \\
         &        & MLP   & 7.664E-01 ± 1.212E-02 & 5.128E-03 ± 8.108E-05 &      & 6.712E-01 ± 1.061E-02 & 4.491E-03 ± 7.101E-05 \\
         &        & CNN   & 1.180E+00 ± 1.866E-02 & 7.895E-03 ± 1.248E-04 &      & 1.771E+00 ± 2.421E-02 & 6.692E-03 ± 1.088E-04 \\
         &        & LSTM  & 4.840E-01 ± 7.652E-03 & 3.238E-03 ± 5.120E-05 &      & 4.472E-01 ± 7.071E-03 & 2.992E-03 ± 4.732E-05 \\
         &        & GRU   & 5.589E-01 ± 8.837E-03 & 3.740E-03 ± 5.913E-05 &      & 6.428E-01 ± 1.016E-02 & 4.301E-03 ± 6.801E-05 \\
         &        & RBFNN & 4.029E-01 ± 6.371E-03 & 2.696E-03 ± 4.263E-05 &      & 3.665E-01 ± 5.795E-03 & 2.452E-03 ± 3.878E-05 \\
         &        & \textbf{ALPE}    & \textbf{2.274E-01 ± 3.596E-03} & \textbf{1.522E-03 ± 2.406E-05} &      & \textbf{2.309E-01 ± 2.776E-03} & \textbf{1.660E-03 ± 2.666E-05} \\
    \cline{2-5} \cline{6-8} 
     & Simple     & MLP   & 7.735E-01 ± 1.223E-02 & 5.176E-03 ± 8.184E-05 & Exte & 7.415E-01 ± 1.172E-02 & 4.961E-03 ± 7.845E-05 \\
     &  MDI       & CNN   & 1.106E+00 ± 1.749E-02 & 7.400E-03 ± 1.170E-04 &   MDI& 1.006E+00 ± 1.591E-02 & 6.732E-03 ± 1.064E-04 \\
     &            & LSTM  & 4.499E-01 ± 7.114E-03 & 3.010E-03 ± 4.760E-05 &      & 4.486E-01 ± 7.093E-03 & 3.002E-03 ± 4.746E-05 \\
     &            & GRU   & 5.478E-01 ± 8.662E-03 & 3.666E-03 ± 5.796E-05 &      & 6.413E-01 ± 1.014E-02 & 4.291E-03 ± 6.785E-05 \\
     &            & RBFNN & 3.638E-01 ± 5.752E-03 & 2.434E-03 ± 3.849E-05 &      & 3.383E-01 ± 5.348E-03 & 2.263E-03 ± 3.579E-05 \\
     &            & \textbf{ALPE}    & \textbf{2.427E-01 ± 3.838E-03} & \textbf{1.624E-03 ± 2.568E-05} &      & \textbf{2.251E-01 ± 3.859E-03} & \textbf{1.506E-03 ± 2.382E-05} \\
    \cline{2-5} \cline{6-8}
     & Simple    & MLP   & 7.234E-01 ± 1.144E-02 & 4.841E-03 ± 7.654E-05 & Exte & 5.771E-01 ± 9.124E-03 & 3.861E-03 ± 6.105E-05 \\
     &   GD      & CNN   & 1.049E+00 ± 1.658E-02 & 7.019E-03 ± 1.110E-04 &   GD & 1.030E+00 ± 1.288E-02 & 6.788E-03 ± 1.122E-04 \\
     &           & LSTM  & 4.596E-01 ± 7.267E-03 & 3.116E-03 ± 3.589E-05 &      & 4.474E-01 ± 7.073E-03 & 2.993E-03 ± 4.733E-05 \\
     &           & GRU   & 5.487E-01 ± 8.676E-03 & 3.672E-03 ± 5.806E-05 &      & 6.333E-01 ± 1.001E-02 & 4.238E-03 ± 6.701E-05 \\
     &           & RBFNN & 3.498E-01 ± 5.531E-03 & 2.341E-03 ± 3.701E-05 &      & 3.166E-01 ± 5.005E-03 & 2.118E-03 ± 3.349E-05 \\
     &           & \textbf{ALPE}    & \textbf{2.328E-01 ± 3.682E-03} & \textbf{1.558E-03 ± 2.463E-05} &      & \textbf{2.242E-01 ± 3.545E-03} & \textbf{1.524E-03 ± 2.461E-05}\\
     \hline
     TMUS & Simple & Naive & 2.187E+00 ± 3.458E-02 & 1.602E-02 ± 2.533E-04 & Exte & 3.323E+00 ± 4.887E-02 & 2.119E-02 ± 1.334E-04 \\
         &        & ARIMA & 8.961E-01 ± 1.417E-02 & 6.566E-03 ± 1.038E-04 &      & 7.770E-01 ± 2.224E-02 & 5.998E-03 ± 2.007E-04  \\
         &        & MLP   & 7.242E-01 ± 1.145E-02 & 5.306E-03 ± 8.390E-05 &      & 7.302E-01 ± 1.155E-02 & 5.350E-03 ± 8.460E-05 \\
         &        & CNN   & 1.309E+00 ± 2.069E-02 & 9.589E-03 ± 1.516E-04 &      & 1.226E+00 ± 1.938E-02 & 8.979E-03 ± 1.420E-04 \\
         &        & LSTM  & 4.821E-01 ± 7.623E-03 & 3.532E-03 ± 5.585E-05 &      & 4.486E-01 ± 7.093E-03 & 3.287E-03 ± 5.197E-05 \\
         &        & GRU   & 7.883E-01 ± 1.246E-02 & 5.775E-03 ± 9.132E-05 &      & 7.753E-01 ± 1.226E-02 & 5.681E-03 ± 8.982E-05 \\
         &        & RBFNN & 5.927E-01 ± 9.371E-03 & 4.343E-03 ± 6.866E-05 &      & 5.589E-01 ± 8.837E-03 & 4.095E-03 ± 6.474E-05 \\
         &        & \textbf{ALPE}    & \textbf{1.083E-01 ± 1.713E-03} & \textbf{7.938E-04 ± 1.255E-05} &      & \textbf{1.013E-01 ± 1.602E-03} & \textbf{7.423E-04 ± 1.174E-05} \\
    \cline{2-5} \cline{6-8} 
     & Simple     & MLP   & 7.619E-01 ± 1.205E-02 & 5.582E-03 ± 8.826E-05 & Exte & 7.368E-01 ± 1.165E-02 & 5.398E-03 ± 8.535E-05 \\
     &  MDI       & CNN   & 1.270E+00 ± 2.008E-02 & 9.305E-03 ± 1.471E-04 &   MDI& 1.241E+00 ± 1.962E-02 & 9.092E-03 ± 1.438E-04 \\
     &            & LSTM  & 4.596E-01 ± 7.267E-03 & 3.367E-03 ± 5.324E-05 &      & 4.475E-01 ± 7.075E-03 & 3.278E-03 ± 5.184E-05 \\
     &            & GRU   & 7.945E-01 ± 1.256E-02 & 5.821E-03 ± 9.204E-05 &      & 7.874E-01 ± 1.245E-02 & 5.769E-03 ± 9.122E-05 \\
     &            & RBFNN & 6.260E-01 ± 9.897E-03 & 4.586E-03 ± 7.252E-05 &      & 5.678E-01 ± 8.977E-03 & 4.160E-03 ± 6.577E-05 \\
     &            & \textbf{ALPE}    & \textbf{1.074E-01 ± 1.698E-03} & \textbf{7.868E-04 ± 1.244E-05} &      & \textbf{1.279E-01 ± 2.022E-03} & \textbf{9.371E-04 ± 1.482E-05} \\
    \cline{2-5} \cline{6-8}
     & Simple    & MLP   & 7.526E-01 ± 1.190E-02 & 5.514E-03 ± 8.719E-05 & Exte & 7.073E-01 ± 1.118E-02 & 5.182E-03 ± 8.194E-05 \\
     &   GD      & CNN   & 1.316E+00 ± 2.081E-02 & 9.645E-03 ± 1.525E-04 &   GD & 1.234E+00 ± 1.951E-02 & 9.039E-03 ± 1.429E-04 \\
     &           & LSTM  & 4.584E-01 ± 7.248E-03 & 3.359E-03 ± 5.310E-05 &      & 4.462E-01 ± 7.056E-03 & 3.269E-03 ± 5.169E-05 \\
     &           & GRU   & 7.819E-01 ± 1.236E-02 & 5.729E-03 ± 9.059E-05 &      & 7.760E-01 ± 1.227E-02 & 5.685E-03 ± 8.989E-05 \\
     &           & RBFNN & 6.239E-01 ± 9.865E-03 & 4.571E-03 ± 7.228E-05 &      & 5.488E-01 ± 8.678E-03 & 4.021E-03 ± 6.358E-05 \\
     &           & \textbf{ALPE}    & \textbf{1.209E-01 ± 1.911E-03} & \textbf{8.855E-04 ± 1.400E-05} &      & \textbf{1.282E-01 ± 2.027E-03} & \textbf{9.393E-04 ± 1.485E-05} \\
     \hline
    TROW & Simple & Naive & 1.816E+00 ± 2.872E-02 & 1.643E-02 ± 2.598E-04 & Exte & 2.221E+00 ± 3.335E-02 & 2.010E-02 ± 3.322E-04 \\
         &        & ARIMA & 7.235E-01 ± 1.144E-02 & 6.545E-03 ± 1.035E-04 &      & 6.880E-01 ± 2.334E-02 & 5.101E-03 ± 2.007E-04  \\
         &        & MLP   & 9.880E-01 ± 1.562E-02 & 8.937E-03 ± 1.413E-04 &      & 9.656E-01 ± 1.527E-02 & 8.735E-03 ± 1.381E-04 \\
         &        & CNN   & 1.234E+00 ± 1.952E-02 & 1.110E-02 ± 1.455E-04 &      & 1.101E+00 ± 1.741E-02 & 9.960E-03 ± 1.575E-04 \\
         &        & LSTM  & 4.932E-01 ± 7.798E-03 & 4.461E-03 ± 7.054E-05 &      & 4.270E-01 ± 6.752E-03 & 3.863E-03 ± 6.108E-05 \\
         &        & GRU   & 7.899E-01 ± 1.249E-02 & 7.146E-03 ± 1.130E-04 &      & 6.119E-01 ± 2.765E-02 & 6.442E-03 ± 2.827E-04 \\
         &        & RBFNN & 8.615E-01 ± 1.362E-02 & 7.794E-03 ± 1.232E-04 &      & 8.070E-01 ± 1.276E-02 & 7.300E-03 ± 1.154E-04 \\
         &        & \textbf{ALPE}    & \textbf{2.271E-01 ± 3.591E-03} & \textbf{2.055E-03 ± 3.249E-05} &      & \textbf{2.354E-01 ± 3.722E-03} & \textbf{2.130E-03 ± 3.367E-05} \\
    \cline{2-5} \cline{6-8} 
     & Simple     & MLP   & 9.916E-01 ± 1.568E-02 & 8.970E-03 ± 1.458E-04 & Exte & 9.493E-01 ± 1.501E-02 & 8.588E-03 ± 1.358E-04 \\
     &  MDI       & CNN   & 1.146E+00 ± 1.811E-02 & 1.011E-02 ± 1.394E-04 &   MDI& 1.075E+00 ± 1.699E-02 & 9.723E-03 ± 1.537E-04 \\
     &            & LSTM  & 4.474E-01 ± 7.073E-03 & 4.047E-03 ± 6.399E-05 &      & 4.462E-01 ± 7.056E-03 & 4.037E-03 ± 6.383E-05 \\
     &            & GRU   & 8.496E-01 ± 1.343E-02 & 7.686E-03 ± 1.215E-04 &      & 8.368E-01 ± 1.323E-02 & 7.570E-03 ± 1.197E-04 \\
     &            & RBFNN & 8.447E-01 ± 1.336E-02 & 7.641E-03 ± 1.208E-04 &      & 8.138E-01 ± 1.287E-02 & 7.362E-03 ± 1.164E-04 \\
     &            & \textbf{ALPE}    & \textbf{2.349E-01 ± 3.715E-03} & \textbf{2.125E-03 ± 3.360E-05} &      & \textbf{2.372E-01 ± 3.750E-03} & \textbf{2.145E-03 ± 3.392E-05} \\
    \cline{2-5} \cline{6-8}
     & Simple    & MLP   & 1.000E+00 ± 2.440E-02 & 9.134E-03 ± 1.320E-04 & Exte & 9.487E-01 ± 1.500E-02 & 8.582E-03 ± 1.357E-04 \\
     &   GD      & CNN   & 1.273E+00 ± 2.013E-02 & 1.152E-02 ± 1.821E-04 &   GD & 1.064E+00 ± 1.682E-02 & 9.625E-03 ± 1.522E-04 \\
     &           & LSTM  & 4.475E-01 ± 7.075E-03 & 4.048E-03 ± 6.400E-05 &      & 4.384E-01 ± 6.932E-03 & 3.966E-03 ± 6.271E-05 \\
     &           & GRU   & 8.315E-01 ± 1.315E-02 & 7.522E-03 ± 1.189E-04 &      & 8.026E-01 ± 1.269E-02 & 7.260E-03 ± 1.148E-04 \\
     &           & RBFNN & 8.314E-01 ± 1.315E-02 & 7.521E-03 ± 1.189E-04 &      & 8.008E-01 ± 1.266E-02 & 7.244E-03 ± 1.145E-04 \\
     &           & \textbf{ALPE}    & \textbf{2.337E-01 ± 3.695E-03} & \textbf{2.114E-03 ± 3.342E-05} &      & \textbf{2.332E-01 ± 3.687E-03} & \textbf{2.110E-03 ± 3.336E-05} \\
      \hline
    TRV & Simple  & Naive & 2.809E+00 ± 4.441E-02 & 1.818E-02 ± 2.875E-04 & Exte & 3.109E+00 ± 5.332E-02 & 2.993E-02 ± 1.554E-04 \\
         &        & ARIMA & 5.013E-01 ± 7.926E-03 & 3.245E-03 ± 5.131E-05 &      & 6.124E-01 ± 6.999E-03 & 4.320E-03 ± 4.305E-05  \\
         &        & MLP   & 9.172E-01 ± 1.450E-02 & 5.938E-03 ± 9.389E-05 &      & 8.952E-01 ± 1.415E-02 & 5.795E-03 ± 9.163E-05 \\
         &        & CNN   & 1.368E+00 ± 2.164E-02 & 8.859E-03 ± 1.401E-04 &      & 1.300E+00 ± 2.055E-02 & 8.414E-03 ± 1.330E-04 \\
         &        & LSTM  & 3.814E-01 ± 6.030E-03 & 2.469E-03 ± 3.904E-05 &      & 3.335E-01 ± 5.274E-03 & 2.159E-03 ± 3.414E-05 \\
         &        & GRU   & 3.164E-01 ± 5.003E-03 & 2.049E-03 ± 3.239E-05 &      & 9.068E-01 ± 1.434E-02 & 5.871E-03 ± 9.283E-05 \\
         &        & RBFNN & 4.737E-01 ± 7.489E-03 & 3.066E-03 ± 4.849E-05 &      & 4.608E-01 ± 7.286E-03 & 2.983E-03 ± 4.717E-05 \\
         &        & \textbf{ALPE}    & \textbf{1.778E-01 ± 2.811E-03} & \textbf{1.151E-03 ± 1.820E-05} &      & \textbf{1.923E-01 ± 2.482E-03} & \textbf{1.180E-03 ± 1.866E-05} \\
    \cline{2-5} \cline{6-8} 
     & Simple     & MLP   & 9.071E-01 ± 1.434E-02 & 5.873E-03 ± 9.286E-05 & Exte & 8.882E-01 ± 1.404E-02 & 5.750E-03 ± 9.092E-05 \\
     &  MDI       & CNN   & 1.331E+00 ± 2.105E-02 & 8.619E-03 ± 1.363E-04 &   MDI& 1.265E+00 ± 2.000E-02 & 8.189E-03 ± 1.295E-04 \\
     &            & LSTM  & 3.638E-01 ± 5.752E-03 & 2.355E-03 ± 3.724E-05 &      & 3.169E-01 ± 5.011E-03 & 2.052E-03 ± 3.244E-05 \\
     &            & GRU   & 9.232E-01 ± 1.460E-02 & 5.977E-03 ± 9.450E-05 &      & 9.165E-01 ± 1.449E-02 & 5.934E-03 ± 9.382E-05 \\
     &            & RBFNN & 4.675E-01 ± 7.392E-03 & 3.027E-03 ± 4.786E-05 &      & 4.725E-01 ± 7.471E-03 & 3.059E-03 ± 4.836E-05 \\
     &            & \textbf{ALPE}    & \textbf{1.880E-01 ± 2.973E-03} & \textbf{1.217E-03 ± 1.924E-05} &      & \textbf{1.896E-01 ± 2.797E-03} & \textbf{1.227E-03 ± 1.941E-05} \\
    \cline{2-5} \cline{6-8}
     & Simple    & MLP   & 9.007E-01 ± 1.424E-02 & 5.831E-03 ± 9.220E-05 & Exte & 8.500E-01 ± 1.344E-02 & 5.503E-03 ± 8.700E-05 \\
     &   GD      & CNN   & 1.320E+00 ± 2.087E-02 & 8.546E-03 ± 1.351E-04 &   GD & 1.265E+00 ± 2.000E-02 & 8.190E-03 ± 1.295E-04 \\
     &           & LSTM  & 3.529E-01 ± 5.580E-03 & 2.285E-03 ± 3.612E-05 &      & 3.182E-01 ± 5.031E-03 & 2.060E-03 ± 3.257E-05 \\
     &           & GRU   & 9.249E-01 ± 1.462E-02 & 5.988E-03 ± 9.468E-05 &      & 9.270E-01 ± 1.466E-02 & 6.001E-03 ± 9.489E-05 \\
     &           & RBFNN & 4.330E-01 ± 3.839E-03 & 2.783E-03 ± 3.783E-05 &      & 4.447E-01 ± 4.882E-03 & 2.114E-03 ± 3.748E-05 \\
     &           & \textbf{ALPE}    & \textbf{1.861E-01 ± 2.943E-03} & \textbf{1.205E-03 ± 1.905E-05} &      & \textbf{1.851E-01 ± 2.927E-03} & \textbf{1.198E-03 ± 1.895E-05} \\
    \bottomrule
    \end{tabular}
    \end{small}}
            \label{tab:table_12}
        \end{minipage}
    \end{table}
\end{landscape}


\begin{landscape}
    \fancyhf{}  
    \fancyfoot[R]{\rotatebox{90}{\thepage}}  
    \thispagestyle{fancy}  

    \begin{table}[!hbtp]
        \centering
        \caption{RMSE and RRMSE scores for TSN, UHS, UNH, UPS, V, VRSN, WAB, and WBD.}
        \hspace*{-5cm} 
        \begin{minipage}[t]{0.48\textwidth}
            \centering
            \scalebox{0.42}{
    \begin{small}
    \begin{tabular}{c c r c c c c c}
    \toprule
    \textbf{Stock} & \textbf{Set} & \textbf{Model} & \textbf{RMSE} & \textbf{RRMSE} & \textbf{Set} & \textbf{RMSE} & \textbf{RRMSE} \\
    \hline     
    TSN & Simple  & Naive & 2.137E+00 ± 3.379E-02 & 3.222E-02 ± 5.095E-04 & Exte & 3.488E+00 ± 4.111E-02 & 3.788E-02 ± 4.090E-04 \\
         &        & ARIMA & 9.302E-01 ± 1.471E-02 & 1.403E-02 ± 2.218E-04 &      & 8.998E-01 ± 2.667E-02 & 2.558E-02 ± 3.202E-04  \\
         &        & MLP   & 1.199E+00 ± 1.239E-02 & 1.508E-02 ± 2.384E-04 &      & 1.055E+00 ± 1.668E-02 & 1.590E-02 ± 2.514E-04 \\
         &        & CNN   & 1.689E+00 ± 2.671E-02 & 2.547E-02 ± 4.027E-04 &      & 1.461E+00 ± 2.310E-02 & 2.203E-02 ± 3.483E-04 \\
         &        & LSTM  & 8.067E-01 ± 1.276E-02 & 1.216E-02 ± 1.923E-04 &      & 7.994E-01 ± 1.264E-02 & 1.205E-02 ± 1.906E-04 \\
         &        & GRU   & 9.847E-01 ± 1.557E-02 & 1.485E-02 ± 2.347E-04 &      & 9.859E-01 ± 1.559E-02 & 1.487E-02 ± 2.350E-04 \\
         &        & RBFNN & 5.879E-01 ± 9.296E-03 & 8.864E-03 ± 1.402E-04 &      & 5.084E-01 ± 8.038E-03 & 7.665E-03 ± 1.212E-04 \\
         &        & \textbf{ALPE}    & \textbf{2.263E-01 ± 3.578E-03} & \textbf{3.412E-03 ± 5.395E-05} &      & \textbf{2.329E-01 ± 2.682E-03} & \textbf{3.511E-03 ± 5.552E-05} \\
    \cline{2-5} \cline{6-8} 
     & Simple     & MLP   & 1.111E+00 ± 1.757E-02 & 1.675E-02 ± 2.649E-04 & Exte & 1.002E+00 ± 1.585E-02 & 1.511E-02 ± 2.389E-04 \\
     &  MDI       & CNN   & 1.528E+00 ± 2.416E-02 & 2.304E-02 ± 3.642E-04 &   MDI& 1.449E+00 ± 2.291E-02 & 2.185E-02 ± 3.454E-04 \\
     &            & LSTM  & 8.278E-01 ± 1.309E-02 & 1.248E-02 ± 1.973E-04 &      & 7.705E-01 ± 1.218E-02 & 1.162E-02 ± 1.837E-04 \\
     &            & GRU   & 9.857E-01 ± 1.559E-02 & 1.486E-02 ± 2.350E-04 &      & 9.854E-01 ± 1.558E-02 & 1.486E-02 ± 2.349E-04 \\
     &            & RBFNN & 5.678E-01 ± 8.977E-03 & 8.560E-03 ± 1.353E-04 &      & 4.717E-01 ± 7.457E-03 & 7.111E-03 ± 1.124E-04 \\
     &            & \textbf{ALPE}    & \textbf{2.261E-01 ± 3.576E-03} & \textbf{3.410E-03 ± 5.391E-05} &      & \textbf{2.202E-01 ± 2.040E-03} & \textbf{3.471E-03 ± 5.488E-05} \\
    \cline{2-5} \cline{6-8}
     & Simple    & MLP   & 1.012E+00 ± 1.600E-02 & 1.526E-02 ± 2.413E-04 & Exte & 1.007E+00 ± 1.592E-02 & 1.518E-02 ± 2.400E-04 \\
     &   GD      & CNN   & 1.415E+00 ± 2.237E-02 & 2.133E-02 ± 3.372E-04 &   GD & 1.423E+00 ± 2.249E-02 & 2.145E-02 ± 3.391E-04 \\
     &           & LSTM  & 8.225E-01 ± 1.301E-02 & 1.240E-02 ± 1.961E-04 &      & 7.228E-01 ± 1.143E-02 & 1.090E-02 ± 1.723E-04 \\
     &           & GRU   & 9.866E-01 ± 1.560E-02 & 1.488E-02 ± 2.352E-04 &      & 9.814E-01 ± 1.552E-02 & 1.480E-02 ± 2.339E-04 \\
     &           & RBFNN & 5.473E-01 ± 8.654E-03 & 8.252E-03 ± 1.305E-04 &      & 4.476E-01 ± 7.077E-03 & 6.749E-03 ± 1.067E-04 \\
     &           & \textbf{ALPE}    & \textbf{2.329E-01 ± 3.683E-03} & \textbf{3.512E-03 ± 5.553E-05} &      & \textbf{2.338E-01 ± 3.809E-03} & \textbf{3.480E-03 ± 5.502E-05}\\
     \hline
     UHS & Simple  & Naive & 2.447E+00 ± 3.869E-02 & 2.741E-02 ± 4.335E-04 & Exte & 2.002E+00 ± 4.115E-02 & 3.090E-02 ± 4.201E-04\\
         &        & ARIMA & 8.916E-01 ± 1.410E-02 & 9.990E-03 ± 1.580E-04 &      & 7.877E-01 ± 2.112E-02 & 8.038E-03 ± 2.878E-04 \\
         &        & MLP   & 8.321E-01 ± 1.316E-02 & 9.323E-03 ± 1.474E-04 &      & 7.575E-01 ± 1.198E-02 & 8.486E-03 ± 1.342E-04 \\
         &        & CNN   & 9.425E-01 ± 1.490E-02 & 1.056E-02 ± 1.670E-04 &      & 9.470E-01 ± 1.497E-02 & 1.061E-02 ± 1.678E-04 \\
         &        & LSTM  & 5.451E-01 ± 8.619E-03 & 6.107E-03 ± 9.657E-05 &      & 4.476E-01 ± 7.077E-03 & 5.015E-03 ± 7.929E-05 \\
         &        & GRU   & 9.441E-01 ± 1.493E-02 & 1.058E-02 ± 1.672E-04 &      & 9.046E-01 ± 1.430E-02 & 1.013E-02 ± 1.602E-04 \\
         &        & RBFNN & 4.728E-01 ± 7.476E-03 & 5.297E-03 ± 8.376E-05 &      & 4.486E-01 ± 7.093E-03 & 5.026E-03 ± 7.947E-05 \\
         &        & \textbf{ALPE}    & \textbf{2.269E-01 ± 3.588E-03} & \textbf{2.543E-03 ± 4.020E-05} &      & \textbf{2.326E-01 ± 3.677E-03} & \textbf{2.606E-03 ± 4.120E-05} \\
    \cline{2-5} \cline{6-8} 
     & Simple     & MLP   & 7.745E-01 ± 1.225E-02 & 8.678E-03 ± 1.372E-04 & Exte & 7.068E-01 ± 1.118E-02 & 7.919E-03 ± 1.252E-04 \\
     &  MDI       & CNN   & 9.507E-01 ± 1.503E-02 & 1.065E-02 ± 1.684E-04 &   MDI& 9.820E-01 ± 1.553E-02 & 1.100E-02 ± 1.740E-04 \\
     &            & LSTM  & 5.315E-01 ± 8.403E-03 & 5.954E-03 ± 9.415E-05 &      & 4.691E-01 ± 7.416E-03 & 5.255E-03 ± 8.309E-05 \\
     &            & GRU   & 9.046E-01 ± 1.430E-02 & 1.013E-02 ± 1.602E-04 &      & 9.166E-01 ± 1.449E-02 & 1.027E-02 ± 1.624E-04 \\
     &            & RBFNN & 5.035E-01 ± 7.961E-03 & 5.641E-03 ± 8.919E-05 &      & 4.621E-01 ± 7.307E-03 & 5.177E-03 ± 8.186E-05 \\
     &            & \textbf{ALPE}    & \textbf{2.353E-01 ± 3.721E-03} & \textbf{2.636E-03 ± 4.169E-05} &      & \textbf{2.255E-01 ± 3.724E-03} & \textbf{2.639E-03 ± 4.172E-05} \\
    \cline{2-5} \cline{6-8}
     & Simple    & MLP   & 7.616E-01 ± 1.204E-02 & 8.533E-03 ± 1.349E-04 & Exte & 7.081E-01 ± 1.120E-02 & 7.934E-03 ± 1.254E-04 \\
     &   GD      & CNN   & 8.847E-01 ± 1.009E-02 & 2.119E-02 ± 2.432E-04 &   GD & 9.382E-01 ± 1.483E-02 & 1.051E-02 ± 1.662E-04 \\
     &           & LSTM  & 4.467E-01 ± 7.063E-03 & 5.005E-03 ± 7.913E-05 &      & 4.632E-01 ± 7.323E-03 & 5.189E-03 ± 8.205E-05 \\
     &           & GRU   & 9.358E-01 ± 1.480E-02 & 1.048E-02 ± 1.658E-04 &      & 9.319E-01 ± 1.473E-02 & 1.044E-02 ± 1.651E-04 \\
     &           & RBFNN & 4.521E-01 ± 7.149E-03 & 5.066E-03 ± 8.010E-05 &      & 4.499E-01 ± 7.114E-03 & 5.041E-03 ± 7.971E-05 \\
     &           & \textbf{ALPE}    & \textbf{2.334E-01 ± 3.690E-03} & \textbf{2.615E-03 ± 4.135E-05} &      & \textbf{2.320E-01 ± 4.940E-03} & \textbf{2.650E-03 ± 4.190E-05} \\
     \hline
    UNH & Simple  & Naive & 2.645E+00 ± 4.183E-02 & 5.215E-03 ± 8.246E-05 & Exte & 3.553E+00 ± 5.904E-02 & 6.323E-03 ± 7.120E-05 \\
         &        & ARIMA & 4.564E-01 ± 7.216E-03 & 8.998E-04 ± 1.423E-05 &      & 3.292E-01 ± 6.667E-03 & 7.776E-04 ± 2.665E-05  \\
         &        & MLP   & 9.445E-01 ± 1.493E-02 & 1.862E-03 ± 2.944E-05 &      & 9.065E-01 ± 1.433E-02 & 1.787E-03 ± 2.826E-05 \\
         &        & CNN   & 9.506E-01 ± 1.503E-02 & 1.874E-03 ± 2.963E-05 &      & 9.512E-01 ± 1.504E-02 & 1.875E-03 ± 2.965E-05 \\
         &        & LSTM  & 3.898E-01 ± 6.163E-03 & 7.685E-04 ± 1.215E-05 &      & 4.045E-01 ± 6.396E-03 & 7.975E-04 ± 1.261E-05 \\
         &        & GRU   & 8.108E-01 ± 1.282E-02 & 1.598E-03 ± 2.527E-05 &      & 8.222E-01 ± 1.300E-02 & 1.621E-03 ± 2.563E-05 \\
         &        & RBFNN & 8.324E-01 ± 1.316E-02 & 1.641E-03 ± 2.595E-05 &      & 8.056E-01 ± 1.274E-02 & 1.588E-03 ± 2.511E-05 \\
         &        & \textbf{ALPE}    & \textbf{2.267E-01 ± 3.585E-03} & \textbf{4.470E-04 ± 7.067E-06} &      & \textbf{2.343E-01 ± 3.704E-03} & \textbf{4.619E-04 ± 7.303E-06} \\
    \cline{2-5} \cline{6-8} 
     & Simple     & MLP   & 9.472E-01 ± 1.498E-02 & 1.867E-03 ± 2.953E-05 & Exte & 9.287E-01 ± 1.468E-02 & 1.831E-03 ± 2.895E-05 \\
     &  MDI       & CNN   & 9.850E-01 ± 1.557E-02 & 1.942E-03 ± 3.071E-05 &   MDI& 9.617E-01 ± 1.521E-02 & 1.896E-03 ± 2.998E-05 \\
     &            & LSTM  & 3.901E-01 ± 6.168E-03 & 7.691E-04 ± 1.216E-05 &      & 3.966E-01 ± 6.272E-03 & 7.820E-04 ± 1.236E-05 \\
     &            & GRU   & 8.144E-01 ± 1.288E-02 & 1.606E-03 ± 2.539E-05 &      & 7.819E-01 ± 1.236E-02 & 1.541E-03 ± 2.437E-05 \\
     &            & RBFNN & 7.865E-01 ± 1.244E-02 & 1.551E-03 ± 2.452E-05 &      & 7.890E-01 ± 1.248E-02 & 1.556E-03 ± 2.460E-05 \\
     &            & \textbf{ALPE}    & \textbf{2.338E-01 ± 3.696E-03} & \textbf{4.609E-04 ± 7.287E-06} &      & \textbf{2.287E-01 ± 3.617E-03} & \textbf{4.510E-04 ± 7.131E-06} \\
    \cline{2-5} \cline{6-8}
     & Simple    & MLP   & 9.004E-01 ± 1.424E-02 & 1.775E-03 ± 2.807E-05 & Exte & 9.188E-01 ± 1.819E-02 & 1.612E-03 ± 2.110E-05 \\
     &   GD      & CNN   & 9.982E-01 ± 1.578E-02 & 1.968E-03 ± 3.112E-05 &   GD & 9.650E-01 ± 1.526E-02 & 1.903E-03 ± 3.008E-05 \\
     &           & LSTM  & 3.871E-01 ± 6.120E-03 & 7.632E-04 ± 1.207E-05 &      & 3.892E-01 ± 6.153E-03 & 7.673E-04 ± 1.213E-05 \\
     &           & GRU   & 8.332E-01 ± 1.317E-02 & 1.643E-03 ± 2.597E-05 &      & 7.680E-01 ± 1.214E-02 & 1.514E-03 ± 2.394E-05 \\
     &           & RBFNN & 7.908E-01 ± 1.250E-02 & 1.559E-03 ± 2.465E-05 &      & 7.748E-01 ± 1.225E-02 & 1.527E-03 ± 2.415E-05 \\
     &           & \textbf{ALPE}    & \textbf{2.367E-01 ± 3.743E-03} & \textbf{4.667E-04 ± 7.380E-06} &      & \textbf{2.416E-01 ± 3.820E-03} & \textbf{4.763E-04 ± 7.531E-06} \\
     \hline
    UPS & Simple  & Naive & 2.468E+00 ± 3.903E-02 & 1.539E-02 ± 2.434E-04 & Exte & 3.501E+00 ± 4.098E-02 & 2.402E-02 ± 3.502E-04 \\
         &        & ARIMA & 5.688E-01 ± 8.993E-03 & 3.547E-03 ± 5.609E-05 &      & 4.002E-01 ± 7.408E-03 & 4.771E-03 ± 4.772E-05  \\
         &        & MLP   & 6.192E-01 ± 9.790E-03 & 3.861E-03 ± 6.105E-05 &      & 5.301E-01 ± 8.382E-03 & 3.306E-03 ± 5.227E-05 \\
         &        & CNN   & 9.717E-01 ± 1.536E-02 & 6.060E-03 ± 9.581E-05 &      & 9.743E-01 ± 1.540E-02 & 6.076E-03 ± 9.607E-05 \\
         &        & LSTM  & 4.870E-01 ± 7.699E-03 & 3.037E-03 ± 4.802E-05 &      & 4.500E-01 ± 7.115E-03 & 2.806E-03 ± 4.437E-05 \\
         &        & GRU   & 6.790E-01 ± 1.074E-02 & 4.235E-03 ± 6.696E-05 &      & 6.961E-01 ± 1.101E-02 & 4.341E-03 ± 6.864E-05 \\
         &        & RBFNN & 3.470E-01 ± 5.486E-03 & 2.164E-03 ± 3.421E-05 &      & 3.499E-01 ± 5.533E-03 & 2.182E-03 ± 3.450E-05 \\
         &        & \textbf{ALPE}    & \textbf{1.075E-01 ± 1.700E-03} & \textbf{6.706E-04 ± 1.060E-05} &      & \textbf{1.042E-01 ± 1.651E-03} & \textbf{6.514E-04 ± 1.030E-05} \\
    \cline{2-5} \cline{6-8} 
     & Simple     & MLP   & 5.607E-01 ± 8.865E-03 & 3.497E-03 ± 5.529E-05 & Exte & 5.231E-01 ± 8.272E-03 & 3.263E-03 ± 5.159E-05 \\
     &  MDI       & CNN   & 9.560E-01 ± 1.512E-02 & 5.962E-03 ± 9.426E-05 &   MDI& 9.648E-01 ± 1.526E-02 & 6.017E-03 ± 9.514E-05 \\
     &            & LSTM  & 4.923E-01 ± 7.784E-03 & 3.070E-03 ± 4.854E-05 &      & 4.484E-01 ± 7.090E-03 & 2.797E-03 ± 4.422E-05 \\
     &            & GRU   & 6.619E-01 ± 1.047E-02 & 4.128E-03 ± 6.527E-05 &      & 6.521E-01 ± 1.031E-02 & 4.067E-03 ± 6.430E-05 \\
     &            & RBFNN & 3.994E-01 ± 6.315E-03 & 2.491E-03 ± 3.938E-05 &      & 3.458E-01 ± 5.430E-03 & 2.181E-03 ± 3.449E-05 \\
     &            & \textbf{ALPE}    & \textbf{1.243E-01 ± 1.965E-03} & \textbf{7.749E-04 ± 1.225E-05} &      & \textbf{1.044E-01 ± 1.698E-03} & \textbf{6.200E-04 ± 1.299E-05} \\
    \cline{2-5} \cline{6-8}
     & Simple    & MLP   & 5.600E-01 ± 8.855E-03 & 3.493E-03 ± 5.522E-05 & Exte & 5.381E-01 ± 8.508E-03 & 3.356E-03 ± 5.306E-05 \\
     &   GD      & CNN   & 9.983E-01 ± 1.578E-02 & 6.226E-03 ± 9.844E-05 &   GD & 9.499E-01 ± 1.502E-02 & 5.924E-03 ± 9.367E-05 \\
     &           & LSTM  & 4.778E-01 ± 7.555E-03 & 2.980E-03 ± 4.711E-05 &      & 4.542E-01 ± 7.182E-03 & 2.833E-03 ± 4.479E-05 \\
     &           & GRU   & 7.178E-01 ± 1.135E-02 & 4.476E-03 ± 7.078E-05 &      & 6.334E-01 ± 1.001E-02 & 3.950E-03 ± 6.246E-05 \\
     &           & RBFNN & 3.619E-01 ± 5.722E-03 & 2.257E-03 ± 3.569E-05 &      & 4.556E-01 ± 4.001E-03 & 2.111E-03 ± 3.337E-05 \\
     &           & \textbf{ALPE}    & \textbf{1.213E-01 ± 1.918E-03} & \textbf{7.566E-04 ± 1.196E-05} &      & \textbf{1.108E-01 ± 1.752E-03} & \textbf{6.909E-04 ± 1.092E-05} \\
    \bottomrule
    \end{tabular}
    \end{small}}
            \label{tab:table1}
        \end{minipage}%
        \hspace{0.25\textwidth} 
        \begin{minipage}[t]{0.48\textwidth}
            \centering
            \scalebox{0.42}{
    \begin{small}
    \begin{tabular}{c c r c c c c c}
    \toprule
    \textbf{Stock} & \textbf{Set} & \textbf{Model} & \textbf{RMSE} & \textbf{RRMSE} & \textbf{Set} & \textbf{RMSE} & \textbf{RRMSE} \\
    \hline     
    V & Simple    & Naive & 2.220E+00 ± 3.510E-02 & 1.237E-02 ± 1.956E-04 & Exte & 3.292E+00 ± 4.210E-02 & 2.873E-02 ± 2.999E-04 \\
         &        & ARIMA & 5.985E-01 ± 9.463E-03 & 3.335E-03 ± 5.274E-05 &      & 4.600E-01 ± 8.533E-03 & 4.441E-03 ± 4.650E-05  \\
         &        & MLP   & 7.239E-01 ± 1.145E-02 & 4.035E-03 ± 6.379E-05 &      & 7.234E-01 ± 1.144E-02 & 4.031E-03 ± 6.374E-05 \\
         &        & CNN   & 9.711E-01 ± 1.535E-02 & 5.412E-03 ± 8.557E-05 &      & 1.055E+00 ± 1.668E-02 & 5.878E-03 ± 9.294E-05 \\
         &        & LSTM  & 5.154E-01 ± 8.149E-03 & 2.872E-03 ± 4.542E-05 &      & 4.887E-01 ± 7.728E-03 & 2.724E-03 ± 4.307E-05 \\
         &        & GRU   & 8.558E-01 ± 1.353E-02 & 4.770E-03 ± 7.541E-05 &      & 8.321E-01 ± 1.316E-02 & 4.637E-03 ± 7.332E-05 \\
         &        & RBFNN & 3.150E-01 ± 4.981E-03 & 1.756E-03 ± 2.776E-05 &      & 3.002E-01 ± 4.747E-03 & 1.673E-03 ± 2.645E-05 \\
         &        & \textbf{ALPE}    & \textbf{2.312E-01 ± 3.656E-03} & \textbf{1.289E-03 ± 2.038E-05} &      & \textbf{2.285E-01 ± 3.613E-03} & \textbf{1.274E-03 ± 2.014E-05} \\
    \cline{2-5} \cline{6-8} 
     & Simple     & MLP   & 7.624E-01 ± 1.205E-02 & 4.249E-03 ± 6.718E-05 & Exte & 7.073E-01 ± 1.118E-02 & 3.942E-03 ± 6.232E-05 \\
     &  MDI       & CNN   & 9.764E-01 ± 1.544E-02 & 5.442E-03 ± 8.604E-05 &   MDI& 9.539E-01 ± 1.508E-02 & 5.317E-03 ± 8.406E-05 \\
     &            & LSTM  & 5.248E-01 ± 8.298E-03 & 2.925E-03 ± 4.624E-05 &      & 4.942E-01 ± 7.814E-03 & 2.724E-03 ± 2.325E-05 \\
     &            & GRU   & 9.077E-01 ± 1.435E-02 & 5.059E-03 ± 7.999E-05 &      & 8.367E-01 ± 1.323E-02 & 4.663E-03 ± 7.373E-05 \\
     &            & RBFNN & 3.117E-01 ± 4.928E-03 & 1.737E-03 ± 2.746E-05 &      & 2.987E-01 ± 4.724E-03 & 1.665E-03 ± 2.633E-05 \\
     &            & \textbf{ALPE}    & \textbf{2.633E-01 ± 4.163E-03} & \textbf{1.467E-03 ± 2.320E-05} &      & \textbf{2.241E-01 ± 3.544E-03} & \textbf{1.249E-03 ± 1.975E-05} \\
    \cline{2-5} \cline{6-8}
     & Simple    & MLP   & 7.601E-01 ± 1.202E-02 & 4.236E-03 ± 6.698E-05 & Exte & 7.158E-01 ± 1.132E-02 & 3.989E-03 ± 6.307E-05 \\
     &   GD      & CNN   & 1.010E+00 ± 1.597E-02 & 5.630E-03 ± 8.902E-05 &   GD & 9.488E-01 ± 1.500E-02 & 5.288E-03 ± 8.361E-05 \\
     &           & LSTM  & 5.021E-01 ± 7.939E-03 & 2.798E-03 ± 4.425E-05 &      & 4.759E-01 ± 7.525E-03 & 2.652E-03 ± 4.194E-05 \\
     &           & GRU   & 8.900E-01 ± 1.407E-02 & 4.960E-03 ± 7.843E-05 &      & 8.383E-01 ± 1.325E-02 & 4.672E-03 ± 7.387E-05 \\
     &           & RBFNN & 3.038E-01 ± 4.804E-03 & 1.693E-03 ± 2.677E-05 &      & 2.831E-01 ± 4.476E-03 & 1.578E-03 ± 2.494E-05 \\
     &           & \textbf{ALPE}    & \textbf{2.209E-01 ± 2.760E-03} & \textbf{1.374E-03 ± 1.009E-05} &      & \textbf{2.263E-01 ± 3.579E-03} & \textbf{1.261E-03 ± 1.995E-05}\\
     \hline
     VRSN & Simple & Naive & 1.982E+00 ± 3.134E-02 & 1.042E-02 ± 1.648E-04 & Exte & 2.550E+00 ± 4.998E-02 & 2.080E-02 ± 2.665E-04 \\
         &        & ARIMA & 7.529E-01 ± 1.190E-02 & 3.959E-03 ± 6.260E-05 &      & 6.775E-01 ± 2.202E-02 & 4.337E-03 ± 7.440E-05  \\
         &        & MLP   & 9.867E-01 ± 1.560E-02 & 5.188E-03 ± 8.204E-05 &      & 9.499E-01 ± 1.502E-02 & 4.995E-03 ± 7.898E-05 \\
         &        & CNN   & 1.387E+00 ± 2.193E-02 & 7.294E-03 ± 1.153E-04 &      & 1.313E+00 ± 2.077E-02 & 6.906E-03 ± 1.092E-04 \\
         &        & LSTM  & 7.370E-01 ± 1.165E-02 & 3.876E-03 ± 6.128E-05 &      & 7.151E-01 ± 1.131E-02 & 3.760E-03 ± 5.946E-05 \\
         &        & GRU   & 8.442E-01 ± 1.335E-02 & 4.439E-03 ± 7.019E-05 &      & 8.362E-01 ± 1.322E-02 & 4.397E-03 ± 6.953E-05 \\
         &        & RBFNN & 8.242E-01 ± 1.303E-02 & 4.334E-03 ± 6.852E-05 &      & 7.901E-01 ± 1.249E-02 & 4.155E-03 ± 6.569E-05 \\
         &        & \textbf{ALPE}    & \textbf{2.279E-01 ± 3.604E-03} & \textbf{1.199E-03 ± 1.895E-05} &      & \textbf{2.364E-01 ± 3.538E-03} & \textbf{1.243E-03 ± 1.966E-05} \\
    \cline{2-5} \cline{6-8} 
     & Simple     & MLP   & 9.702E-01 ± 1.534E-02 & 5.102E-03 ± 8.067E-05 & Exte & 9.494E-01 ± 1.501E-02 & 4.993E-03 ± 7.894E-05 \\
     &  MDI       & CNN   & 1.238E+00 ± 1.957E-02 & 6.510E-03 ± 1.029E-04 &   MDI& 1.150E+00 ± 1.819E-02 & 6.050E-03 ± 9.566E-05 \\
     &            & LSTM  & 7.158E-01 ± 1.132E-02 & 3.764E-03 ± 5.951E-05 &      & 7.082E-01 ± 1.120E-02 & 3.724E-03 ± 5.888E-05 \\
     &            & GRU   & 8.373E-01 ± 1.324E-02 & 4.403E-03 ± 6.962E-05 &      & 7.964E-01 ± 1.259E-02 & 4.188E-03 ± 6.622E-05 \\
     &            & RBFNN & 8.148E-01 ± 1.288E-02 & 4.284E-03 ± 6.774E-05 &      & 7.747E-01 ± 1.225E-02 & 4.074E-03 ± 6.441E-05 \\
     &            & \textbf{ALPE}    & \textbf{2.351E-01 ± 3.717E-03} & \textbf{1.236E-03 ± 1.955E-05} &      & \textbf{2.234E-01 ± 2.998E-03} & \textbf{1.302E-03 ± 2.802E-05} \\
    \cline{2-5} \cline{6-8}
     & Simple    & MLP   & 9.609E-01 ± 1.519E-02 & 5.053E-03 ± 7.990E-05 & Exte & 9.452E-01 ± 1.495E-02 & 4.970E-03 ± 7.859E-05 \\
     &   GD      & CNN   & 1.193E+00 ± 1.886E-02 & 6.274E-03 ± 9.920E-05 &   GD & 1.121E+00 ± 1.772E-02 & 5.893E-03 ± 9.318E-05 \\
     &           & LSTM  & 7.101E-01 ± 1.123E-02 & 3.734E-03 ± 5.904E-05 &      & 7.065E-01 ± 1.247E-02 & 3.715E-03 ± 5.874E-05 \\
     &           & GRU   & 8.493E-01 ± 1.343E-02 & 4.466E-03 ± 7.061E-05 &      & 8.193E-01 ± 1.295E-02 & 4.308E-03 ± 6.812E-05 \\
     &           & RBFNN & 8.260E-01 ± 1.306E-02 & 4.344E-03 ± 6.868E-05 &      & 7.755E-01 ± 1.226E-02 & 4.078E-03 ± 6.448E-05 \\
     &           & \textbf{ALPE}    & \textbf{2.341E-01 ± 3.702E-03} & \textbf{1.231E-03 ± 1.947E-05} &      & \textbf{2.337E-01 ± 2.695E-03} & \textbf{1.229E-03 ± 1.943E-05} \\
     \hline
    WAB & Simple & Naive & 2.003E+00 ± 3.167E-02 & 2.429E-02 ± 3.841E-04 & Exte & 3.112E+00 ± 4.665E-02 & 3.767E-02 ± 4.008E-04 \\
        &        & ARIMA & 1.174E+00 ± 1.856E-02 & 1.424E-02 ± 2.251E-04 &      & 2.200E+00 ± 2.909E-02 & 2.565E-02 ± 3.476E-04  \\
        &        & MLP   & 1.372E+00 ± 2.169E-02 & 1.664E-02 ± 2.631E-04 &      & 1.109E+00 ± 1.753E-02 & 1.345E-02 ± 2.127E-04 \\
        &        & CNN   & 1.631E+00 ± 2.580E-02 & 1.979E-02 ± 3.129E-04 &      & 1.541E+00 ± 2.437E-02 & 1.870E-02 ± 2.956E-04 \\
        &        & LSTM  & 8.235E-01 ± 1.302E-02 & 9.988E-03 ± 1.579E-04 &      & 8.374E-01 ± 1.324E-02 & 1.016E-02 ± 1.606E-04 \\
        &        & GRU   & 1.270E+00 ± 2.008E-02 & 1.540E-02 ± 2.436E-04 &      & 1.001E+00 ± 1.583E-02 & 1.214E-02 ± 1.920E-04 \\
        &        & RBFNN & 8.440E-01 ± 1.335E-02 & 1.024E-02 ± 1.619E-04 &      & 8.301E-01 ± 1.313E-02 & 1.007E-02 ± 1.592E-04 \\
        &        & \textbf{ALPE}    & \textbf{2.851E-01 ± 4.508E-03} & \textbf{3.458E-03 ± 5.468E-05} &      & \textbf{2.829E-01 ± 4.372E-03} & \textbf{3.431E-03 ± 5.425E-05} \\
    \cline{2-5} \cline{6-8} 
     & Simple     & MLP   & 1.230E+00 ± 1.925E-02 & 1.492E-02 ± 2.359E-04 & Exte & 1.001E+00 ± 1.582E-02 & 1.214E-02 ± 1.919E-04 \\
     &  MDI       & CNN   & 1.634E+00 ± 2.584E-02 & 1.982E-02 ± 3.134E-04 &   MDI& 1.632E+00 ± 3.344E-02 & 1.979E-02 ± 3.129E-04 \\
     &            & LSTM  & 8.368E-01 ± 1.323E-02 & 1.015E-02 ± 1.605E-04 &      & 8.444E-01 ± 1.335E-02 & 1.024E-02 ± 1.619E-04 \\
     &            & GRU   & 1.213E+00 ± 1.918E-02 & 1.471E-02 ± 2.326E-04 &      & 1.006E+00 ± 1.590E-02 & 1.220E-02 ± 1.929E-04 \\
     &            & RBFNN & 8.375E-01 ± 1.324E-02 & 1.016E-02 ± 1.606E-04 &      & 7.929E-01 ± 1.254E-02 & 9.617E-03 ± 1.521E-04 \\
     &            & \textbf{ALPE}    & \textbf{2.830E-01 ± 4.475E-03} & \textbf{3.433E-03 ± 5.428E-05} &      & \textbf{2.876E-01 ± 4.548E-03} & \textbf{3.489E-03 ± 5.516E-05} \\
    \cline{2-5} \cline{6-8}
     & Simple    & MLP   & 1.054E+00 ± 1.667E-02 & 1.279E-02 ± 2.022E-04 & Exte & 1.006E+00 ± 1.590E-02 & 1.220E-02 ± 1.929E-04 \\
     &   GD      & CNN   & 1.453E+00 ± 2.297E-02 & 1.762E-02 ± 2.787E-04 &   GD & 1.415E+00 ± 2.237E-02 & 1.716E-02 ± 2.713E-04 \\
     &           & LSTM  & 8.484E-01 ± 1.341E-02 & 1.029E-02 ± 1.627E-04 &      & 8.258E-01 ± 1.306E-02 & 1.002E-02 ± 1.584E-04 \\
     &           & GRU   & 1.114E+00 ± 1.812E-02 & 1.412E-02 ± 2.233E-04 &      & 1.180E+00 ± 1.866E-02 & 1.432E-02 ± 2.263E-04 \\
     &           & RBFNN & 8.196E-01 ± 1.296E-02 & 9.942E-03 ± 1.572E-04 &      & 7.754E-01 ± 1.226E-02 & 9.405E-03 ± 1.487E-04 \\
     &           & \textbf{ALPE}    & \textbf{2.857E-01 ± 4.518E-03} & \textbf{3.466E-03 ± 5.480E-05} &      & \textbf{2.828E-01 ± 4.402E-03} & \textbf{3.430E-03 ± 5.424E-05} \\
     \hline
    WBD & Simple & Naive & 2.997E+00 ± 4.739E-02 & 2.577E-01 ± 4.075E-03 & Exte & 3.984E+00 ± 5.888E-02 & 3.501E-01 ± 5.129E-03 \\
        &        & ARIMA & 1.374E+00 ± 2.172E-02 & 1.181E-01 ± 1.868E-03 &      & 2.004E+00 ± 3.101E-02 & 2.008E-01 ± 2.786E-03  \\
        &        & MLP   & 1.378E+00 ± 2.179E-02 & 1.185E-01 ± 1.874E-03 &      & 1.246E+00 ± 1.969E-02 & 1.071E-01 ± 1.693E-03 \\
        &        & CNN   & 1.968E+00 ± 3.111E-02 & 1.692E-01 ± 2.675E-03 &      & 1.732E+00 ± 2.739E-02 & 1.490E-01 ± 2.355E-03 \\
        &        & LSTM  & 9.916E-01 ± 1.568E-02 & 8.526E-02 ± 1.348E-03 &      & 9.457E-01 ± 1.495E-02 & 8.132E-02 ± 1.286E-03 \\
        &        & GRU   & 1.454E+00 ± 2.299E-02 & 1.250E-01 ± 1.977E-03 &      & 1.338E+00 ± 2.116E-02 & 1.151E-01 ± 1.820E-03 \\
        &        & RBFNN & 9.881E-01 ± 1.562E-02 & 8.496E-02 ± 1.343E-03 &      & 8.889E-01 ± 1.405E-02 & 7.643E-02 ± 1.208E-03 \\
        &        & \textbf{ALPE}    & \textbf{3.423E-01 ± 5.412E-03} & \textbf{2.943E-02 ± 4.653E-04} &      & \textbf{4.486E-01 ± 7.093E-03} & \textbf{3.857E-02 ± 6.099E-04} \\
    \cline{2-5} \cline{6-8} 
     & Simple     & MLP   & 1.183E+00 ± 1.870E-02 & 1.017E-01 ± 1.608E-03 & Exte & 1.002E+00 ± 1.585E-02 & 8.020E-02 ± 2.884E-03 \\
     &  MDI       & CNN   & 1.673E+00 ± 2.645E-02 & 1.438E-01 ± 2.274E-03 &   MDI& 1.698E+00 ± 2.685E-02 & 1.460E-01 ± 2.309E-03 \\
     &            & LSTM  & 9.962E-01 ± 1.575E-02 & 8.566E-02 ± 1.354E-03 &      & 9.488E-01 ± 1.500E-02 & 8.158E-02 ± 1.290E-03 \\
     &            & GRU   & 1.416E+00 ± 2.239E-02 & 1.217E-01 ± 1.925E-03 &      & 1.297E+00 ± 2.051E-02 & 1.115E-01 ± 1.763E-03 \\
     &            & RBFNN & 9.601E-01 ± 1.518E-02 & 8.256E-02 ± 1.305E-03 &      & 8.957E-01 ± 1.416E-02 & 7.701E-02 ± 1.218E-03 \\
     &            & \textbf{ALPE}    & \textbf{3.573E-01 ± 5.649E-03} & \textbf{3.072E-02 ± 4.857E-04} &      & \textbf{4.349E-01 ± 6.876E-03} & \textbf{3.739E-02 ± 5.912E-04} \\
    \cline{2-5} \cline{6-8}
     & Simple    & MLP   & 1.003E+00 ± 1.585E-02 & 8.411E-02 ± 1.102E-03 & Exte & 1.301E+00 ± 2.057E-02 & 1.119E-01 ± 1.769E-03 \\
     &   GD      & CNN   & 1.703E+00 ± 2.693E-02 & 1.464E-01 ± 2.315E-03 &   GD & 1.597E+00 ± 2.525E-02 & 1.373E-01 ± 2.172E-03 \\
     &           & LSTM  & 9.855E-01 ± 1.558E-02 & 8.474E-02 ± 1.340E-03 &      & 9.587E-01 ± 1.516E-02 & 8.244E-02 ± 1.303E-03 \\
     &           & GRU   & 1.410E+00 ± 2.229E-02 & 1.212E-01 ± 1.916E-03 &      & 1.295E+00 ± 2.048E-02 & 1.114E-01 ± 1.761E-03 \\
     &           & RBFNN & 9.441E-01 ± 1.493E-02 & 8.117E-02 ± 1.283E-03 &      & 9.116E-01 ± 1.441E-02 & 7.838E-02 ± 1.239E-03 \\
     &           & \textbf{ALPE}    & \textbf{5.173E-01 ± 8.179E-03} & \textbf{4.448E-02 ± 7.400E-04} &      & \textbf{3.891E-01 ± 5.152E-03} & \textbf{3.346E-02 ± 5.290E-04} \\
    \bottomrule
    \end{tabular}
    \end{small}}
            \label{tab:table_13}
        \end{minipage}
    \end{table}
\end{landscape}


\begin{landscape}
    \fancyhf{}  
    \fancyfoot[R]{\rotatebox{90}{\thepage}}  
    \thispagestyle{fancy}  

    \begin{table}[!hbtp]
        \centering
        \caption{RMSE and RRMSE scores for WEC, WMT, WRB, XOM, XYL, ZBH, ZBRA, and ZTS.}
        \hspace*{-5cm} 
        \begin{minipage}[t]{0.48\textwidth}
            \centering
            \scalebox{0.42}{
    \begin{small}
    \begin{tabular}{c c r c c c c c}
    \toprule
    \textbf{Stock} & \textbf{Set} & \textbf{Model} & \textbf{RMSE} & \textbf{RRMSE} & \textbf{Set} & \textbf{RMSE} & \textbf{RRMSE} \\
    \hline     
    WEC & Simple & Naive & 2.211E+00 ± 3.496E-02 & 2.424E-02 ± 3.833E-04 & Exte & 1.377E+00 ± 2.205E-02 & 3.229E-02 ± 4.776E-04 \\
        &        & ARIMA & 5.001E-01 ± 6.446E-03 & 4.540E-03 ± 5.763E-05 &      & 5.004E-01 ± 6.665E-03 & 4.875E-03 ± 6.909E-05  \\
        &        & MLP   & 4.809E-01 ± 7.079E-03 & 4.059E-03 ± 7.763E-05 &      & 5.135E-01 ± 8.119E-03 & 5.630E-03 ± 8.903E-05 \\
        &        & CNN   & 1.282E+00 ± 2.026E-02 & 1.405E-02 ± 2.222E-04 &      & 1.179E+00 ± 1.864E-02 & 1.293E-02 ± 2.044E-04 \\
        &        & LSTM  & 4.257E-01 ± 6.730E-03 & 4.667E-03 ± 7.380E-05 &      & 4.253E-01 ± 6.724E-03 & 4.663E-03 ± 7.373E-05 \\
        &        & GRU   & 5.523E-01 ± 8.733E-03 & 6.057E-03 ± 9.577E-05 &      & 4.519E-01 ± 7.146E-03 & 4.956E-03 ± 7.835E-05 \\
        &        & RBFNN & 3.423E-01 ± 5.411E-03 & 3.753E-03 ± 5.934E-05 &      & 3.152E-01 ± 5.004E-03 & 3.470E-03 ± 5.487E-05 \\
        &        & \textbf{ALPE}    & \textbf{1.370E-01 ± 2.166E-03} & \textbf{1.502E-03 ± 2.375E-05} &      & \textbf{1.001E-01 ± 1.583E-03} & \textbf{1.098E-03 ± 1.736E-05} \\
    \cline{2-5} \cline{6-8} 
     & Simple     & MLP   & 4.583E+01 ± 7.246E-03 & 5.025E-03 ± 7.945E-05 & Exte & 5.212E-01 ± 8.241E-03 & 5.716E-03 ± 9.037E-05 \\
     &  MDI       & CNN   & 1.222E+00 ± 1.932E-02 & 1.340E-02 ± 2.118E-04 &   MDI& 1.010E+00 ± 1.597E-02 & 1.108E-02 ± 1.751E-04 \\
     &            & LSTM  & 4.238E-01 ± 6.700E-03 & 4.647E-03 ± 7.347E-05 &      & 4.251E-01 ± 6.722E-03 & 4.662E-03 ± 7.371E-05 \\
     &            & GRU   & 5.421E-01 ± 8.571E-03 & 5.944E-03 ± 9.399E-05 &      & 4.852E-01 ± 7.672E-03 & 5.321E-03 ± 8.413E-05 \\
     &            & RBFNN & 3.186E-01 ± 5.038E-03 & 3.494E-03 ± 5.524E-05 &      & 3.446E-01 ± 5.448E-03 & 3.779E-03 ± 5.974E-05 \\
     &            & \textbf{ALPE}    & \textbf{1.239E-01 ± 1.959E-03} & \textbf{1.358E-03 ± 2.148E-05} &      & \textbf{1.028E-01 ± 1.625E-03} & \textbf{1.127E-03 ± 1.782E-05} \\
    \cline{2-5} \cline{6-8}
     & Simple    & MLP   & 5.482E-01 ± 8.667E-03 & 6.011E-03 ± 9.504E-05 & Exte & 4.237E-01 ± 5.109E-03 & 4.909E-03 ± 7.763E-05 \\
     &   GD      & CNN   & 1.261E+00 ± 1.993E-02 & 1.382E-02 ± 2.186E-04 &   GD & 1.202E+00 ± 1.900E-02 & 1.318E-02 ± 2.084E-04 \\
     &           & LSTM  & 4.230E-01 ± 6.689E-03 & 4.639E-03 ± 7.335E-05 &      & 4.255E-01 ± 6.727E-03 & 4.666E-03 ± 7.377E-05 \\
     &           & GRU   & 4.594E-01 ± 7.264E-03 & 5.038E-03 ± 7.966E-05 &      & 5.054E-01 ± 7.991E-03 & 5.542E-03 ± 8.763E-05 \\
     &           & RBFNN & 3.499E-01 ± 5.532E-03 & 3.837E-03 ± 6.066E-05 &      & 3.163E-01 ± 5.001E-03 & 3.469E-03 ± 5.484E-05 \\
     &           & \textbf{ALPE}    & \textbf{1.151E-01 ± 1.820E-03} & \textbf{1.262E-03 ± 1.996E-05} &      & \textbf{1.011E-01 ± 1.482E-03} & \textbf{1.097E-03 ± 1.735E-05}\\
     \hline
     WMT & Simple & Naive & 2.630E+00 ± 4.159E-02 & 2.019E-02 ± 3.192E-04 & Exte & 3.554E+00 ± 3.007E-02 & 4.001E-02 ± 4.633E-04 \\
        &        & ARIMA & 9.213E-01 ± 1.457E-02 & 7.071E-03 ± 1.118E-04 &      & 8.677E-01 ± 2.337E-02 & 6.113E-03 ± 2.763E-04 \\
        &        & MLP   & 3.639E-01 ± 5.753E-03 & 2.793E-03 ± 4.416E-05 &      & 3.170E-01 ± 5.012E-03 & 2.401E-03 ± 3.436E-05 \\
        &        & CNN   & 7.701E-01 ± 1.218E-02 & 5.911E-03 ± 9.346E-05 &      & 7.747E-01 ± 1.225E-02 & 5.946E-03 ± 9.401E-05 \\
        &        & LSTM  & 3.368E-01 ± 5.325E-03 & 2.585E-03 ± 4.087E-05 &      & 3.340E-01 ± 5.001E-03 & 2.546E-03 ± 4.025E-05 \\
        &        & GRU   & 9.230E-01 ± 1.459E-02 & 7.084E-03 ± 1.120E-04 &      & 8.321E-01 ± 1.316E-02 & 6.386E-03 ± 1.010E-04 \\
        &        & RBFNN & 3.016E-01 ± 4.768E-03 & 2.315E-03 ± 3.660E-05 &      & 2.829E-01 ± 4.473E-03 & 2.171E-03 ± 3.433E-05 \\
        &        & \textbf{ALPE}    & \textbf{1.475E-01 ± 2.332E-03} & \textbf{1.132E-03 ± 1.790E-05} &      & \textbf{1.615E-01 ± 2.554E-03} & \textbf{1.240E-03 ± 1.960E-05} \\
    \cline{2-5} \cline{6-8} 
     & Simple     & MLP   & 3.501E+01 ± 5.535E-03 & 2.687E-03 ± 4.248E-05 & Exte & 3.180E-01 ± 5.028E-03 & 2.441E-03 ± 3.859E-05 \\
     &  MDI       & CNN   & 7.738E-00 ± 1.224E-02 & 5.939E-03 ± 9.391E-05 &   MDI& 7.921E-01 ± 1.252E-02 & 6.080E-03 ± 9.613E-05 \\
     &            & LSTM  & 3.570E-01 ± 5.644E-03 & 2.740E-03 ± 4.332E-05 &      & 3.202E-01 ± 5.063E-03 & 2.458E-03 ± 3.886E-05 \\
     &            & GRU   & 8.393E-00 ± 1.327E-02 & 6.442E-03 ± 1.019E-04 &      & 8.242E-00 ± 1.303E-02 & 6.326E-03 ± 1.030E-04 \\
     &            & RBFNN & 2.999E-01 ± 4.742E-03 & 2.302E-03 ± 3.639E-05 &      & 2.832E-01 ± 4.478E-03 & 2.174E-03 ± 3.437E-05 \\
     &            & \textbf{ALPE}    & \textbf{1.596E-01 ± 2.523E-03} & \textbf{1.225E-03 ± 1.936E-05} &      & \textbf{1.572E-01 ± 2.486E-03} & \textbf{1.207E-03 ± 1.908E-05} \\
    \cline{2-5} \cline{6-8}
     & Simple    & MLP   & 3.368E-01 ± 5.325E-03 & 2.585E-03 ± 4.087E-05 & Exte & 3.607E-01 ± 5.339E-03 & 2.546E-03 ± 4.026E-05 \\
     &   GD      & CNN   & 7.593E+00 ± 1.201E-02 & 5.828E-03 ± 9.214E-05 &   GD & 7.826E-00 ± 1.237E-02 & 6.007E-03 ± 9.497E-05 \\
     &           & LSTM  & 3.184E-01 ± 5.034E-03 & 2.443E-03 ± 3.863E-05 &      & 3.299E-01 ± 5.111E-03 & 2.546E-03 ± 4.025E-05 \\
     &           & GRU   & 8.375E-01 ± 1.324E-02 & 6.428E-03 ± 1.016E-04 &      & 8.382E-01 ± 1.325E-02 & 6.433E-03 ± 1.017E-04 \\
     &           & RBFNN & 2.998E+00 ± 4.741E-03 & 2.301E-03 ± 3.639E-05 &      & 2.828E-01 ± 4.472E-03 & 2.171E-03 ± 3.432E-05 \\
     &           & \textbf{ALPE}    & \textbf{1.574E-01 ± 2.489E-03} & \textbf{1.208E-03 ± 1.911E-05} &      & \textbf{1.653E-01 ± 2.613E-03} & \textbf{1.268E-03 ± 2.005E-05} \\
     \hline
    WRB & Simple & Naive & 1.172E+00 ± 1.852E-02 & 1.811E-02 ± 2.863E-04 & Exte & 2.089E+00 ± 2.113E-02 & 2.667E-02 ± 3.887E-04 \\
        &        & ARIMA & 2.323E-01 ± 3.674E-03 & 3.591E-03 ± 5.678E-05 &      & 3.555E-01 ± 4.101E-03 & 4.225E-03 ± 4.909E-05  \\
        &        & MLP   & 4.839E-01 ± 7.651E-03 & 7.480E-03 ± 1.183E-04 &      & 4.210E-01 ± 6.656E-03 & 6.507E-03 ± 1.029E-04 \\
        &        & CNN   & 9.332E-01 ± 1.475E-02 & 1.442E-02 ± 2.281E-04 &      & 8.543E-01 ± 1.351E-02 & 1.321E-02 ± 2.088E-04 \\
        &        & LSTM  & 2.849E-01 ± 4.504E-03 & 4.403E-03 ± 6.962E-05 &      & 2.452E-01 ± 3.877E-03 & 3.790E-03 ± 5.992E-05 \\
        &        & GRU   & 3.064E-01 ± 4.844E-03 & 4.735E-03 ± 7.487E-05 &      & 2.401E-01 ± 3.796E-03 & 3.711E-03 ± 5.868E-05 \\
        &        & RBFNN & 3.810E-01 ± 6.024E-03 & 5.889E-03 ± 9.311E-05 &      & 2.306E-01 ± 3.172E-03 & 3.301E-03 ± 4.903E-05 \\
        &        & \textbf{ALPE}    & \textbf{2.254E-01 ± 3.564E-03} & \textbf{3.485E-03 ± 5.510E-05} &      & \textbf{2.094E-01 ± 3.311E-03} & \textbf{3.037E-03 ± 5.118E-05} \\
    \cline{2-5} \cline{6-8} 
     & Simple     & MLP   & 4.474E+01 ± 7.073E-03 & 6.915E-03 ± 1.093E-04 & Exte & 3.917E-01 ± 6.193E-03 & 6.055E-03 ± 9.573E-05 \\
     &  MDI       & CNN   & 8.800E+00 ± 1.391E-02 & 1.360E-02 ± 2.151E-04 &   MDI& 8.296E+01 ± 1.312E-02 & 1.282E-02 ± 2.028E-04 \\
     &            & LSTM  & 2.739E-01 ± 4.331E-03 & 4.234E-03 ± 6.694E-05 &      & 1.883E-01 ± 2.977E-03 & 2.911E-03 ± 4.602E-05 \\
     &            & GRU   & 2.717E-00 ± 4.296E-03 & 4.200E-03 ± 6.641E-05 &      & 2.127E-00 ± 3.363E-03 & 3.287E-03 ± 5.198E-05 \\
     &            & RBFNN & 2.804E-01 ± 3.168E-03 & 3.907E-03 ± 4.897E-05 &      & 2.416E-01 ± 3.819E-03 & 3.734E-03 ± 5.904E-05 \\
     &            & \textbf{ALPE}    & \textbf{2.595E-01 ± 4.104E-03} & \textbf{3.412E-03 ± 6.343E-05} &      & \textbf{1.745E-01 ± 2.759E-03} & \textbf{2.497E-03 ± 4.265E-05} \\
    \cline{2-5} \cline{6-8}
     & Simple    & MLP   & 4.257E-01 ± 6.731E-03 & 6.581E-03 ± 1.040E-04 & Exte & 3.606E-01 ± 5.702E-03 & 5.574E-03 ± 8.813E-05 \\
     &   GD      & CNN   & 8.946E+00 ± 1.414E-02 & 1.383E-02 ± 2.186E-04 &   GD & 8.374E+00 ± 1.324E-02 & 1.294E-02 ± 2.047E-04 \\
     &           & LSTM  & 2.406E-01 ± 3.805E-03 & 3.719E-03 ± 5.881E-05 &      & 2.425E-01 ± 3.835E-03 & 3.749E-03 ± 5.927E-05 \\
     &           & GRU   & 2.419E-01 ± 3.825E-03 & 3.739E-03 ± 5.912E-05 &      & 2.509E-01 ± 3.224E-03 & 3.760E-03 ± 5.945E-05 \\
     &           & RBFNN & 2.399E-01 ± 2.738E-03 & 3.760E-03 ± 5.946E-05 &      & 2.372E-01 ± 3.181E-03 & 3.410E-03 ± 4.917E-05 \\
     &           & \textbf{ALPE}    & \textbf{1.851E-01 ± 2.927E-03} & \textbf{2.862E-03 ± 4.525E-05} &      & \textbf{2.055E-01 ± 3.249E-03} & \textbf{3.076E-03 ± 5.022E-05} \\
     \hline
    XOM & Simple & Naive & 1.702E+00 ± 2.692E-02 & 1.888E-02 ± 2.985E-04 & Exte & 2.876E+00 ± 3.443E-02 & 2.544E-02 ± 3.080E-04 \\
        &        & ARIMA & 6.600E-01 ± 1.044E-02 & 7.319E-03 ± 1.157E-04 &      & 7.833E-01 ± 2.454E-02 & 6.535E-03 ± 2.120E-04  \\
        &        & MLP   & 6.556E-01 ± 1.037E-02 & 7.270E-03 ± 1.149E-04 &      & 5.486E-01 ± 8.674E-03 & 6.083E-03 ± 9.618E-05 \\
        &        & CNN   & 1.449E+00 ± 2.291E-02 & 1.607E-02 ± 2.540E-04 &      & 1.096E+00 ± 1.732E-02 & 1.215E-02 ± 1.921E-04 \\
        &        & LSTM  & 5.769E-01 ± 9.121E-03 & 6.397E-03 ± 1.011E-04 &      & 4.032E-01 ± 6.043E-03 & 4.964E-03 ± 7.849E-05 \\
        &        & GRU   & 8.279E-01 ± 1.309E-02 & 9.181E-03 ± 1.452E-04 &      & 6.876E-01 ± 1.087E-02 & 7.625E-03 ± 1.206E-04 \\
        &        & RBFNN & 4.608E-01 ± 7.286E-03 & 5.110E-03 ± 8.079E-05 &      & 4.475E-01 ± 7.075E-03 & 4.962E-03 ± 7.845E-05 \\
        &        & \textbf{ALPE}    & \textbf{2.990E-01 ± 4.727E-03} & \textbf{3.315E-03 ± 5.242E-05} &      & \textbf{2.358E-01 ± 3.729E-03} & \textbf{2.615E-03 ± 4.135E-05} \\
    \cline{2-5} \cline{6-8} 
     & Simple     & MLP   & 6.324E+01 ± 9.999E-03 & 7.013E-03 ± 1.109E-04 & Exte & 5.490E-01 ± 8.681E-03 & 6.088E-03 ± 9.626E-05 \\
     &  MDI       & CNN   & 1.333E+00 ± 2.108E-02 & 1.478E-02 ± 2.337E-04 &   MDI& 1.114E+01 ± 1.762E-02 & 1.236E-02 ± 1.954E-04 \\
     &            & LSTM  & 5.027E-01 ± 7.948E-03 & 5.574E-03 ± 8.814E-05 &      & 4.329E-01 ± 6.845E-03 & 4.800E-03 ± 7.590E-05 \\
     &            & GRU   & 7.741E+00 ± 1.224E-02 & 8.584E-03 ± 1.357E-04 &      & 7.616E+00 ± 1.204E-02 & 8.445E-03 ± 1.335E-04 \\
     &            & RBFNN & 4.342E-01 ± 6.865E-03 & 4.814E-03 ± 7.612E-05 &      & 4.443E+01 ± 7.024E-03 & 4.926E-03 ± 7.789E-05 \\
     &            & \textbf{ALPE}    & \textbf{2.129E-01 ± 3.366E-03} & \textbf{2.361E-03 ± 3.733E-05} &      & \textbf{1.771E+01 ± 2.800E-03} & \textbf{1.964E-03 ± 3.105E-05} \\
    \cline{2-5} \cline{6-8}
     & Simple    & MLP   & 5.877E-01 ± 9.292E-03 & 6.517E-03 ± 1.030E-04 & Exte & 6.557E-01 ± 1.037E-02 & 7.271E-03 ± 1.150E-04 \\
     &   GD      & CNN   & 9.942E+00 ± 1.572E-02 & 1.102E-02 ± 1.743E-04 &   GD & 1.001E+00 ± 1.583E-02 & 1.110E-02 ± 1.755E-04 \\
     &           & LSTM  & 4.704E-01 ± 7.437E-03 & 5.216E-03 ± 8.247E-05 &      & 3.627E-01 ± 5.735E-03 & 4.022E-03 ± 6.360E-05 \\
     &           & GRU   & 7.071E-01 ± 1.118E-02 & 7.841E-03 ± 1.240E-04 &      & 5.125E-01 ± 1.000E-02 & 7.013E-03 ± 1.109E-04 \\
     &           & RBFNN & 3.975E+00 ± 6.285E-03 & 4.408E-03 ± 6.970E-05 &      & 3.695E-01 ± 5.842E-03 & 4.097E-03 ± 6.478E-05 \\
     &           & \textbf{ALPE}    & \textbf{2.435E-01 ± 3.850E-03} & \textbf{2.700E-03 ± 4.269E-05} &      & \textbf{1.567E-01 ± 2.477E-03} & \textbf{1.737E-03 ± 2.747E-05} \\
    \bottomrule
    \end{tabular}
    \end{small}}
            \label{tab:table1}
        \end{minipage}%
        \hspace{0.25\textwidth} 
        \begin{minipage}[t]{0.48\textwidth}
            \centering
            \scalebox{0.42}{
    \begin{small}
    \begin{tabular}{c c r c c c c c}
    \toprule
    \textbf{Stock} & \textbf{Set} & \textbf{Model} & \textbf{RMSE} & \textbf{RRMSE} & \textbf{Set} & \textbf{RMSE} & \textbf{RRMSE} \\
    \hline     
    XYL & Simple & Naive & 1.001E+00 ± 1.583E-02 & 1.135E-02 ± 1.794E-04 & Exte & 2.223E+00 ± 3.775E-02 & 2.181E-02 ± 3.773E-04 \\
        &        & ARIMA & 8.259E-01 ± 1.306E-02 & 9.360E-03 ± 1.480E-04 &      & 7.220E-01 ± 2.448E-02 & 8.776E-03 ± 2.579E-04  \\
        &        & MLP   & 1.414E+00 ± 2.236E-02 & 1.603E-02 ± 2.534E-04 &      & 1.020E+00 ± 1.436E-02 & 1.133E-02 ± 1.792E-04 \\
        &        & CNN   & 9.873E-01 ± 1.561E-02 & 1.119E-02 ± 1.769E-04 &      & 9.366E-01 ± 1.481E-02 & 1.062E-02 ± 1.678E-04 \\
        &        & LSTM  & 5.730E-01 ± 9.059E-03 & 6.494E-03 ± 1.027E-04 &      & 6.528E-01 ± 1.032E-02 & 7.398E-03 ± 1.170E-04 \\
        &        & GRU   & 7.690E-01 ± 1.216E-02 & 8.715E-03 ± 1.378E-04 &      & 7.411E-01 ± 1.172E-02 & 8.399E-03 ± 1.328E-04 \\
        &        & RBFNN & 9.484E-01 ± 1.500E-02 & 1.075E-02 ± 1.700E-04 &      & 4.167E-01 ± 6.589E-03 & 4.723E-03 ± 7.467E-05 \\
        &        & \textbf{ALPE}    & \textbf{2.749E-01 ± 4.347E-03} & \textbf{3.116E-03 ± 4.926E-05} &      & \textbf{1.368E-01 ± 2.164E-03} & \textbf{1.551E-03 ± 2.452E-05} \\
    \cline{2-5} \cline{6-8} 
     & Simple     & MLP   & 1.370E+00 ± 2.166E-02 & 1.553E-02 ± 2.455E-04 & Exte & 1.072E+00 ± 1.694E-02 & 1.215E-02 ± 1.920E-04 \\
     &  MDI       & CNN   & 9.217E+00 ± 1.457E-02 & 1.045E-02 ± 1.652E-04 &   MDI& 9.286E+01 ± 1.468E-02 & 1.052E-02 ± 1.664E-04 \\
     &            & LSTM  & 6.508E-01 ± 1.029E-02 & 7.375E-03 ± 1.166E-04 &      & 4.632E-01 ± 7.324E-03 & 5.250E-03 ± 8.300E-05 \\
     &            & GRU   & 6.808E+00 ± 1.076E-02 & 7.716E-03 ± 1.220E-04 &      & 7.905E+00 ± 1.250E-02 & 8.959E-03 ± 1.417E-04 \\
     &            & RBFNN & 6.582E-01 ± 1.041E-02 & 7.460E-03 ± 1.179E-04 &      & 3.300E-01 ± 4.702E-03 & 3.983E-03 ± 5.682E-05 \\
     &            & \textbf{ALPE}    & \textbf{2.289E-01 ± 3.619E-03} & \textbf{2.594E-03 ± 4.102E-05} &      & \textbf{1.887E-01 ± 2.984E-03} & \textbf{2.139E-03 ± 3.382E-05} \\
    \cline{2-5} \cline{6-8}
     & Simple    & MLP   & 1.139E+00 ± 1.801E-02 & 1.291E-02 ± 2.042E-04 & Exte & 1.010E+00 ± 1.597E-02 & 1.145E-02 ± 1.810E-04 \\
     &   GD      & CNN   & 1.242E+00 ± 1.964E-02 & 1.408E-02 ± 2.226E-04 &   GD & 8.426E+00 ± 1.332E-02 & 9.549E-03 ± 1.510E-04 \\
     &           & LSTM  & 5.936E-01 ± 9.385E-03 & 6.727E-03 ± 1.064E-04 &      & 6.422E-01 ± 1.015E-02 & 7.279E-03 ± 1.151E-04 \\
     &           & GRU   & 7.999E-01 ± 1.265E-02 & 9.066E-03 ± 1.433E-04 &      & 6.344E-01 ± 1.003E-02 & 7.190E-03 ± 1.137E-04 \\
     &           & RBFNN & 4.478E-01 ± 7.080E-03 & 5.075E-03 ± 8.024E-05 &      & 3.384E-01 ± 5.351E-03 & 3.836E-03 ± 5.352E-05 \\
     &           & \textbf{ALPE}    & \textbf{1.530E-01 ± 2.420E-03} & \textbf{1.735E-03 ± 2.743E-05} &      & \textbf{1.234E-01 ± 1.952E-03} & \textbf{1.399E-03 ± 2.212E-05}\\
     \hline
     ZBH & Simple  & Naive  & 2.988E+00 ± 4.725E-02 & 1.665E-02 ± 2.633E-04 & Exte & 3.996E+00 ± 3.551E-02 & 2.991E-02 ± 3.090E-04 \\
         &        & ARIMA  & 9.911E-01 ± 1.567E-02 & 5.524E-03 ± 8.734E-05 &      & 8.699E-01 ± 2.339E-02 & 4.110E-03 ± 7.337E-05 \\
         &        & MLP    & 1.350E+00 ± 2.135E-02 & 7.526E-03 ± 1.190E-04 &      & 1.045E+00 ± 1.653E-02 & 5.825E-03 ± 9.210E-05 \\
         &        & CNN    & 1.457E+00 ± 2.304E-02 & 8.121E-03 ± 1.284E-04 &      & 1.487E+00 ± 2.352E-02 & 8.290E-03 ± 1.311E-04 \\
         &        & LSTM   & 9.181E-01 ± 1.452E-02 & 5.117E-03 ± 8.090E-05 &      & 8.955E-01 ± 1.416E-02 & 4.991E-03 ± 7.891E-05 \\
         &        & GRU    & 1.101E+00 ± 1.741E-02 & 6.136E-03 ± 9.702E-05 &      & 1.032E+00 ± 1.632E-02 & 5.754E-03 ± 9.098E-05 \\
         &        & RBFNN  & 5.651E-01 ± 8.934E-03 & 3.149E-03 ± 4.979E-05 &      & 4.914E-01 ± 7.769E-03 & 2.738E-03 ± 4.330E-05 \\
         &        & \textbf{ALPE}     & \textbf{3.040E-01 ± 4.806E-03} & \textbf{1.694E-03 ± 2.678E-05} &      & \textbf{2.813E-01 ± 4.448E-03} & \textbf{1.568E-03 ± 2.479E-05} \\
    \cline{2-5} \cline{6-8} 
     & Simple     & MLP   & 1.198E+00 ± 1.894E-02 & 6.677E-03 ± 1.056E-04 & Exte    & 9.552E-01 ± 1.510E-02 & 5.323E-03 ± 8.417E-05 \\
     &   MDI      & CNN   & 1.502E+00 ± 2.375E-02 & 8.372E-03 ± 1.324E-04 & MDI     & 1.580E+00 ± 2.445E-02 & 7.904E-03 ± 1.250E-04 \\
     &            & LSTM  & 9.117E-01 ± 1.442E-02 & 5.081E-03 ± 8.034E-05 &         & 8.946E-01 ± 1.414E-02 & 4.986E-03 ± 7.883E-05 \\
     &            & GRU   & 1.060E+00 ± 1.676E-02 & 5.906E-03 ± 9.338E-05 &         & 1.016E+00 ± 1.606E-02 & 5.662E-03 ± 8.952E-05 \\
     &            & RBFNN & 5.669E-01 ± 8.963E-03 & 3.159E-03 ± 4.995E-05 &         & 4.509E-01 ± 6.009E-03 & 2.507E-03 ± 3.964E-05 \\
     &            & \textbf{ALPE}    & \textbf{2.987E-01 ± 4.723E-03} & \textbf{1.665E-03 ± 2.632E-05} &         & \textbf{2.788E-01 ± 4.408E-03} & \textbf{1.554E-03 ± 2.457E-05} \\
    \cline{2-5} \cline{6-8}
     & Simple     & MLP   & 1.101E+00 ± 1.742E-02 & 6.139E-03 ± 9.706E-05 & Exte    & 9.665E-01 ± 1.528E-02 & 5.387E-03 ± 8.517E-05 \\
     &    GD      & CNN   & 1.695E+00 ± 2.680E-02 & 9.447E-03 ± 1.494E-04 & GD      & 1.463E+00 ± 2.314E-02 & 8.156E-03 ± 1.290E-04 \\
     &            & LSTM  & 8.957E-01 ± 1.416E-02 & 4.992E-03 ± 7.893E-05 &         & 9.019E-01 ± 1.426E-02 & 5.027E-03 ± 7.948E-05 \\
     &            & GRU   & 1.001E+00 ± 1.582E-02 & 5.577E-03 ± 8.817E-05 &         & 9.962E-01 ± 1.575E-02 & 5.552E-03 ± 8.779E-05 \\
     &            & RBFNN & 5.397E-01 ± 8.533E-03 & 3.008E-03 ± 4.756E-05 &         & 4.608E-01 ± 7.286E-03 & 2.568E-03 ± 4.061E-05 \\
     &            & \textbf{ALPE}   & \textbf{2.936E-01 ± 4.642E-03} & \textbf{1.636E-03 ± 2.587E-05} &         & \textbf{2.795E-01 ± 4.420E-03} & \textbf{1.558E-03 ± 2.463E-05} \\
    \hline
    ZBRA & Simple & Naive & 1.085E+00 ± 1.715E-02 & 4.114E-03 ± 6.504E-05 & Exte & 2.887E+00 ± 2.776E-02 & 5.101E-03 ± 7.990E-05 \\
         &        & ARIMA & 5.435E-01 ± 8.594E-03 & 2.061E-03 ± 3.259E-05 &      & 4.703E-01 ± 7.443E-03 & 3.874E-03 ± 4.335E-05  \\
         &        & MLP   & 5.395E-01 ± 8.531E-03 & 2.046E-03 ± 3.235E-05 &      & 5.549E-01 ± 8.774E-03 & 2.104E-03 ± 3.327E-05 \\
         &        & CNN   & 5.621E-01 ± 8.887E-03 & 2.131E-03 ± 3.370E-05 &      & 5.646E-01 ± 8.927E-03 & 2.141E-03 ± 3.385E-05 \\
         &        & LSTM  & 4.609E-01 ± 7.287E-03 & 1.748E-03 ± 2.763E-05 &      & 4.785E-01 ± 7.566E-03 & 1.815E-03 ± 2.869E-05 \\
         &        & GRU   & 5.650E-01 ± 8.933E-03 & 2.142E-03 ± 3.387E-05 &      & 5.319E-01 ± 8.410E-03 & 2.017E-03 ± 3.189E-05 \\
         &        & RBFNN & 4.903E-01 ± 7.752E-03 & 1.859E-03 ± 2.940E-05 &      & 4.162E-01 ± 6.581E-03 & 1.578E-03 ± 2.496E-05 \\
         &        & \textbf{ALPE}    & \textbf{2.361E-01 ± 3.734E-03} & \textbf{8.955E-04 ± 1.416E-05} &      & \textbf{2.424E-01 ± 3.832E-03} & \textbf{9.190E-04 ± 1.453E-05} \\
    \cline{2-5} \cline{6-8} 
     & Simple     & MLP   & 5.317E-01 ± 8.407E-03 & 2.016E-03 ± 3.188E-05 & Exte & 5.215E-01 ± 8.246E-03 & 1.978E-03 ± 3.127E-05 \\
     &  MDI       & CNN   & 5.638E-01 ± 8.914E-03 & 2.138E-03 ± 3.380E-05 &   MDI& 5.588E-01 ± 8.835E-03 & 2.119E-03 ± 3.350E-05 \\
     &            & LSTM  & 4.247E-01 ± 6.715E-03 & 1.611E-03 ± 2.546E-05 &      & 3.958E-01 ± 6.258E-03 & 1.501E-03 ± 2.373E-05 \\
     &            & GRU   & 5.515E-01 ± 8.719E-03 & 2.091E-03 ± 3.306E-05 &      & 5.501E-01 ± 5.802E-03 & 2.077E-03 ± 3.284E-05 \\
     &            & RBFNN & 3.336E-01 ± 5.275E-03 & 1.265E-03 ± 2.000E-05 &      & 3.409E-01 ± 5.664E-03 & 1.412E-03 ± 2.282E-05 \\
     &            & \textbf{ALPE}    & \textbf{2.427E-01 ± 3.837E-03} & \textbf{9.202E-04 ± 1.455E-05} &      & \textbf{2.419E-01 ± 3.825E-03} & \textbf{9.174E-04 ± 1.451E-05} \\
    \cline{2-5} \cline{6-8}
     & Simple    & MLP   & 5.153E-01 ± 8.147E-03 & 1.954E-03 ± 3.089E-05 & Exte & 5.325E-01 ± 8.420E-03 & 2.019E-03 ± 3.193E-05 \\
     &   GD      & CNN   & 5.794E-01 ± 9.161E-03 & 2.197E-03 ± 3.474E-05 &   GD & 5.625E-01 ± 8.894E-03 & 2.133E-03 ± 3.373E-05 \\
     &           & LSTM  & 4.539E-01 ± 7.177E-03 & 1.721E-03 ± 2.722E-05 &      & 4.247E-01 ± 6.715E-03 & 1.610E-03 ± 2.546E-05 \\
     &           & GRU   & 5.458E-01 ± 7.352E-03 & 2.077E-03 ± 3.284E-05 &      & 5.477E-01 ± 8.660E-03 & 2.077E-03 ± 3.284E-05 \\
     &           & RBFNN & 4.144E-01 ± 6.552E-03 & 1.571E-03 ± 2.484E-05 &      & 3.761E-01 ± 5.947E-03 & 1.426E-03 ± 2.255E-05 \\
     &           & \textbf{ALPE}    & \textbf{2.431E-01 ± 3.843E-03} & \textbf{9.217E-04 ± 1.457E-05} &      & \textbf{2.410E-01 ± 3.520E-03} & \textbf{9.139E-04 ± 1.445E-05} \\
    \hline
    ZTS & Simple  & Naive & 1.423E+00 ± 2.249E-02 & 9.530E-03 ± 1.507E-04 & Exte & 2.171E+00 ± 3.372E-02 & 8.181E-03 ± 2.638E-04 \\
         &        & ARIMA & 6.446E-01 ± 1.019E-02 & 4.318E-03 ± 6.827E-05 &      & 5.432E-01 ± 2.837E-02 & 5.935E-03 ± 5.761E-05  \\
         &        & MLP   & 7.384E-01 ± 1.168E-02 & 4.947E-03 ± 7.821E-05 &      & 5.359E-01 ± 8.474E-03 & 3.590E-03 ± 5.677E-05 \\
         &        & CNN   & 9.487E-01 ± 1.500E-02 & 6.355E-03 ± 1.005E-04 &      & 8.254E-01 ± 1.305E-02 & 5.529E-03 ± 8.743E-05 \\
         &        & LSTM  & 4.160E-01 ± 6.577E-03 & 2.787E-03 ± 4.406E-05 &      & 4.249E-01 ± 6.718E-03 & 2.846E-03 ± 4.500E-05 \\
         &        & GRU   & 9.784E-01 ± 1.547E-02 & 6.554E-03 ± 1.046E-04 &      & 8.203E-01 ± 1.297E-02 & 5.495E-03 ± 8.689E-05 \\
         &        & RBFNN & 3.811E-01 ± 6.026E-03 & 2.553E-03 ± 4.037E-05 &      & 3.529E-01 ± 5.580E-03 & 2.364E-03 ± 3.738E-05 \\
         &        & \textbf{ALPE}    & \textbf{3.073E-01 ± 4.858E-03} & \textbf{2.058E-03 ± 3.255E-05} &      & \textbf{2.346E-01 ± 3.709E-03} & \textbf{1.572E-03 ± 2.485E-05} \\
    \cline{2-5} \cline{6-8} 
     & Simple     & MLP   & 6.271E-01 ± 9.916E-03 & 4.201E-03 ± 3.496E-05 & Exte & 5.487E-01 ± 8.675E-03 & 3.675E-03 ± 5.811E-05 \\
     &  MDI       & CNN   & 8.509E-01 ± 1.345E-02 & 5.700E-03 ± 9.013E-05 &   MDI& 8.944E-01 ± 1.414E-02 & 5.992E-03 ± 9.473E-05 \\
     &            & LSTM  & 4.182E-01 ± 6.612E-03 & 2.802E-03 ± 4.430E-05 &      & 4.224E-01 ± 6.679E-03 & 2.830E-03 ± 4.475E-05 \\
     &            & GRU   & 7.896E-01 ± 1.248E-02 & 5.290E-03 ± 8.364E-05 &      & 7.747E-01 ± 1.225E-02 & 5.190E-03 ± 8.206E-05 \\
     &            & RBFNN & 4.954E-01 ± 7.833E-03 & 3.319E-03 ± 5.247E-05 &      & 3.166E-01 ± 5.005E-03 & 2.121E-03 ± 3.353E-05 \\
     &            & \textbf{ALPE}    & \textbf{2.549E-01 ± 4.031E-03} & \textbf{1.708E-03 ± 2.700E-05} &      & \textbf{2.341E-01 ± 2.481E-03} & \textbf{1.568E-03 ± 2.479E-05} \\
    \cline{2-5} \cline{6-8}
     & Simple    & MLP   & 5.637E-01 ± 8.913E-03 & 3.776E-03 ± 5.971E-05 & Exte & 5.258E-01 ± 8.314E-03 & 3.522E-03 ± 5.569E-05 \\
     &   GD      & CNN   & 8.944E-01 ± 1.414E-02 & 5.992E-03 ± 9.473E-05 &   GD & 9.778E-01 ± 1.546E-02 & 6.550E-03 ± 1.032E-04 \\
     &           & LSTM  & 4.166E-01 ± 6.587E-03 & 2.791E-03 ± 4.413E-05 &      & 4.236E-01 ± 6.698E-03 & 2.442E-03 ± 2.101E-05 \\
     &           & GRU   & 7.506E-01 ± 1.187E-02 & 5.029E-03 ± 7.951E-05 &      & 7.702E-01 ± 1.218E-02 & 5.160E-03 ± 8.159E-05 \\
     &           & RBFNN & 3.761E-01 ± 5.947E-03 & 2.520E-03 ± 3.984E-05 &      & 4.725E-01 ± 7.471E-03 & 3.165E-03 ± 5.005E-05 \\
     &           & \textbf{ALPE}    & \textbf{2.331E-01 ± 3.685E-03} & \textbf{1.561E-03 ± 2.469E-05} &      & \textbf{2.244E-01 ± 2.755E-03} & \textbf{1.090E-03 ± 2.393E-05} \\

    \bottomrule
    \end{tabular}
    \end{small}}
            \label{tab:table_14}
        \end{minipage}
    \end{table}
\end{landscape}







\end{document}